\documentclass[11pt]{article}
    
\usepackage{lettrine}    
\usepackage{amsmath} 
\usepackage{amsfonts}
\usepackage{amssymb}
\usepackage{booktabs}
\usepackage{bm}
\usepackage[usenames, dvipsnames]{color} 
\usepackage{dsfont}
\usepackage{graphicx}
\usepackage{epstopdf}
\usepackage[utf8]{inputenc}
\usepackage{lscape}
\usepackage{natbib}
\usepackage{multirow}
\usepackage{setspace}
\usepackage[capposition=top]{floatrow}
\usepackage[margin=0.9in]{geometry}
\usepackage{subfloat}
\usepackage{caption}
\usepackage{xr}
\usepackage[caption=false]{subfig}
\usepackage{sectsty}
\usepackage{titlesec}
\usepackage{titletoc}
\usepackage[bottom,hang,flushmargin]{footmisc}
\usepackage{longtable}
\usepackage{colortbl}

\sectionfont{\fontsize{14}{15}\selectfont}
\subsectionfont{\fontsize{13}{15}\selectfont}
\titlespacing\section{0pt}{12pt plus 0pt minus 0pt}{0pt plus 0pt minus 0pt}
\titlespacing\subsection{0pt}{12pt plus 0pt minus 0pt}{0pt plus 0pt minus 0pt}

\newcommand{\appendixpagenumbering}{
  \pagenumbering{arabic}
  \renewcommand{\thepage}{A\arabic{page}}
}

\usepackage[usenames,dvipsnames]{xcolor} 
\usepackage[colorlinks = true,
            linkcolor = Rhodamine,
            urlcolor  = Rhodamine,
            citecolor = MidnightBlue,
            anchorcolor = MidnightBlue]{hyperref}

\title{\textbf{Schools as a Safety-net: The Impact of School Closures and Reopenings on Rates of Reporting of Violence Against Children}\thanks{Clarke and Larroulet gratefully acknowledge funding from ANID of the Government of Chile, grant number COVID0593, and Clarke additionally acknowledges research funding from FEN, University of Chile. We are very grateful to Patricia Gonz\'alez and Carmen Puga from the Superintendency of Crime Prevention of the Ministry of the Interior for help in accessing administrative data, and extremely useful discussion relating to criminal reporting data, to Francisca Espinoza from the Ministry of Education for her help in clarifying the data, and to Angelina Orellana for data access and support. Data and replication materials are available at \href{https://github.com/Daniel-Pailanir/childrenSchools}{https://github.com/Daniel-Pailanir/childrenSchools}.}} 
\author{Damian Clarke\thanks{University of Chile, MIPP \& IZA.  .  Contact email: \href{damian.clarke@uchile.cl}{damian.clarke@uchile.cl}.}
  \and Pilar Larroulet\thanks{Pontificia Universidad Cat\'olica de Chile \& Viodemos. Contact email: \href{plarroul@uc.cl}{plarroul@uc.cl}.}
  \and Daniel Paila\~nir\thanks{University of Chile. Contact email: \href{dpailanir@fen.uchile.c}{dpailanir@fen.uchile.cl}.}
  \and Daniela Quintana\thanks{University of Chile. Contact email: \href{dquintanaa@fen.uchile.cl}{dquintanaa@fen.uchile.cl}.}
}

\date{\today}

\begin{document}
\captionsetup[figure]{list=no}
\captionsetup[table]{list=no}

\renewcommand{\thefootnote}{\arabic{footnote}}
\renewcommand{\footnotesize}{\small} 
\setcounter{footnote}{0} 
\thispagestyle{empty}
\maketitle

\begin{spacing}{1.2}
\begin{abstract} 
Ongoing school closures and gradual reopenings have been occurring since the beginning of the COVID-19 pandemic.  One substantial cost of school closure is breakdown in channels of reporting of violence against children, in which schools play a considerable role. There is, however, little evidence documenting how widespread such a breakdown in reporting of violence against children has been, and scant evidence exists about potential recovery in reporting as schools re-open.  We study all formal criminal reports of violence against children occurring in Chile up to December 2021, covering physical, psychological, and sexual violence.  This is combined with administrative records of school re-opening, attendance, and epidemiological and public health measures.  We observe sharp declines in violence reporting at the moment of school closure across all classes of violence studied. Estimated reporting declines range from -17\% (rape), to -43\% (sexual abuse). While reports rise with school re-opening, recovery of reporting rates is slow. Conservative projections suggest that reporting gaps remained into the final quarter of 2021, nearly two years after initial school closures.  Our estimates suggest that school closure and incomplete re-opening resulted in around 2,800 `missing' reports of intra-family violence, 2,000 missing reports of sexual assault, and 230 missing reports of rape against children, equivalent to between 10-25 weeks of reporting in baseline periods. The immediate and longer term impacts of school closures account for between 40-70\% of `missing' reports in the post-COVID period.
\end{abstract}

\noindent\emph{Keywords:} Violence against children; Domestic violence; Sexual violence; Pandemic Recovery; Schooling systems.
    
\thispagestyle{empty}
\setlength{\baselineskip}{1.3\baselineskip} 
\newpage 
\end{spacing}
\begin{spacing}{1.3}
\section{Introduction}
After more than two years of global pandemic, it is key to understand how to recover, and build resilient social systems designed to provide well-being for all members of society. The costs of the COVID-19 pandemic and necessary public health responses have been noted in multitude dimensions \citep{GassmanPinesetal2022,Hansenetal2022,Chandra2022}, from early life up to old age \citep{Singhetal2020,Vahia2020}. In this paper we focus particularly on the well-being of children, and in understanding how formal educational systems have recovered a specific function---that of identifying and reporting violence against children. We seek to understand how school re-openings and associated challenges have allowed for recovery of rates of reporting of violence against children, which have been clearly observed to plummet with school closures \citep{Baronetal2020,CHPR2020,Prettyman2021}, despite concerns that violence may have actually risen \citep{Pereda2020}, and consider the implications of this in a post-pandemic world. 

Violence against children has long-term consequences in terms of educational attainment and earnings \citep{CurriSpatz2010}, mental health \citep{widom2007}, and antisocial behaviors \citep{smith1995,thornberry2001}. An early detection of maltreatment could mitigate those negative effects by leading to timely interventions which may be more effective at altering abusive behavior \citep{fitzpatrick2020}. Schools play a key role in this regard, identifying early signs of abuse and maltreatment and channeling these cases into the justice and child protection systems \citep{fitzpatrick2020}. Research conducted in the context of COVID-19 has confirmed this key bridging role played by teachers and educational professionals, documenting significant declines in reporting following school closure \citep{Baronetal2020,Barbozaetal2021,Prettyman2021,Bullingeretal2021,Rapoportetal2021, takaku2021}. 

Importantly, this decrease came despite concerns about a potential rise in victimization of children and adolescents, given the confluence of risk factors associated with child maltreatment, such as parental unemployment and economic stress, parental burnout, limited sources of social support \citep{bullinger2022evaluating, Pereda2020, Lindoetal2018}, and the significant increase in the time spent together \citep{Lindoetal2018,fitzpatrick2020}. Indeed, surveys conducted among parents \citep{rodriguez2021}, teachers \citep{vermeulen2022}, and social-service professionals \citep{bullinger2022evaluating} report an increase in family conflict, harsh parenting, and child neglect during the first months of the COVID-19 pandemic. Thus, undetected cases may have increased substantially above baseline levels, with long-term consequences should these undetected victims of violence remain undetected over time.

In this paper, we extend the relatively nascent literature on the negative consequences of school closures by analyzing i) how school closure contributed to a decline in reporting over a substantial time period, and distinguishing between different types of violence including physical, sexual and psychological violence, and ii) determining whether that decline was reverted once schools re-opened, and if so over what time-frame and under what conditions. By analyzing the recovery of reporting, we are able to shed light on the relevance of in-person interactions for the identification of maltreatment.  While a small number of existing studies have clearly documented reporting declines with school closure \citep{Baronetal2020,CHPR2020,Prettyman2021}, what is new here is an extension into a considerably broader range of classes of violence including sexual violence, consideration of a longer period extending into pandemic recovery, the first consideration of the subsequent impact of school re-openings, the use of complete and nationally comparable administrative data which resolves a number of substantial measurement concerns noted in the literature \citep{Bullingeretal2021}, and bounds estimation to incorporate potential increases in violence above baseline rates.

\section{Study Setting}
This study is based on Chile. We observe each administrative record of a criminal report occurring in the country between 2010 and 2021.  We focus specifically on the pre-COVID, school closure, and school reopening period, spanning January 2019-December 2021. We examine all reports of crimes against children classified as intra-family violence (or domestic violence), sexual abuse, and rape. 

While schools closed worldwide as an early reaction to the spread of COVID-19, they remained closed for particularly long periods in Latin America \citep{Economist2021,UNICEF2021}, potentially increasing extant inequalities and interrupting three decades of educational improvement in the continent \citep{WorldBank2021}. In March 2021, still close to 60\% of Latin American school-age children were affected by school closure \citep{UNICEF2021}, having lost more days of schools than any other region in the world \citep{WorldBank2021}. Similarly, in Chile, schools closed nationally on March 16\textsuperscript{th}, 2020, and started to open--although very gradually--during August of 2020. By March 2021 only 25\% of all schools had some kind of in-person education, increasing to 98\% by December of 2021, the end of the school year. Student attendance, however, remained under 50\% of the total number of students. For these reasons, Chile provides a particular opportunity to analyze long-term impacts of school closures and the specific contribution of in-person activities to the reporting of violence against children.

A key component of our study design is related to Chile's pandemic response and recovery, in which there is very fine-grained (municipal level) variation in school opening statuses, lockdown measures, and COVID-19 case rates (additional discussion in Supplementary Information, Section \ref{SIscn:context}).  We leverage temporal variation in school closure policies to isolate impacts of school reporting channels on violence against children, combining high quality administrative records of crime reporting, educational opening and attendance, and very local epidemiological conditions at each moment in time.  We document impacts of school closure and re-openings using recent advances in observational methods based on temporal and geographic variation in policy exposures, consider dynamic impacts and placebo tests, and conduct projections and sensitivity testing to estimate under-reporting, and post-COVID recovery.  Our results, laid out below, suggest sequelae from school closures have declined over time, but still persist almost 2 years following initial closure decisions.

\section{Materials and Methods}
\subsection{Materials}
We collect administrative data from a number of national Ministries in Chile covering a range of factors.  These data, which are generally available at the individual or municipal$\times$day level, are aggregated consistently to the level of municipality$\times$week, covering each of Chile's 346 municipalities over the period of (at least) January 2019 to December 2021.  We additionally hand-compiled daily data on municipal-level lockdown status from public announcements made over the period.
We describe these variables and their sources briefly here.  Additional details on each variable, their formal definitions, and summary measures are provided in Supplementary Information.

\paragraph{Reporting of Crimes Against Children}
From Chile's Ministry of the Interior we collect full official microdata records on all victims of crime known to the police occurring between January 2010 -- December 2021 for crimes classified as intra-family violence against children, sexual abuse of children, and rape of children.  These data are available at the individual level, registering the victim's age, sex, and official classification of the crime.  

\paragraph{School Opening Status}
From the Ministry of Education, we collect administrative data on official records of dates of school closures and reopening from August 2020 to December 2021. These data cover each of the 10,847 and 10,875 schools in the country (respectively) for 2020 and 2021, both public and private.
The Ministry also provided administrative records on official enrollment for each school over the entire period of study, and attendance for a specific time-period (July-December, 2021).

\paragraph{Other measures}
We compiled data on COVID-19 infection rates and death rates, testing rates, and test positivity rates in each municipality and week from open repositories maintained by the Chilean Ministry of Science, and hand-collected daily records of municipal-level lockdown  statuses from public reports made by the Ministry of Health.
Data on municipal-level population by age and sex is collected from the National Institute of Statistics, and a Municipal Development Index was collected from \citet{Gattinietal2014}.

\subsection{Methods}
We have two principal aims in this study.  The first is to estimate changes in rates of reporting in crimes against children owing to school closure and reopening, and the second is to estimate counterfactual outcomes in the absence of school closures.  We lay these out briefly below.  Additional details are provided in SI, section \ref{scn:methodsSI}.

For the first of these aims we estimate linear regression models following:
\begin{eqnarray}
\label{eqn:2wayFE}
Reporting_{mt} &=& \alpha + \beta School\ Closure_{mt} + \gamma Schools\ Reopen_{mt} \nonumber \\ 
&& + WoY_t + Municipal_m +  \bm{X}^\prime_{mt}\bm{\Gamma}+\varepsilon_{mt}
\end{eqnarray}
where reporting refers to rates of child victimization in crimes classified as intra-family violence and sexual violence against children.  In each case, reporting is measured as victims per 100,000 children residing in each municipality $m$ and week $t$.  The measure of school closure, $School\ Closure_{mt}$ equals 1 if all schools in the municipality have been ordered closed by official decree and remain closed, or 0 otherwise. The school reopening measure, $Schools\ Reopen_{mt}$ takes the value of 1 if municipality $m$ has re-opened any schools in week $t$, following their initial closure, otherwise takes the value of 0.\footnote{Thus, by definition, $School\ Closure_{mt}=0$ everywhere prior to the official school closure decree on March 16, 2020.  Similarly, 
 $School\ Reopen_{mt}=0$ prior to the official school re-opening decree on August 17, 2020.}  In alternative models, in place of the binary measure of any schools being open, a continuous variable is used capturing the proportion of students whose schools have re-opened in municipality $m$ and week $t$.  

We include a series of fixed effects and time-varying controls, $\bm{X}^\prime_{mt}$ to capture potentially confounding factors.  In baseline models, we include week of year fixed effects for each of the 52 weeks of the year to capture temporal regularities in reporting within each year, as well as municipal-level fixed effects to capture all time-invariant municipal-specific unobservable factors.  In most demanding specifications, we include controls for (a) the epidemiological circumstances in each municipality and week, specifically rates of COVID infection, rates of testing, and test positivity; and (b) the lockdown status of each municipality which has been documented to additionally proxy mobility \citep{Bhalotraetal2021}.  

The model is estimated by Ordinary Least Squares, and standard errors are clustered by municipality. Given the specification of the model, the estimated parameter $\widehat\beta$ captures the change in reporting when comparing periods of school closure with baseline (pre-closure) periods, while $\widehat\gamma$ captures the change in reporting when comparing periods of school opening with baseline (pre-closure) periods.  In SI section \ref{scn:methodsSI} we lay out necessary assumptions for these parameters to have a causal interpretation, provide partial tests of these assumptions, and discuss a number of alternative tests related to concerns owing to time-varying policy adoption \citep{GB2021,dCDH2020}, and interactions with attendance levels.

We conduct counterfactual analysis which consists of estimating post school closure hypothetical reporting rates based on historical trends, which generate expected reporting in the absence of school closure and other pandemic-related events.  This is:
\begin{equation}
    \label{eqn:counterfactual}
    \widehat{Reporting}^{post}_{mt} = \widehat{\alpha}^{pre}+\widehat{WoY}^{pre}_t + \widehat{Municipal}^{pre}_m + \widehat{f(t)}^{pre},
\end{equation}
where projected reporting per 100,000 minors is estimated off pre-closure averages in each municipality and week of year, as well as flexible temporal trends $\widehat{f(t)}^{pre}$ chosen optimally as laid out below.  We then consider baseline reporting differentials as
\begin{equation}
\label{eqn:Diff1}
\text{Difference}^{post}_{mt} = Reporting^{post}_{mt}-\widehat{Reporting}^{post}_{mt},
\end{equation}
based on observed (actual) reporting, and projected baseline reporting from equation \ref{eqn:counterfactual}. The absolute number of ``missing cases'' are estimated as:
\begin{equation}
\label{eqn:Difference}
\text{Reporting Differential}_t=\sum_{m=1}^{346} \text{Difference}_{mt}^{post}\times \frac{Population_{mt}}{100,000}.
\end{equation}

These reporting differentials consider total aggregate changes in the post-COVID period.  To consider the contribution of school closure and reopening to these reporting changes, we generate another counterfactual projection, following equation \ref{eqn:counterfactual}, however now controlling entirely for the contribution of school closures and reopening:
\begin{eqnarray}
    \label{eqn:counterfactual2}
    \widehat{Reporting}^{post}_{mt}\Big|_{SO} &=& \widehat{\alpha}^{pre}+\widehat{WoY}^{pre}_t + \widehat{Municipal}^{pre}_m + \widehat{f(t)}^{pre} \nonumber \\ && + \widehat\delta \text{School Opening}_{mt}.
\end{eqnarray}
Here School Opening$_{mt}$ ($SO$) records the total proportion of students whose school is open (1 in pre-closure periods, 0 in full closure, and a positive proportion in the post-reopening period).  Following identical procedures from \ref{eqn:Diff1}, we can estimate reporting differentials absent the effect of school closure, and infer the relative importance of schools in declines in reporting.  Additional details related to this procedure, as well as sensitivity to inclusion of additional controls, is discussed in SI section \ref{scn:methodsSI}.

Models to estimate counterfactuals in equation \ref{eqn:counterfactual}, as well as the length of ``pre'' school closure periods are chosen based on optimal selection procedures, as models which minimize mean squared prediction errors.  Full discussion of this procedure, sensitivity to alternative modeling consideration, as well as analyses where counterfactual reporting allows for growth in baseline rates is provided in SI section \ref{scn:methodsSI}.  Inference procedures are consistently based on blocked bootstrap resampling over municipalities (see SI section \ref{scn:methodsSI}).

\section{Results}
\subsection{Impacts of School Closure and Reopening on Violence Reporting}
\begin{figure}[h!]
\begin{center}
\caption{Temporal Trends -- Crimes Reported Against Children, School Closure, and Epidemiological Measures}
\label{fig:trends}
\subfloat[Reports of Intra-family Violence Against Children\label{fig:pa}]{%
\includegraphics[width=0.33\textwidth]{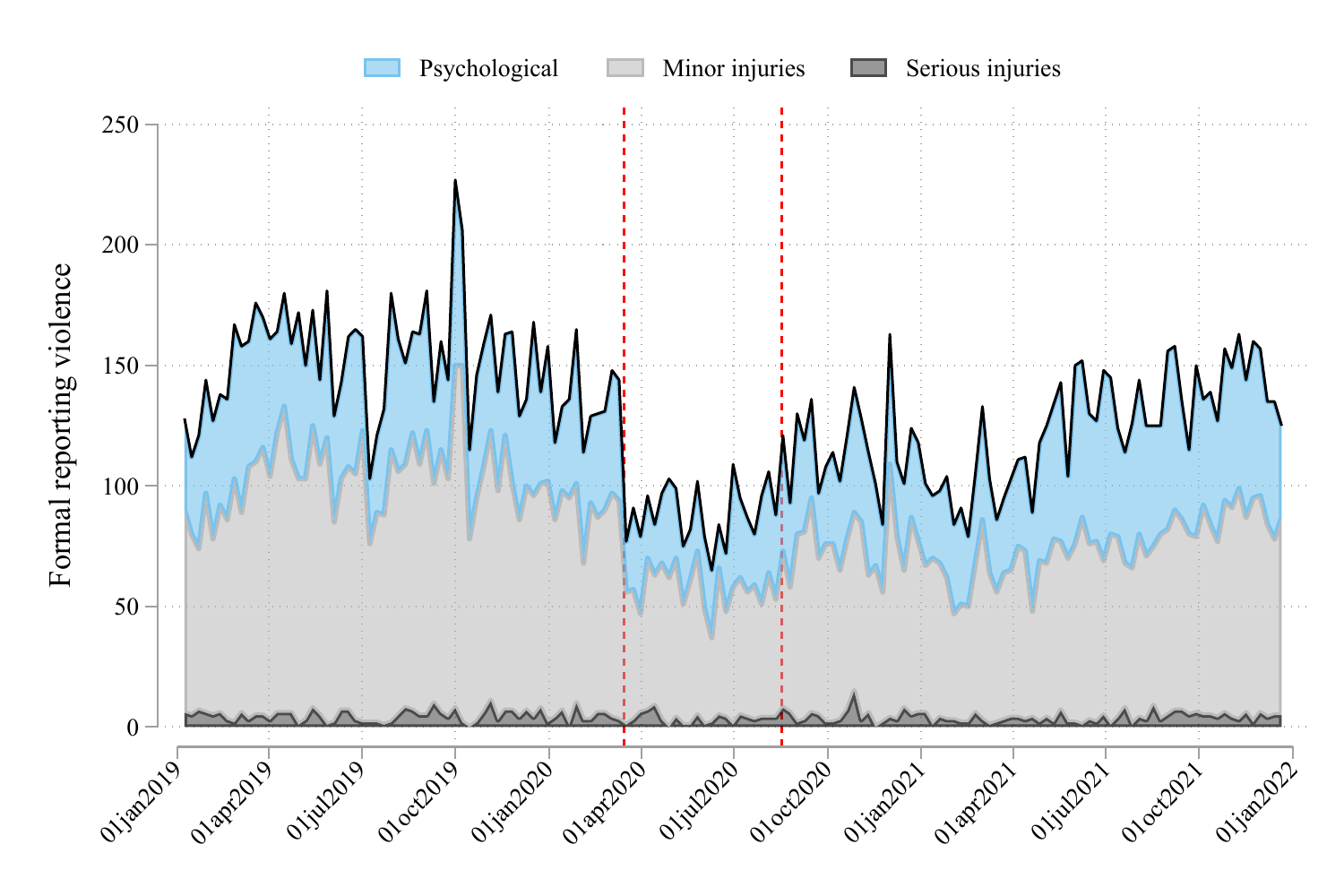}%
}
\subfloat[Reports of Sexual Assault Against Children\label{fig:pb}]{%
\includegraphics[width=0.33\textwidth]{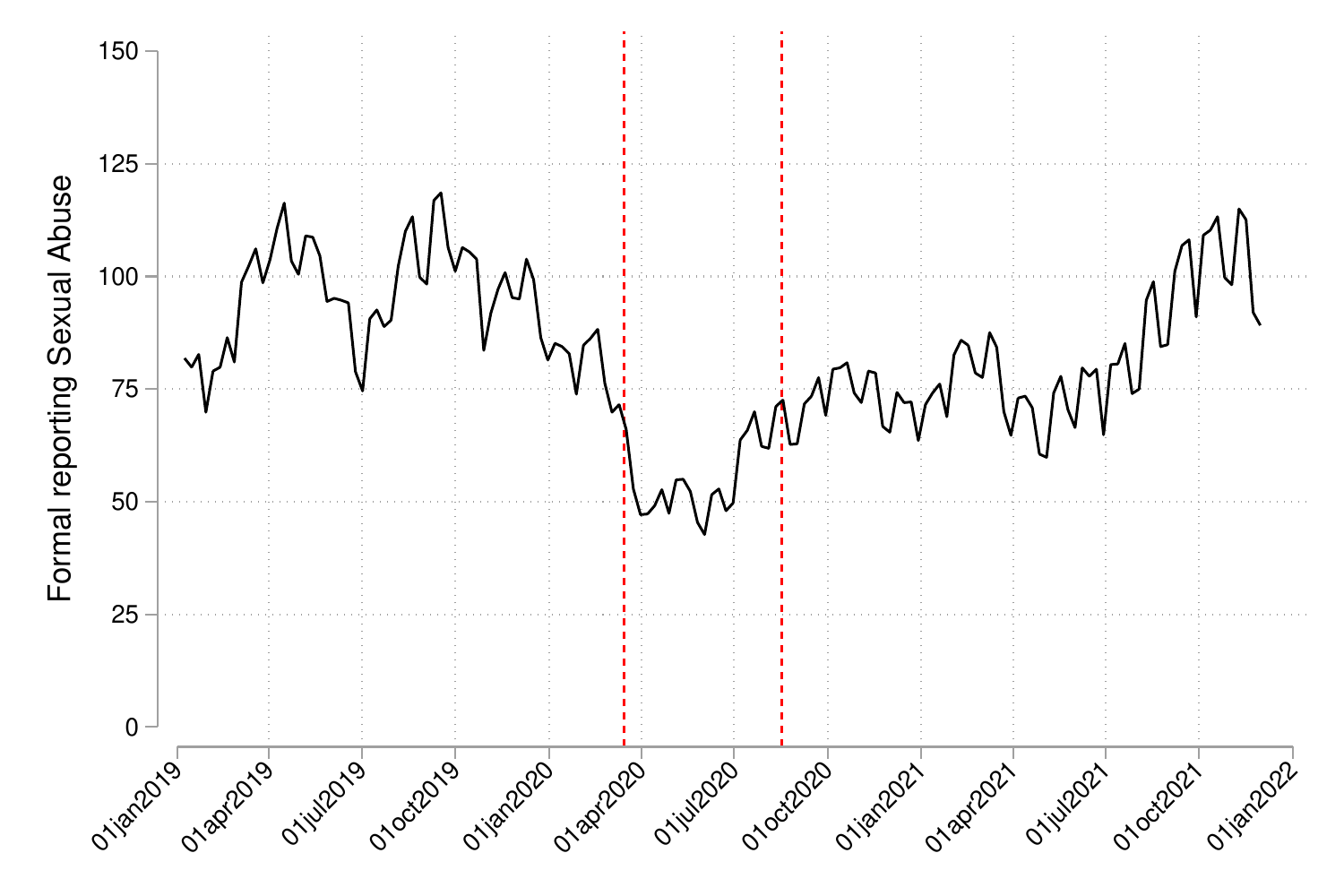}%
}
\subfloat[Reports of Rape Against Children\label{fig:pc}]{%
\includegraphics[width=0.33\textwidth]{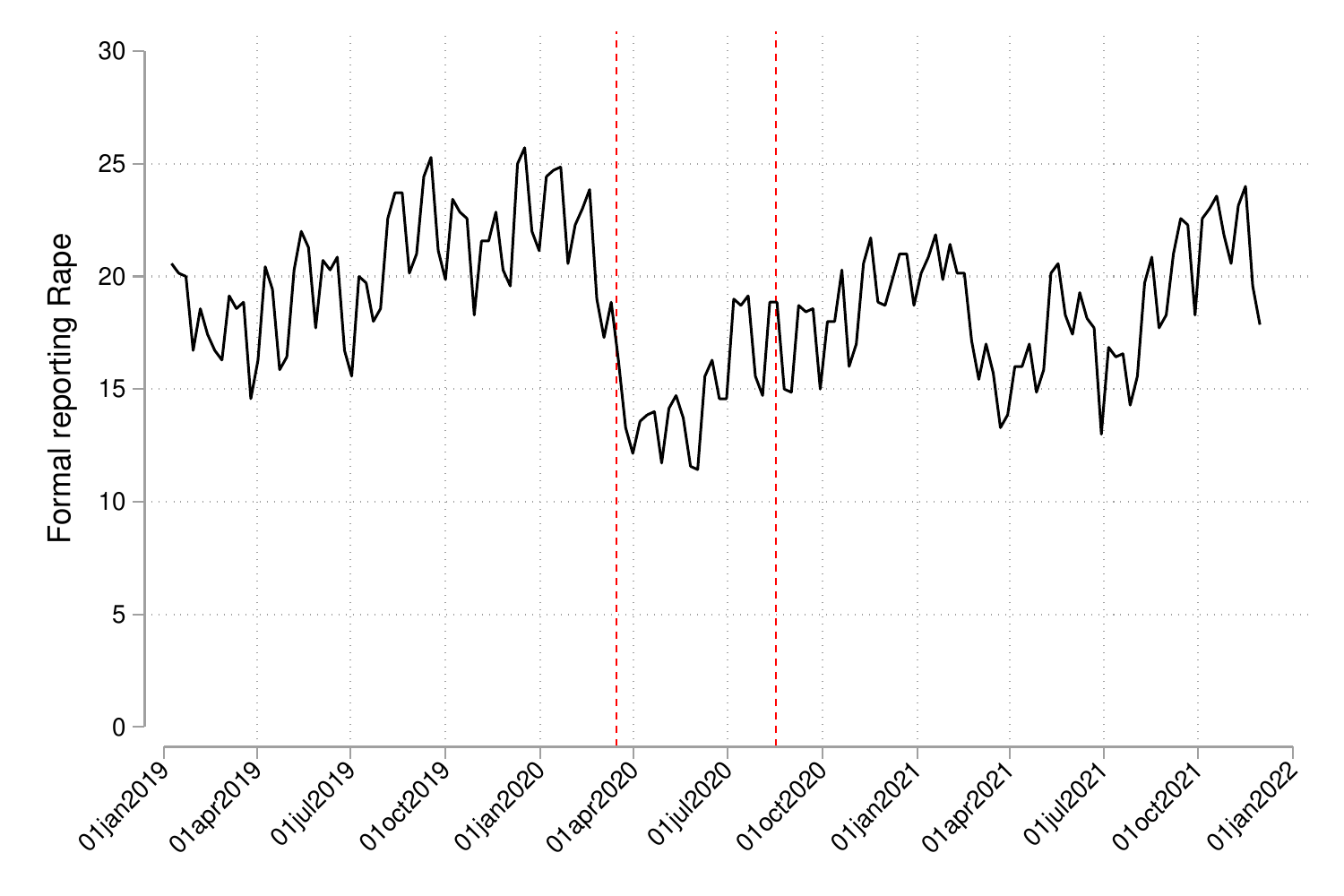}%
}\\
\subfloat[School Closure/Reopening\label{fig:pd}]{%
 \includegraphics[width=0.33\textwidth]{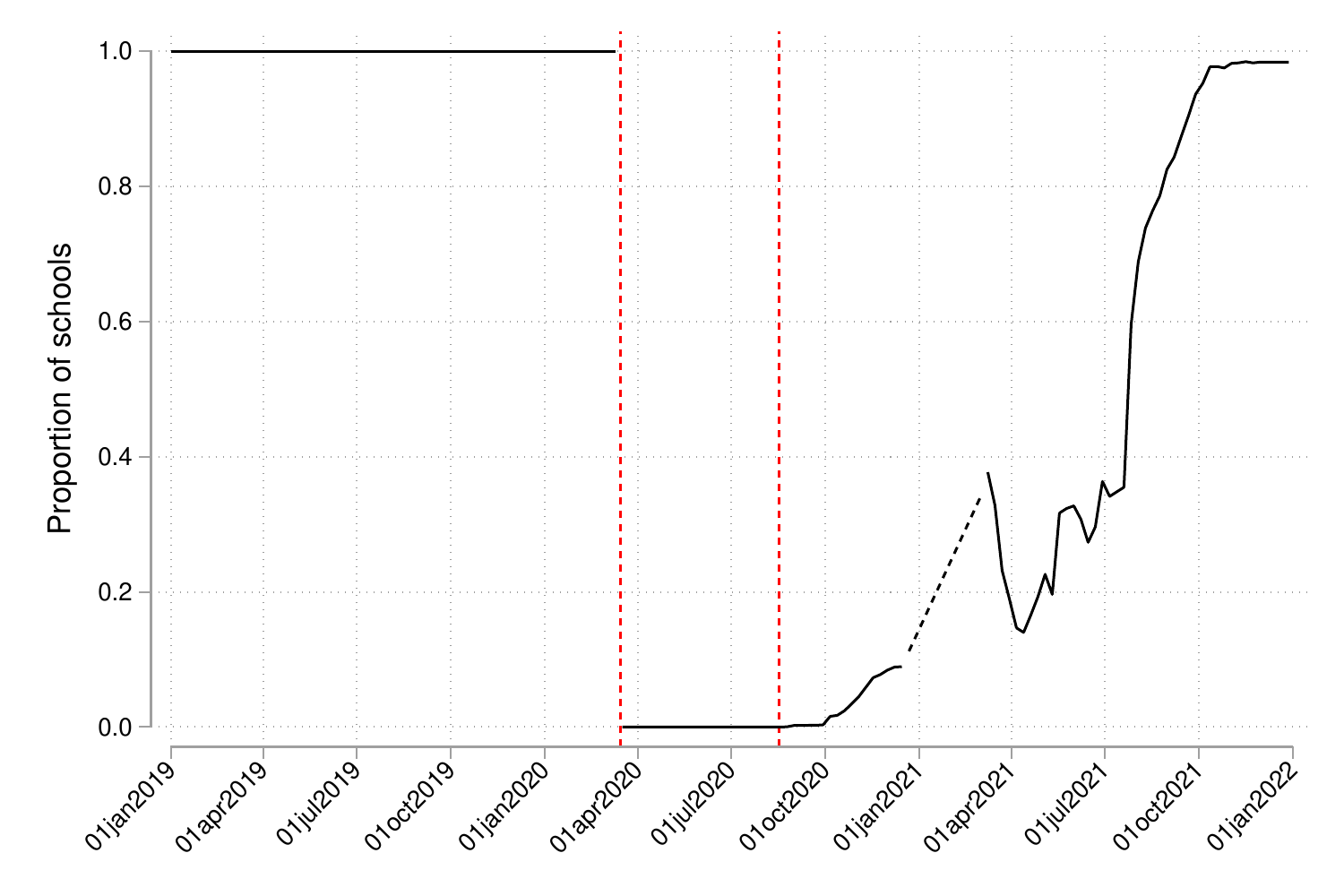}%
}
\subfloat[Lockdowns\label{fig:pe}]{%
\includegraphics[width=0.33\textwidth]{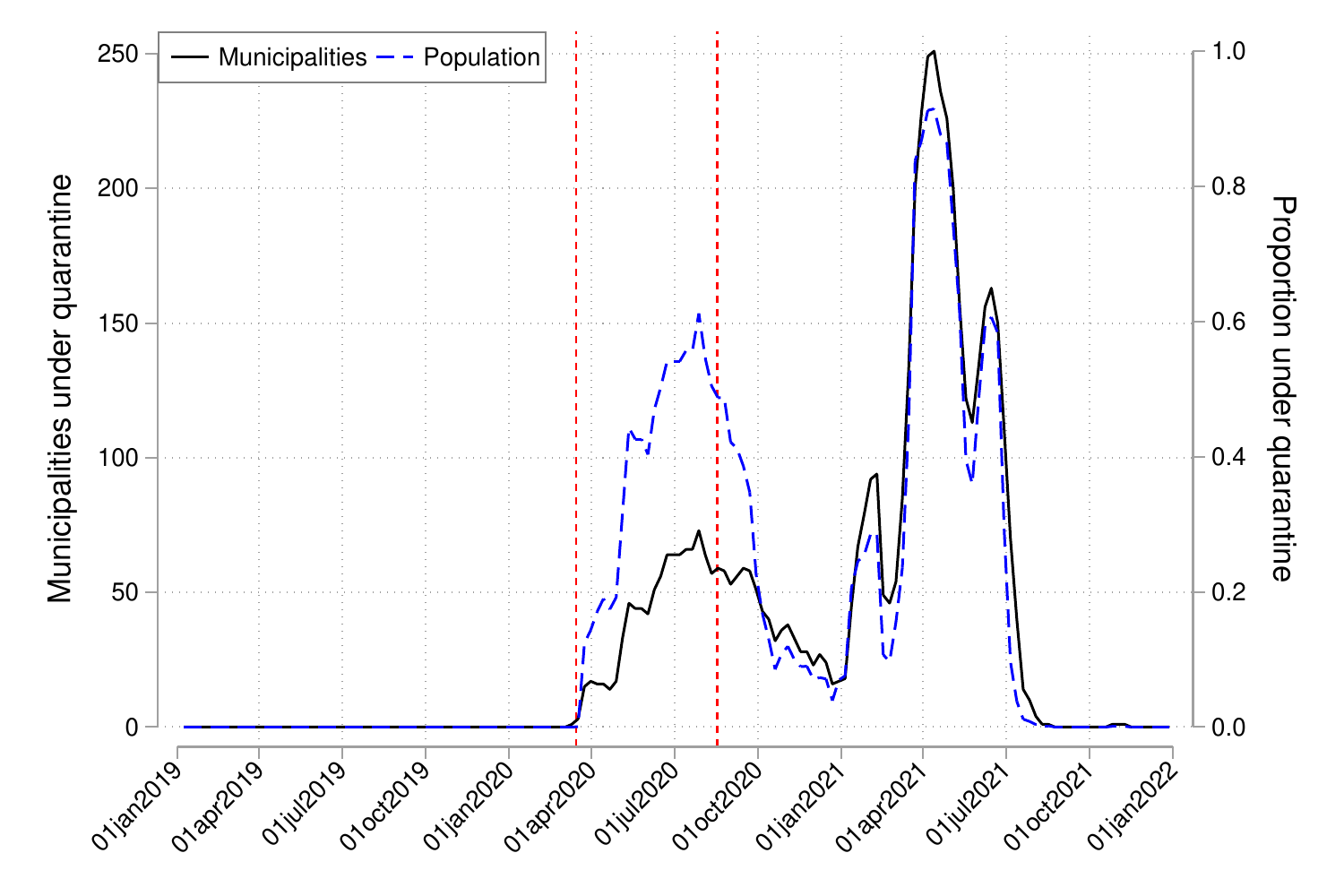}%
}
\subfloat[COVID Cases\label{fig:pf}]{%
 \includegraphics[width=0.33\textwidth]{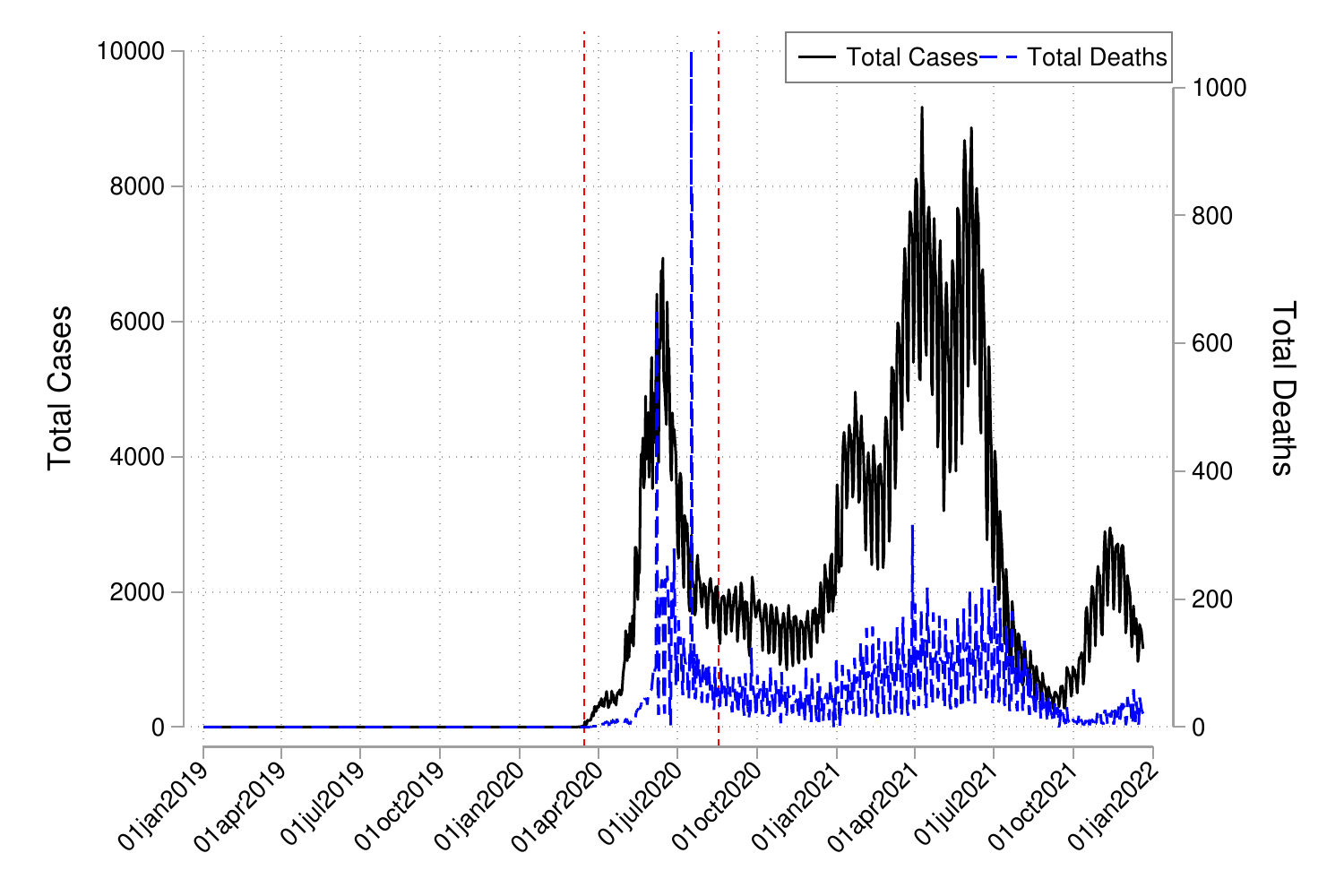}%
}
\end{center}
\floatfoot{\textbf{Notes to Fig.\ \ref{fig:trends}}: Each panel presents weekly measures of crime reporting against children (panels (a)-(c)), proportion of schools open (panel (d)), formal lockdown measures (panel (e)), and COVID cases and deaths.  Trends are reported between Jan.\ 1, 2019-Dec.\ 31, 2021, and are all based on official administrative records maintained by the Ministry for the Interior, Ministry of Education, Ministry of Health, or Ministry of Science (refer to Supplementary Information).  In panel (e) the number of municipalities under lockdown (left-hand axis) can be at most 346 (the total number of municipalities), while the proportion of the Chilean population under lockdown is plotted on right-hand axis.  The first vertical dashed line indicates the day which schools were ordered shut by the central government, while the second vertical dashed line indicates the day which schools were officially allowed to re-open.}
\end{figure}

Figure \ref{fig:trends} documents trends in reporting of intra-family violence (hereafter DV, for Domestic Violence), sexual assault, and rape against children, school closures, and the epidemiological and public health response in Chile.  These are displayed during the pre-pandemic period, initial arrival, and later pandemic; specifically, January 1, 2019-December 31, 2021.  The first case of COVID-19 was documented in Chile on March 3, 2020, and formal lockdowns were first put in place on March 26, 2020, reaching a considerable proportion of the population with their implementation in the entire capital of Santiago, on May 15, 2020. Schools were closed nation-wide on March 16, 2020 (for additional context of the strategies adopted in Chile, refer to SI, section \ref{SIscn:context}). Note that there is considerable temporal and spatial variation in lockdowns (which vary by municipality, and were implemented and revoked multiple times each week by the Ministry of Health), the spread of COVID-19 in the country, and school re-opening. School closures are the only event which were decreed country-wide on a single date, with both lockdowns and school re-openings occurring at different moments in each of the country's 346 municipalities (for details of school reopening, refer to SI Figures \ref{SIfig:Chile}-\ref{SIfig:RM}).

Figure \ref{fig:trends} shows large declines in reporting of violence against children in weekly aggregates occurring sharply with the closure of schools on March 16, 2020.  This is apparent even when comparing to much longer time series of data, see SI Figure \ref{SIfig:longtrendVIF}.  Total cases fall from approximately 150 to 75 criminal reports per week in the case of DV against children, from around 90 to around 50 in the case of sexual assault against minors, and from around 20 to around 12 in the case of rape against minors.  Complaints generally stayed at these reduced rates during the period in which schools were completely closed before moving in the direction of baseline rates as schools began to gradually reopen (refer to panel (d)). Note that declines in  formal reporting precede the large increases in population under lock-down and the first wave of COVID-19 cases, suggesting school closures itself may be a relevant channel.  This is studied formally in models below.\footnote{In the case of sexual assault and rape, reporting procedures result in an over-representation of cases officially reported as occurring on the 1\textsuperscript{st} day of the month, despite actually occurring later in the month.  In Figure \ref{fig:trends} these cases are smoothed to occur later in the month, however in SI Figure \ref{SIfig:unsmoothed} we note that patterns are similar if using original uncorrected data.  We document later in the paper that results are not sensitive to whether smoothing these data or not.}

Table \ref{tab:estimates} estimates observed changes in rates of reporting of violence against children when schools close and when schools re-open compared with rates prevailing at baseline (pre school closure).  In each case, violence (DV, sexual abuse, or rape) is measured as cases per 100,000 minors in municipality by week cells, and baseline means are displayed at the base of each panel.  School Closure is a binary measure taking the value of 1 when schools close in March 2020, and remaining at 1 until schools reopen in the municipality, while school reopening is a measure set to zero when schools have not yet re-opened, switching to 1 when at least one school reopens (panel A) or a continuous measure capturing the proportion of students whose schools have reopened (panel B).  Thus, each of the coefficients on School Closure and School Reopening capture relative changes compared to the baseline (pre-school closure) period, with negative values implying declines in rates of violence reporting compared with rates prevailing in the pre-closure period.

\begin{landscape}
\begin{table}[h!]
    \caption{Modelled Impacts of School Closure and Re-opening on Reporting of Violence Against Children}
    \label{tab:estimates}
    \centering
    \begin{tabular}{lccccccccc} \toprule
    & \multicolumn{3}{c}{Intra-family Violence} &  \multicolumn{3}{c}{Sexual Abuse} &  \multicolumn{3}{c}{Rape} \\ \cmidrule(r){2-4}\cmidrule(r){5-7}\cmidrule(r){8-10}
    & (1) & (2) & (3) & (4) & (5) & (6) & (7) & (8) & (9)  \\ \midrule 
    \multicolumn{10}{l}{\textbf{Panel A:} Binary Re-opening Measure} \\
    School Closure      &      -1.393***&      -1.589***&      -1.342***&      -0.819***&      -0.949***&      -1.014***&      -0.114***&      -0.099***&      -0.086** \\
                    &     (0.104)   &     (0.126)   &     (0.145)   &     (0.067)   &     (0.080)   &     (0.095)   &     (0.024)   &     (0.027)   &     (0.037)   \\
School Reopening    &      -0.771***&      -0.843***&      -0.892***&      -0.306***&      -0.352***&      -0.622***&      -0.047*  &      -0.050*  &      -0.085*  \\
                    &     (0.127)   &     (0.131)   &     (0.195)   &     (0.070)   &     (0.073)   &     (0.112)   &     (0.027)   &     (0.028)   &     (0.044)   \\
 \\
    Test of $\beta=\gamma$ (p-value) &     0.000 &     0.000 &     0.002  &     0.000 &     0.000 &     0.000 &     0.006  &     0.082  &     0.964     \\
Observations   &    54,214 &    54,214 &    54,214  &    53,179 &    53,179 &    53,179 &    53,179  &    53,179  &    53,179  \\
Baseline Mean   &     4.302 &     4.302 &     4.302  &     2.677 &     2.677 &     2.677  &     0.582  &     0.582 &     0.582 \\
\\
    \midrule
    \multicolumn{10}{l}{\textbf{Panel B:} Continuous Re-opening Measure} \\
    School Closure      &      -1.202***&      -1.432***&      -0.940***&      -0.692***&      -0.855***&      -0.691***&      -0.104***&      -0.088***&      -0.048   \\
                    &     (0.093)   &     (0.119)   &     (0.114)   &     (0.060)   &     (0.075)   &     (0.077)   &     (0.021)   &     (0.025)   &     (0.030)   \\
School Reopening    &      -0.556***&      -0.725***&      -0.261   &      -0.014   &      -0.189** &      -0.048   &      -0.041   &      -0.038   &      -0.024   \\
                    &     (0.174)   &     (0.171)   &     (0.199)   &     (0.093)   &     (0.095)   &     (0.117)   &     (0.038)   &     (0.039)   &     (0.046)   \\
 \\
    Test of $\beta=\gamma$ (p-value) &     0.000 &     0.000 &     0.000  &     0.000 &     0.000 &     0.000 &     0.078  &     0.198  &     0.598    \\
Observations   &    54,214 &    54,214 &    54,214  &    53,179 &    53,179 &    53,179 &    53,179  &    53,179  &    53,179  \\
Baseline Mean   &     4.302 &     4.302 &     4.302  &     2.677 &     2.677 &     2.677  &     0.582  &     0.582 &     0.582 \\
\\
    \midrule
    Municipal \& WoY FEs     &  & Y & Y &  & Y & Y &  & Y & Y \\
    Lockdown \& Epidemiological controls    &  &   & Y &  &   & Y &  &   & Y \\
    \bottomrule
    \multicolumn{10}{p{24.0cm}}{{\footnotesize \textbf{Notes to Table \ref{tab:estimates}}: Each column presents coefficients and standard errors from separate weighted linear regression models with alternative sets of covariates.  Outcomes are consistently measured as the number of violence reports per 100,000 children per municipality and week, for each week between January 1, 2019, and December 31, 2021.  Each column estimates equation \ref{eqn:2wayFE}, where School Closure is a binary indicator for periods in which schools are closed due to national decree, and Schools Reopening is a measure capturing the schools having reopened. Panel A measures re-opening as an indicator of at least one school in a municipality being open, while panel B measures re-opening as the proportion of students in the municipality whose school has re-opened.  `Test of $\beta=\gamma$' refers to coefficients in equation \ref{eqn:2wayFE}, ie the equality of estimates on School Closure and Schools Reopening.  Cluster-robust standard errors allowing for arbitrary correlations of unobservable shocks over time within each municipality are presented in parentheses below coefficient estimates.   $^{***}$ p$<0.01$; $^{**}$ p$<0.05$; $^{*}$ p$<0.10$.}}
    \end{tabular}
\end{table}
\end{landscape}

Columns (1), (4) and (7) document models which simply capture changes in the time series rates of reporting, while columns (2)-(3), (5)-(6) and (8)-(9) report estimates from two-way fixed effect (FE) models with or without controls, which capture all time-invariant municipal-specific factors as municipality FEs, and cyclical components as week of year FEs.  As we lay out in the Methods section and SI section \ref{scn:methodsSI}, identifying assumptions to causally estimate effects are that conditional on included controls, no other relevant events occur in affected municipalities at precisely the same moment as policy changes (parallel-trend style assumptions).  We point to suggestive evidence in favour of such assumptions in the Discussion section.

Focusing on baseline two-way FE models in Panel A, we estimate that school closure results in declines in reporting of DV by approximately 1.6 per 100,000 children per week, compared to a baseline 4.3 cases per 100,000 children.  This is in line with the sharp declines observed graphically (Fig \ref{fig:trends}). Once schools reopen, cases are observed to decline by `only' 0.84 cases per 100,000 children.  This suggests that cases sharply decline upon school closure and increase upon school reopening\footnote{We formally test the difference between coefficients on closure and re-opening, and find clear evidence to suggest that even though reporting is lower than in the baseline period, it is considerably higher than during the period of full closure ($p$-value<0.01 in table footer).}, but that this increase upon reopening is not sufficient to recover baseline rates of violence reporting.  Similar patterns, albeit it with different magnitudes, are observed in the case of sexual abuse, and rape against children.  In the case of sexual abuse, initial declines are estimated as 0.95 fewer cases per 100,000, while post-opening declines are more moderate, at 0.35 fewer cases per 100,000 children, while in the case of rape, these values are estimated at declines of 0.10 (closure) and 0.05 (re-opening).  In each case, these values are substantial when compared with baseline rates.  In the case of DV and sexual abuse against children substantively similar results are observed even when conditioning on each municipality's lockdown status as well as rates of COVID infection, testing and test positivity, suggesting that these results do not simply capture changes owing to municipal circumstances beyond school closures. In the case of rape, which is the most infrequent outcome and hence least powered outcome, while we still observe reductions in rates of criminal reporting both during school closure and re-opening, we can no longer conclude that reporting rates \emph{increase} when moving from closure to re-opening.\footnote{Indeed, estimates are virtually identical at around $-0.085$, resulting in a $p$-value$=0.964$ on the test of equality of coefficients on School Closure and Schools Reopening.}

Results documented in Panel A of Table \ref{tab:estimates} are based on binary measures of first school reopening, however this may underplay the importance of re-opening, particularly in municipalities with many schools where reopening occurred only gradually.  Panel B thus re-estimates with a measure of the continuous proportion of students whose school were re-opened, which varies between 0 (no students with open schools) to 1 (all students with an open school). It is important to note here that all coefficients on Schools Reopening are thus cast as the effect of moving all children back into school.  In reality, this occurred only substantially after first reopening\footnote{For example, by September 2021, 84.4\% of students' schools were reopened, by October 2021, this value had reached 96.6\%, and by December 2021 this value had reached 98.6\%.} Here, in the case of DV reporting we observe that if school reopening does indeed become complete, rates of reporting are estimated to no longer be statistically significantly below baseline rates.  Estimates fully conditioning on time-varying controls in column (3) suggest that while complaints would still be 0.26 per 100,000 children lower than in baseline periods, we cannot formally rule out that the confidence interval of this estimate contains 0, or a return to pre-closure rates of reporting.\footnote{The results between panels A and B are consistent, in that panel A refers to a binary measure of municipal reopening (regardless of the proportion of students whose schools had reopened), while panel B refers to continuous measures of students whose schools had re-opened, and estimates are interpreted as the impact of moving from 0 (full closure) to 1 (full opening), with few municipal by week cells observed with full re-opening.}  As in panel A, we consistently observe sharp reporting declines with initial school closures.

In the case of sexual abuse and rape, we once again observe sharp declines in reporting upon school closure, and evidence to suggest that re-opening may provide recovery in rates of complaints, at least compared to baseline levels (later in the paper we consider counterfactual outcomes in which rates of violence against children actually increased), once re-opening reaches 100\%.  In columns (6), (8) and (9) we observe that when schools fully re-opened, we cannot reject that rates of reporting would have been the same as in pre-closure periods, with all point estimates being slightly negative, but again not statistically distinguishable from zero.

However, even when schools re-opened, not all students went back to in-person activities (SI section \ref{SIscn:context}). For example, by November, 2021, while almost all schools had resumed some form of in-person activities, 30\% of students were still observed to not attend even one day per week. If, as we argue, identifying violence requires time spent in school \citep{fitzpatrick2020}, the lack of attendance will inevitably result in a more limited recovery. We explore these patterns by interacting the school reopening measures with the proportion of attendance of students that were attending at least one day a week.\footnote{As data on attendance is not observed for all periods (refer to SI section \ref{SIscn:data}, these models are only estimated for periods in which all data are available, and reported in Supplementary Information.} The results indicate that criminal complaints recover more rapidly when student attendance is higher, which is consistent with in-person interactions being a key mechanism to increase reporting (SI Tables \ref{SItab:attendanceDV}-\ref{SItab:attendanceRape}).

A number of additional results are displayed in SI; refer to discussion in SI section \ref{SIscn:results}.  In SI Figures \ref{SIfig:eventCloseDV}-\ref{SIfig:eventOpenRa} we present event studies documenting full dynamic effects and quite flat pre-trends (with one exception where cyclicality is observed), providing support to parallel trend style assumptions, given that units are at least observed to be trending in similar directions prior to school closure/reopening events.  Secondly, these results hold when eliminating months of summer vacations when schools are closed (SI Table \ref{SItab:novacations}), and are virtually unchanged if we use raw measures of sexual abuse and rape with over-reporting on the first day of each month rather than smoothed measures (SI Table \ref{SItab:unadjusted}).  If disaggregating DV by specific classifications, we observe results are both largest in magnitude and proportion when considering moderate physical violence, followed by psychological violence, and smallest or insignificant when considering serious physical violence, consistent with these more serious cases being captured by authorities even when schools are closed \citep{loiseau2021} (SI Table \ref{SItab:subDV}). The interpretation of estimates of school re-opening hold when accounting for potential heterogeneous impacts over time and staggered roll-out of school re-opening (SI Figure \ref{SIfig:GB} and Table \ref{SItab:2WayWeights}). Finally, we use data from a local agency in charge of child protection to examine how different institutional channels changed during this period (refer to SI section \ref{SIscn:data}). Descriptive results show that the number of referrals coming from school sharply decrease once schools closed, and they recovered, but not to pre-pandemic levels, during the second semester of 2021 (SI Figure \ref{SIfig:OPD} and Table \ref{SItab:OPDdesc}). We also observed a decrease in cases reported by the health system, particularly during the first months of the pandemic.

\paragraph{Heterogeneity of Policy Impacts}

\begin{figure}[htpb!]
\begin{center}
\caption{Modelled Impact of School Closures and Openings: Demographic and Socio-Economic Variation}
\label{fig:heterogeneity}
\subfloat[Intra-family Violence\vspace{-2mm}]{%
\includegraphics[width=0.46\textwidth]{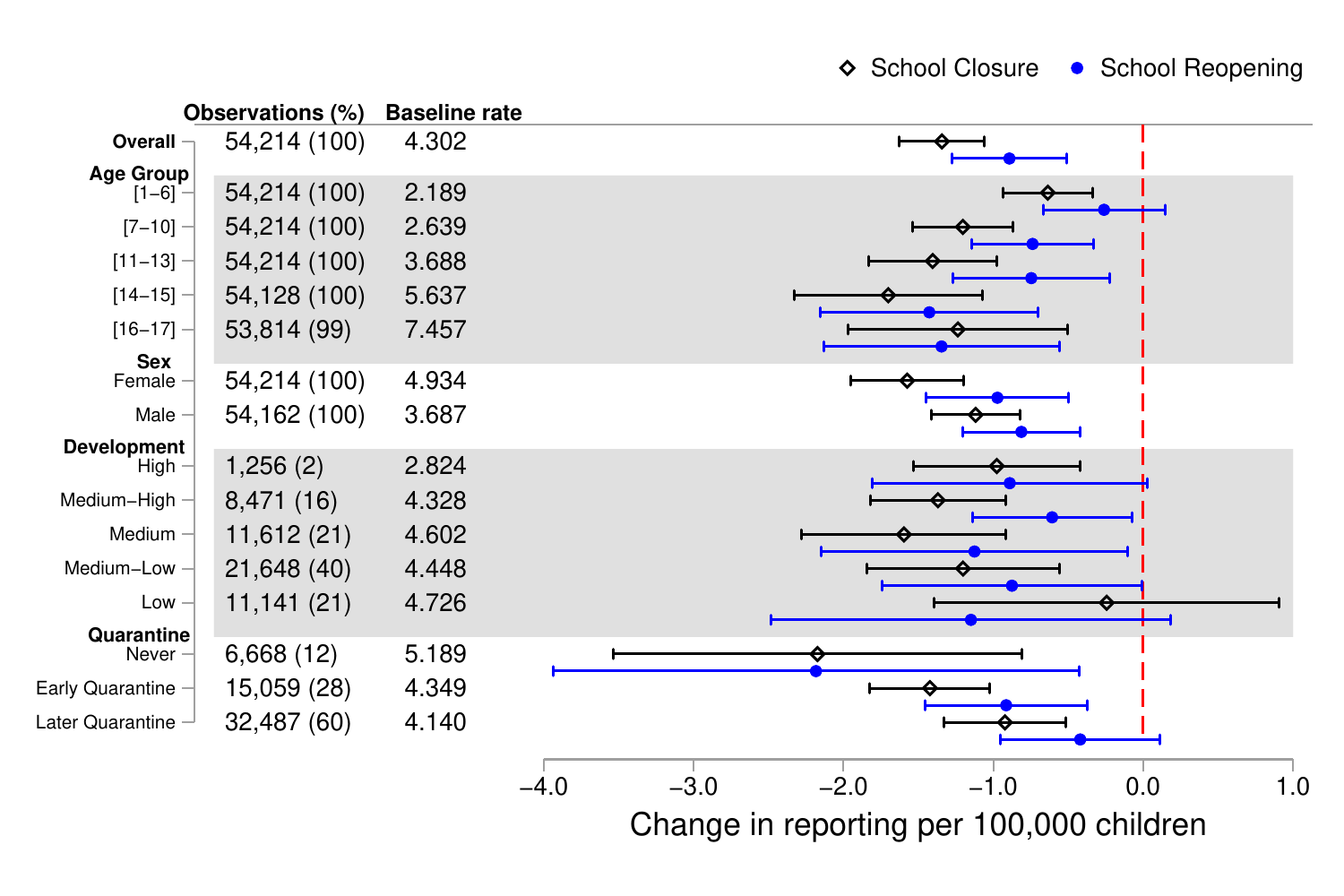}%
}\\
\subfloat[Sexual Abuse\vspace{-2mm}]{%
\includegraphics[width=0.46\textwidth]{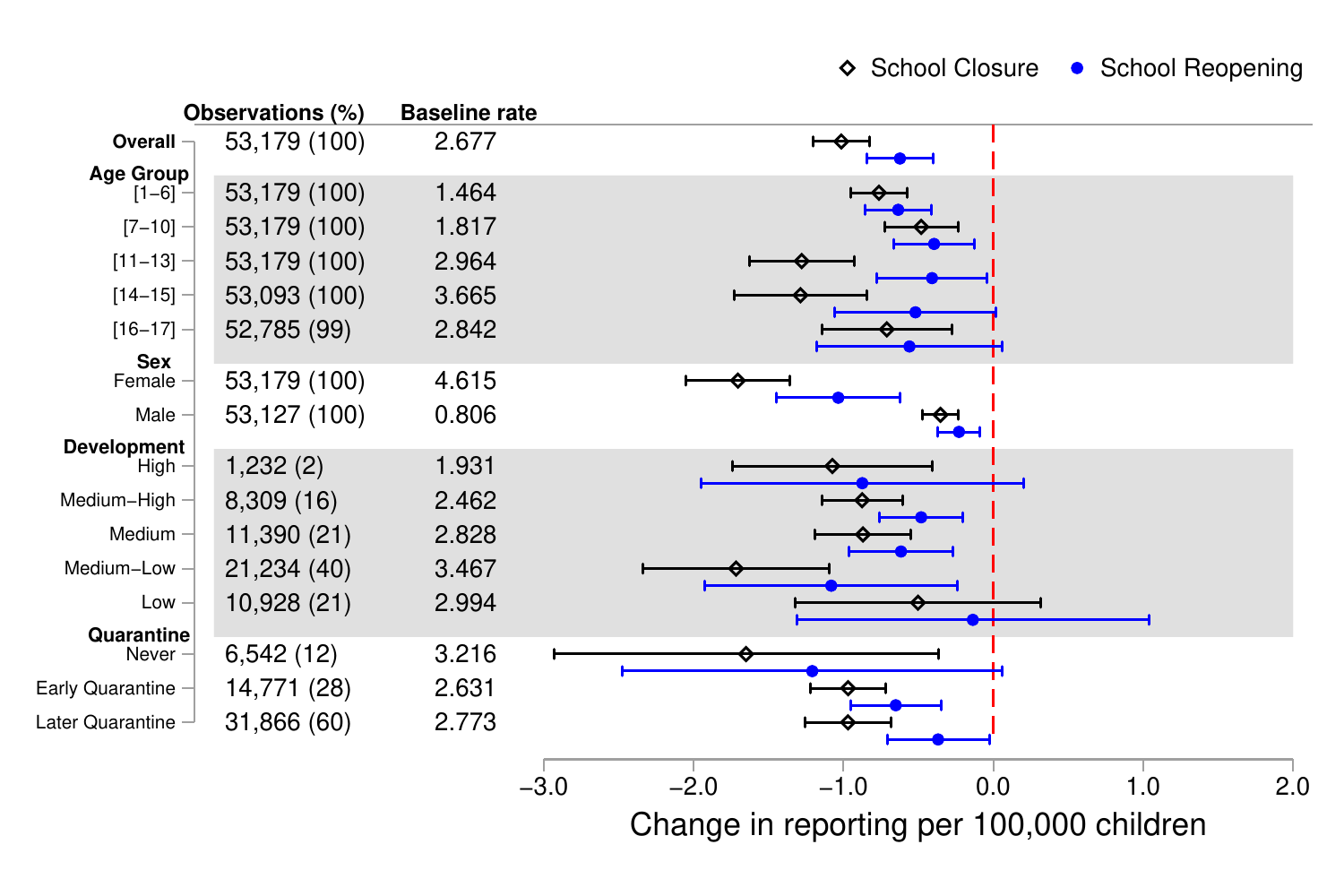}
}\\
\subfloat[Rape\vspace{-2mm}]{%
\includegraphics[width=0.46\textwidth]{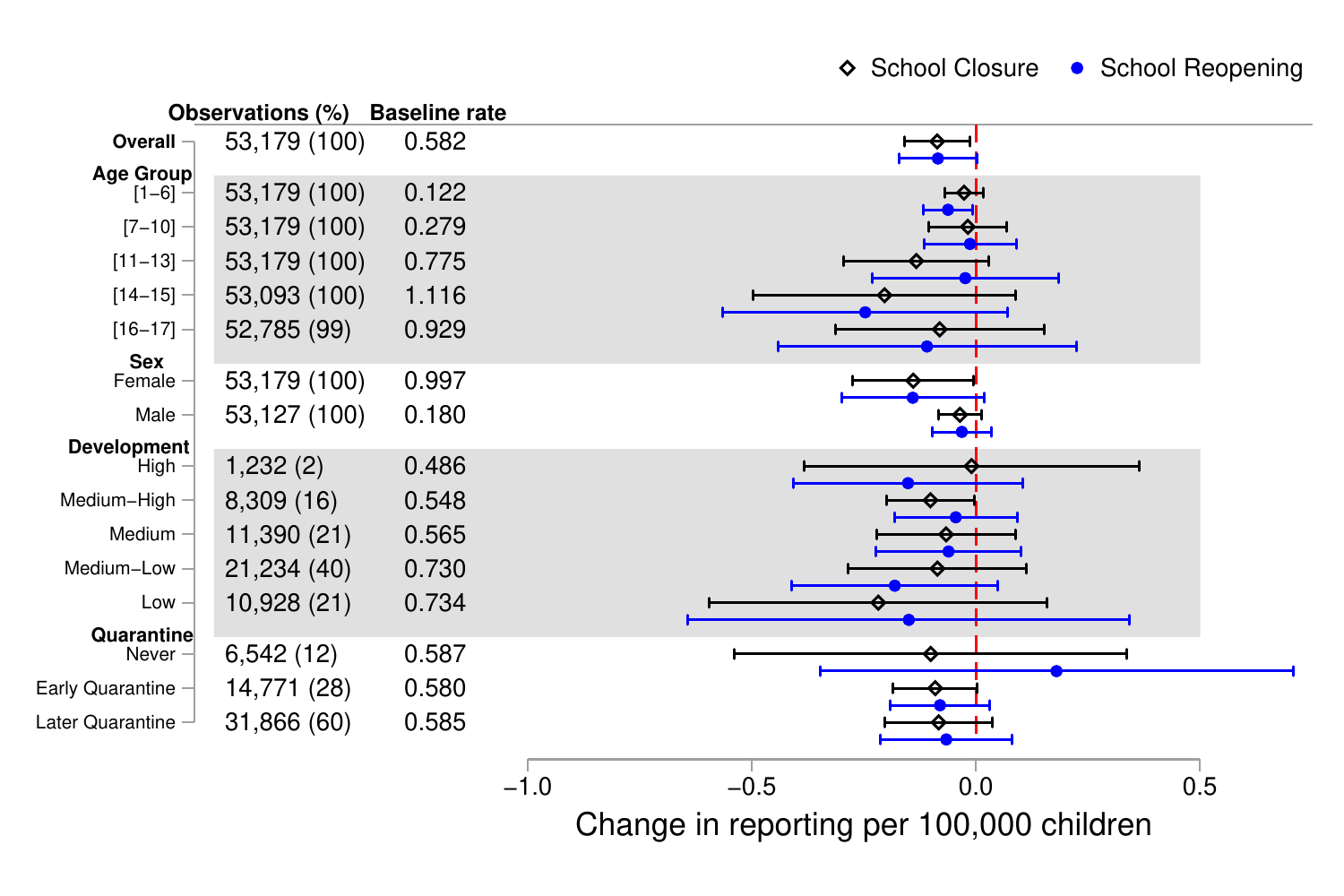}%
}
\end{center}
\floatfoot{\textbf{Notes to Fig.\ \ref{fig:heterogeneity}}: Estimates (diamonds and circles) and 95\% confidence intervals are displayed for models corresponding to sub-groups indicated on the vertical axis.  Estimates for each group correspond to coefficients on School Closure (diamonds), and Schools Reopening (circles), following equation \ref{eqn:2wayFE}, with full time-varying controls. In each case, estimates are based on the population or municipality-specific estimation sample, and estimates are consistently weighted by the population of the estimation sample.  The total number of municipality by week observations are indicated in 
``Observations", with the \% referring to the percent of the full sample of municipality by week cells.  Baseline (pre-2020) rates per 100,000 individuals of each group are displayed as ``Baseline rate".}
\end{figure}

Figure \ref{fig:heterogeneity} documents variation in estimates from Table \ref{tab:estimates} within different sectors of the population.  In each case point estimates are reported along with 95\% CIs, based on two-way FE models with time-varying controls (models with graduated controls are documented in SI Figures \ref{SIfig:heterogeneity1} and \ref{SIfig:heterogeneity2}).  Estimates signalled with hollow diamonds present impacts of school closure, while filled circles present impacts of re-opening.  Variation in observed estimates allows us to consider how and where schools act as a safety net in cases of reporting of violence against children, and additionally, in the case of variation in quarantine status, provide a partial test of identifying assumptions as we can observe impacts independent of formal government policy responses.

Consider first the impact by age.  In sub-figure (a) we observe a `backwards J' pattern, in which the impact of school closure is largest in absolute magnitude among children in their mid teenage years, lower among older teens, and lowest among younger children.  This pattern is consistent with schools acting to channel complaints most frequently for children above 6 who are most connected to school systems, and the fact that infants may have better access to other channels, such as the health system (refer to SI Table \ref{SItab:channelsOPD} for a descriptive example of this). The role of schools is also reduced when children are older (16-17), and potentially more empowered to make their own complaints \citep{ortiz2021}. Broadly similar impacts of closure by age are observed in cases of sexual violence in panels (b) and (c).  In the case of re-opening, we observe largest recoveries in rates of reporting among younger children, in line with increases in attendance patterns in these groups (SI, Figures \ref{SIfig:AttendanceAll}-\ref{SIfig:AttendanceGroups}). Indeed, significant increases in reporting are observed only among individuals aged 13 and under in the case of DV, and are concentrated among younger individuals in the case of sexual abuse (SI Figure \ref{SIfig:heterogeneity3}).

We report estimates by each municipality's lockdown status, as an early lockdown (March 16-August 30, 2020), late lockdown (September 1, 2020 or after), or no lockdown area. In each of these three cases we observe sharp declines in rates of complaints for DV and sexual abuse (though noisier estimate for rape) suggesting that these results do not simply capture reductions in movement owing to lockdowns (or a `lock-down' effect), but rather transversal effects of school closure on violence reporting, observed across all municipality types.  Similarly, with the case of re-opening, we generally observe that declines in reporting are substantially reversed, regardless of a municipality's lockdown status.

\subsection{Estimated under-reporting based on counterfactual projections}
\begin{figure}[h!]
\begin{center}
\caption{Reporting, Projected Reporting, and Under-reporting Under Various Counterfactual Assumptions}
\label{fig:counterfactuals}
\textbf{Panel A: Simple Counterfactual (Time Only)} \\
\subfloat[Intra-family Violence]{%
\includegraphics[width=0.33\textwidth]{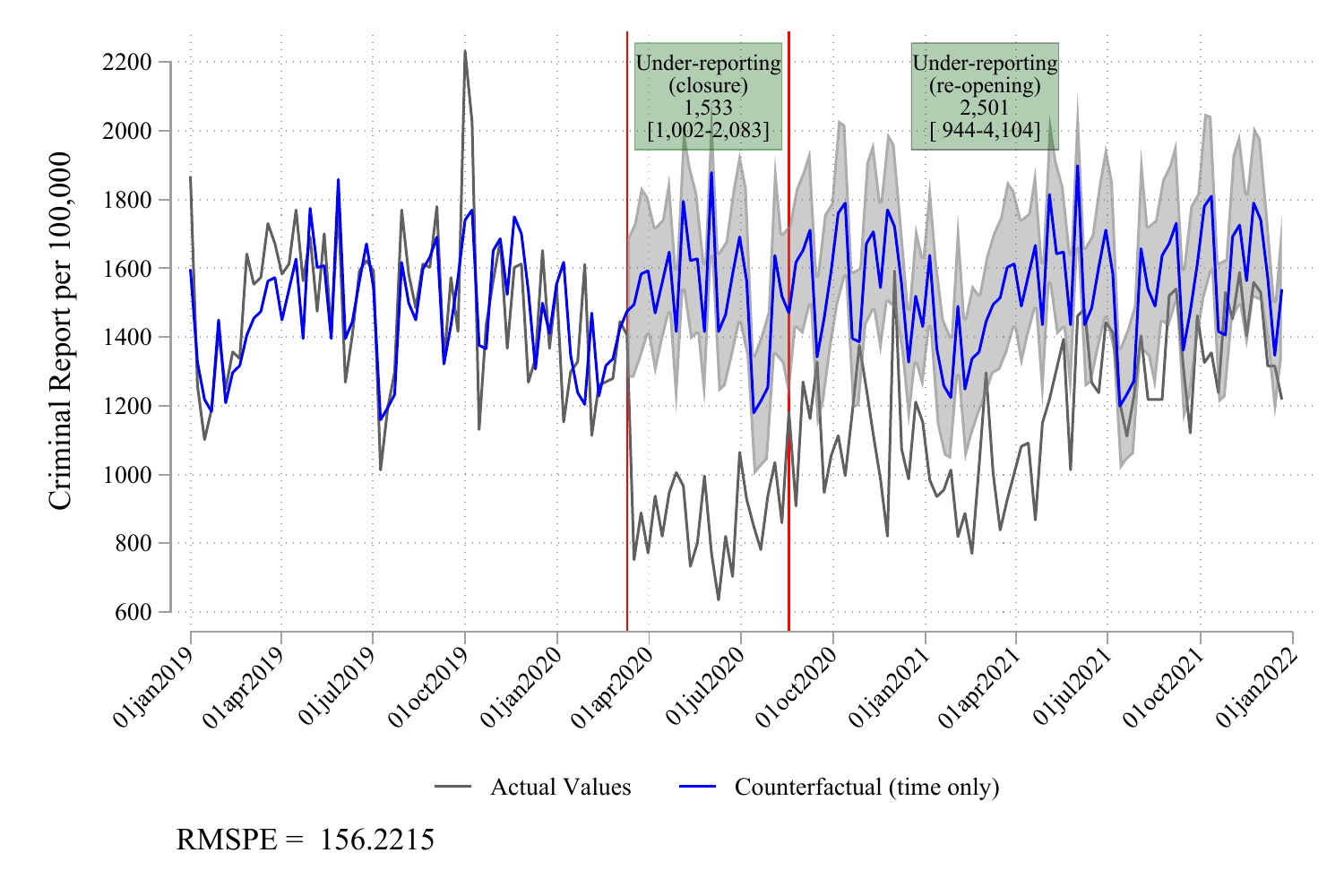}%
}
\subfloat[Sexual Abuse]{%
\includegraphics[width=0.33\textwidth]{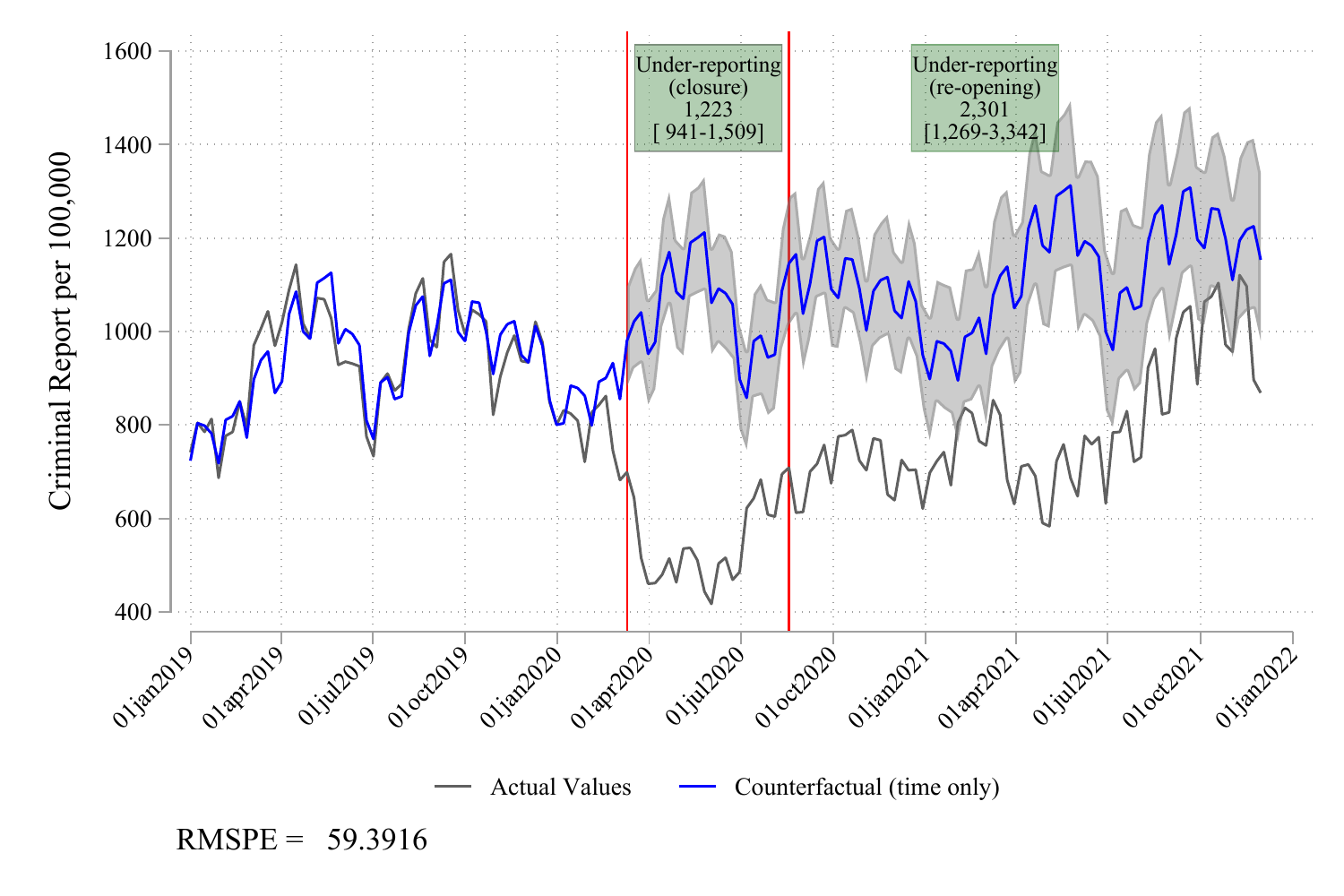}%
}
\subfloat[Rape]{%
\includegraphics[width=0.33\textwidth]{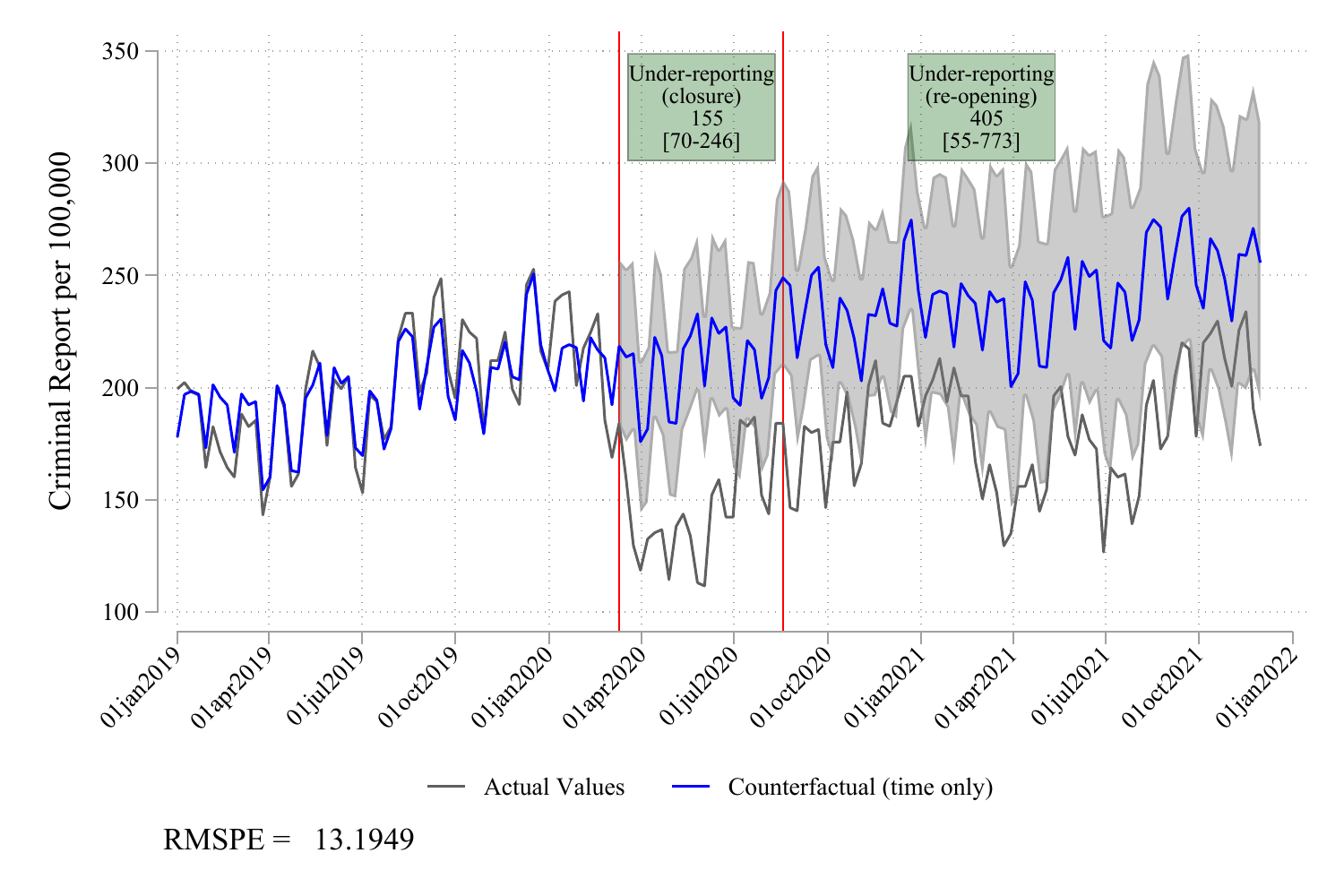}%
}
\\
\textbf{Panel B: Counterfactual (No School Channel)} \\
\subfloat[Intra-family Violence]{%
\includegraphics[width=0.33\textwidth]{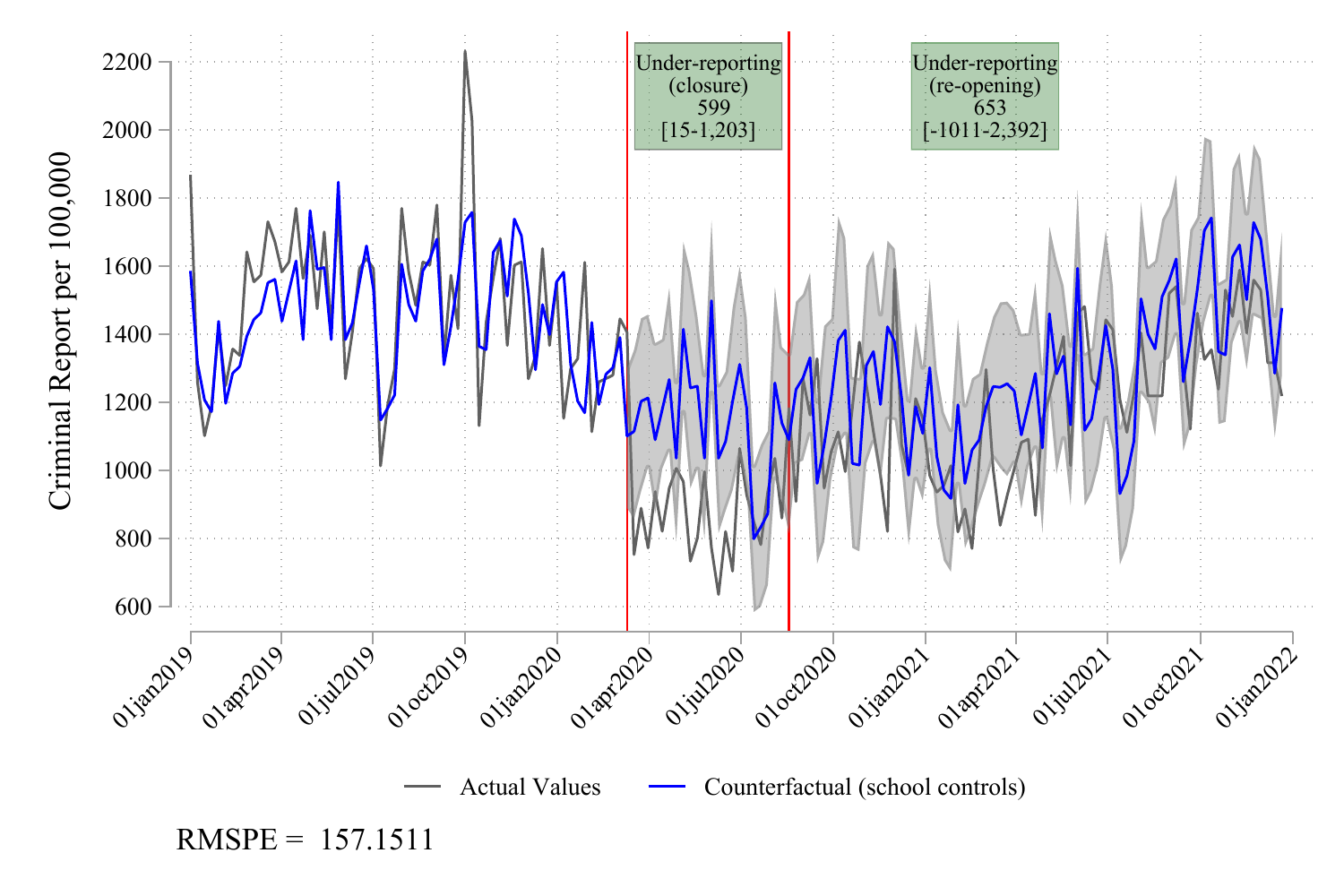}%
}
\subfloat[Sexual Abuse]{%
\includegraphics[width=0.33\textwidth]{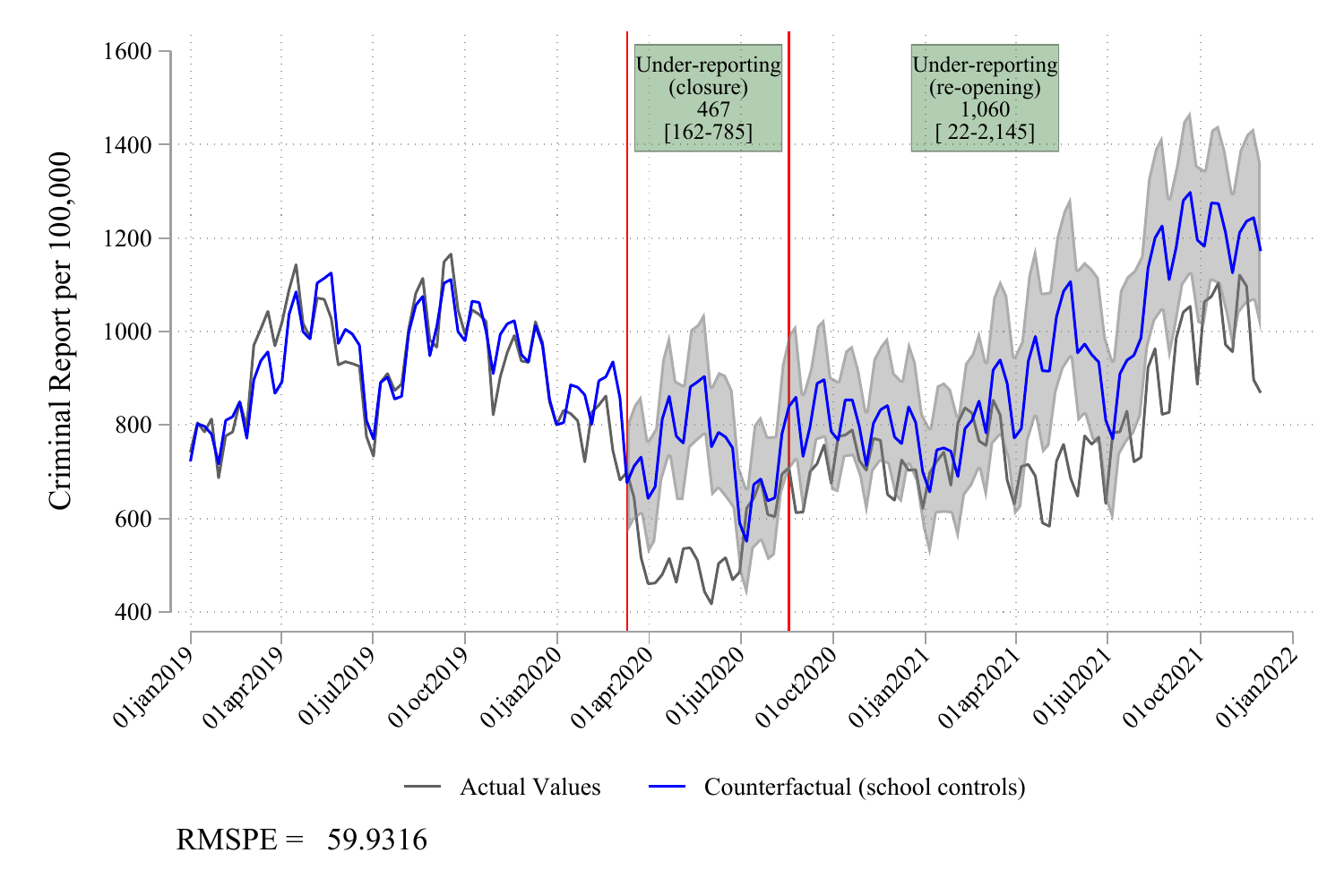}%
}
\subfloat[Rape]{%
\includegraphics[width=0.33\textwidth]{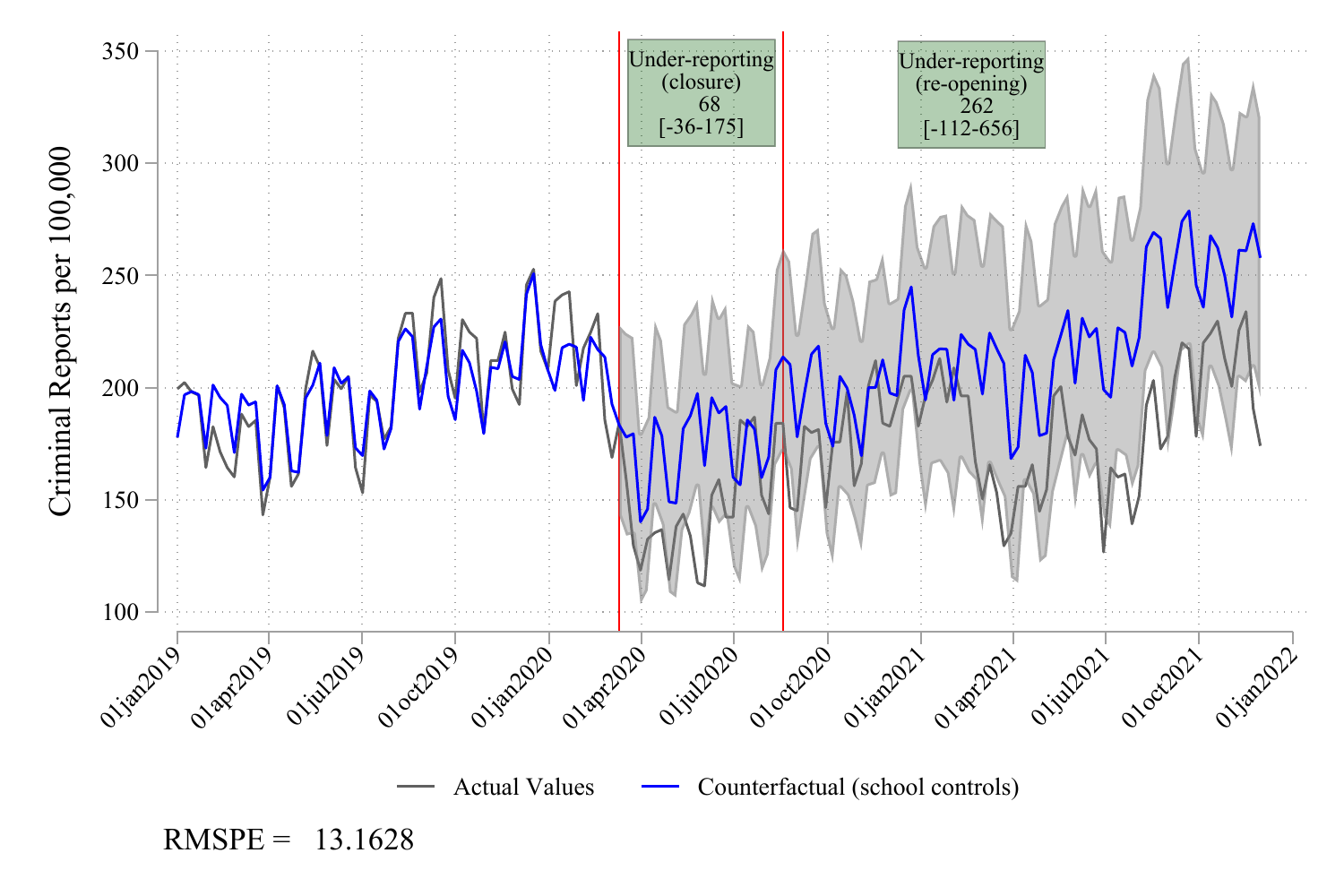}%
}\\
\textbf{Panel C: Projected Under-reporting} \\
\subfloat[Intra-family Violence]{%
\includegraphics[width=0.33\textwidth]{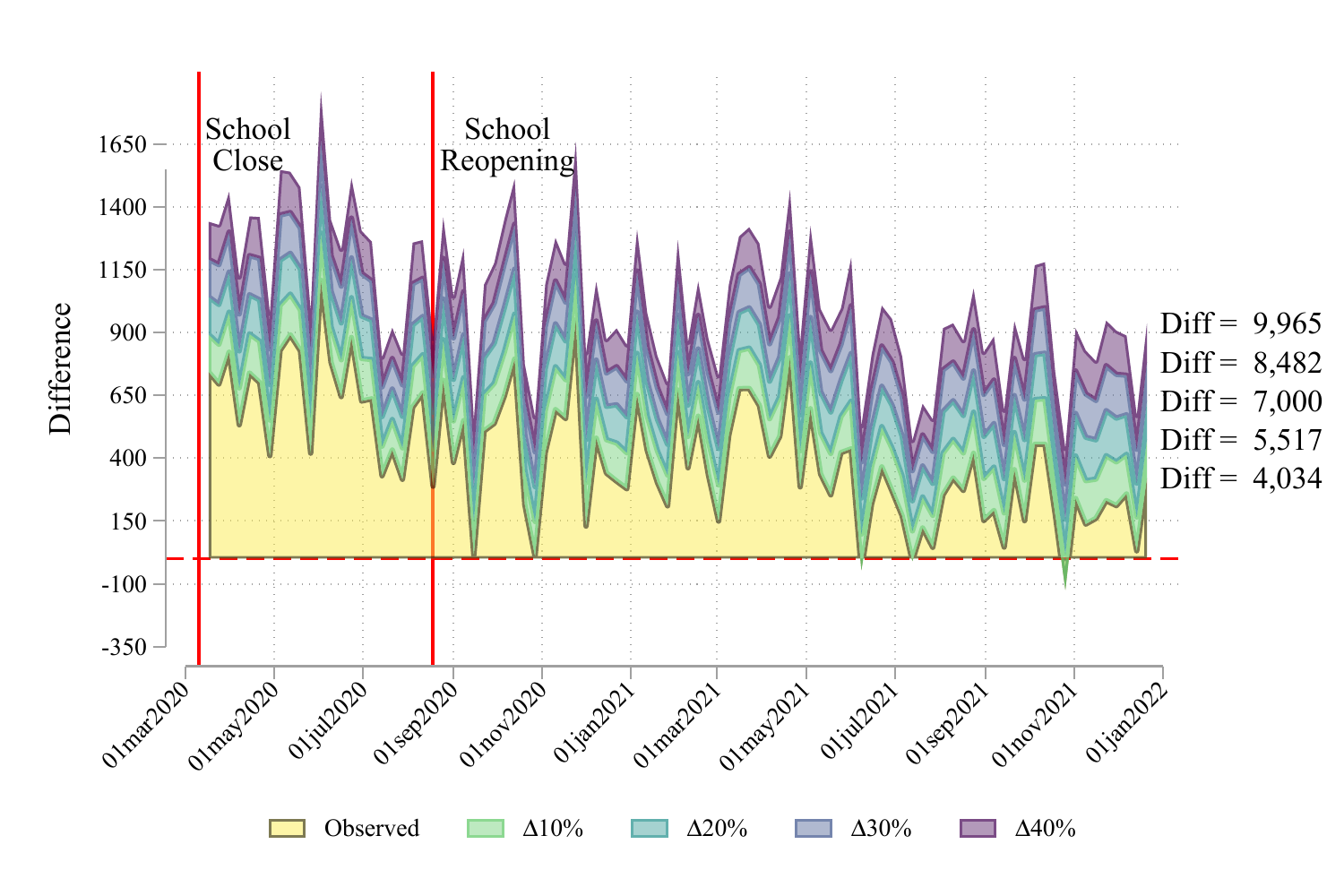}%
}
\subfloat[Sexual Abuse]{%
\includegraphics[width=0.33\textwidth]{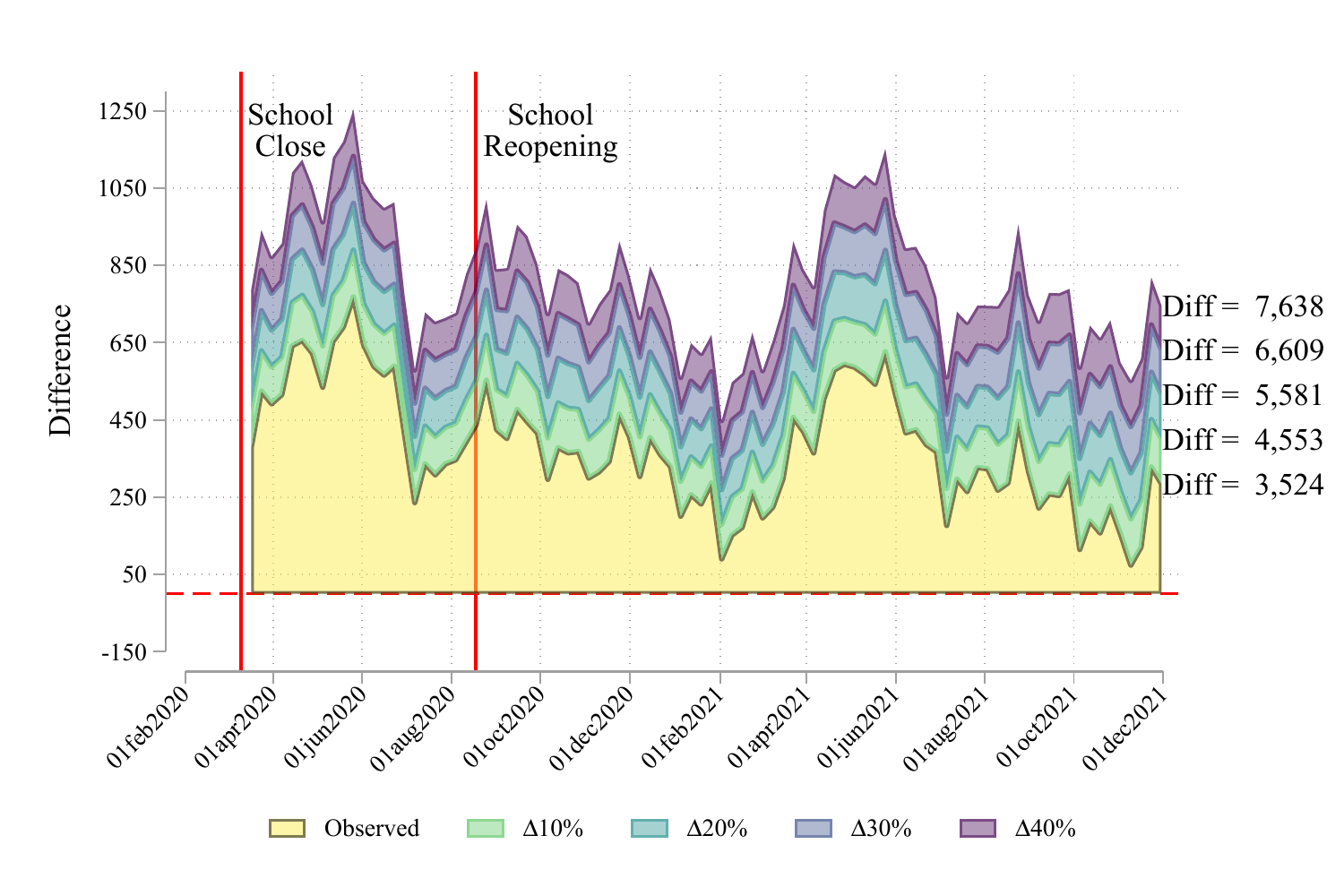}%
}
\subfloat[Rape]{%
\includegraphics[width=0.33\textwidth]{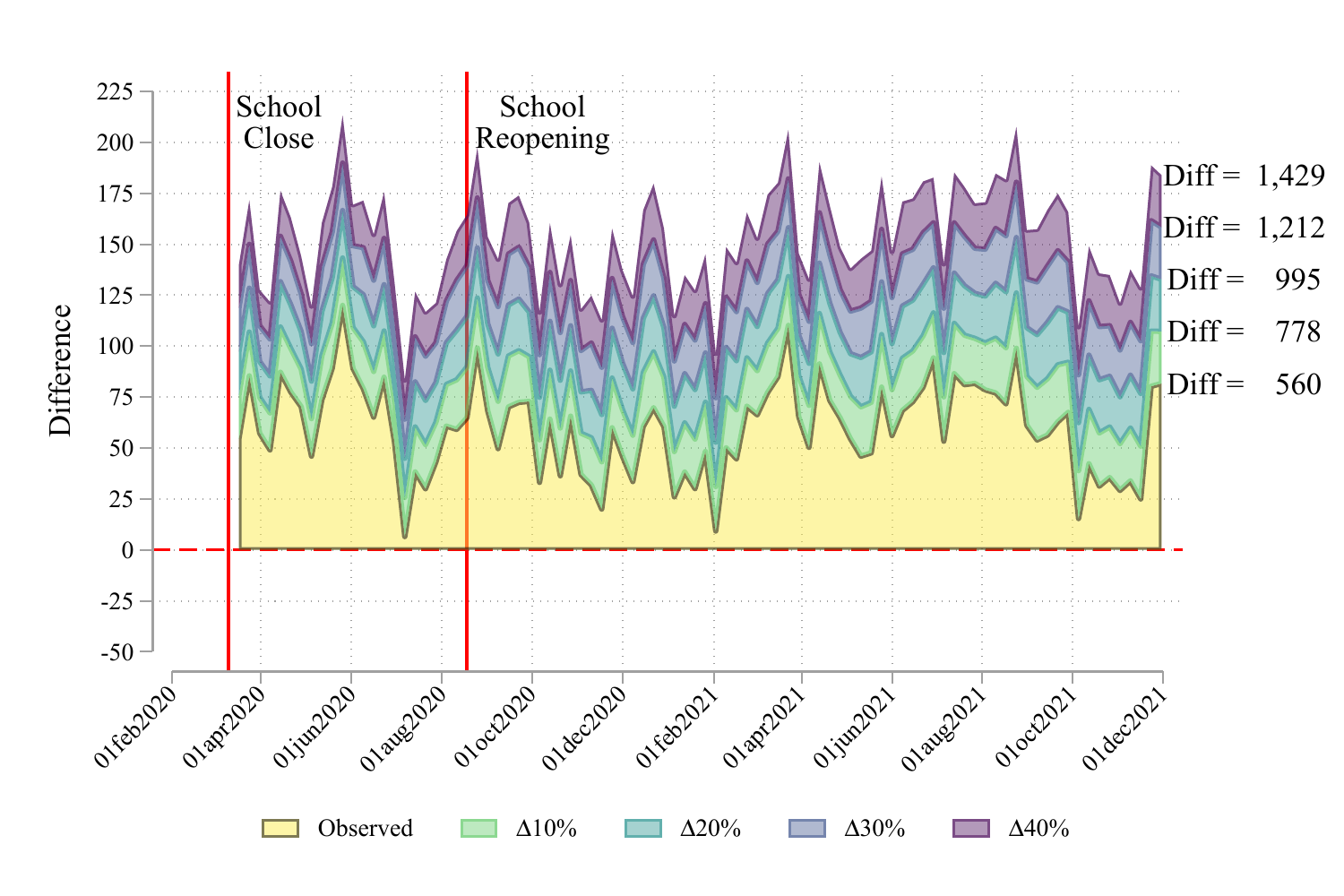}%
}
\end{center}
\floatfoot{\textbf{Notes to Fig.\ \ref{fig:counterfactuals}}: Subfigures (a)-(c) document actual reporting (grey line) and projected counterfactuals (blue line), with 95\% bootstrap confidence intervals for (a) DV, (b) sexual abuse, and (c) rape against children, following equation \ref{eqn:counterfactual}.  Counterfactuals are estimated using optimal temporal trends (linear in subfigure (a), and quadratic in (b) and (c), and pre-pandemic prediction periods, with root mean squared prediction errors displayed in the bottom left corner of each panel.  Subfigures (d)-(f) document identical counterfactual procedures, but now `switching off' the school closure and reopening channel, following equation \ref{eqn:counterfactual2}.  Aggregate differences between real and projected reporting for the closure and reopening period along with bootstrapped 95\% CIs are displayed in green squares, and week by week reporting differentials are displayed in subfigures (g)-(i), along with sensitivity testing following SI equation \ref{SIeqn:Diff1}, in which rates of projected violence are allowed to increase in the post-pandemic period.}
\end{figure}

To understand the impact of school closure, as well as the dynamics of recovery, we conduct counterfactual projections, presented in Figure \ref{fig:counterfactuals}.  In panel A we document simple projections: how would complaints of violence against children perform if simply projecting optimally-chosen cyclical (week of year) and temporal trends forward estimated off the pre-pandemic period.  In each case, we observe that such projections perform well in predicting in-sample (2019) and out-of-sample (2020), up until the week of school closure.  We then observe sharp declines when comparing actual complaints (thin grey line), to counterfactual predictions (thick blue line).  We observe that over time, these lines nearly converge, with actual reporting nearly reaching counterfactual projections, though this convergence is slow, only approaching predicted levels by around the fourth quarter of 2021, over a year after the first schools were re-opened.  Indeed, when mapping estimated differences between real and projected complaints, we estimate that in the school closure period 1,533 (95\% CI: 1,002--2,083) cases of DV against children were not reported, 1,223 (95\% CI: 941--1,509) cases of sexual abuse against children were not reported, and 155 (95\% CI: 70--246) cases of rape against children were not reported.  Somewhat similar values are observed in the post-reopening period, but these are estimated over a much longer period.  In weekly terms, we estimate that reductions are generally much larger during school closure than school reopening, at 64 versus 36 per week in the case of DV, 51 versus 33 in the case of sexual abuse, and 7 versus 6 in the case of rape.

Counterfactual projections in Panel A are simply based on optimally chosen historical cyclical and temporal trends, however we extend these in a number of ways.  In Panel B, we consider alternative projections, however now `turning off' the school reporting channel.  This is, we consider the counterfactual outcomes based on the projections from panel A, but also concentrating out effects of school closure and re-opening (refer to Methods).  If school closure accounts for the full difference in reporting, we would expect that these counterfactual and observed trends would now entirely overlap. While we observe large movements, we do not observe that school closure can explain away \emph{the entirety} of under-reported cases in these projections. Comparing panel (a) to panel (d), we observe that school closure can explain away 934 of the 1,533 estimated `missing' DV reports during the school closure period, and that partial school closure can account for 1,848 of the 2,501 `missing' DV reports during the school opening period.  In the case of sexual abuse and rape, we observe similar patterns, with school closure explaining between 41\% (rape) to 57\% (sexual abuse) of the drop in reporting observed over the entire period.  In SI Figure \ref{SIfig:counterfactualsV4} we document that these results are broadly similar if additionally controlling for COVID case rates, testing and positivity rates, as well as municipality lockdown status, suggesting that school closure plays a substantial and substantive role in crime reporting, which was missing upon closures.

These counterfactual trends are optimally estimated as laid out in SI, section \ref{scn:methodsSI}, and we document robustness to alternative counterfactual modelling processes in SI Figures \ref{SIfig:counterfactuals2015}-\ref{SIfig:counterfactualsQuadratic}.  Total estimated under-reporting across a number of scenarios is documented in SI Figures \ref{SIfig:Under-reporting}-\ref{SIfig:Under-reportingNoSchool}, and shows projections are stable across a range of models, particularly so in the case of DV reporting, which concentrates the largest number of `missing' criminal complaints.

Finally, in panel C of Figure \ref{fig:counterfactuals} we consider alternative projections where rather than assuming that historical and cyclical trends predict counterfactual (no COVID-19 related school closures) outcomes, we assume that violence may actually have increased. As discussed above and in wider literature, policy responses to the pandemic may conceivably result in increases in the risk of violence against children \citep{Pereda2020}. Children experience higher risk of violence at home, and the most likely perpetrator are parents and other family members \citep{valenzuela2022,fitzpatrick2020}.\footnote{Our data also confirm this fact: over time, the most likely place where sexual abuse and rape against children occur is at home (SI Figure \ref{SIfig:partestrends}). And domestic violence happens, by definition, inside the family environment, with over 80\% of the cases taking place at home.} With stay-at-home orders in place and schools closed, families spent more time together, increasing the opportunities for violence, and decreasing the interactions with non-family members and other sources of social support. Additionally, child maltreatment is more likely to occur under situations of economic strain, financial hardship, mental stress, and family conflict \citep{bullinger2022evaluating,Lindoetal2018,rodriguez2021}. As displayed in Figure \ref{SIfig:othertrends}, the COVID pandemic increased exposure to all these factors, making an increase in violence against children a most likely hypothesis. Therefore, the models presented in Panels A and B could be read as the lower bound of reports missing due to school closure.

We do not know, however, the magnitude of the increase. A survey conducted with US parents in the first weeks of the pandemic documents that 20\% of the parents report hitting or spanking their child in the past 2 weeks, with 5\% reporting doing so more often than usual, while 1 in 4 recognizing an increase in conflict during that time \citep{lee2021}. Similarly estimates accounting for factors associated with violence point to increases of up to 29\% in referrals of violence against children in a specific US-setting \citep{Prettyman2021}, while an increase of around 10\% in the use of women's DV shelters in Chile was observed as a consequence of lockdowns \citep{Bhalotraetal2021}, pointing to an increases in violence within the household based on objective measures.  In panel C, rather than projecting counterfactual and actual outcomes, we document the difference between counterfactuals and actual reporting under alternative assumptions of increases in underlying violence by \{10, 20, 30, 40\}\%. This can be considered a bounding exercise, given that we do not know by how much true violence rates may have increased. If true rates of violence had actually increased by 10\% above trend, rather than 4,034 unreported cases of DV against children in aggregate, this would rise to 5,517, with broadly similar proportional changes in the case of sexual abuse (from 3,524 to 4,553) and rape (from 560 to 778).  Sensitivity of these estimates are displayed in SI Figures \ref{SIfig:counterfactuals2015}-\ref{SIfig:counterfactualsQuadratic}, and additional discussion of sensitivity is provided in SI, section \ref{SIscn:results}.

\section{Discussion}
The importance of schools, and the impact of their closure during the COVID-19 pandemic has been noted across a range of outcomes including learning loss \citep{Engzelletal2021,Angristetal2021}, child mental health \citep{Vineretal2022} and inequality \citep{Agostinelli2022,VANLANCKER2020e243}.  However, the results documented here make clear (a) that schools play a substantive role as a safety net in cases of violence against children, and (b) that recovering this channel has required substantial time.

An important contextual detail of this study setting is that in Chile there was significant variation in the implementation of lockdowns, infection rates, and school re-openings.  Thus, we observe municipalities with very different epidemiological and public health profiles at the moment of school closures. The fact that sharp declines in reporting are observed in all settings suggests that it is unlikely that these owe to other (non-school) channels. In general, the substantial temporal and geographic variation of school opening allows us to examine the plausibility of identifying causal effects in this observational setting (SI, section \ref{scn:methodsSI}).  Across all outcomes considered, we observe substantial changes precisely at the moment of closures and re-opening, rather than prior to policy shifts (SI Figures \ref{SIfig:eventCloseDV}-\ref{SIfig:eventCloseRa}), suggesting that it is indeed changes in the availability of in person contact between students, teachers and other education professionals which drives large changes in reporting of violence against children.

Limited in-person interaction may also explain the persistent effect of school closure. While schools were mandated to close nation-wide in March, 2020, the decision to open and how to resume in-person activities was a school-level decision (SI section \ref{SIscn:context}). While only 10\% of schools had some in-person activity by December 2020, the opening process evolved at an increasing--but still gradual--pace, with 31\% of schools opened by the end of the fall semester of 2021 and 98\% by the end of that academic year (Figure \ref{fig:trends}). 
Beyond the school status, attendance remained voluntary up to March 2022, with only between 35\% and 55\% of students attending school each day by the end of the academic year of 2021 \citep{claro2021}. Low attendance rates limit the ability of teachers to detect signs of abuse, and could explain ongoing declines in reporting. These results are in line with previous evidence that show that additional time spent in school leads to an increase in reporting of child maltreatment \citep{fitzpatrick2020}.

While school opening results in increased rates of contagion in schools \citep{TupperColijn2021}, and schools are still being closed as a prophylactic measure in many countries \citep{Haleettal2020}, contagion in educational systems can be reduced if taking adequate avoidance measures \citep{MACARTNEY2020807,Tupperetal2020}. In contrast, the results of this study suggest that continued use of school closures imposes potentially significant costs on child well-being, even if \emph{only} considering reduced rates of violence reporting, and that those costs remain over time, suggesting that such factors should be accounted for when weighing up the costs of school closure decisions.

These results have several implications. First, teachers and professionals could be trained and more staff can be hired in order to better identify violence against children, even after time has passed, as schools with more and better trained personnel have higher chances of identifying maltreatment \citep{Baronetal2020, fitzpatrick2020}. In the post-COVID era, this decision could promote recovery in schools' capacities to observe and channel victims of violence to legal and child protection systems. Second, the results confirm the relevance of in-person interactions for detecting cases of violence against children. Policies that encourage school attendance or generate alerts in cases of frequent absenteeism, may have an impact not only on the chances of school dropout--with all its negative consequences \citep{Mussidaetal2019}--but would also result in a higher likelihood of identifying maltreatment

Finally, developing and implementing alternative types of reporting channels for children who experience victimization may help in times when the `school reporting channel' is not available. For example, the use of text messages or other private messaging services may provide children with an accessible and safer way of seeking help, while additionally being robust to weakened ties between children and schooling systems \citep{ortiz2021}. 

All told, these results confirm that schools act as a social safety net for children, detecting and formalising complaints for violence which otherwise may be left undetected. They do so in so far as schools provide opportunities for in-person interactions with teachers and school personnel who are able to identify signals of maltreatment and report such cases to the relevant authorities. Thus, their role in protecting children is likely substantially interrupted as schools remain closed, or attendance remains low. And, while our results suggest the this school reporting channel could be recovered after periods of closure--due to holidays, weather, or future pandemics, this recovery occurs slowly, and certainly does not appear to suggest a spike in reporting upon re-opening which would be consistent with `missing' cases being channeled into the criminal system with a lag.

\renewcommand{\footnotesize}{\footnotesize}
\clearpage
\end{spacing}
\begin{spacing}{1}
\bibliographystyle{chicago}
\bibliography{pnas-sample}
\end{spacing}

\clearpage
\newgeometry{margin=0.5in}
\begin{spacing}{1.3}

\newpage

\appendix
\setcounter{page}{1}
\renewcommand{\thepage}{A\arabic{page}}

\begin{center}
{\Large Supplementary Information -- Not for Print}\\
\textbf{Schools as a Safety-net: The Impact of School Closures and Reopenings on Rates of Reporting of Violence Against Children} \\
Damian Clarke, Pilar Larroulet, Daniel Paila\~nir \& Daniela Quintana
\end{center}

\end{spacing}


\section{Additional Context}

\appendixpagenumbering

\label{SIscn:context}
\subsection{COVID Responses, Formal Lockdowns and Case Measurement}
The first COVID case was identified in Chile on March 3, 2020. While initially arriving to Chile from Europe and Asia, COVID cases expanded quickly with local transmission, with around 7000-8000 cases per day in mid-June, 2020 at the peak of the first-wave of the pandemic in Chile (see Figure 1, panel (f)). 
Since the arrival of the first cases, a number of social distancing responses were adopted by the Chilean government, such as an early ban of any large gathering, the closure of schools and universities, the mandatory use of face-masks in public, and the implementation of formal lock-downs \citep{tariq2021transmission}.

The first mandatory lockdown was put in place on March 28, 2020 \citep{tariq2021transmission}.  The particularity of Chile is that lock-downs were defined at the national level by the national Ministry of Health (MoH), though implemented at the municipal level (Figure 1, panel (e)). Chile is divided into 346 municipalities, which in urban settings are smaller than a city, though in rural settings can cover various towns. Therefore, two neighboring municipalities may have had different lockdown statuses based on the MoH assessment of the need for lockdown. The assessment was mostly based on the case growth and the risk of contagion, although there was no declared metric, making the exact time of lockdown hard to predict for a specific municipality \citep{Bhalotraetal2021,lee2021}.\footnote{A video on the dynamic nature of lockdown imposition in the Metropolitan Region of Santiago is available here: \href{https://www.damianclarke.net/resources/quarantines.gif}{https://www.damianclarke.net/resources/quarantines.gif}} Moreover, formal lockdowns were strictly enforced, with police personnel conducting spot checks and citizens--with the exception of essential workers--allowed to leave their house only twice a week for three hours with a specific permit. In fact, the implementation of lockdowns led to a sharp drop in mobility (about 35\%) \citep{Bhalotraetal2021}, beyond the decline already observed after schools were closed \citep{bennett2021}. Similarly, the decision to exit lockdown was determined by the MoH. 

The vast majority of cases of COVID-19 in Chile are detected by PCR tests, and are recorded in a public reporting system known as EPIVIGILA. Mandatory reporting of all PCR tests conducted in cases of suspected contagion with the SARS-CoV-2 virus (hereafter COVID-19), as well as automated reporting of positive tests (in which the individual's municipality of residence is registered) ensures that we have a highly local measure of testing, test positivity, and diagnosed COVID-19 cases.  This reporting covers all public and private health care providers, and the EPIVIGILA system was in place before the COVID-19 pandemic, meaning that information on case measurement is available from the earliest days of the pandemic.  Healthcare providers are obliged by the national Sanitary Code to report all cases of certain mandated transmissible diseases \citep{MinSal2022}, though individual identities and exact addresses are only known by the central government given laws which protect private information of individuals.  As such, all epidemiological data used in this paper is based on municipal by week aggregates, as discussed in section \ref{SIscn:data} below. 

Additional discussion of Chile's pandemic response is provided in broader literature, see for example \cite{Menaetal2021,GilUndurraga2020,tariq2021transmission,Bhalotraetal2021,Castilloetal2021,li2022}.

\subsection{Educational Responses}
\label{SIsscn:educResponse}
As one of the earliest governmental responses to COVID, all K-12 schools nationwide were mandated to close on March 16, 2020. Schools moved to online education. However, data collected by the Ministry of Education make clear the unequal access to education: only 31\% of the parents surveyed report their child had a personal electronic device with which to connect to classes and a similar percentage reported that they had a good internet connection.  

In July 2020, the government implemented a ``step by step'' strategy (in Spanish, the \emph{Paso a Paso} policy) that considered five phases of gradual opening of each municipality--from full lockdown to no restrictions \citep{tariq2021transmission}. Under this scheme, schools located in municipalities that were not under full lockdown (Phase 1) were allowed--though not mandated--to resume in-person education in August 2020. In order to do so, they needed to follow the protocols established by the Ministry of Education. For example, students needed to wear masks, and classrooms should guarantee physical distancing between students, along with adequate ventilation. In municipalities which were not classified as Phase 1 (and hence not under complete lockdown), thus, the decision to open and how to resume in-person activities was a school-level decision. For example, some schools adopted a gradual return, establishing shifts with in-person vs.\ remote schooling, or opening only some days during the week. Attendance remained voluntary up to March 2021.   By December 2020, only 10\% of schools had had some in-person activity, although this was mostly part time and with low attendance \citep{claro2021}. In March 2021, with the start of a new school year, 32\% of the schools resumed some in-person activities, with all schools being required by the Ministry of Education to develop a plan for a safe but ideally in-person academic year. Early vaccination of teachers and educational personnel was prioritized as part of the plan. However, as the cases increased and full lockdowns were put back in place (see Figure 1, panel (e)), most schools remained or moved back into remote education, with a steady increase in in-person activities starting with the spring semester, which started in July, 2021. That same month, the Ministry of Health allowed schools based in municipalities under Phase 1 to re-open, given the start of the vaccination process among school children \citep{jara2021}. By the end of the academic year of 2021, 98\% of schools had some in person activity in place \citep{claro2021}. Most of the schools, however, opened either in shorter school days or alternating days/weeks among the students in order to fulfill the requirements of social distancing. 

Low-income schools were less likely to open and opened, on average, for less days, but the school's administration type\footnote{The Chilean school system comprises private schools, public schools (administered by municipalities until 2017, where some public schools have been transferred to a new governmental organization \citep{canales2022}), and private state-subsidized schools \citep{bellei2021}. The later represents over 50\% of the total number of students, while another 35\% are enrolled in public schools.} was the most significant predictor of opening earlier during 2021 and explained all the observed socioeconomic differences in opening probabilities \citep{canales2022}. Particularly, schools administrated by the municipality were the least likely to open, regardless of the political affiliation of the mayor. And, while rates of re-opening among different school types converged by September 2021, large differences remained between private schools and other schools (public and state-subsidized private schools) in terms of percentage of days open and rates of attendance. For example, as reported by \cite{claro2021}, in November, 2021, while about 70\% of the students in private education attended at least one day a week in person, only 40\% of those studying in public and private subsidized schools did so. 

A key element of school opening decisions, at least for the analysis in this paper discussed in section \ref{scn:methodsSI} below, is that school opening decisions seem highly unlikely to be related to changes in rates of DV or sexual violence against children.  As we lay out further in section \ref{scn:methodsSI}, an identifying concern would occur if schools which opted to re-open were those which were in municipalities in which systematically different changes in rates of DV or sexual violence were occurring, for example if schools were more likely to open when rates of reporting were increasing, or were more likely to open when rates of reporting were decreasing.  In practice the precise moment of school re-opening for each school appeared to depend more on the school's abilities to meet the required criteria indicated by the Ministry of Education. 

\section{Underlying Data Sources}
\label{SIscn:data}
We compile administrative data from different governmental and public sources, all fine-grained at the temporal (day or week) and geographic level (at the individual, school or municipal level).  In principal analyses, all data are aggregated to the municipal$\times$week level.  We describe these various data sources below in the format originally accessed to generate municipal by week cells.

\subsection{Crime Reporting}
Reports of Violence Against Children come from police information that is reported to Chile's Ministry of the Interior. A single observation is provided for each victim, along with demographic characteristics and details of the crime, such as municipality of occurrence and the type of crime.  We requested information from the Ministry of the Interior on all victims of crimes reported to the police between January 2010 -- December 2021. Specifically, we requested full data for those victims that i) were under 18 years old at the time of the offense, and ii) had been victims of crimes classified as intra-familiar violence, sexual abuse, or rape. Regarding the former, the police distinguish between psychological violence, moderate physical violence, and serious physical violence. The data also provided information about the victim's age and sex, and more detailed classification of the offense. As the specific date of the crime is recorded, those dates were used to generate weekly 
rates of reports for each type of offense in each municipality.  Rates were generated for each class of violence (intra-family violence, sexual abuse, and rape), as well as in sub-groups by age and by sex.  Rates are consistently generated using populations by municipality which are available from Chile's National Statistics Agency (INE) for each age and sex.

These principal data contain one line for each victim of crime, and so in certain circumstances may include more than one victim for a single crime event.  In data on victimhood, administrative records do not contain information on the exact context of the crime (i.e., inside the house, or in a public place).  While in our main analysis we wish to determine rates of victimization, and as such work with data on all crime victims known to police, in supplementary figures we also work with another administrative database with a single line for each crime known to police (irrespective of the number of victims).  While this event-based database allows us to explore the precise location of the crime (specifically as occurring inside the house, or in a public place), it does not account for all the potential victims involved in one crime, nor provides precise demographic information about the victims (such as sex and exact age), and as such we simply use this in Supplementary Analyses.

\subsection{Educational Information}
From the Ministry of Education we obtain administrative data on records of dates of school closures and reopening for each of the 10,847 schools in the country in 2020 and 10,875 schools in the country in 2021.  This covers all schools (both public and private) excluding those which only provide adult education.  These data provide a weekly record of whether the school was officially open to receive students for in person instruction.  The data are available at the following link \href{www.https://datosabiertos.mineduc.cl/}{www.https://datosabiertos.mineduc.cl/} covering months of October 2020-December 2021, and are supplemented by two months of records provided in transparency requests from the Ministry of Education for the months of August and September, 2020, which are not available in public repositories.  The Ministry also provided a database with the number of children attending in-person education every month per school, recorded in four categories: 1) at least once during that month; 2) between one and five days; 3) between six and ten days; 4) More than ten days. 

While school-level reopening data is available over the entire period under study, school-level information on attendance was only available in a consistent format between July and December, 2021, and as such analysis based on school level attendance measures is included only as supplementary results. When working with attendance, we calculate proportional measures of attendance in each school, which can be generated from the attendance database, as well as an additional publicly available administrative record of the number of children registered in each school in each academic year.

\subsection{COVID-19}
Data of the dates of entry and exit of the confinement of each municipality was prepared by hand by the authors based on daily televised reports made by the Ministry of Health, and open repositories of the Ministry of Science of Chile. The data recorded the exact date when mandatory lockdown started and the date when it was formally lifted. These data are available from March 2020 to December 2021, and are available at \href{https://github.com/Daniel-Pailanir/Cuarentenas}{https://github.com/Daniel-Pailanir/Cuarentenas} and \href{https://github.com/MinCiencia/Datos-COVID19/tree/master/output/producto74}{https://github.com/MinCiencia/Datos-COVID19/tree/master/output/producto74}.  Official records from the Ministry of Science were maintained from approximately August 2020.  Earlier records were announced in highly viewed public announcements made by the Minister of Health or Under-Secretary of Health.  Thus, official records available from the Ministry of Science are complemented by our hand-collected records of every lockdown status in all periods from March-August 2020.
Data on COVID-19 infection, all PCR tests, PCR test positivity and COVID-related deaths come from open repositories from the Chilean Ministry of Science. The data provide information for each of the 346 municipalities, and are generally recorded at a frequency of at least weekly, generally with multiple records in each municipality and week.  These are consistently measured based on the EPIVIGILA system, and in principal analyses we aggregate these to municipal by week records of test positivity (proportion of PCR tests with a positive result), number of PCR tests per total population, and number of COVID-19 cases per total population.

\subsection{One Municipal's Complete Records}
We collaborated with the Children's Rights Protection Office (OPD for its initials in Spanish) of a large municipality in the capital city of Santiago.\footnote{In the interests of privacy, we cannot reveal the name of this municipality.  For this reason, this data is used only as Supplementary Information.} OPDs are the institutions in charge of the prevention and early detection of children's rights violations, and either refer cases or provide interventions when a violation has been identified \citep{StutzinVallejos2018}.

This local OPD collects data on all cases which they cover, including information about the type of right that has been violated, the gender and age of the victim, and--importantly for the analysis here--the institution that reported the case to the OPD. They shared this information with us subject to a privacy agreement, and we harmonized these data for the period of 2019-2021, for use in a number of descriptive supplementary analyses to examine in a specific case how institutional channels changed surrounding the time of school closures in this particular context.  

\section{Supplementary Methods}
\label{scn:methodsSI}
\subsection{Two-way Fixed Effect Models}
We estimate the impact of school closure and school reopening in a two-way fixed effect model.  We consider school closure and school reopening which vary by municipality and by time, and separately include fixed effects for all time-specific factors common in the entire country (eg regular fluctuations in reporting within the year), and all municipal-specific time invariant factors, or factors which evolve slowly enough that they are unlikely to vary under our study period of 2019-2021, for example demographic factors which may affect rates of violence.  To quantify effects of school closures and reopenings, we begin by estimating the following two-way fixed effect model based on a balanced panel at the municipality$\times$week level:
\begin{equation}
\label{SIeqn:2wayFE}
Reporting_{mt} = \alpha + \beta School\ Closure_{mt} + \gamma Schools\ Reopen_{mt} + WoY_t + Municipal_m +  \bm{X}^\prime_{mt}\bm{\Gamma}+\varepsilon_{mt}.
\end{equation}
Here, $Reporting_{mt}$ refers to crime reporting of violence against children (individuals aged under 18 years), and is consistently expressed as reporting per 100,000 minors.  $School\ Closure_{mt}$ and $Schools\ Reopen_{mt}$ refer to the status of schools in municipality $m$ and week $t$, which are consistently coded relative to the pre-school closure period.  That is, $School\ Closure_{mt}$ moves from 0 to 1 sharply at the week schools are closed in the country (week of March 16, 2020), and then remains at 1 until schools reopen in each municipality $m$, at which point it moves back to 0. And the measure $Schools\ Reopen_{mt}$ is set at zero for the entire pre-closure, and closure period, and switches to 1 (or in alternative specifications, to the proportion of students with schools reopen) precisely when schools reopen in the municipality.  Thus, given that in the post-school closure period, either $School\ Closure_{mt}=1$, or $Schools\ Reopen_{mt}=1$, but never both, coefficients $\beta$ and $\gamma$, are both interpreted as changes in rates of reporting compared to the \emph{pre-school closure period}.  Below we discuss the formal interpretation of these coefficients, and identifying assumptions.  

In equation \ref{SIeqn:2wayFE}, $School\ Closure_{mt}$ switches sharply from 0 to 1 in a specific week $t$.  We thus include 52 week of year fixed effects, as
$WoY_t$, these are separately identified, and allow us to capture all common factors associated with particular weeks of years in all years under study.  346 municipality-specific fixed effects are included as $Municipal_m$.  The vector of time-varying controls $\bm{X}_{mt}$ is discussed below, and standard errors are estimated such that the unobserved stochastic error term can be arbitrarily correlated within each municipality over time, allowing, for example, for temporal dependence in municipal-level shocks \citep{CGM2008}.

The estimands of interest on $School\ Closure_{mt}$ and $Schools\ Reopen_{mt}$ are interpreted as the observed changes in rates of reporting, holding constant week of year and municipal fixed effects.  This is, then, the mean change in outcomes with school closure, when comparing with rates in the same municipality and week of year in years in which the school indicator is equal to 0 (pre-COVID periods).  Formally, in the case of school closure, the estimate simply captures: 
\[
\widehat\beta = E[Reporting_{mt}|School\ Closure_{mt}=1,\bm{W}_{mt}]-E[Reporting_{mt}|School\ Closure_{mt}=0,\bm{W}_{mt}]
\]
where conditioning variables $\bm{W}_{mt}$ include aforementioned fixed effects as well as, potentially, time-varying controls $\bm{X}_{ct}$ discussed below.  Similarly, in the case of school reopening, when considering binary measures of reopening, $\widehat\gamma$ captures estimated mean differences conditional on week of year and municipal FEs, and any time-varying controls:
\[
\widehat\gamma = E[Reporting_{mt}|Schools\ Reopen_{mt}=1,\bm{W}_{mt}]-E[Reporting_{mt}|Schools\ Reopen_{mt}=0,\bm{W}_{mt}].
\]
In the case of continuous models in School reopening, $\gamma$ refers to marginal changes in rates of reporting compared with rates in the same municipality and week of year, conditional on time-varying controls: 
\[  
\widehat\gamma = \frac{\partial E[Reporting_{mt}|\bm{W}_{mt}]}{\partial Schools\ Open_{mt}}.
\]

Causal identification requires that---conditional on factors specific to each week of the year, and each municipality---additional unobserved factors are not correlated with school closure and re-opening.  We note that this is equivalent to a `parallel trends' assumption, given the nature of the two-way FE model.  In the absence of school closing and reopening decisions, rates of violence reporting in each municipality would have followed parallel trends to rates occurring in the same moment in previous years. The key identifying concern is that estimated impacts of school closures may actually owe to epidemiological factors related to COVID-19, or other policy changes such as lockdowns.  When examining school closures, this concern is minimized given that closures occurred prior to sharp increases in rates of infection, and prior to the announcement of formal lockdowns, however this is likely more relevant in the case of municipal level reopenings.\footnote{We also consider a test which examines results stratifying by whether a municipality was ever under lockdown, further dismissing such concerns.} For this reason, time-varying controls $\bm{X}_{mt}$ are included in equation \ref{SIeqn:2wayFE}, which consist of rates of COVID infection, rates of COVID testing, and test positivity.  Similarly, we control for the existence of a formal lockdown in each period.  A key aspect of this study is that we observe considerable temporal and spatial variation in the degree to which municipalities were affected by COVID, were under lockdown, and had mobility changes.  We are interested in examining the common shock of school closure, and staggered re-opening which occurred throughout the country in municipalities with considerably different epidemiological and public health responses. In all cases, we document models with and without the inclusion of controls, as movement of coefficients upon the inclusion of controls allows us to consider the likelihood that unobservable factors may actually account for observed effects \citep{Altonjietal2005}.  

Regressions are consistently estimated by weighted ordinary least squares, where weights refer to the population of individuals under the age of 18 in the municipality.  This allows us to conduct inference assigning equal weight to individuals, rather than assigning equal weights to municipalities which have considerable variation in populations.

In the case of estimated impacts of school reopenings which occurred in a time-varying fashion, and where impacts of reopening are likely to be heterogeneous over time, estimates may fail to recover average treatment effects given that already treated units act as control units in periods in which their treatment status does not change \citep{GB2021,dCDH2020}.  In SI Additional Results we provide the decomposition proposed by Goodman-Bacon \citep{GB2021,GBetal2019}, allowing us to document the relatively low concern of this bias in this particular setting.  

\paragraph{Interactions with School-level Attendance}
In Supplementary results, we seek to test whether rates of reporting increase more upon school re-opening when rates of attendance are higher.  If in-person contact with educational staff is key for violence to be detected and reported, criminal complaints would be expected to rise as rates of attendance increase.  To test this, we estimate interactive two-way FE models, as laid out below.
\begin{eqnarray}
\label{SIeqn:Attendance}
Reporting_{mt} &=& \alpha + \beta School\ Closure_{mt} + \gamma Schools\ Reopen_{mt} + \delta (Schools\ Reopen_{mt})\times Attendance_{mt} \\ && + WoY_t + Municipal_m +  \bm{X}^\prime_{mt}\bm{\Gamma}+\upsilon_{mt} \nonumber.
\end{eqnarray}
All elements of this model follow those in equation \ref{SIeqn:2wayFE}, however here we additionally include an interaction between school reopening, and rates of attendance.  This model can only be estimated for a shorter time period in which attendance data is consistently available (refer to SI, section \ref{SIscn:data}), and as such, when estimating this model, it is consistently displayed alongside a version which omits the interaction term to consider baseline effects in this particular sub-sample.  All other estimating details follow those laid out above in equation \ref{SIeqn:2wayFE}.

\subsection{Event Study Methods}
As a supporting test of assumptions, and given the importance of considering dynamics in this setting \citep{Goodman-Bacon_Marcus_2020}, we provide as supporting results event study methods, which consider rates of reporting in the lead up to, and following changes in, school closure or school reopening policies.  Such methods allow for the consideration of whether observed changes in policies actually \emph{do} emerge only following the implementation of said policies, in which case leads to the event (pre-event estimates) act as placebo tests, and lags to the event (post-event estimates) allow for the consideration of the emergence of dynamics \citep{Autor2003,ClarkeTS2021}.   This consists of estimating:
\begin{equation}
\label{SIeqn:event}
Reporting_{mt} = \alpha + \sum_{j=2}^J \beta_j(\text{Lead}_j)_{mt} + \sum_{k=0}^K \gamma_k(\text{Lag}_k)_{mt} + WoY_t + Municipal_m +  \bm{X}^\prime_{mt}\bm{\xi}+\eta_{mt},
\end{equation}
where leads and lags are binary variables indicating that a given municipality was a given number of periods away from the event of interest in the respective time period. Specifically, in the case of school closure:
\begin{eqnarray}
(Lead_J)_{mt}&=&1[t\leq Schools Close_{mt}-J]\nonumber \\
(Lead_j)_{mt}&=&1[t = Schools Close_{mt}-j] \ \forall\ j\in\{2,\ldots,J-1\} \nonumber \\
(Lag_k)_{mt}&=&1[t = Schools Close_{mt}+k] \ \forall\ k\in\{0,\ldots,K-1\} \nonumber \\
(Lag_K)_{mt}&=&1[t\geq Schools Close_{mt}+K].\nonumber 
\end{eqnarray}
Event studies for school closure and school re-opening are estimated separately, where in the case of school re-opening, identical lags and leads are considered relative to re-opening, rather than closure.  All other details of this event study follow equation \ref{SIeqn:2wayFE}. In each case, event studies are estimated at the level of the week, and the omitted baseline category refers to one week prior to the adoption of school closure or re-opening.  When considering event studies for school closure, we consider up to $J=$60 leads (60 weeks prior to school closure) and $K=$20 lags, given that after 20 lags, school reopenings begin to occur, which potentially contaminate further lags.  Inversely, in the case of re-openings, we consider $J=$20 leads (20 weeks prior to re-opening), as this covers periods in which schools were entirely closed, and $K=$40 weeks post reopening, as the majority of municipalities are observed over the entirely of this time horizon.  Given that final leads and lags accumulate for all periods greater than this time, in the case of the 20\textsuperscript{th} lead in school re-opening event studies, this will include pre-school closure periods, and similarly, the 20\textsuperscript{th} lag in school closure event studies will include post-school reopening periods.  Once again, standard errors are clustered by municipality, and population weights are consistently used.

\subsection{Counterfactual Projections}
While we wish to consider aggregate changes in rates surrounding closure and re-opening, a key consideration is how trends in reporting would have evolved in the absence of school closures and reopenings.  Based on such counterfactual projections, we can consider the differences between actual and projected reporting rates, and additionally, what proportion of these differences can be explained by the school closure and reopening channels.  

Using data on reporting incidence per 100,000 individuals aged under 18 for each of the three outcomes discussed above, we estimate such counterfactual trends from observable (pre-COVID) data as: 
\begin{equation}
    \label{SIeqn:counterfactual}
    \widehat{Reporting}^{post}_{mt} = \widehat{\alpha}^{pre}+\widehat{WoY}^{pre}_t + \widehat{Municipal}^{pre}_m + \widehat{f(t)}^{pre},
\end{equation}
where projected reporting in the post school closure and re-opening period in municipality $m$ and week $t$, denoted $\widehat{Reporting}^{post}_{mt}$ is estimated by projecting pre-closure averages in each municipality and week of year, as well as flexible temporal trends $\widehat{f(t)}^{pre}$.  Estimated week of year fixed effects $\widehat{WoY}^{pre}_t$ allow us to capture cyclical (within year) variation, while flexible time trends allow us to capture secular changes in reporting over time.  We discuss the selection of prediction periods and modeling of secular trends below. A key factor of this counterfactual projection is that it estimates all coefficients and fixed effects entirely off pre-COVID data, allowing for the projection of such trends into the post-COVID period, abstracting from the actual effects of COVID and school closure on violence reporting. 

Based on real and projected trends, we calculate differences between real and `expected' reporting rates in the absence of COVID and school closures.  This is simply:
\begin{equation}
\label{SIeqn:Diff1}
\text{Difference}^{post}_{mt} = Reporting^{post}_{mt}-\widehat{Reporting}^{post}_{mt}
\end{equation}
Note that here, we can calculate a difference \emph{for each} municipality and week, thus allowing a fine-grained consideration of how school re-opening allows for recovery, or lack thereof, to reporting trends in the pre-COVID world.  The quantity $\text{Difference}^{post}_{mt}$ thus captures the reporting shortfall (or excedent if positive), measured in cases per 100,000 individuals aged under 18, as the actual rate of reporting observed, compared with the expected rate of reporting based on counterfactual projections. In the body of the paper, we consider total differences in reporting measured as \emph{absolute case} differentials at a national level in each time period $t$ by converting these per capita reporting differentials into total reporting differentials, and aggregating over the entire country.  This is: 
\begin{equation}
\label{SIeqn:Difference}
\text{Reporting Differential}_t=\sum_{m=1}^{346} \text{Difference}_{mt}^{post}\times \frac{Population_{mt}}{100,000}.
\end{equation}
As we calculate the Reporting Differential at each period $t$, we can observe in a dynamic way how this evolves, considering both pre- and post-school reopening periods.

Finally, while these projections allow us to consider reporting differentials between real and counterfactual reporting, they do not allow us to determine the contribution of school closures and openings to this observed differential.  Thus, to consider this differential, we re-estimate counterfactual reporting, however now controlling additionally for the channel of school closures and re-openings.  This is, we calculate a counterfactual reporting projection, based on secular and cyclical trends in the pre-COVID period, but also accounting for the decline in reporting owing to school closures.  This is calculated as:  
\begin{equation}
    \label{SIeqn:counterfactual2}
    \widehat{Reporting}^{post}_{mt}\Big|_{SO} = \widehat{\alpha}^{pre}+\widehat{WoY}^{pre}_t + \widehat{Municipal}^{pre}_m + \widehat{f(t)}^{pre} + \widehat\delta \text{School Opening}_{mt},
\end{equation}
where all details follow equation \ref{SIeqn:counterfactual}, but we additionally control for $\text{School Opening}_{mt}$ ($SO$), which takes the value of 1 while schools are fully open, 0 while schools are fully closed, and then the proportion of students whose school is re-open upon school re-opening periods.  We follow identical procedures in calculating reporting differentials from equation \ref{SIeqn:Difference}, however now conditional on School Opening channels, and document relative movements when calculating the unconditional counterfactual in \ref{SIeqn:counterfactual} and the conditional counterfactual in \ref{SIeqn:counterfactual2} to estimate the proportional contribution of school closures to reporting differentials.  In Supplementary Information, section \ref{SIscn:results} we additionally document robustness to the additional inclusion of time-varying epidemiological and lockdown controls discussed previously.

\paragraph{Model Selection}
Model selection following \ref{SIeqn:counterfactual} requires the specification of the secular time term $f(t)^{pre}$, and additionally the selection of periods over which $pre$ parameters should be estimated.  In the case of secular time trends, we consider three alternative cases.  The first is a case with no trend, equivalent to specifying $f(t)^{pre}=0$, thus simply using week of year fixed effects to capture temporal dynamics.  The second is a case with a linear trend in time equivalent to specifying $f(t)^{pre}=\alpha t^{pre}$, and projecting forward any pre-existing trends when calculating $\widehat{Reporting}^{post}_{mt}$.  And the third is a case with a quadratic trend in time, namely $f(t)^{pre}=\alpha_1 t^{pre} + \alpha_2(t^{pre})^2$, projecting   forward any pre-existing non-linear trends when calculating $\widehat{Reporting}^{post}_{mt}$.  

As we observe pre-COVID data over a long time horizon (refer to SI Figure \ref{SIfig:longtrendVIF}), we can estimate $\widehat{Reporting}^{post}_{mt}$ using a range of time windows. We consider a number of options, beginning in 2015, 2016, 2017 or 2018.  The benefit of using a shorter time horizon is that it may provide us a more adequate estimation of cyclical trends in working with week of year fixed effects in years closer to the time period of interest, while the benefit of working with longer timer horizon is that it may provide a more adequate estimation of secular trends in observing macro changes in reporting rates over a longer time horizon.  

While we consider projections using each time period as an estimation window, and each functional form to capture secular trends, we report as our main model that which provides the best fit over the final pre-COVID period of 2019, and January, February of 2020. This model is chosen (for each of the three outcomes studied) as the models which minimizes the following Root Mean Squared Prediction Error (RMSPE), where for ease of notation we write 2019, whereas in practice we additionally use the first two months of 2020:
\[
RMSPE = \sqrt{\left(Reporting^{2019}_{mt}-\widehat{Reporting}^{2019}_{mt}\right)^2}.
\]
In SI Additional Results, we provide full results documenting the sensitivity of reporting differentials to each of the potential alternative models, additionally reporting the RMSPE in each case.

\paragraph{Sensitivity Testing}
This counterfactual activity assumes that we can infer projections based on levels and trends in reporting prior to the arrival of COVID-19 and associated school closures.  However, a broad stream of literature \citep{Evansetal2020,Bullingeretal2020,Bhalotraetal2021,ErtenKeskin2022,Pereda2020} suggests that violence may have in fact increased, suggesting that counterfactuals may actually under-estimate the true expected reporting if rates of reporting had been maintained constant. We thus conduct additional sensitivity testing, where Reporting Differentials are calculated under less conservative assumptions, where we project that violence, and hence violence reporting, would have actually increased in the post COVID period.  This consists of increasing counterfactual reporting by fixed rates, for example by 10\%, as below: 
\begin{equation}
\label{SIeqn:Diff1}
\text{Difference}^{post,\Delta 10\%}_{mt} = Reporting^{post}_{mt}-\left(\widehat{Reporting}^{post}_{mt}\right)\times 1.10.
\end{equation}
We consider a wide range of sensitivity values ranging from 0 to as much as a 40\% increase.  We display Reporting Differentials following \ref{SIeqn:counterfactual} under these alternative sensitivity assumptions.

\paragraph{Inference}
Finally, in conducting inference on these projections, we must take account of the fact that counterfactual outcomes are estimated based on observed data, and hence are subject to sampling uncertainty inherent in this estimation procedure.  To conduct inference, we undertake a block bootstrap procedure, resampling over Chile's 346 municipalities to maintain the time-series dependence within each municipality.  This bootstrap inference procedure consists of the following:
\begin{enumerate}
\item Set $B$=250, and $b$=1
\item Generate a bootstrap resample of data, resampling over each of Chile's 346 municipalities. Use this sample to re-estimate:  
\begin{enumerate}
    \item $\widehat{Reporting}^{post,b^{*}}_{mt}$,
    \item $\text{Difference}^{post,b^{*}}_{mt}$, and
    \item $\text{Reporting Differential}_t^{b^{*}}$
\end{enumerate}
\item Save these bootstrap resampled parameters as element $b$ in a vector of $B$ elements.
\item Replace $b=b+1$
\item If $b=B$ end.  Else, return to step 2.
\end{enumerate}
With bootstrap resampled vectors in hand for $\widehat{Reporting}^{post,b^{*}}_{mt}$, we can calculate a 95\% confidence interval (CI) for $\widehat{Reporting}^{post}_{mt}$ as the centiles 2.5 and 97.5 of the corresponding bootstrap resampled vector $\widehat{Reporting}^{post,b^{*}}_{mt}$.  Identical 95\% CIs are calculated for $\text{Difference}^{post}_{mt}$ and $\text{Reporting Differential}_t$, and are reported along with point estimates.

\subsection{Computational Implementation}
Computational implementations are conducted in Stata MP, version 15.0.  Along with standard Stata routines, we use a number of contributed commands for econometric analyses.  These are \texttt{bacondecomp} command to estimate Goodman-Bacon's 2 way FE decomposition \citep{GB2021,GBetal2019}, \texttt{twowayfeweights} to consider the proportion of negative weights in 2 way FE models following \citep{dCDH2020,2wayStata2020} and the \texttt{eventdd} command to estimate event study models \citep{ClarkeTS2020}.

\section{Additional Results}
\label{SIscn:results}
\subsection{Descriptive Statistics}
\label{SIscn:sumstats}
Descriptive statistics of all dependent variables, independent variables and covariates are presented in SI Table \ref{SItab:sumstats}.  These are displayed over the full period of study, from January 1 2019--December 31 2021. Municipality by year cells document substantial variation in rates of DV against children and sexual abuse and rape against children, all of which are observed to have substantial standard deviations.  Of all violence types, reports of DV are highest, at 3.65 per 100,000 children per municipality by week, followed by values of 2.33 per 100,000 in the case of sexual abuse, and 0.5 per municipality per week in the case of rape.  Panels B and C document variation in measures of school closure, reopening and attendance, and the epidemiological situation of each municipality by week cell.

Trends in these variables over the period of study are documented in Figure 1 of the paper. Longer trends in each of these variables are documented in SI Figure \ref{SIfig:longtrendVIF}.  Panel (a) documents quite clear flat trends in rates of domestic violence against children from 2015 up until 2020, prior to large declines with school closures.  This is observed in each of the sub-classes of crimes classified as DV against children (psychological violence, minor injuries, or serious injuries).  In the case of sexual abuse and rape of minors, similar sharp declines are observed with school closures however, increasing secular trends are observed from 2015 onwards, especially in the case of sexual abuse, which increases from around 75 cases per 100,000 minors per week in 2015-2017, to around 90 cases per 100,000 minors per week in 2019 and 2020.  In the case of sexual assault and rape, spikes in reporting generally occur given inexact dates of reporting within each month.  Where the precise date is not recorded, the crime is recorded in microdata on the 1\textsuperscript{st} day of each month.  In Figure \ref{SIfig:unsmoothed}, both original data, as well as smoothed data used in the body of the paper are displayed.  In SI Additional results, we document that estimates are not sensitive to using original (unsmoothed) data.

Finally, the geographic variation in school reopening is displayed in SI Figure \ref{SIfig:Chile} (for all of Chile), and SI Figure \ref{SIfig:RM} for the Metropolitan region of Santiago.  Here we observe considerable variation of initial school re-opening dates.  This is also observed \emph{within} municipalities in the capital of Santiago, despite the fact that contagion rates were generally quite similar, particularly in the central municipalities.

\subsection{Event Study Models}
\label{SIsscn:eventstudies}
Event study models are displayed in Figures \ref{SIfig:eventCloseDV}-\ref{SIfig:eventOpenRa}.  These are displayed for each of the three main outcomes, and for each of school closure and then school opening.  Generally speaking, we observe flat pre-event leads prior to closure or opening events, and then changes which are sudden in the case of closures, or more gradual in the case of reopening.  This is consistent with a sharp `switching off' of the school reporting channel when schools close, and a more gradual recovery of the channel, consistent with lags in times between children returning to school, and in interaction with children and educational professionals.  In one case, that of rape and school closure (Fig.\ \ref{SIfig:eventCloseRa}), we note variation in trends in the run-up to school closure.  Rather than being consistent with violations of pre-trends, this variation corresponds to cyclical variation in rates of rape-reporting observed clearly in Fig.\ 1 of the main text (in line with lower rates of rape reports in winter and summer school vacation periods, even in the pre-school closure period).  We note, additionally, that even despite this variation, rates of rape reporting are observed to be at their lowest immediately following school closure.

\subsection{Two-way FE Diagnostics}
\label{SIsscn:twowayFEdiagnose}
In the case of estimated impacts of school re-opening, staggered adoption of re-opening implies that when estimating two-way FE models, units which have re-opened in the past and not changed their re-opening status in a given period will be viewed as equivalent to control units in regression models \citep{GB2021,dCDH2020}. In cases where treatment effects are heterogeneous, this can lead to two-way FE estimates being considerably different to the underlying average treatment effect of interest (no such concern exists for school closure given that adoption occurs at a single moment in time).  In Fig. \ref{SIfig:GB} and Tab.\ \ref{SItab:2WayWeights} we consider whether this is likely to be problematic for our global estimates documented in Tab.\ 1 of the main paper.  In Fig. \ref{SIfig:GB} we observe that this does not appear to be a significant issue.  We note that here, in general grey ``x" marks which capture estimated impacts in a $2\times2$ DD setting are reasonably closely clustered around the red line indicating the single-coefficient estimate, and, additionally,  these units---which compare already treated units with not yet treated units---are those which take the majority of weights in the aggregate estimate.  In Tab. \ref{SItab:2WayWeights}  we present summary values of weights and estimates for each of the two groups which form the aggregate estimate.  We observe that nearly 80\% of the estimate is generated from comparisons of interest between already treated and not-yet treated municipalities, while only around 20\% of the estimate is generated off later-treated to already treated comparisons.  In both the cases of DV against children and rape, effects are observed to be large and negative in the case of the prior estimate, and small or slightly positive in the latter estimate, consistent with the (negative) impact of school closure compared to the baseline period shrinking over time. In the case of sexual abuse, both estimates are observed to be negative, consistent with negative effects of school closures compared with pre-closure periods.  When considering the decomposition proposed by \cite{dCDH2020}, we find that no units are assigned a negative weight, which is where concerns may be most serious given that treatment effects may be mis-signed.  All told, these results suggest that in the case of school re-opening where adoption is staggered, the two-way FE estimates do not suffer from substantial problems flagged in \cite{dCDH2020,GB2021}.

\subsection{Attendance Interactions}
\label{SIsscn:attendanceModels}
Tables \ref{SItab:attendanceDV}-\ref{SItab:attendanceRape} present attendance interaction models described in equation \ref{SIeqn:Attendance}.  Here coefficients on School Reopening are interpreted as reporting declines when schools reopen, but when attendance is 0, while coefficients on School Reopening $\times$ Attendance document how reporting declines are moderated as attendance increases. Across all tables, we observe evidence consistent with declines in reporting in the post school reopening period (compared with pre-school closure), but these declines become less acute as attendance increases.  While these are always significant in baseline models, when including controls, although attendance interactions are consistently positive, these are at times not sufficiently precise to rule out null effects.  When examining gradients in attendance in table footers we see that attendance appears to be a relevant mediator, for example in Tab.\ \ref{SItab:attendanceSA} column 6, reports of sexual abuse against minors are observed to decline by 1.0 per 100,000 minors when schools close, compared to 0.59 per 100,000 when schools open but attendance is only at the 25\textsuperscript{th} percentile, 0.33 per 100,000 at the 50\textsuperscript{th} percentile, 0.14 per 100,000 at the 75\textsuperscript{th} percentile, and return to baseline rates (or 0.03 per 100,000 above baseline rates) at the 90\textsuperscript{th} percentile.

\subsection{Sensitivity to Alternative Counterfactual Projections}
\label{SIsscn:projectionSensitivity}
Projections in Fig 3 of the text consider counterfactual projections generated by optimally selecting the number of pre-COVID years on which to estimate temporal trends, as well as the polynomial degree with which to estimate long-term trends.  Figures \ref{SIfig:counterfactualsV4}-\ref{SIfig:counterfactuals2015} present a number of alternative counterfactual simulations where models are limited in certain ways, while Figures \ref{SIfig:Under-reporting}-\ref{SIfig:Under-reportingNoSchool} document estimated under-reporting across 12 different counterfactual models (using each of 4 potential pre-COVID baseline periods, and using no trends, linear, or quadratic trends).  In general, across Figures \ref{SIfig:counterfactualsV4}-\ref{SIfig:counterfactuals2015} we see that predictions in pre-school closure trends are often substantially better in the main figure of the paper rather than more restricted models, while Figures \ref{SIfig:Under-reporting}-\ref{SIfig:Under-reportingNoSchool} suggest that broadly speaking, under-reporting predictions are robust to alternative models.  This is particularly clear in the case of DV against children, where under each of the 12 models considered, point estimates and confidence intervals are similar.  The one case where there is slightly more variation is that of sexual abuse, given that sexual abuse appears to be following an upward trend prior to school closures, so if this trend is not included in counterfactual modeling, we observe relatively lower projected under-reporting differentials.  Our preferred models (included in the paper), do a considerably better job in explaining pre-school closure trends, though we note that if such a trend had not considered during the school closure and re-opening period, under-reporting would be lower, as laid out in Figures \ref{SIfig:counterfactualsNoTrend}, \ref{SIfig:Under-reporting} and \ref{SIfig:Under-reportingNoSchool}.

\subsection{Additional Robustness Tests}
\label{SIsscn:robustness}
In Table \ref{SItab:novacations} we observe that results are robust to removing months of January and February (school summer vacations) with final point estimates being substantively similar to those presented in Table 1 of main analysis.  Similarly, Table \ref{SItab:unadjusted} shows that if instead of smoothing measures of rape and sexual abuse which are over-reported on the first day of each month, we use the original un-smoothed data, all conclusions hold.

\clearpage

\section*{Appendix Figures and Tables}
\setcounter{figure}{0}
\renewcommand{\thefigure}{A\arabic{figure}}
\setcounter{table}{1}
\renewcommand{\thetable}{A\arabic{table}}

\begin{figure}[htpb!]
\begin{center}
\caption{Administrative Records of School Re-opening Across Chile}
\label{SIfig:Chile}
\includegraphics[width=\textwidth]{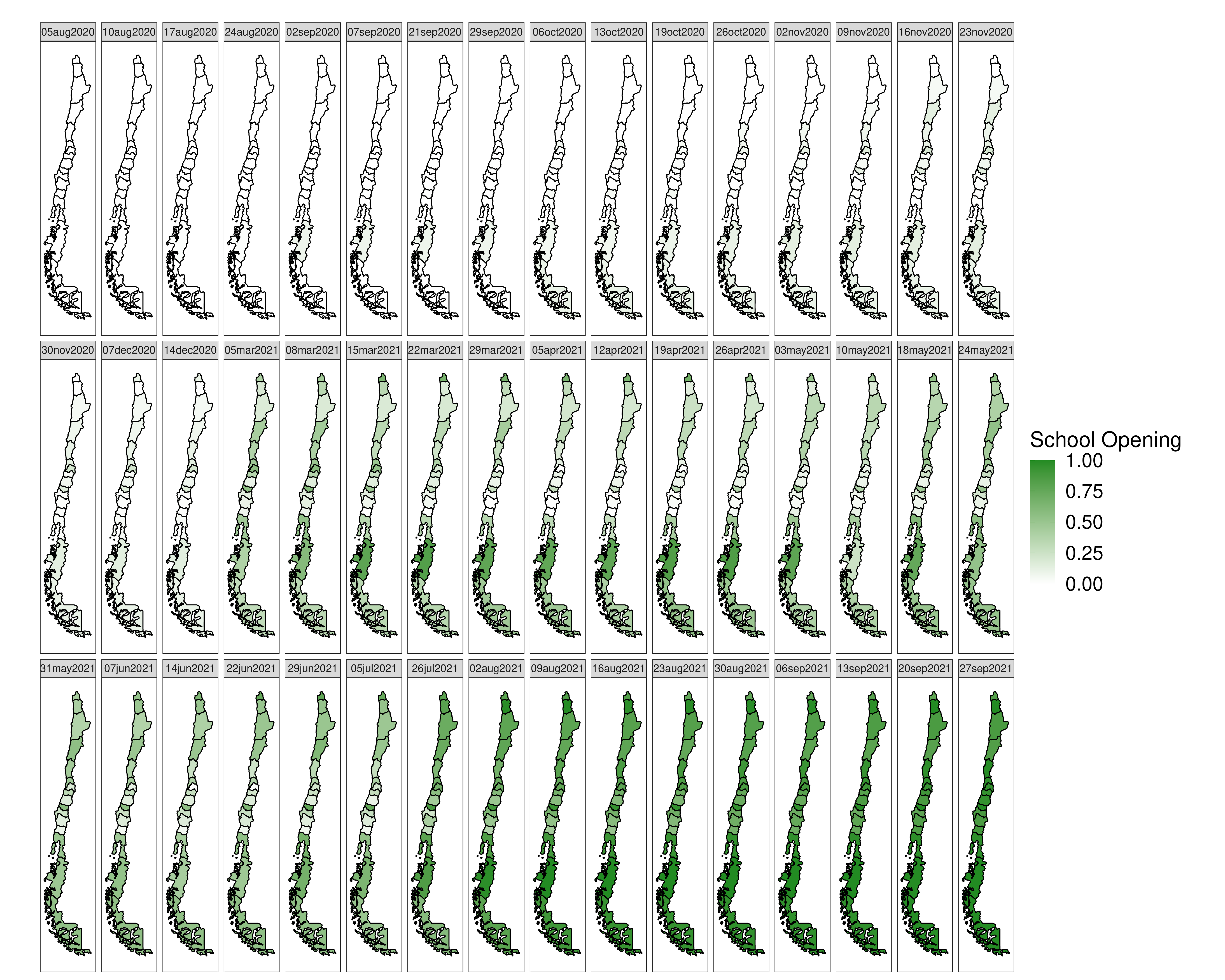}%
\end{center}
\floatfoot{\textbf{Notes to Fig.\ \ref{SIfig:Chile}}: School opening is displayed across the entire of country of Chile for weeks beginning 5 August 2020 (the week of first re-opening is 17 August 2020), up to 27 September 2021.  After this date, schools were nearly entirely reopened (see Figure 1).  Proportion of schools re-opened are displayed at the regional level for each of Chile's 16 regions, and refers to the proportion of all students whose school is reopened based on administrative records of student enrollment, and school reopenings.}
\end{figure}

\begin{figure}
\begin{center}
\caption{Administrative Records of School Re-opening Across Chile's Metropolitan Region}
\label{SIfig:RM}
\includegraphics[width=\textwidth]{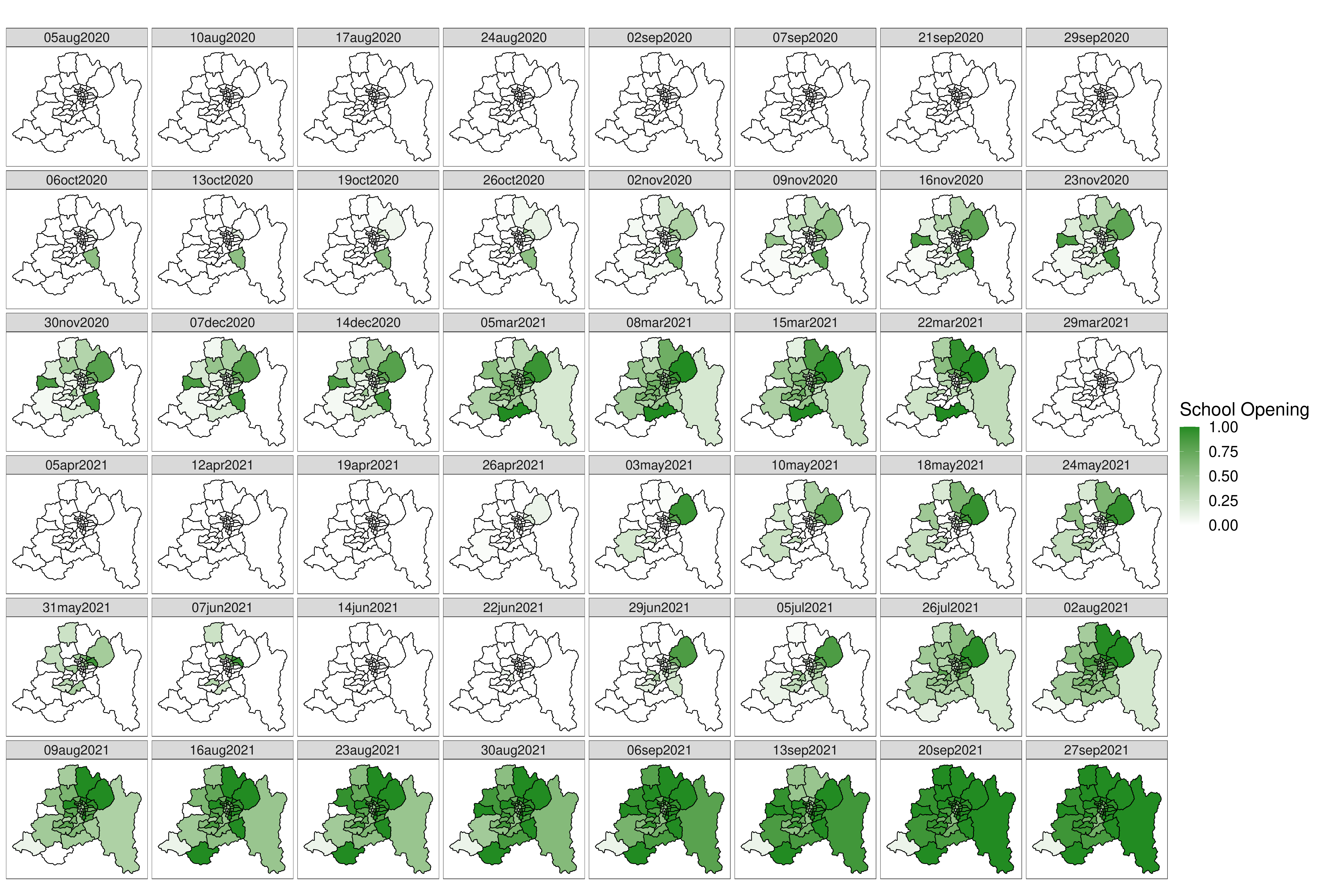}%
\end{center}
\floatfoot{\textbf{Notes to Fig.\ \ref{SIfig:RM}}: School opening is displayed across the Metropolitan Region of Santiago (the capital of Chile) for weeks beginning 5 August 2020 (the week of first re-opening is 17 August 2020), and ending 27 September 2021.  After this date, schools were nearly entirely reopened (see Figure 1).  Proportion of schools re-opened are displayed at the municipal level for each of the Metropolitan Region's 32 municipalities (of the total of 346 municipalities in the country), and refers to the proportion of all students whose school is reopened based on administrative records of student enrollment, and school reopenings.}
\end{figure}

\begin{figure}[tbhp]
\begin{center}
\caption{Extended Trends: Intra-family violence, Sexual Abuse and Rape Against Minors}
\label{SIfig:longtrendVIF}
\subfloat[Intra-family Violence]{%
\includegraphics[width=0.5\textwidth]{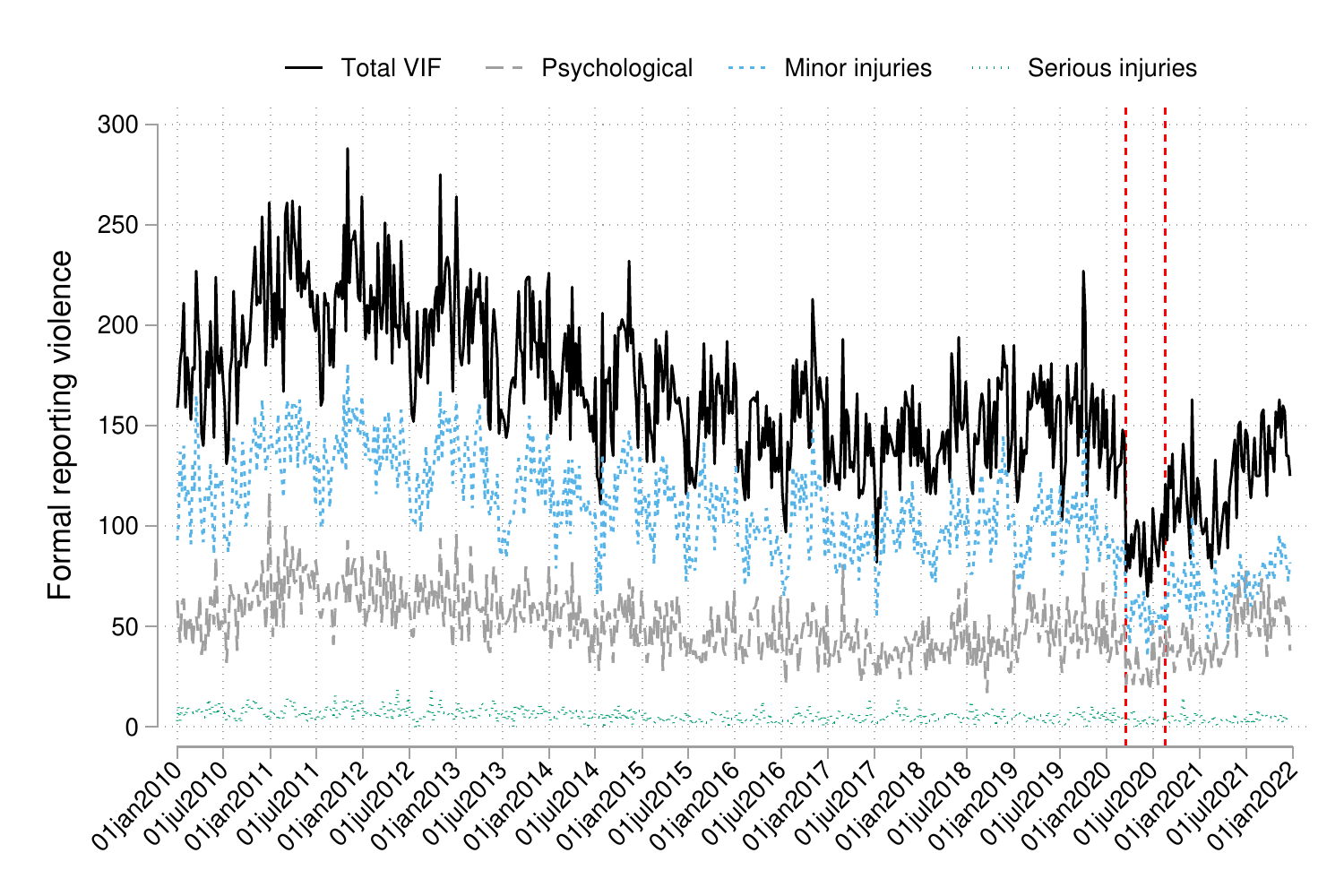}%
} \\
\subfloat[Sexual Abuse]{%
 \includegraphics[width=0.5\textwidth]{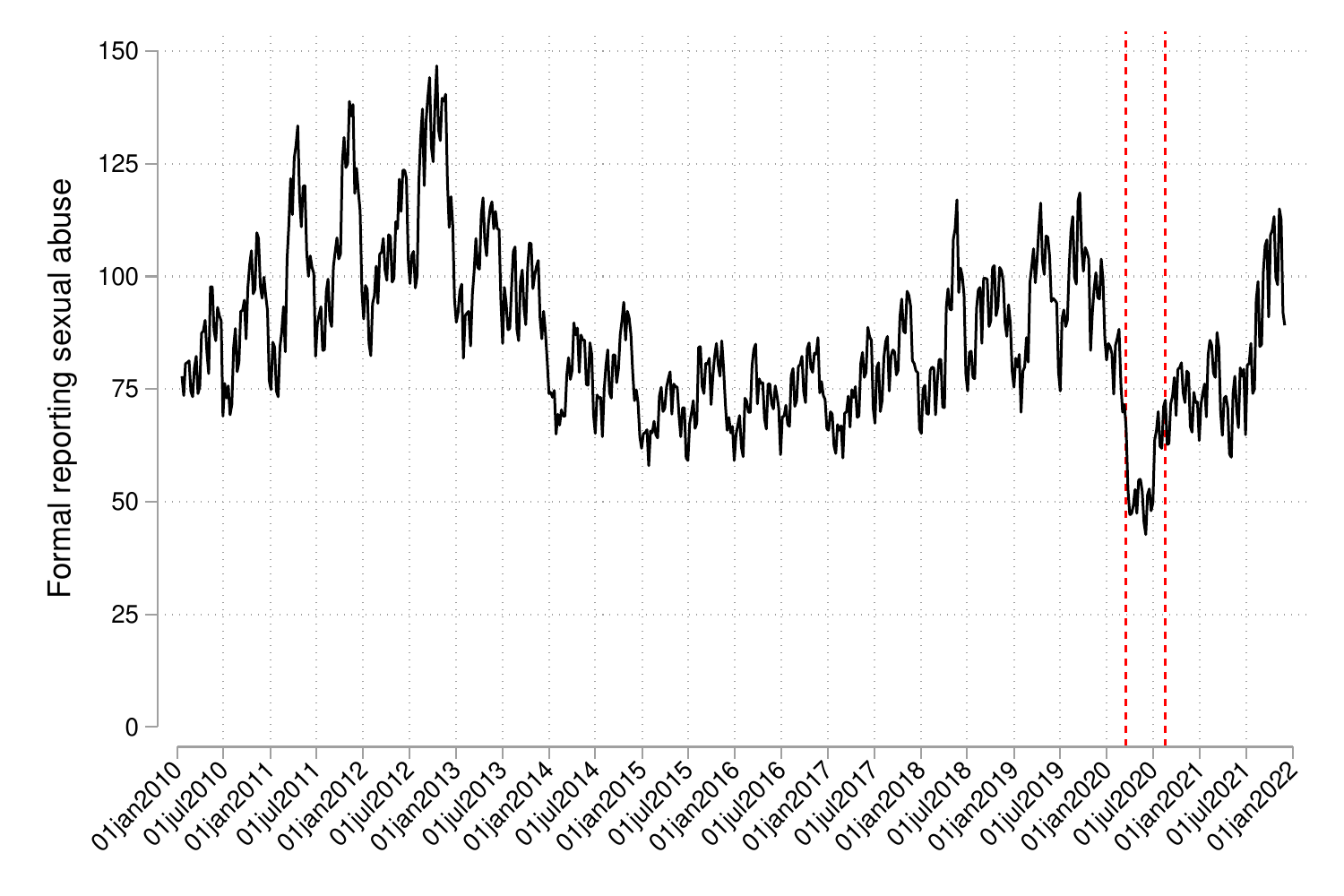}%
} \\
\subfloat[Rape]{%
\includegraphics[width=0.5\textwidth]{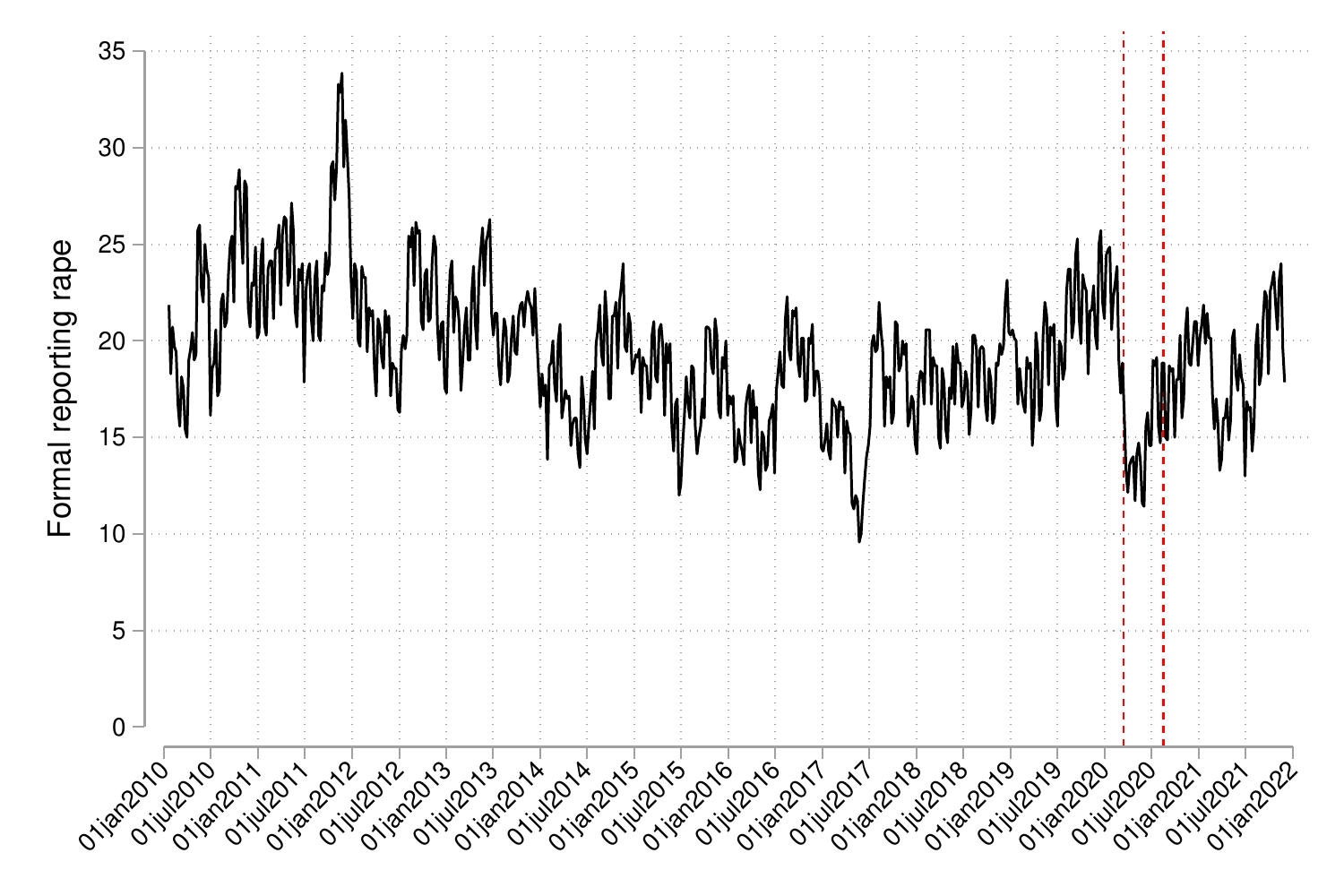}%
}
\\
\end{center}
\floatfoot{\textbf{Notes to Fig.\ \ref{SIfig:longtrendVIF}}: Trends show the weekly  number of formal complaints received by police related to DV (panel (a)), sexual abuse (panel (b)), and rape (panel (c)) against individuals aged under 18 years.  Here, longer trends in outcomes are documented, dating from January 1, 2010 to 31 December, 2021.  In main analysis, the period of January 1, 2019 to 31 December, 2021 is used.  Vertical red lines denote school closures and the date of first reopening.  Panel (a) additionally breaks down total DV (black solid line), into complaints classified as psychological violence, minor injuries, or serious injuries.}
\end{figure}

\begin{figure}[htpb!]
\begin{center}
\caption{Temporal Trends -- Sexual abuse and Rape against Minors, Smoothed and Unsmoothed Outcomes}
\label{SIfig:unsmoothed}
\subfloat[Reporting of sexual assault against minors]{%
\includegraphics[width=0.49\textwidth]{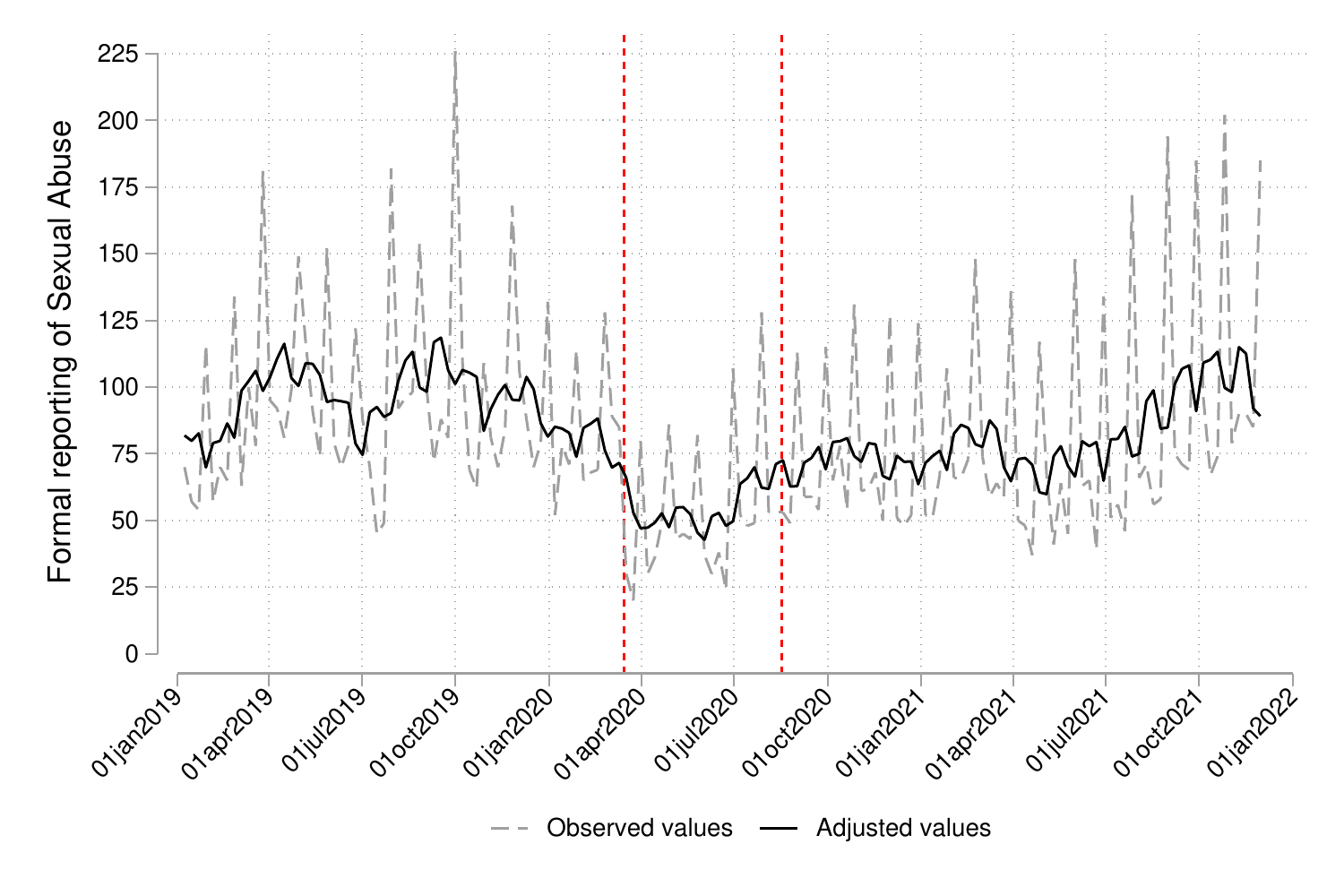}%
}
\subfloat[Reporting of rape against minors]{%
\includegraphics[width=0.49\textwidth]{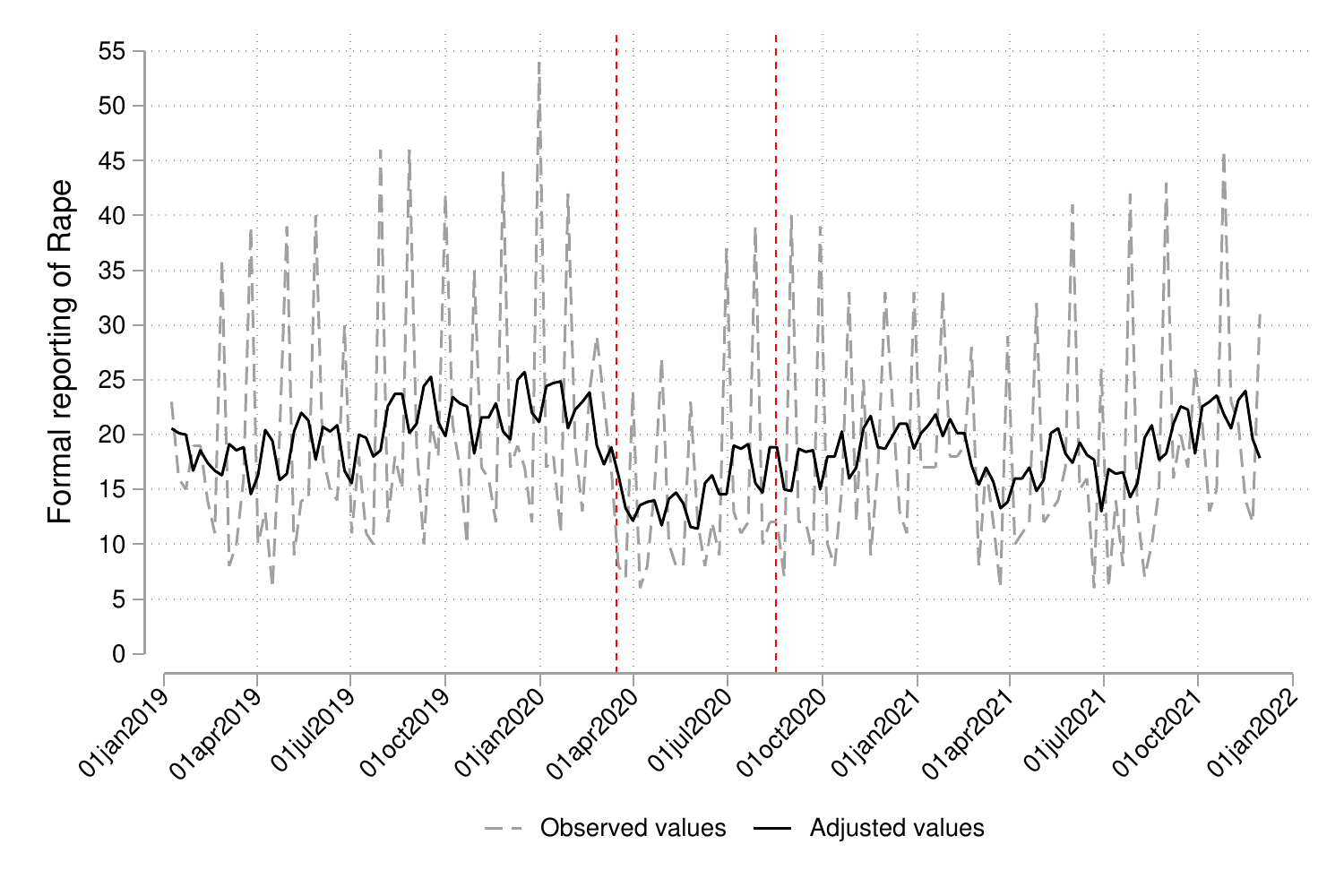}%
}
\end{center}
\floatfoot{\textbf{Notes to Fig.\ \ref{SIfig:unsmoothed}}: Trends show the total number of cases of Sexual assault (panel (a)), and rape (panel (b)) against minors according to original records which tend to over-assign dates as the first day of each month, and smoothed records where these over assigned cases have been uniformly reassigned in each municipality within each month.  In principal analysis smoothed values (solid black line) are used, as these are closer to the actual occurrence of crimes.  In SI Additional Results, we document results using original unsmoothed measures.}
\end{figure}

\begin{figure}[tbhp]
\begin{center}
\caption{Event Study Estimates of School Closure on Domestic Violence Against Children}
\label{SIfig:eventCloseDV}
\subfloat[No controls\label{fig:eventDV1}]{%
\includegraphics[width=0.45\textwidth]{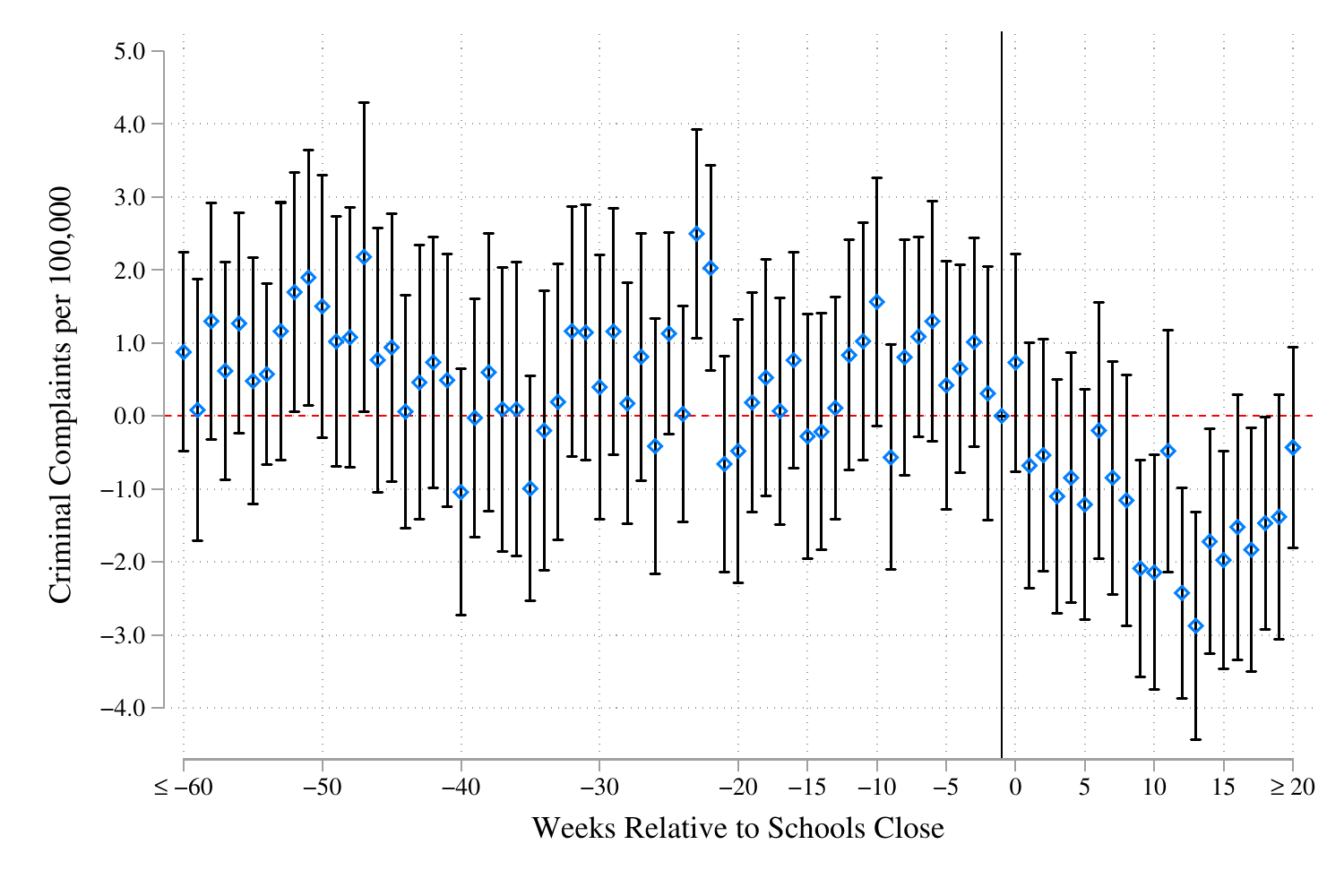}%
}\\
\subfloat[Quarantine control\label{fig:eventDV2}]{%
 \includegraphics[width=0.45\textwidth]{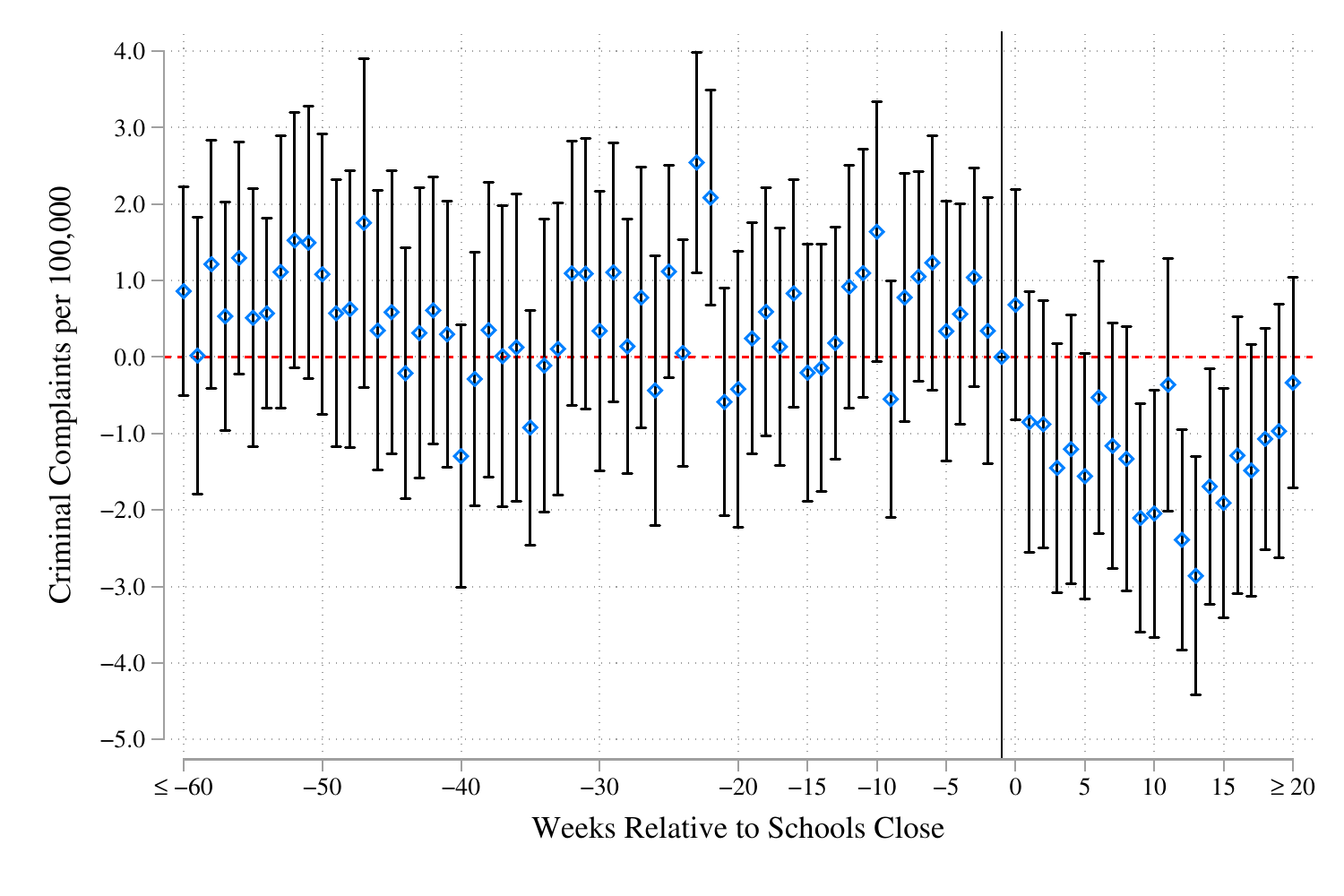}%
}\\
\subfloat[Quarantine and COVID controls\label{fig:eventDV3}]{%
\includegraphics[width=0.45\textwidth]{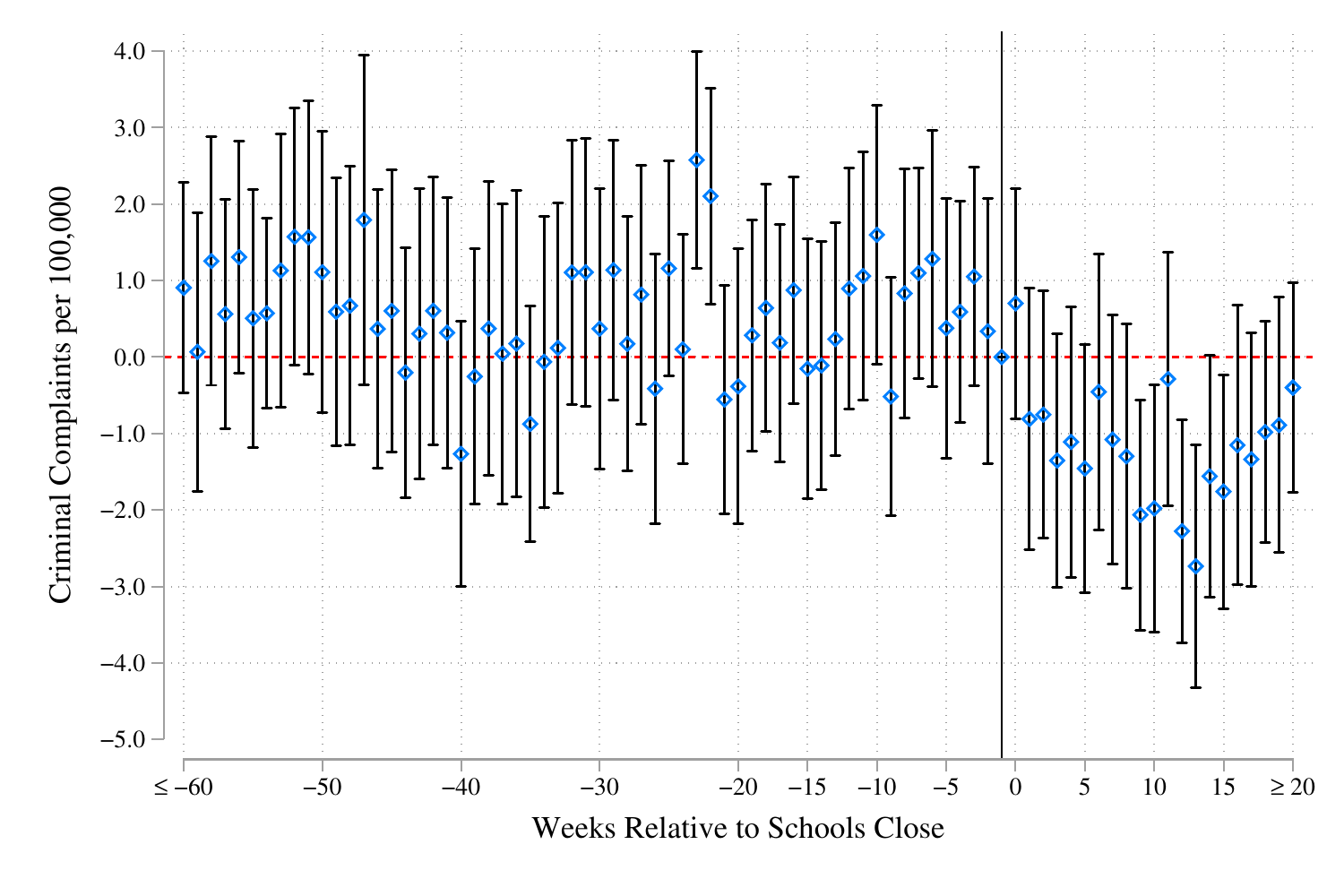}%
}
\end{center}
\floatfoot{\textbf{Notes to Fig.\ \ref{SIfig:eventCloseDV}}: Event studies are documented as described in equation \ref{SIeqn:event}.  Hollow blue diamonds display point estimates, and error bars denote 95\% CIs.  Here the `event' occurring at time 0 refers to school closure, with period -1 (one week prior to school closure) included as the omitted base period.  The outcome is cases of DV against minors per 100,000 minors.  Post-closure lags are included up to 20 weeks post-closure, as this is when first re-openings begin to occur.  Graduated controls are included in panels (b)-(c), indicated in plot titles. All other details follow those described in equation \ref{SIeqn:event}.}
\end{figure}

\begin{figure}[tbhp]
\begin{center}
\caption{Event Study Estimates of School Re-opening on Domestic Violence Against Children}
\label{SIfig:eventOpenDV}
\subfloat[No controls\label{fig:eventDV5}]{%
\includegraphics[width=0.45\textwidth]{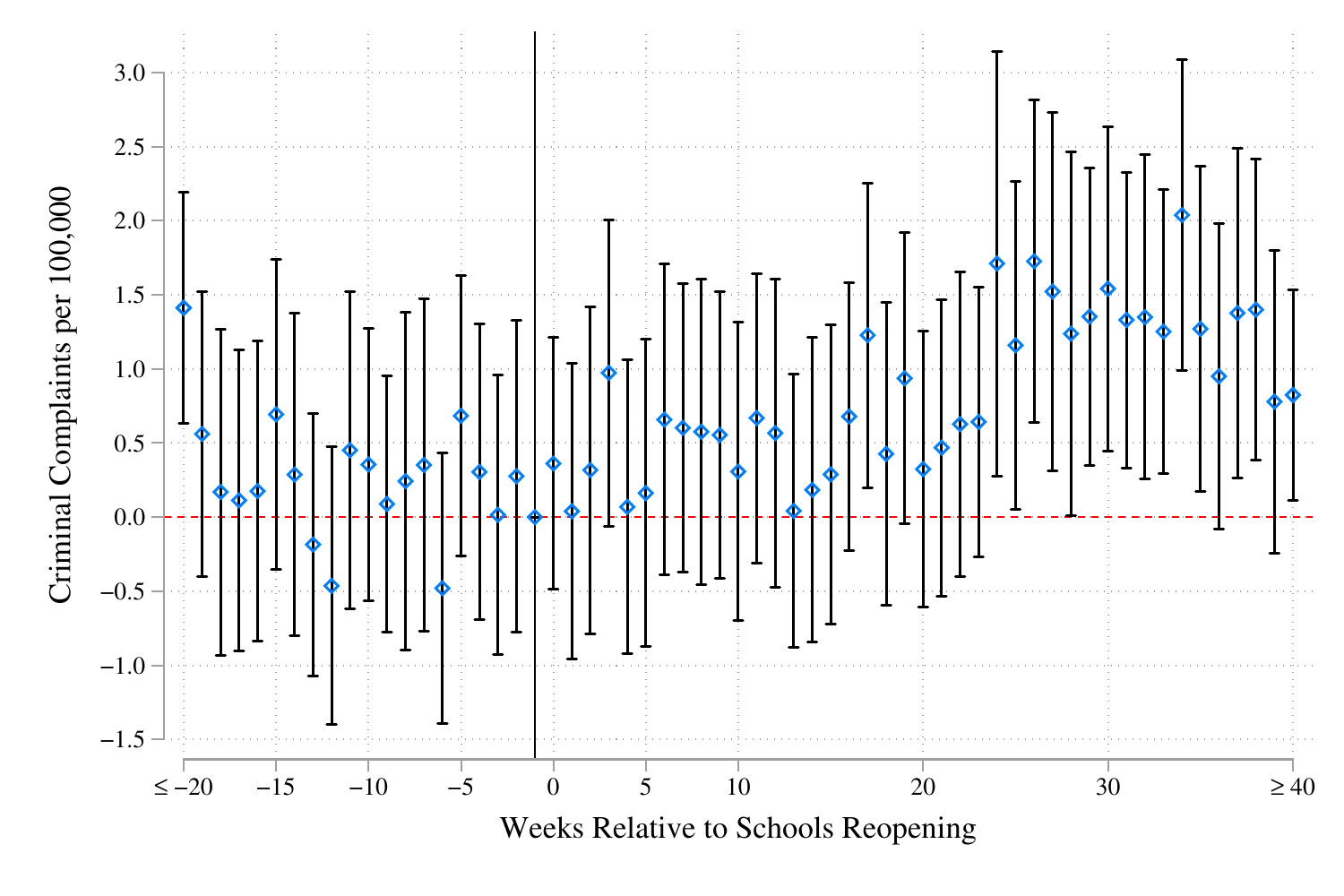}%
}\\
\subfloat[Quarantine control\label{fig:eventDV6}]{%
 \includegraphics[width=0.45\textwidth]{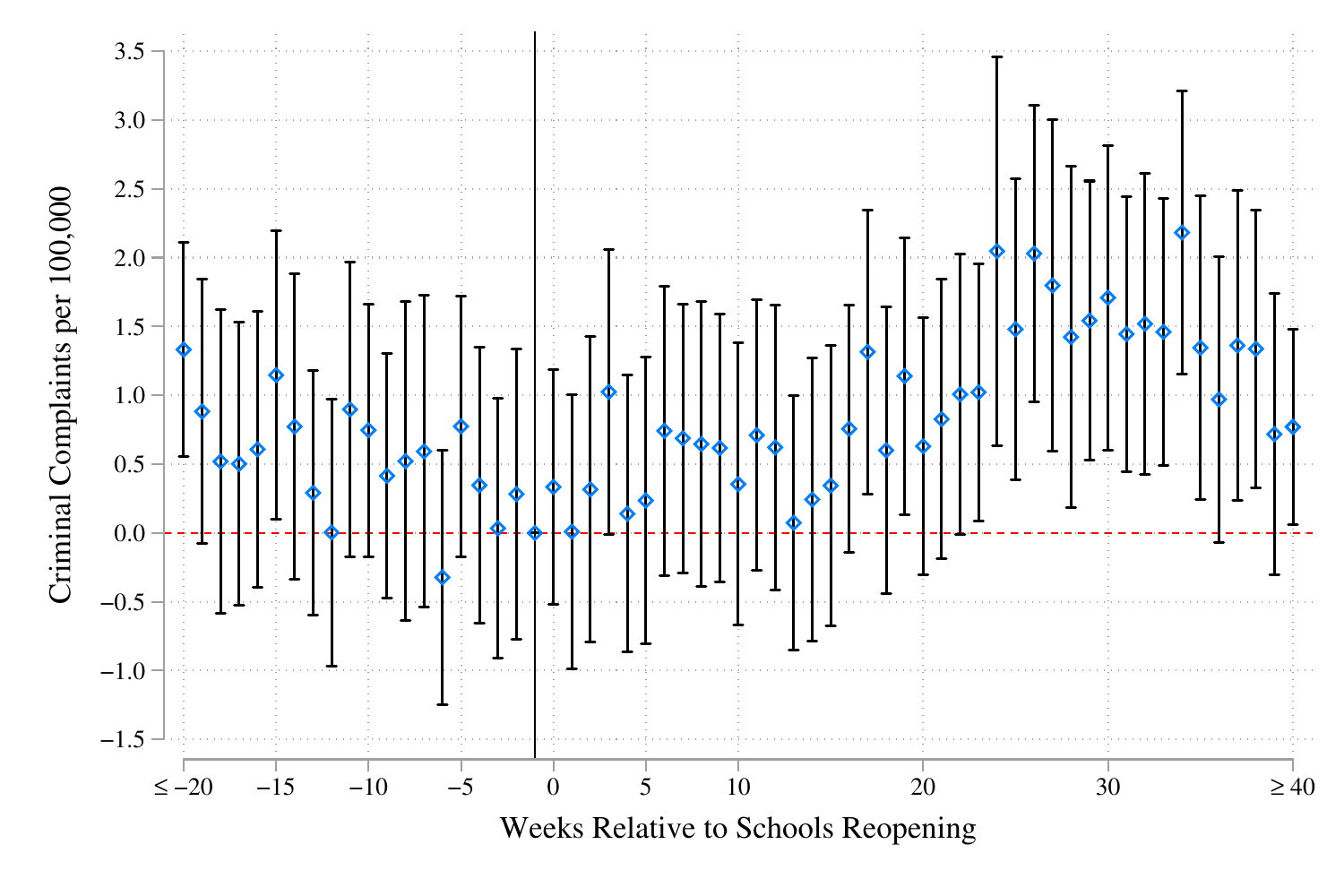}%
}\\
\subfloat[Quarantine and COVID controls\label{fig:eventDV7}]{%
\includegraphics[width=0.45\textwidth]{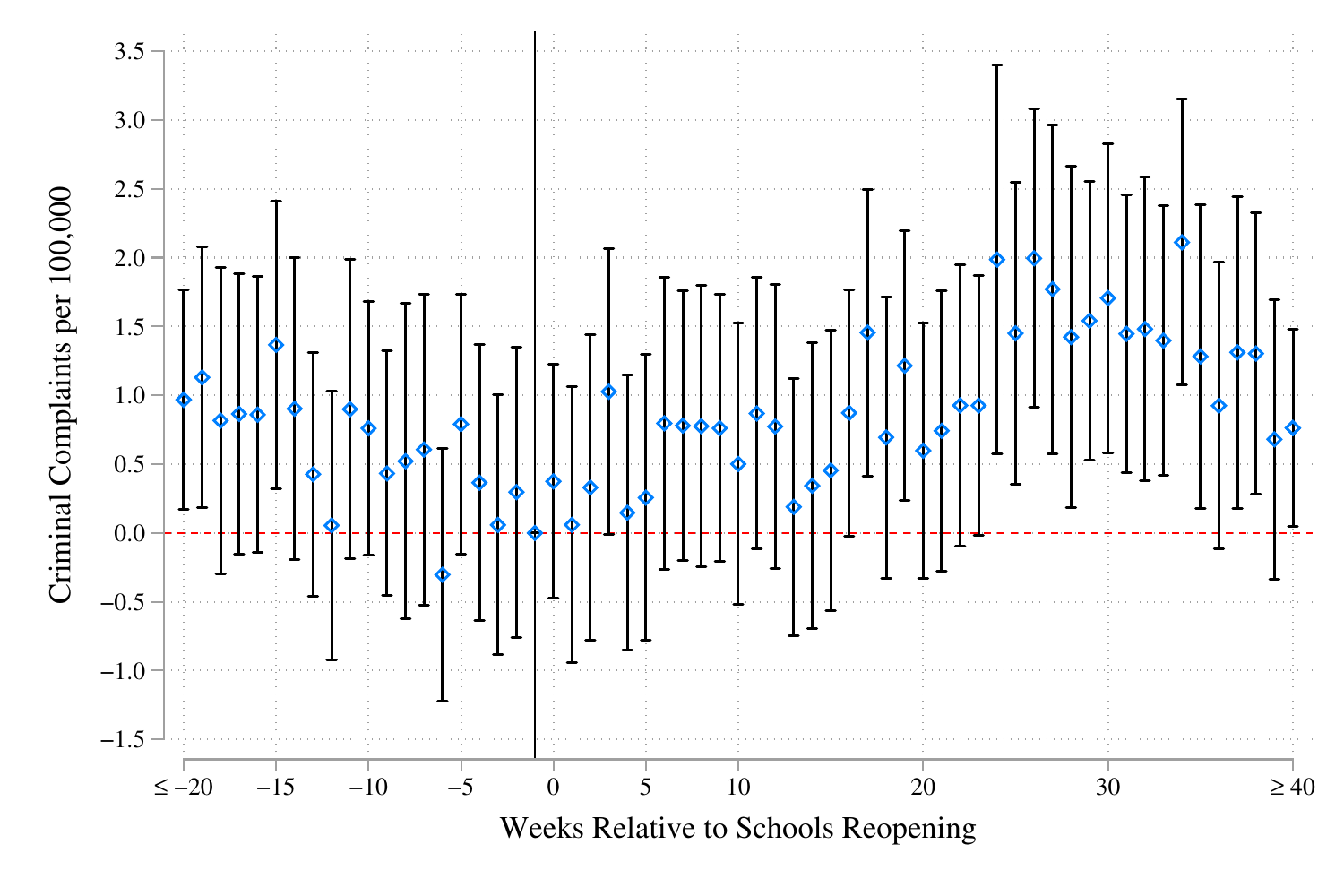}%
}
\end{center}
\floatfoot{\textbf{Notes to Fig.\ \ref{SIfig:eventOpenDV}}: Event studies are documented as described in equation \ref{SIeqn:event}.  Hollow blue diamonds display point estimates, and error bars denote 95\% CIs.  Here the `event' occurring at time 0 refers to school reopening, with period -1 (one week prior to school re-opening) included as the omitted base period.  The outcome is cases of DV against minors per 100,000 minors.  Pre-reopening leads are included up to 20 weeks pre-reopening, as beyond this point is when schools had not yet closed.  Graduated controls are included in panels (b)-(c), indicated in plot titles. All other details follow those described in equation \ref{SIeqn:event}.}
\end{figure}

\begin{figure}[tbhp]
\begin{center}
\caption{Event Study Estimates of School Closure on Sexual Abuse Against Children}
\label{SIfig:eventCloseSA}
\subfloat[No controls\label{fig:eventSA}]{%
\includegraphics[width=0.45\textwidth]{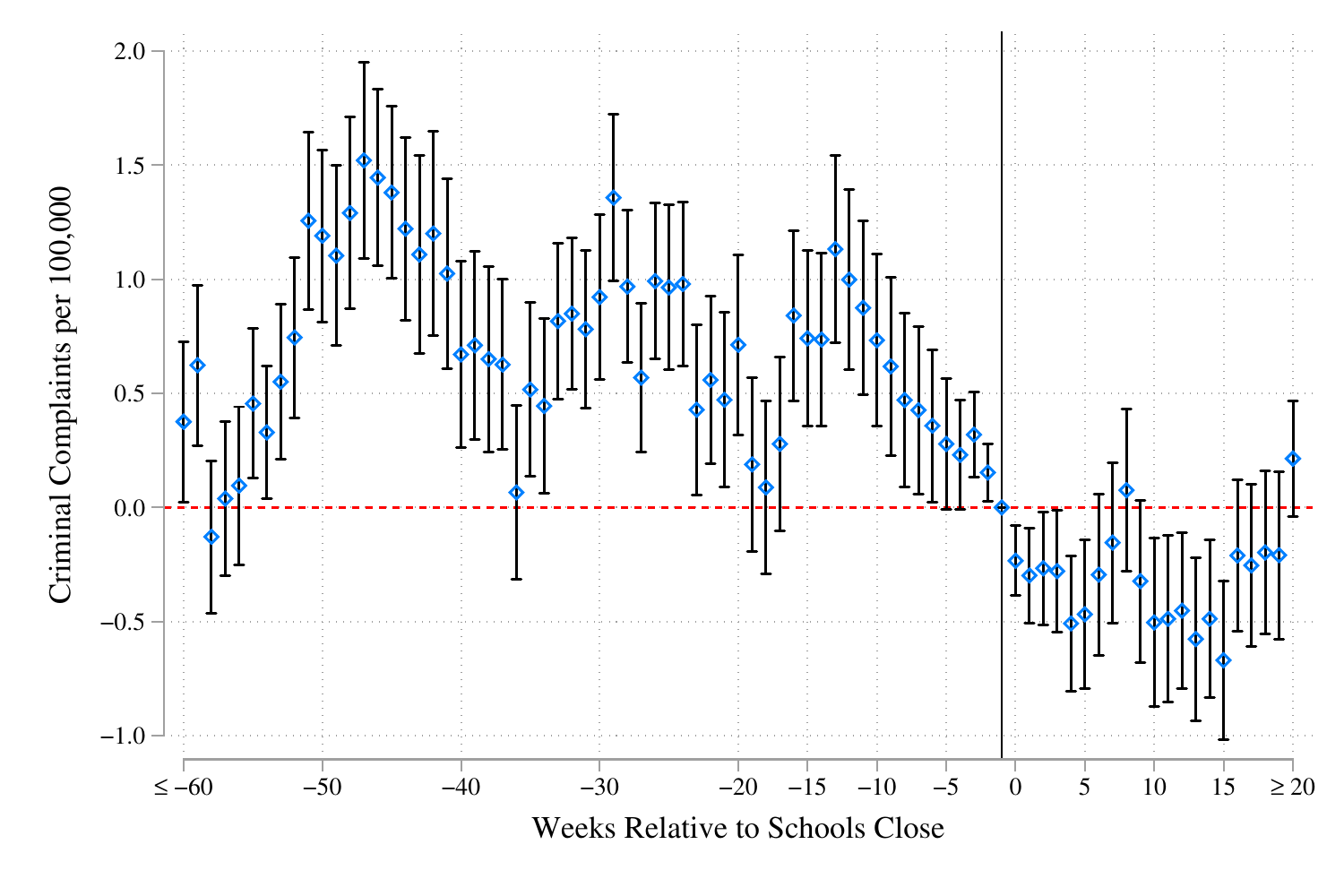}%
}\\
\subfloat[Quarantine control\label{fig:eventSA2}]{%
 \includegraphics[width=0.45\textwidth]{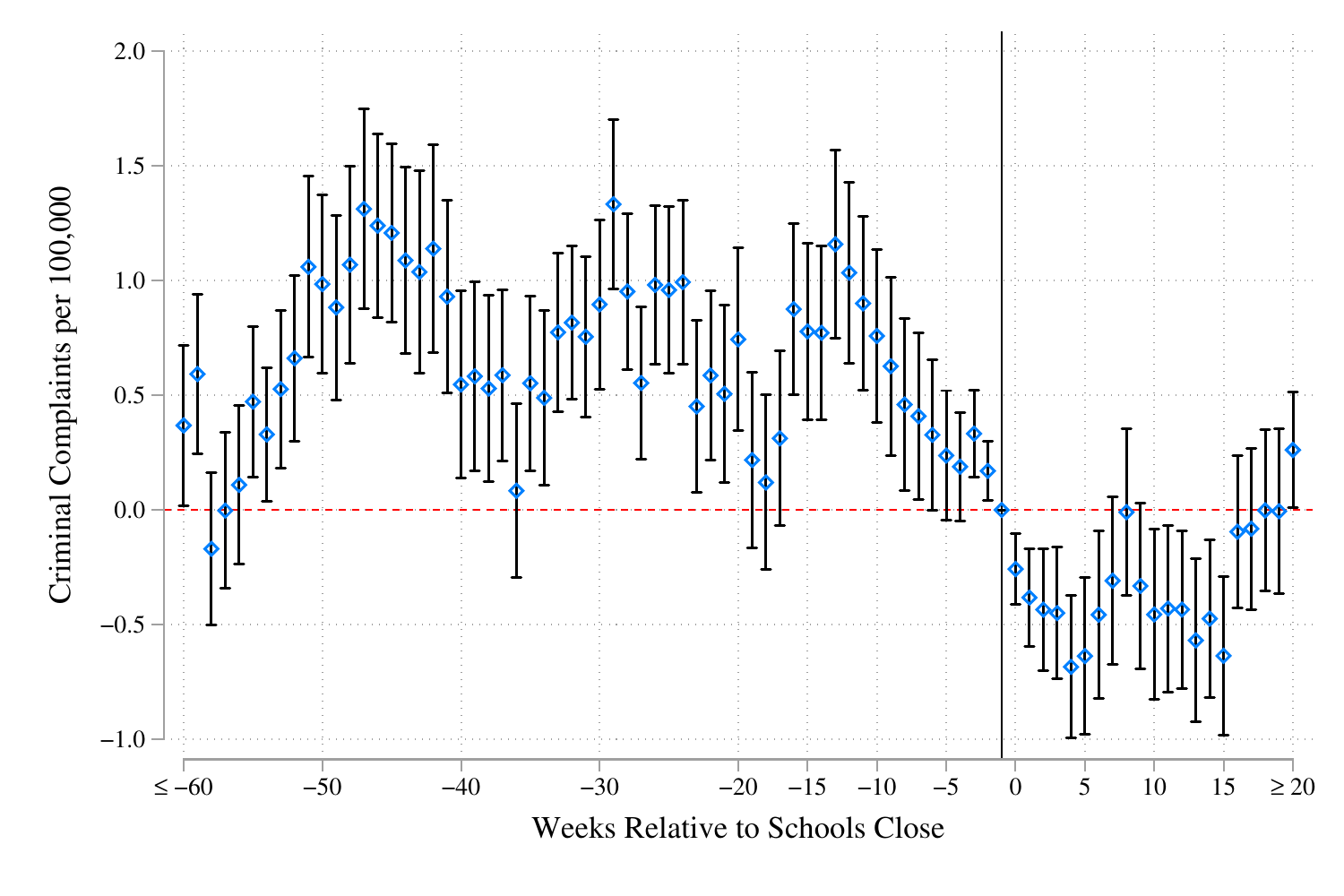}%
}\\
\subfloat[Quarantine and COVID controls\label{fig:eventSA3}]{%
\includegraphics[width=0.45\textwidth]{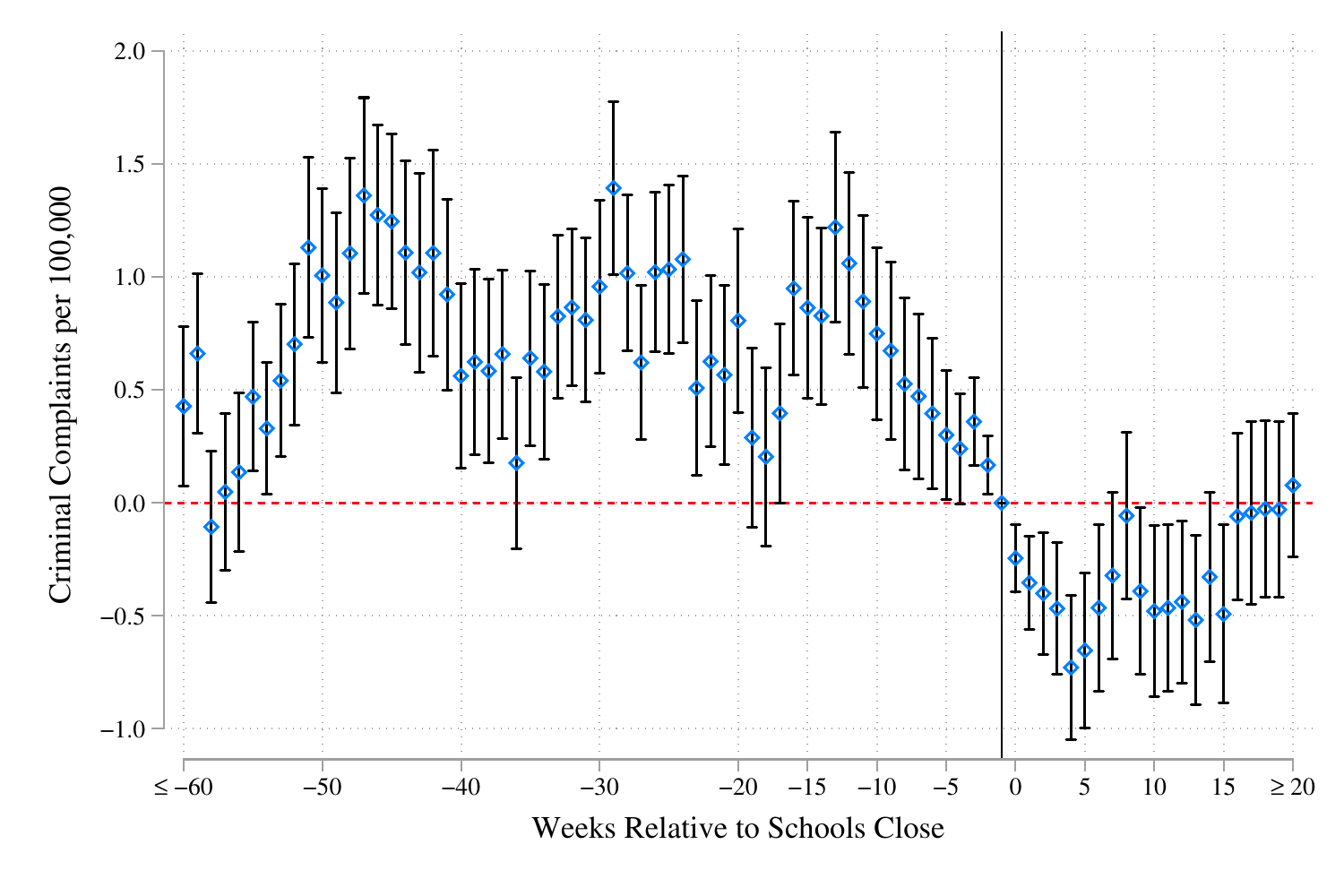}%
}
\end{center}
\floatfoot{\textbf{Notes to Fig.\ \ref{SIfig:eventCloseSA}}: Event studies are documented as described in equation \ref{SIeqn:event}.  Hollow blue diamonds display point estimates, and error bars denote 95\% CIs.  Here the `event' occurring at time 0 refers to school closure, with period -1 (one week prior to school closure) included as the omitted base period.  The outcome is cases of sexual abuse against minors per 100,000 minors.  Post-closure lags are included up to 20 weeks post-closure, as this is when first re-openings begin to occur.  Graduated controls are included in panels (b)-(c), indicated in plot titles. All other details follow those described in equation \ref{SIeqn:event}.}
\end{figure}

\begin{figure}[tbhp]
\begin{center}
\caption{Event Study Estimates of School Reopening on Sexual Abuse Against Children}
\label{SIfig:eventOpenSA}
\subfloat[No controls\label{fig:eventSA5}]{%
\includegraphics[width=0.45\textwidth]{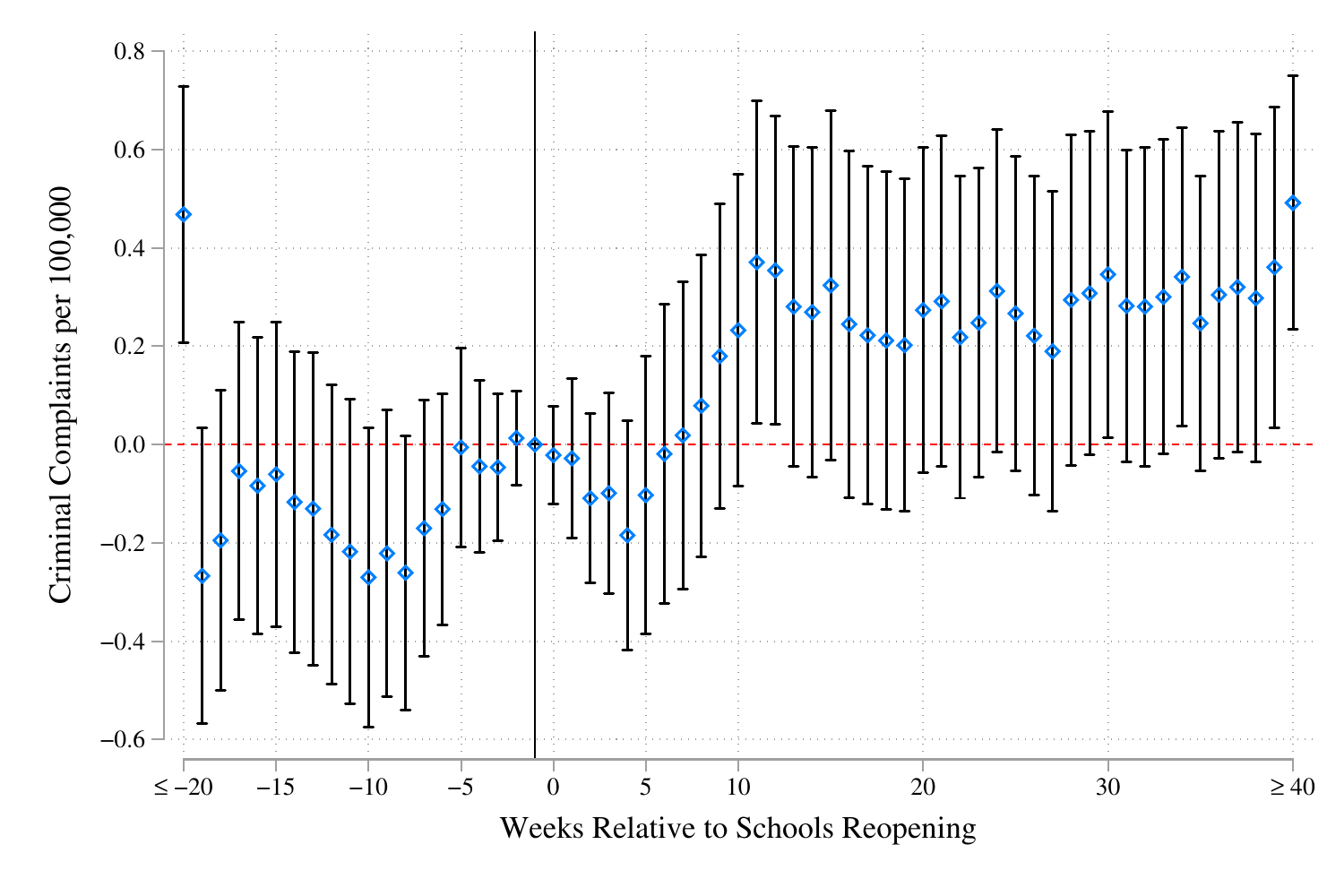}%
}\\
\subfloat[Quarantine control\label{fig:eventSA6}]{%
 \includegraphics[width=0.45\textwidth]{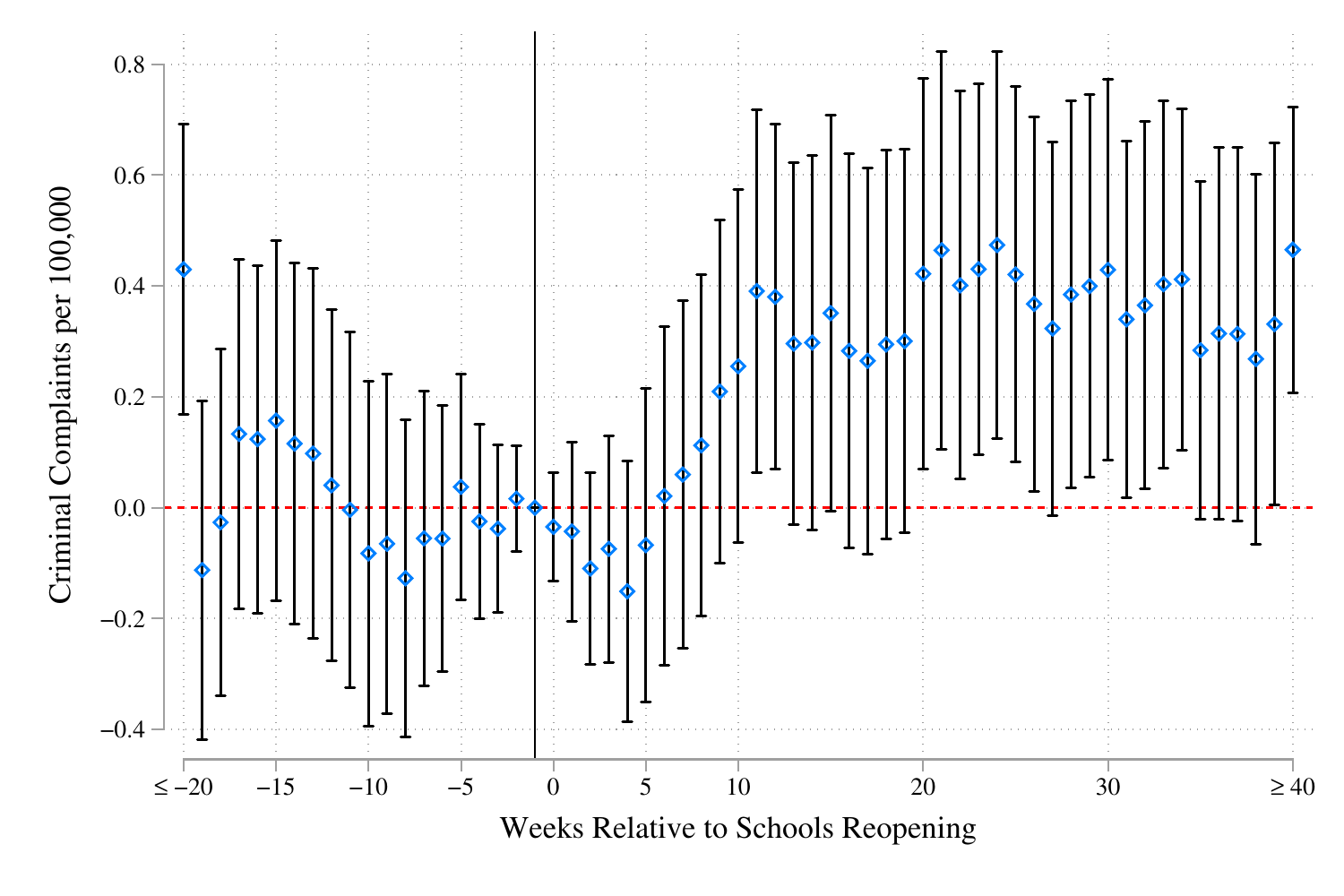}%
}\\
\subfloat[Quarantine and COVID controls\label{fig:eventSA7}]{%
\includegraphics[width=0.45\textwidth]{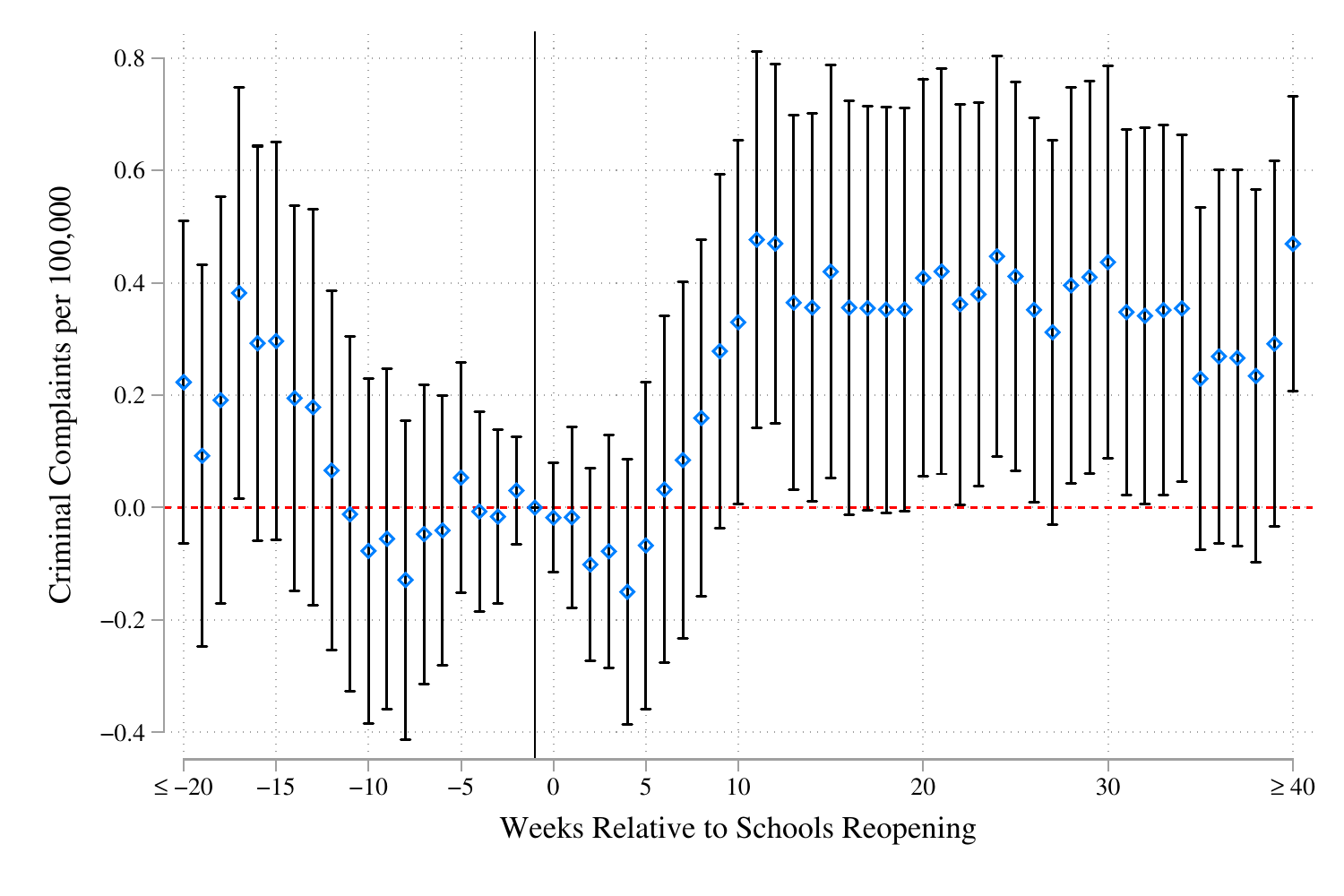}%
}
\end{center}
\floatfoot{\textbf{Notes to Fig.\ \ref{SIfig:eventOpenSA}}: Event studies are documented as described in equation \ref{SIeqn:event}.  Hollow blue diamonds display point estimates, and error bars denote 95\% CIs.  Here the `event' occurring at time 0 refers to school reopening, with period -1 (one week prior to school re-opening) included as the omitted base period.  The outcome is cases of sexual abuse against minors per 100,000 minors.  Pre-reopening leads are included up to 20 weeks pre-reopening, as beyond this point is when schools had not yet closed.  Graduated controls are included in panels (b)-(c), indicated in plot titles. All other details follow those described in equation \ref{SIeqn:event}.}
\end{figure}

\begin{figure}[tbhp]
\begin{center}
\caption{Event Study Estimates of School Closure on Rape Against Children}
\label{SIfig:eventCloseRa}
\subfloat[No controls\label{fig:eventRa1}]{%
\includegraphics[width=0.45\textwidth]{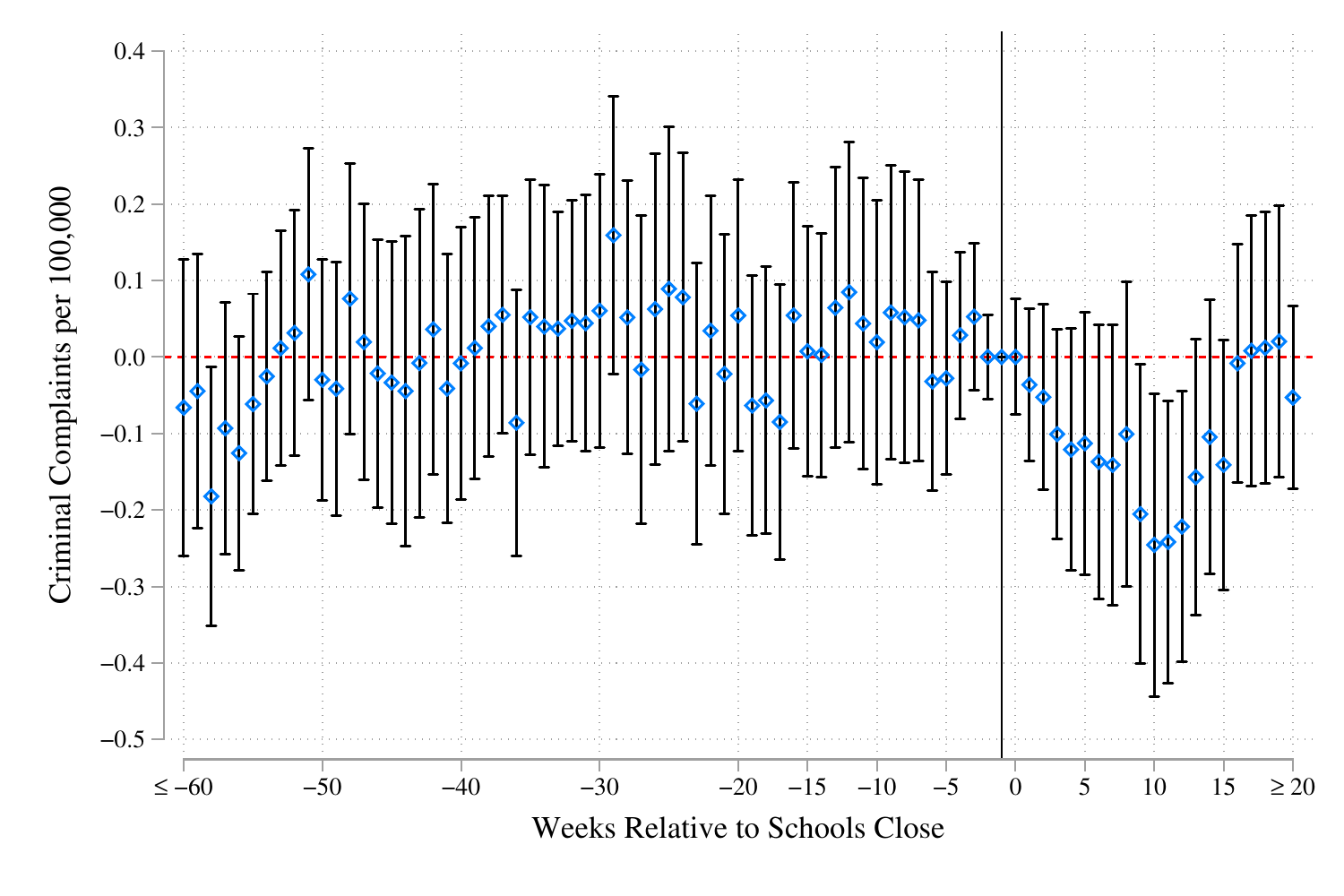}%
}\\
\subfloat[Quarantine control\label{fig:eventRa2}]{%
 \includegraphics[width=0.45\textwidth]{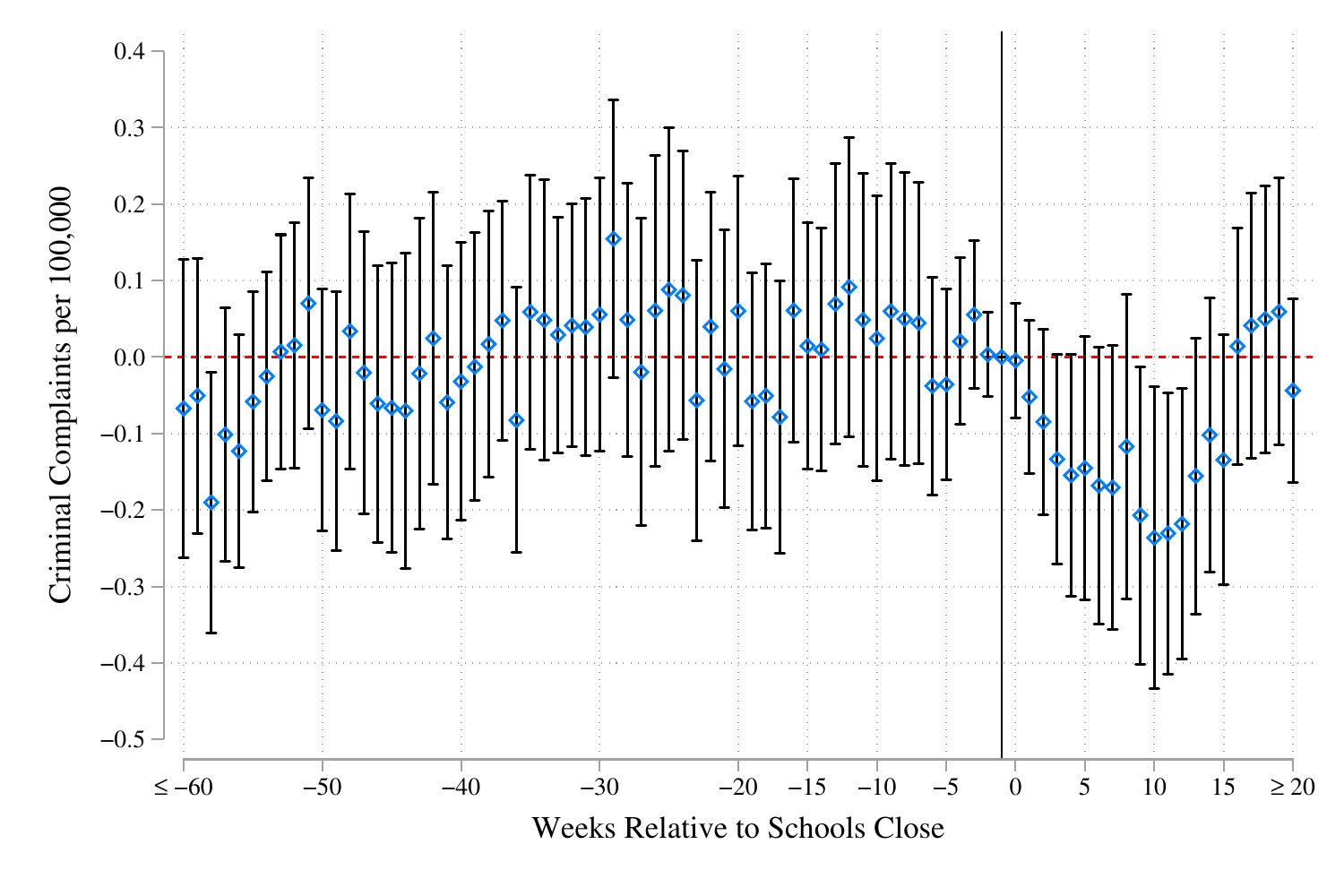}%
}\\
\subfloat[Quarantine and COVID controls\label{fig:eventRa3}]{%
\includegraphics[width=0.45\textwidth]{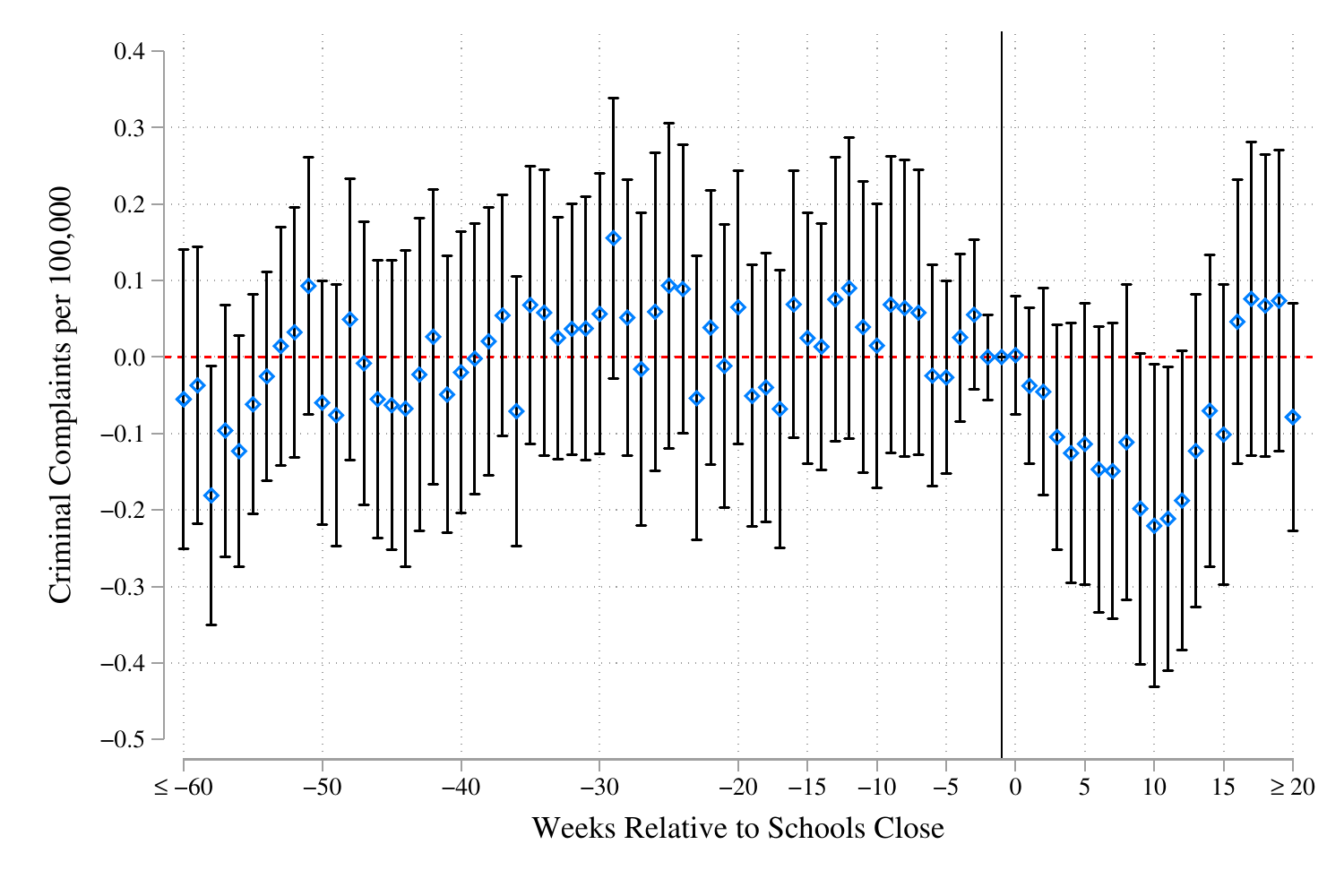}%
}
\end{center}
\floatfoot{\textbf{Notes to Fig.\ \ref{SIfig:eventCloseRa}}: Event studies are documented as described in equation \ref{SIeqn:event}.  Hollow blue diamonds display point estimates, and error bars denote 95\% CIs.  Here the `event' occurring at time 0 refers to school closure, with period -1 (one week prior to school closure) included as the omitted base period.  The outcome is cases of rape against minors per 100,000 minors.  Post-closure lags are included up to 20 weeks post-closure, as this is when first re-openings begin to occur.  Graduated controls are included in panels (b)-(c), indicated in plot titles. All other details follow those described in equation \ref{SIeqn:event}.}
\end{figure}

\begin{figure}[tbhp]
\begin{center}
\caption{Event Study Estimates of School Reopening on Rape Against Children}
\label{SIfig:eventOpenRa}
\subfloat[No controls\label{fig:eventRa5}]{%
\includegraphics[width=0.45\textwidth]{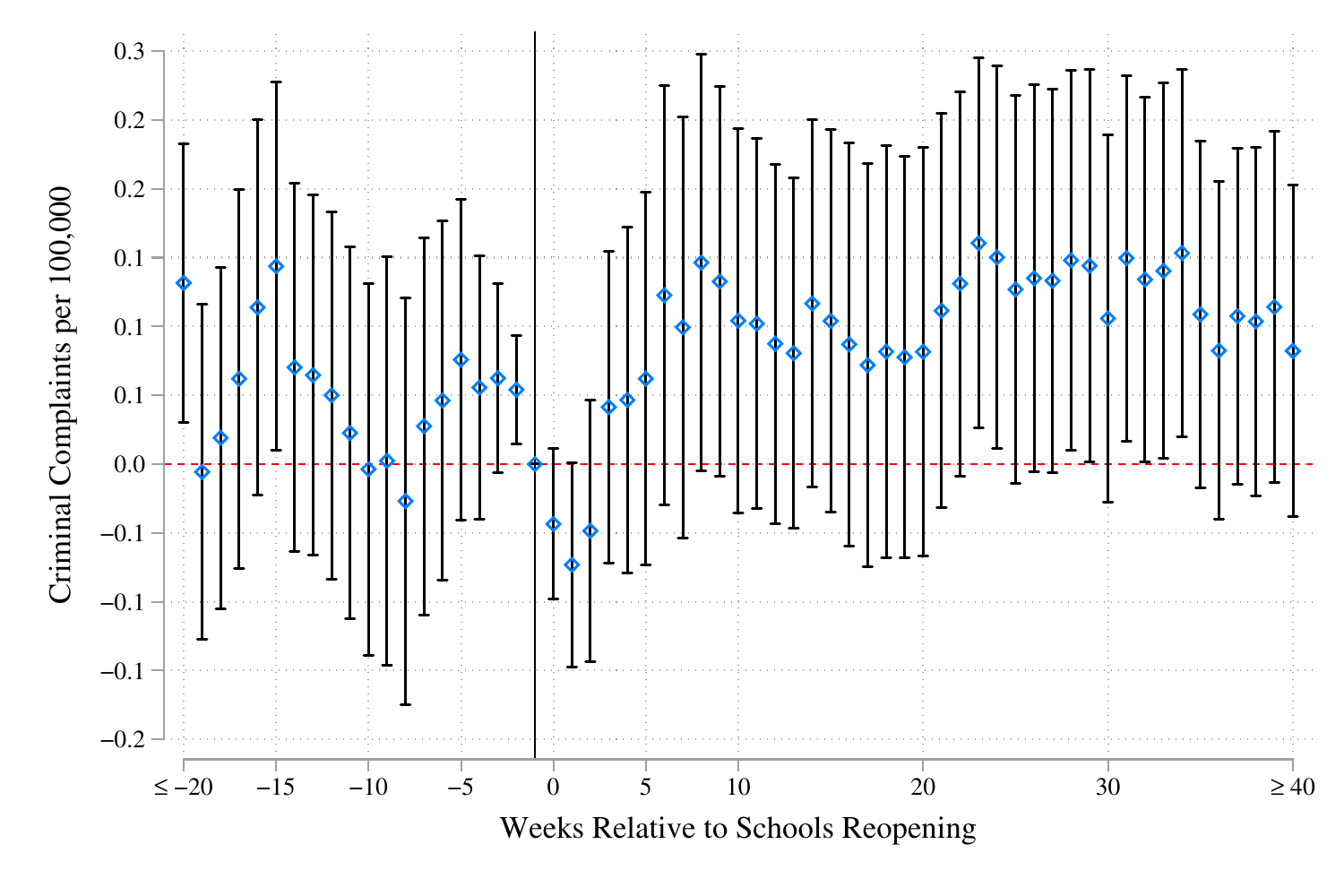}%
}\\
\subfloat[Quarantine control\label{fig:eventRa6}]{%
 \includegraphics[width=0.45\textwidth]{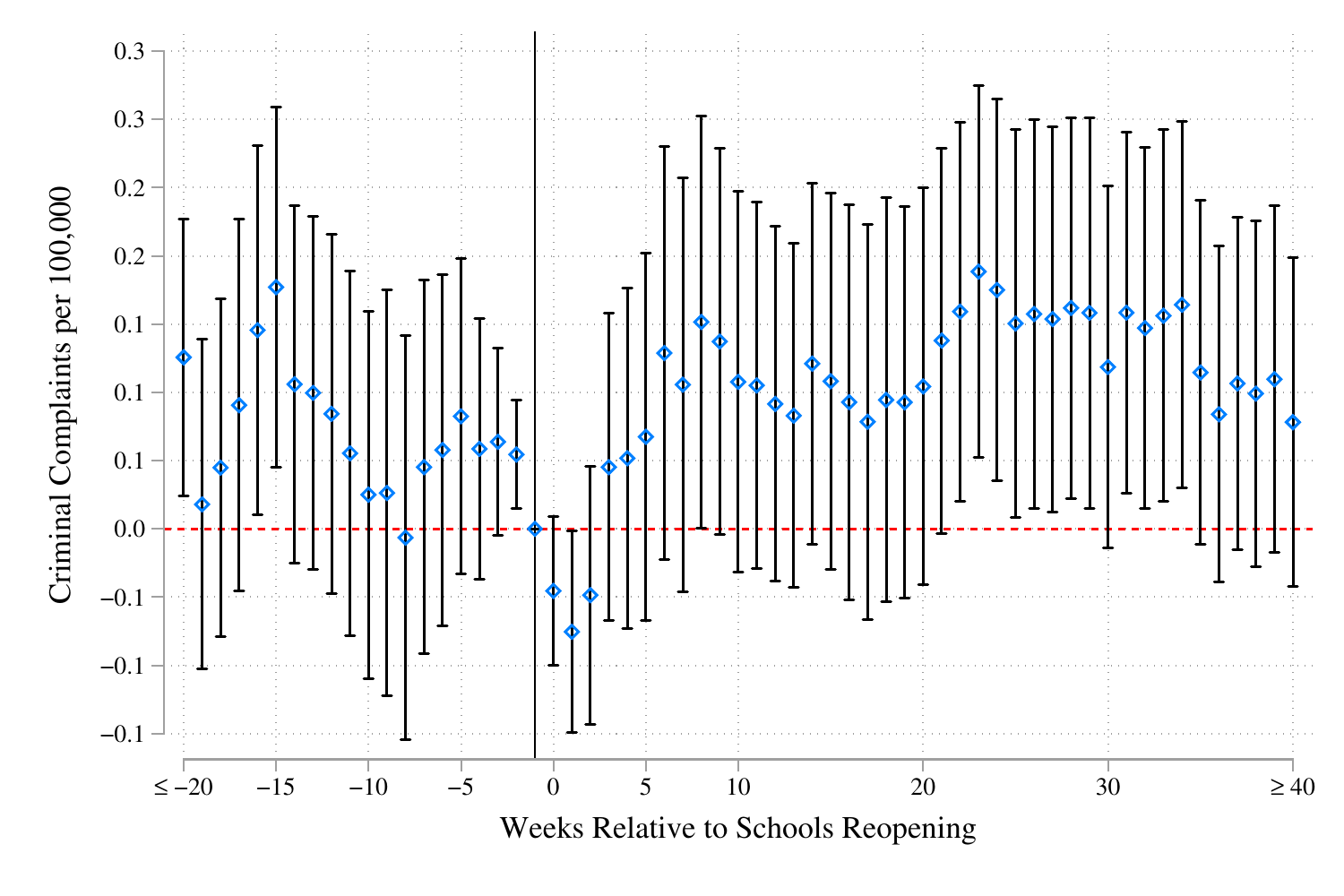}%
}\\
\subfloat[Quarantine and COVID controls\label{fig:eventRa7}]{%
\includegraphics[width=0.45\textwidth]{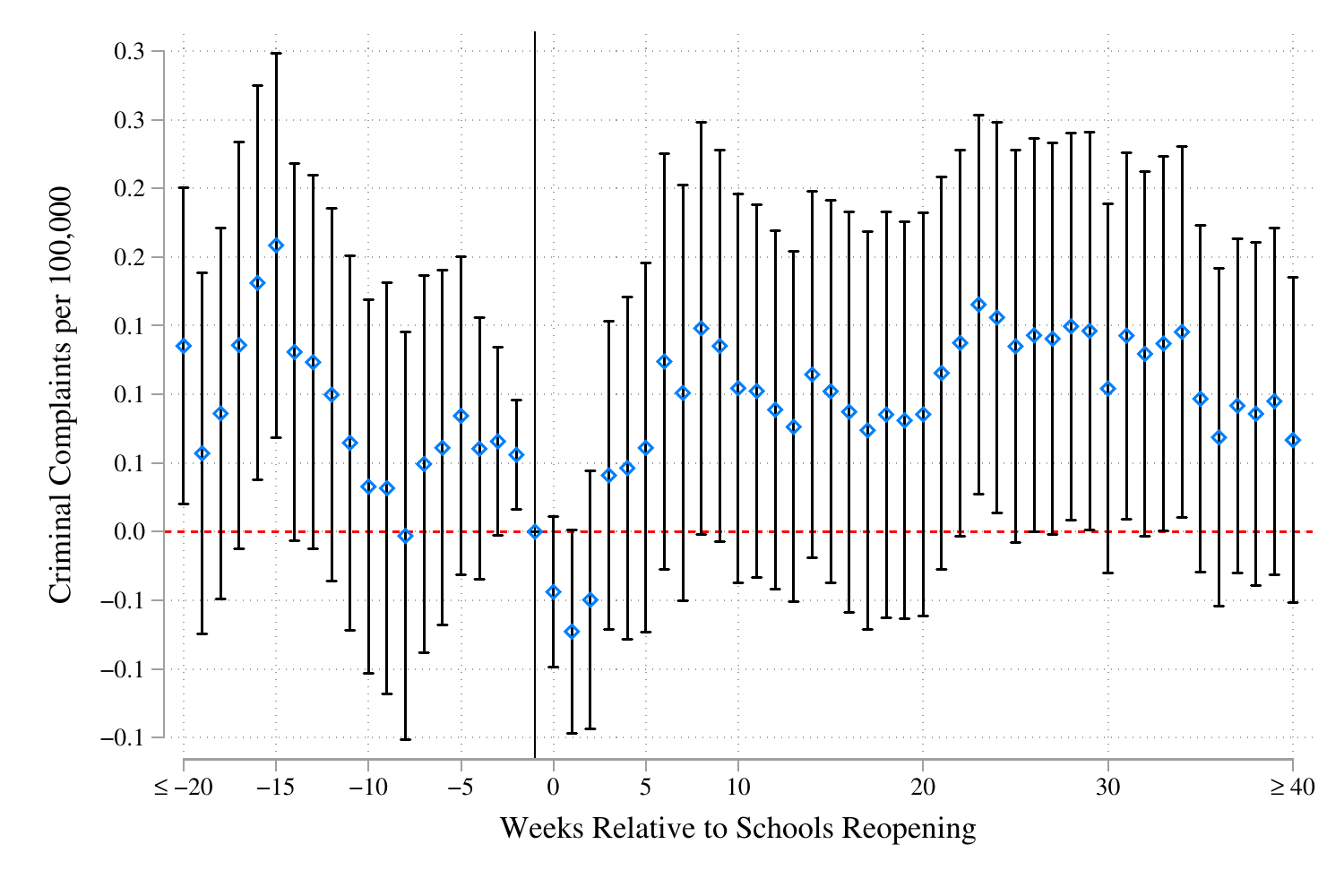}%
}
\end{center}
\floatfoot{\textbf{Notes to Fig.\ \ref{SIfig:eventOpenRa}}: Event studies are documented as described in equation \ref{SIeqn:event}.  Hollow blue diamonds display point estimates, and error bars denote 95\% CIs.  Here the `event' occurring at time 0 refers to school reopening, with period -1 (one week prior to school re-opening) included as the omitted base period.  The outcome is cases of rape against minors per 100,000 minors.  Pre-reopening leads are included up to 20 weeks pre-reopening, as beyond this point is when schools had not yet closed.  Graduated controls are included in panels (b)-(c), indicated in plot titles. All other details follow those described in equation \ref{SIeqn:event}.}
\end{figure}

\begin{figure*}[htpb!]
\begin{center}
\caption{Weights and 2$\times$2 Double-Difference Estimates in Two-way Fixed Effect Models}
\label{SIfig:GB}
\subfloat[Intra-family Violence\label{SIfig:GB_V}]{%
\includegraphics[width=0.5\textwidth]{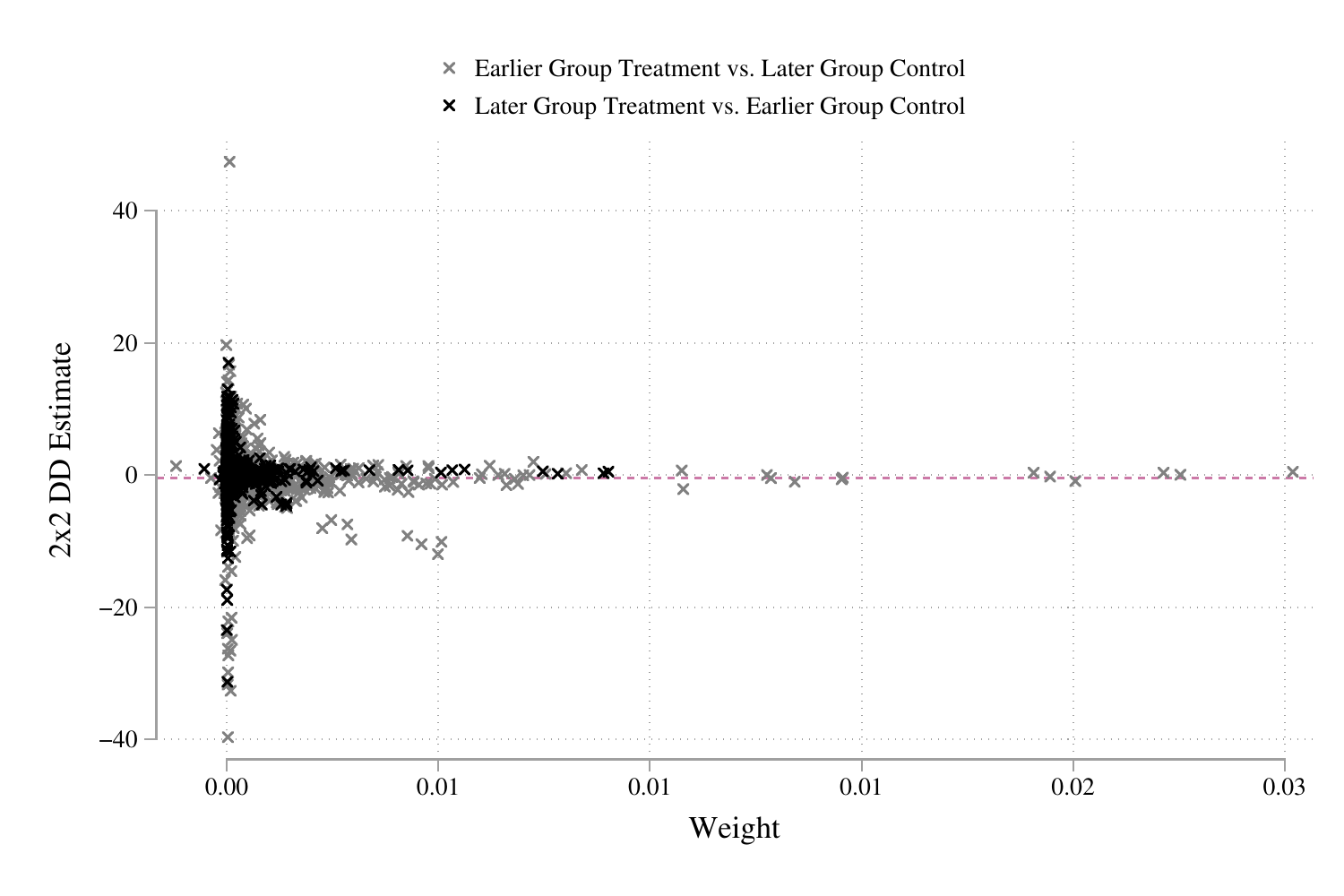}%
}\\
\subfloat[Sexual Abuse\label{SIfig:GB_SA}]{%
 \includegraphics[width=0.5\textwidth]{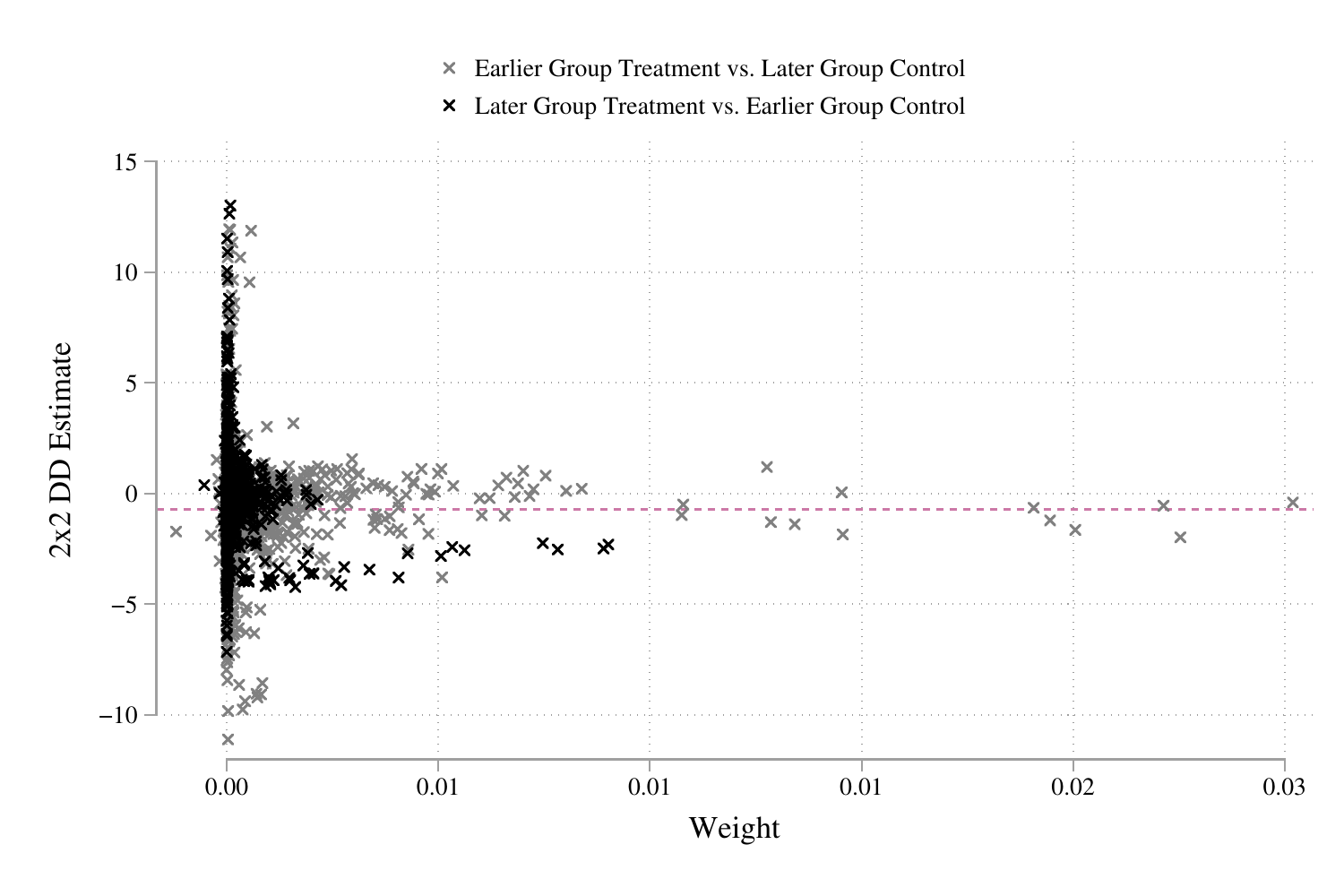}%
}\\
\subfloat[Rape\label{SIfig:GB_R}]{%
\includegraphics[width=0.5\textwidth]{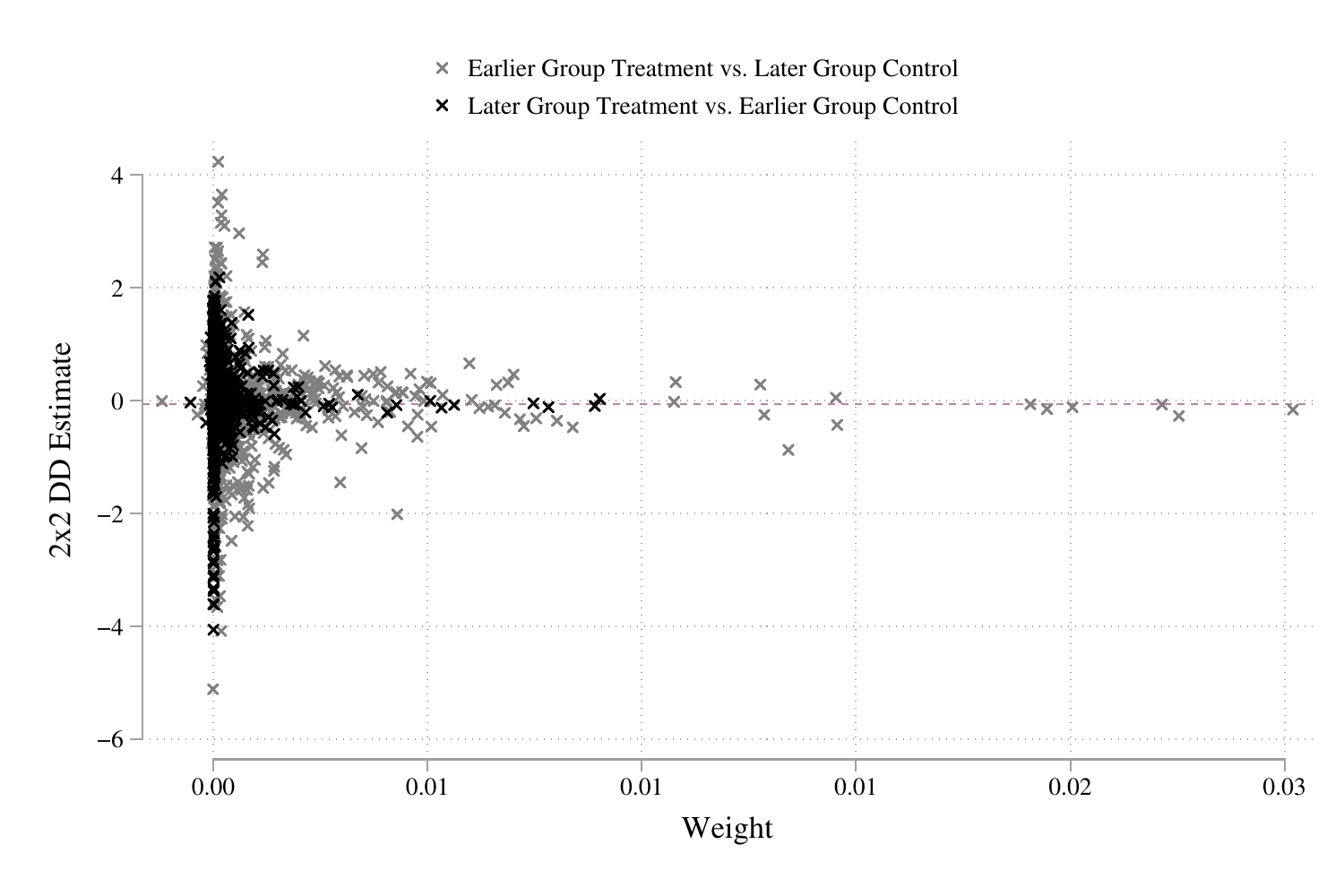}%
}
\end{center}
\vspace{-5mm}
\floatfoot{\textbf{Notes to Fig.\ \ref{SIfig:GB}}: Plots document the double-difference decomposition laid out by \cite{GB2021} to decompose single coefficient estimates on School Reopening displayed in Table 1 of the paper.  Here, each cross displays the proportional weight of each municipal$\times$week switching estimate (horizontal axis), as well as the DD estimate for each switching pair (vertical axis), compared with the single-coefficient estimate (dotted horizontal line).  Grey crosses represent individual estimates based on the (desired) comparison of treated to not yet treated units, while black crosses represent comparisons of later switchers to earlier switchers.}
\end{figure*}

\begin{figure*}[tbhp]
\begin{center}
\caption{Actual Reporting Channels Reported by a Single Child Protection Office in a Large Municipality}
\label{SIfig:OPD}
\subfloat[Violence Reporting by Reporting Channel\label{fig:OPD1}]{%
\includegraphics[width=0.5\textwidth]{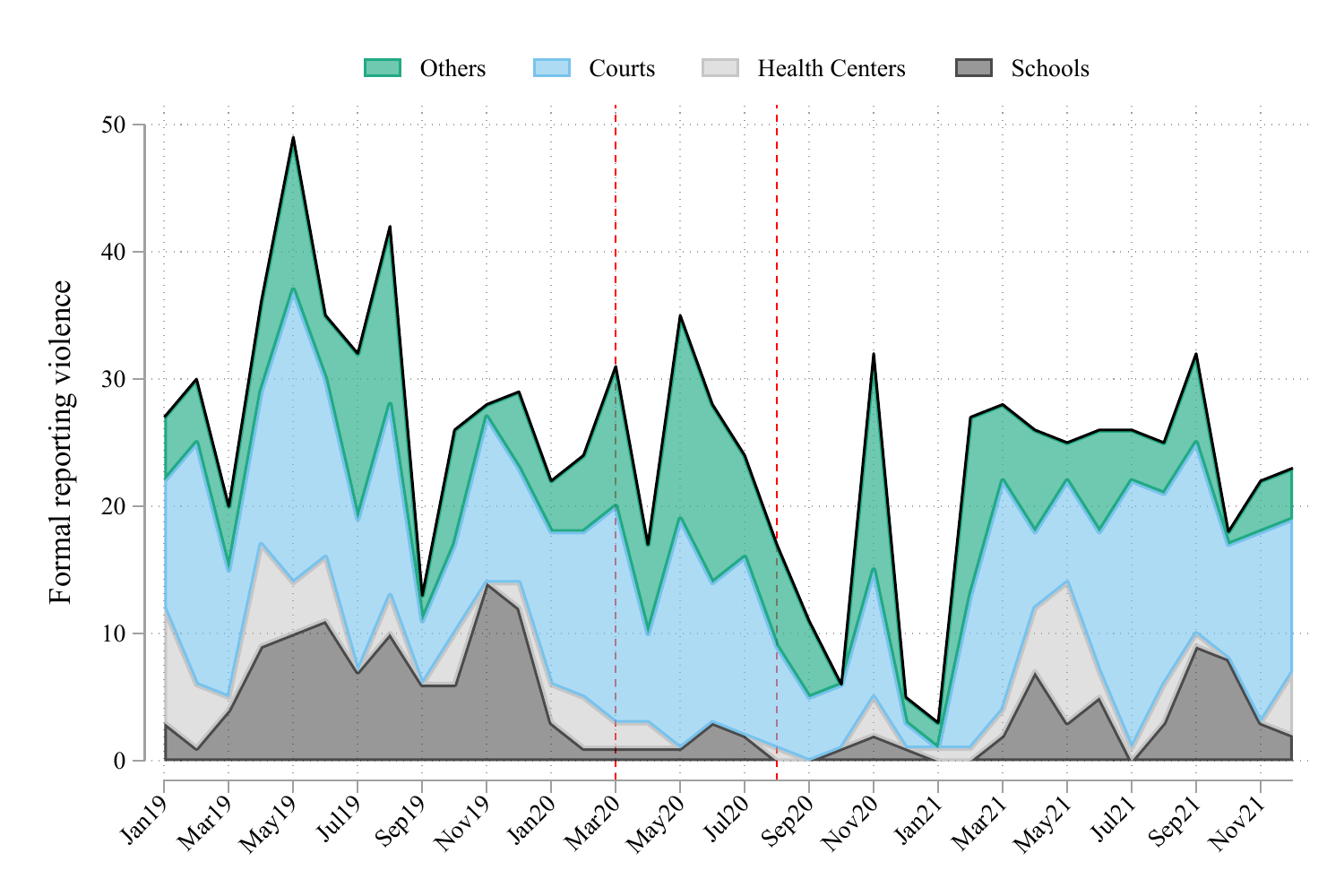}%
}
\subfloat[Percentage of Violence Reporting by Reporting Channel\label{fig:OPD2}]{%
 \includegraphics[width=0.5\textwidth]{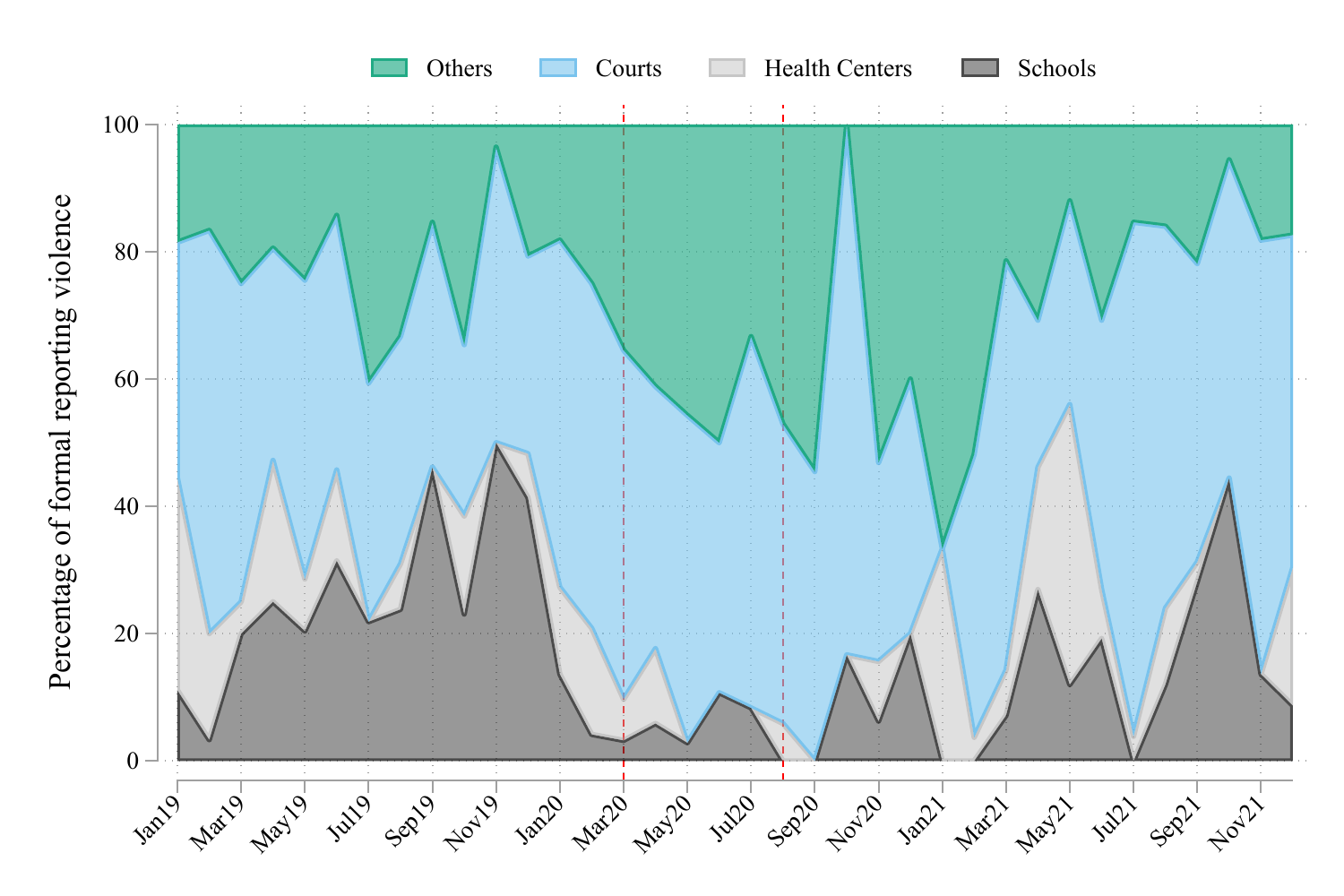}%
}
\end{center}
\floatfoot{\textbf{Notes to Fig.\ \ref{SIfig:OPD}}: All reports of violence received by the child protection office of a large municipality in Santiago are displayed.  These are broken down by reporting channels as from schools, from municipal health care centres, from courts and from other sources.  Total numbers of cases by month in this municipality are displayed in the left-hand panel, and absolute proportions are displayed in the right-hand panel.  Vertical dotted lines represent dates of school closure and school reopening in the municipality.}
\end{figure*}

\begin{figure}[t!]
\begin{center}
\caption{Alternative Specifications of Demographic and Socio-economic Variation in Closure/Reopening (No controls)}
\label{SIfig:heterogeneity1}
\subfloat[Intra-family Violence]{%
\includegraphics[width=0.44\textwidth]{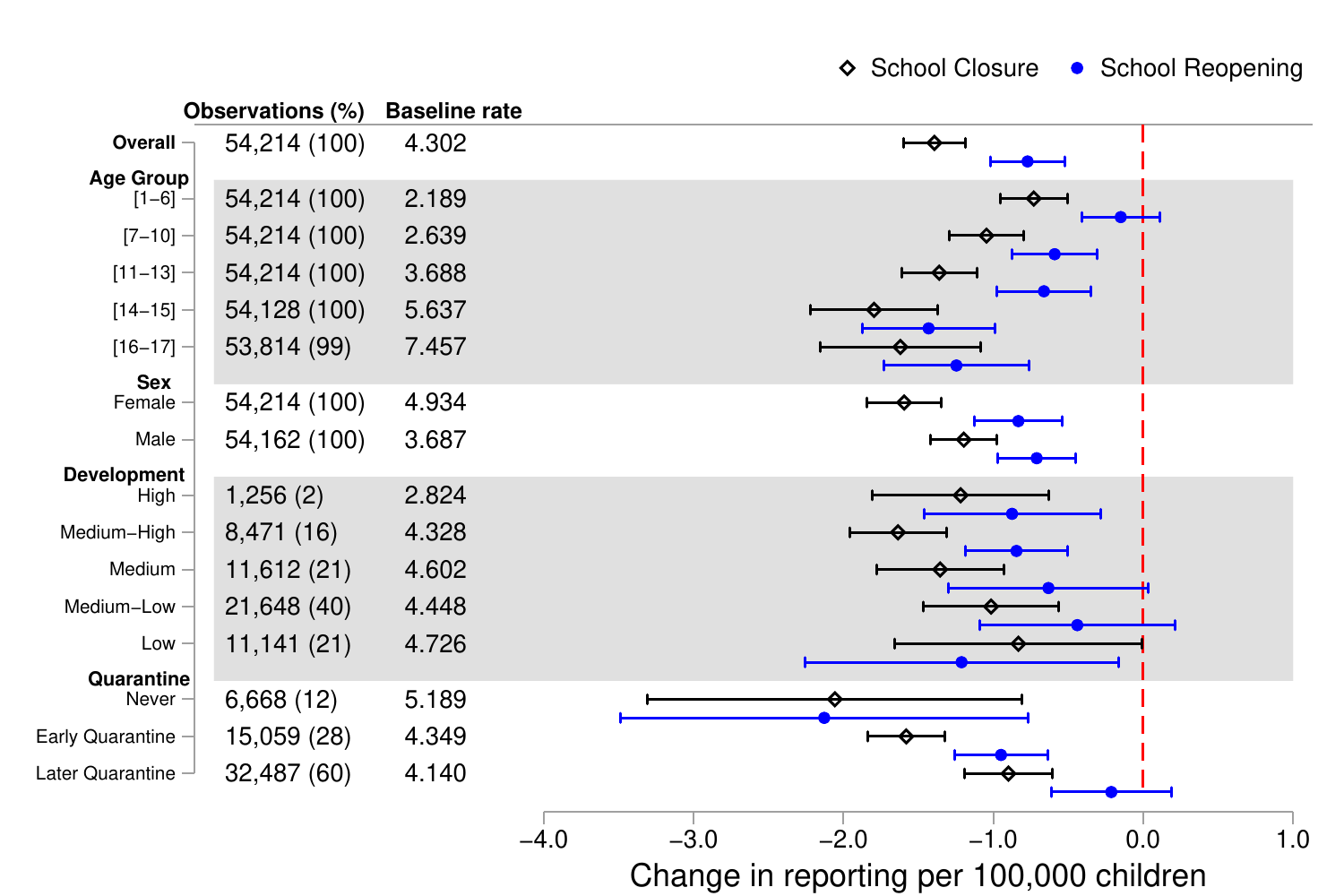}%
}\\
\subfloat[Sexual Abuse]{%
\includegraphics[width=0.44\textwidth]{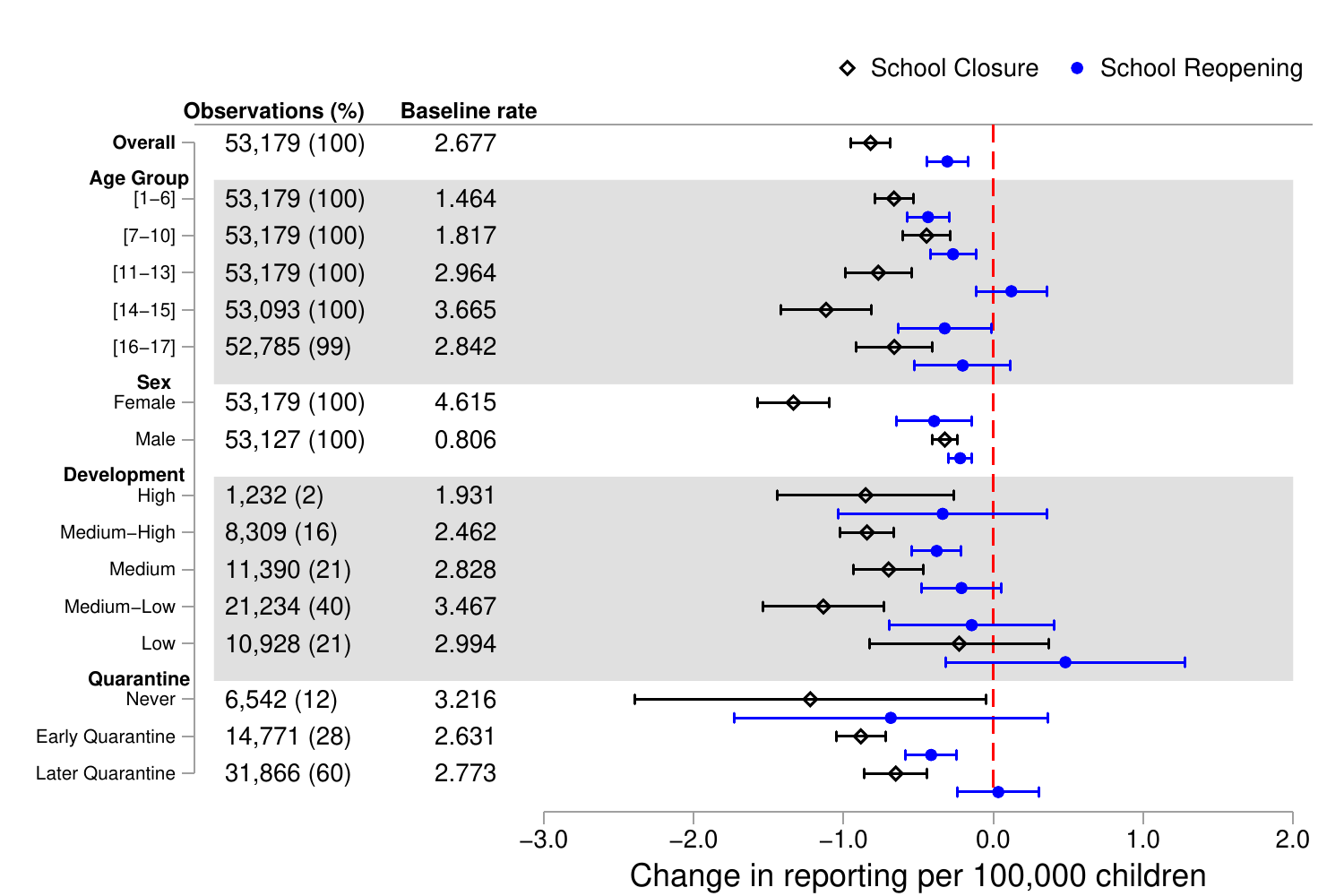}%
}\\
\subfloat[Rape]{%
\includegraphics[width=0.44\textwidth]{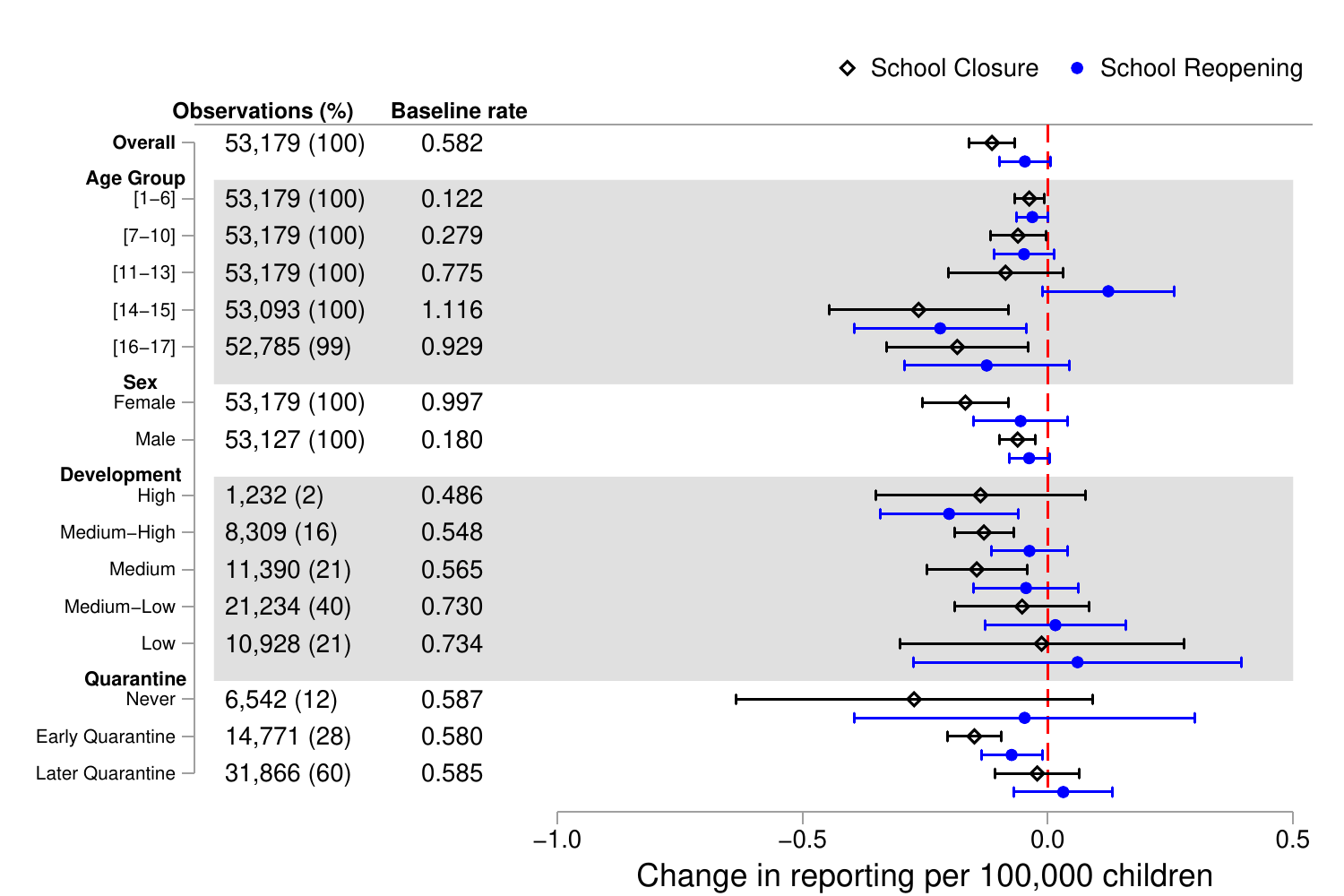}%
}
\end{center}
\floatfoot{\textbf{Notes to Fig.\ \ref{SIfig:heterogeneity1}}: Figure replicates results of Figure 2 of the main analysis, however here omitting all controls.  This corresponds to heterogeneity in estimates of columns (1), (4) and (7) of Table 1  (in panels (a), (b) and (c) respectively).  Refer to Notes to Figure 2 for further details.}
\end{figure}
    
\begin{figure}[t!]
\begin{center}
\caption{Alternative Specifications of Demographic and Socio-economic Variation in Closure/Reopening (Baseline FEs)}
\label{SIfig:heterogeneity2}
\subfloat[Intra-family Violence]{%
\includegraphics[width=0.44\textwidth]{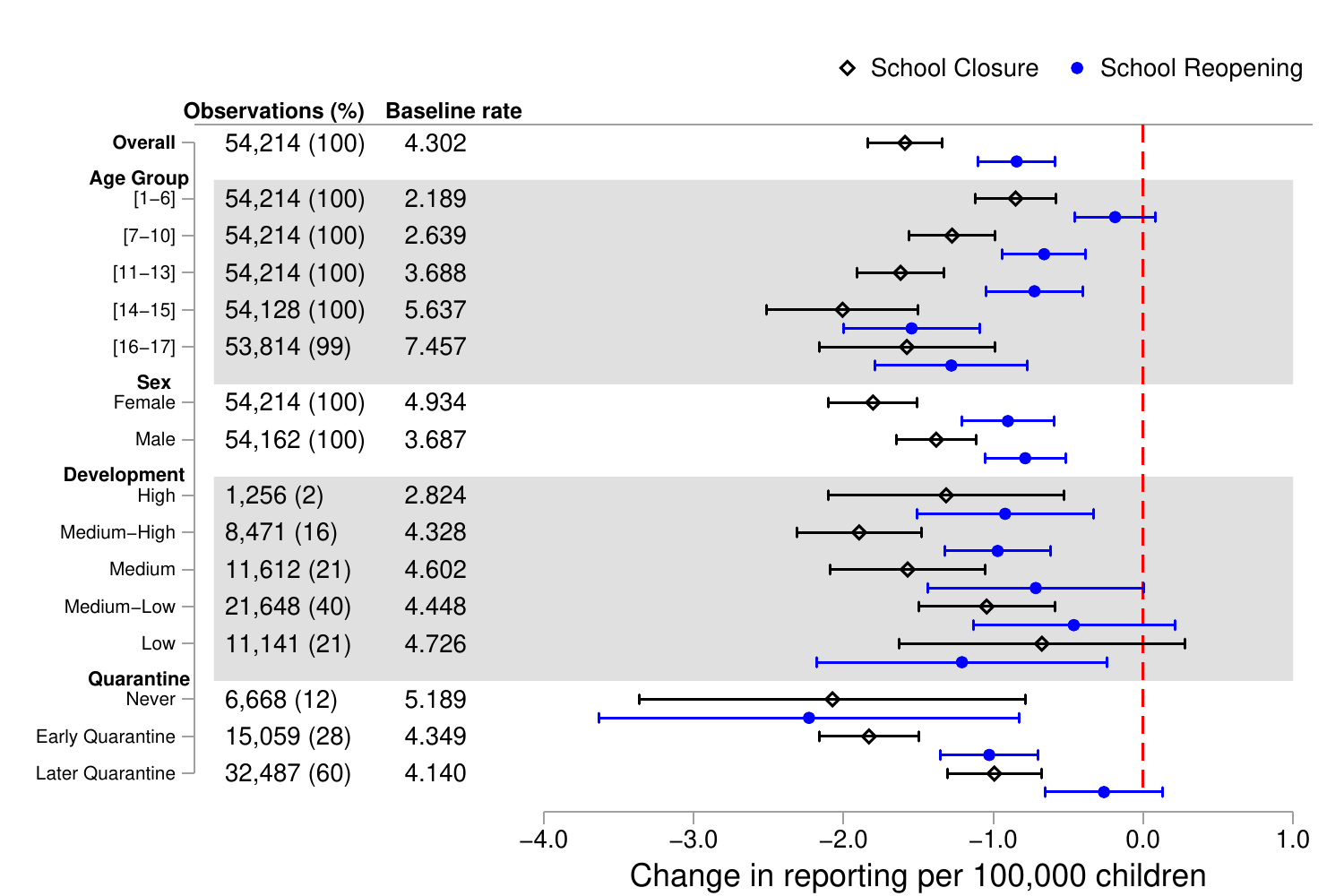}%
}\\
\subfloat[Sexual Abuse]{%
\includegraphics[width=0.44\textwidth]{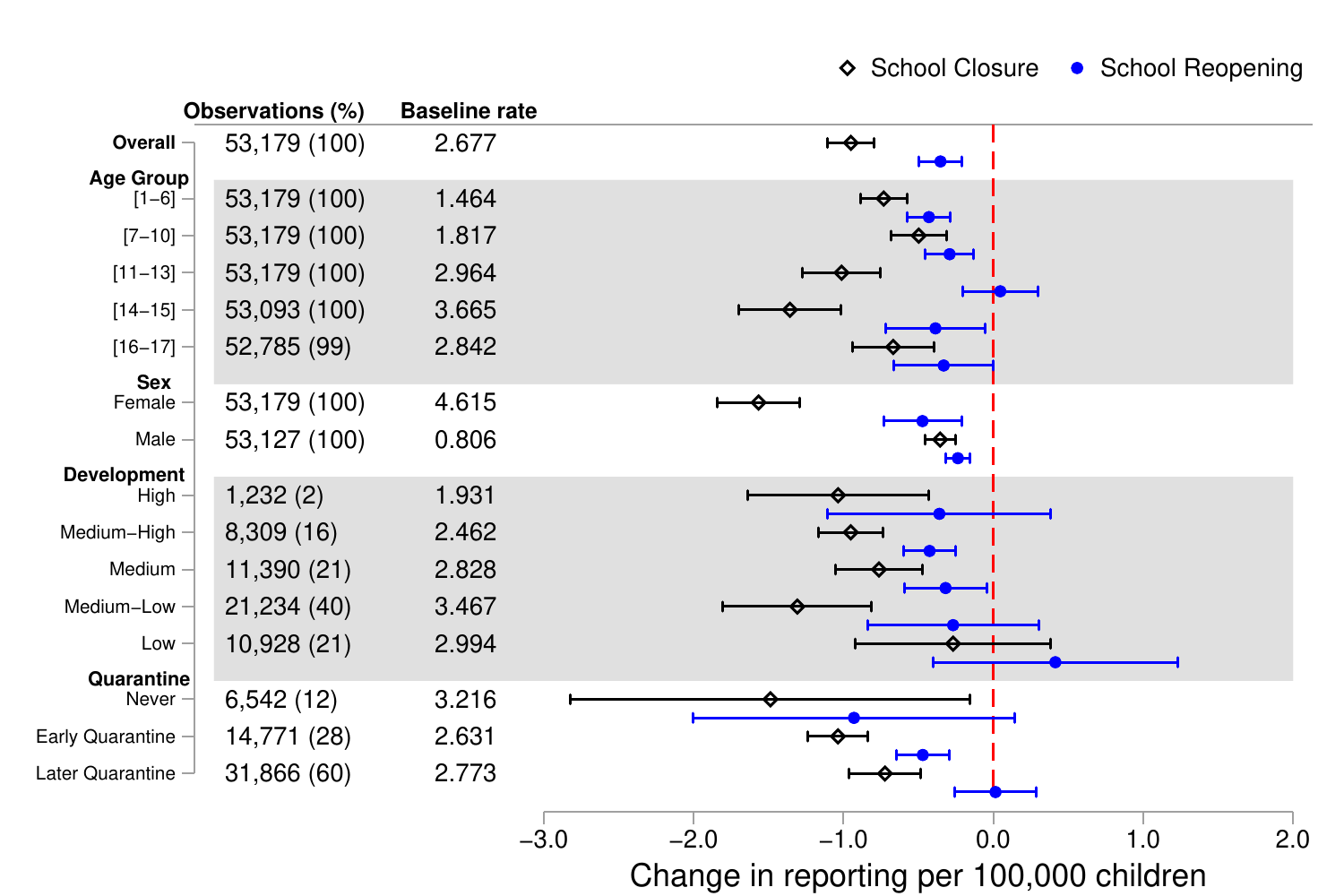}%
}\\
\subfloat[Rape]{%
\includegraphics[width=0.44\textwidth]{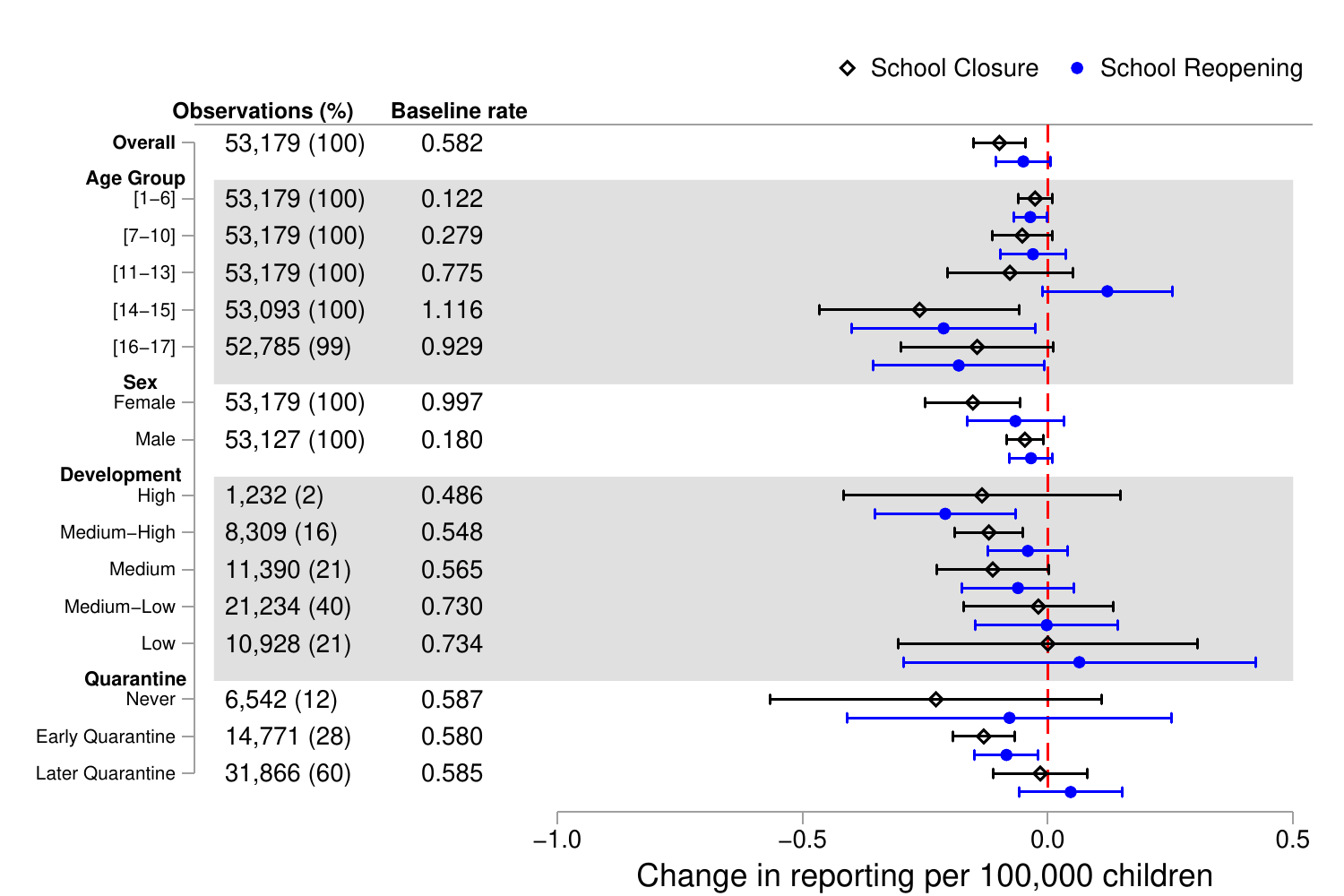}%
}
\end{center}
\floatfoot{\textbf{Notes to Fig.\ \ref{SIfig:heterogeneity2}}: Figure replicates results of Figure 2 of the main analysis, however here controlling only for municipal and week of year FEs.  This corresponds to heterogeneity in estimates of columns (2), (5) and (7) of Table 1  (in panels (a), (b) and (c) respectively).  Refer to Notes to Figure 2 for further details.}
\end{figure}

\begin{figure}[tbhp]
\begin{center}
\caption{School Attendance Proportions by Education Level in School Reopening Periods}
\label{SIfig:AttendanceAll}
\includegraphics[width=1\textwidth]{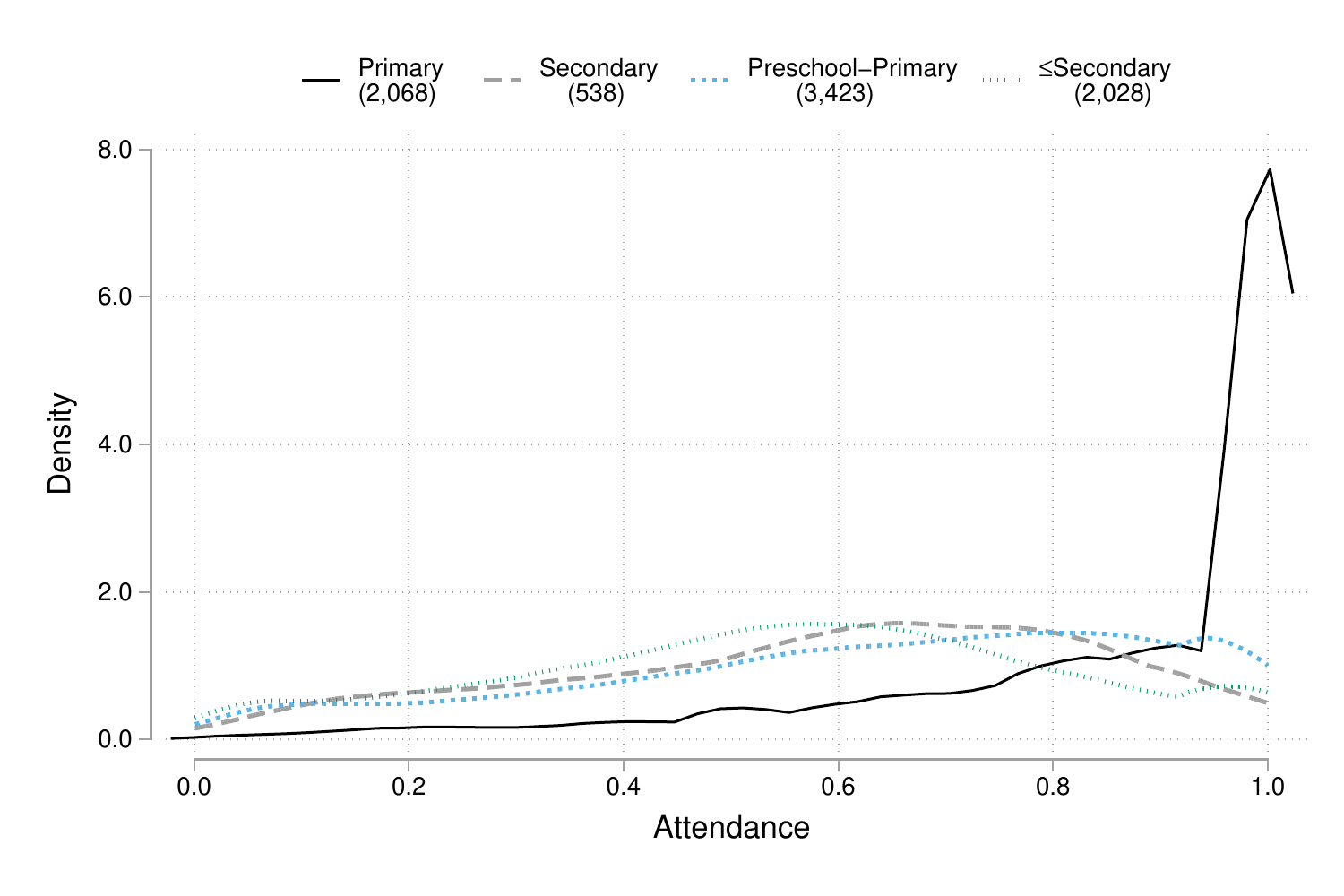}\\
\end{center}
\floatfoot{\textbf{Notes to Fig.\ \ref{SIfig:AttendanceAll}}: Kernel densities of the proportion of individuals observed to attend school at least one day in each month are displayed by school type.  Here densities are documented over all schools in each month for which attendance data is available and in which any attendance occurs.  Values of 1 imply that all students attend school (at least 1 day) in a given school in a given month.  Schools are stratified by level as Primary (only covering primary classes), Secondary (only covering Secondary classes), Preschool-Primary (covering both pre-school and primary, but not secondary), and $\leq$ Secondary refers to schools which have secondary as well as earlier levels (eg secondary and primary levels).  Numbers below school types refer to the total number of each type of school in the country (a number of schools are not included here, as these are registered for adult learning, or as special education, and potentially cover all grades).}
\end{figure}

\begin{figure}[t!]
\begin{center}
\caption{School Attendance Proportion by Education Level and by Month in School Reopening Periods}
\label{SIfig:AttendanceGroups}
\subfloat[Primary level]{%
\includegraphics[width=0.5\textwidth]{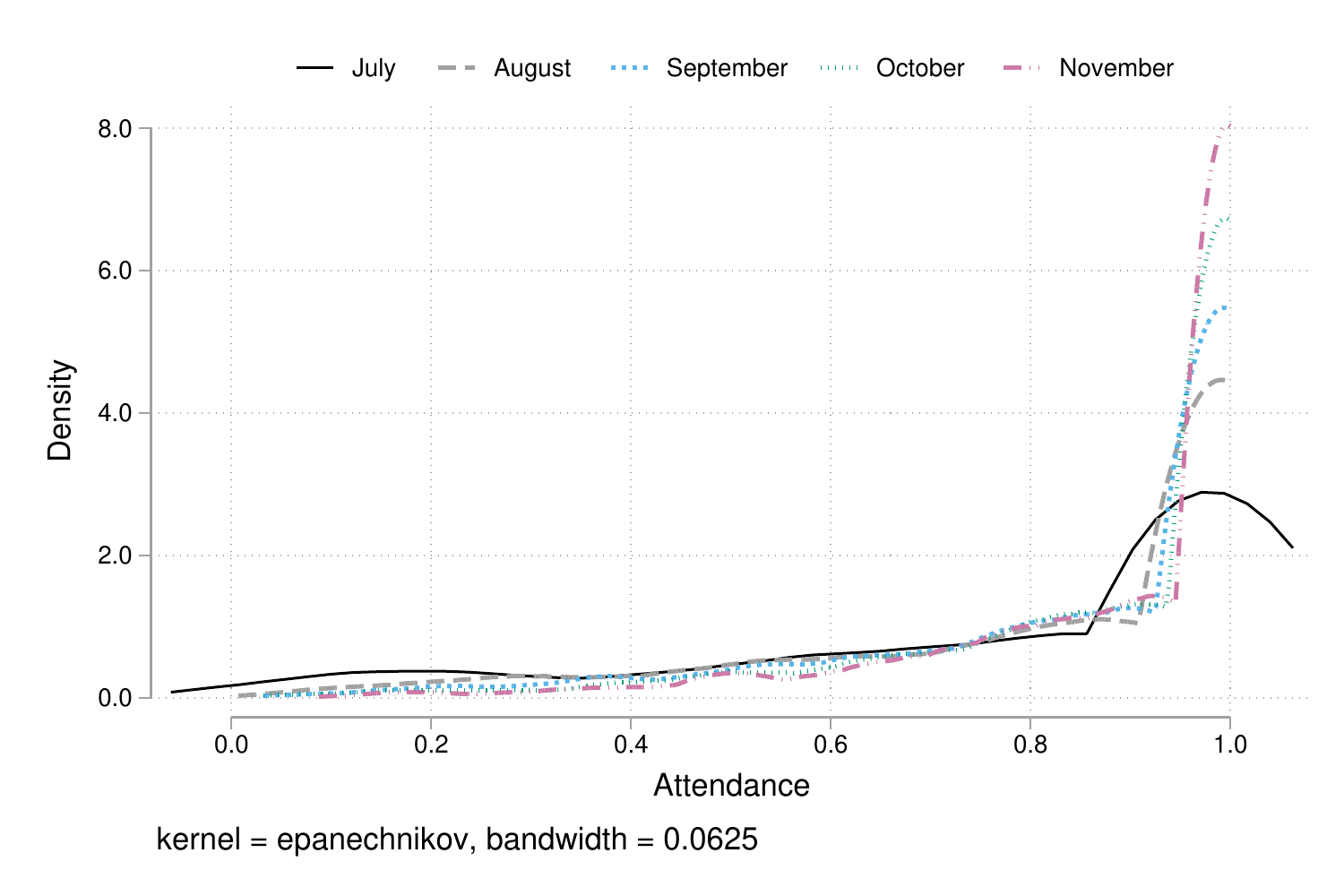}%
}
\subfloat[Secondary level]{%
\includegraphics[width=0.5\textwidth]{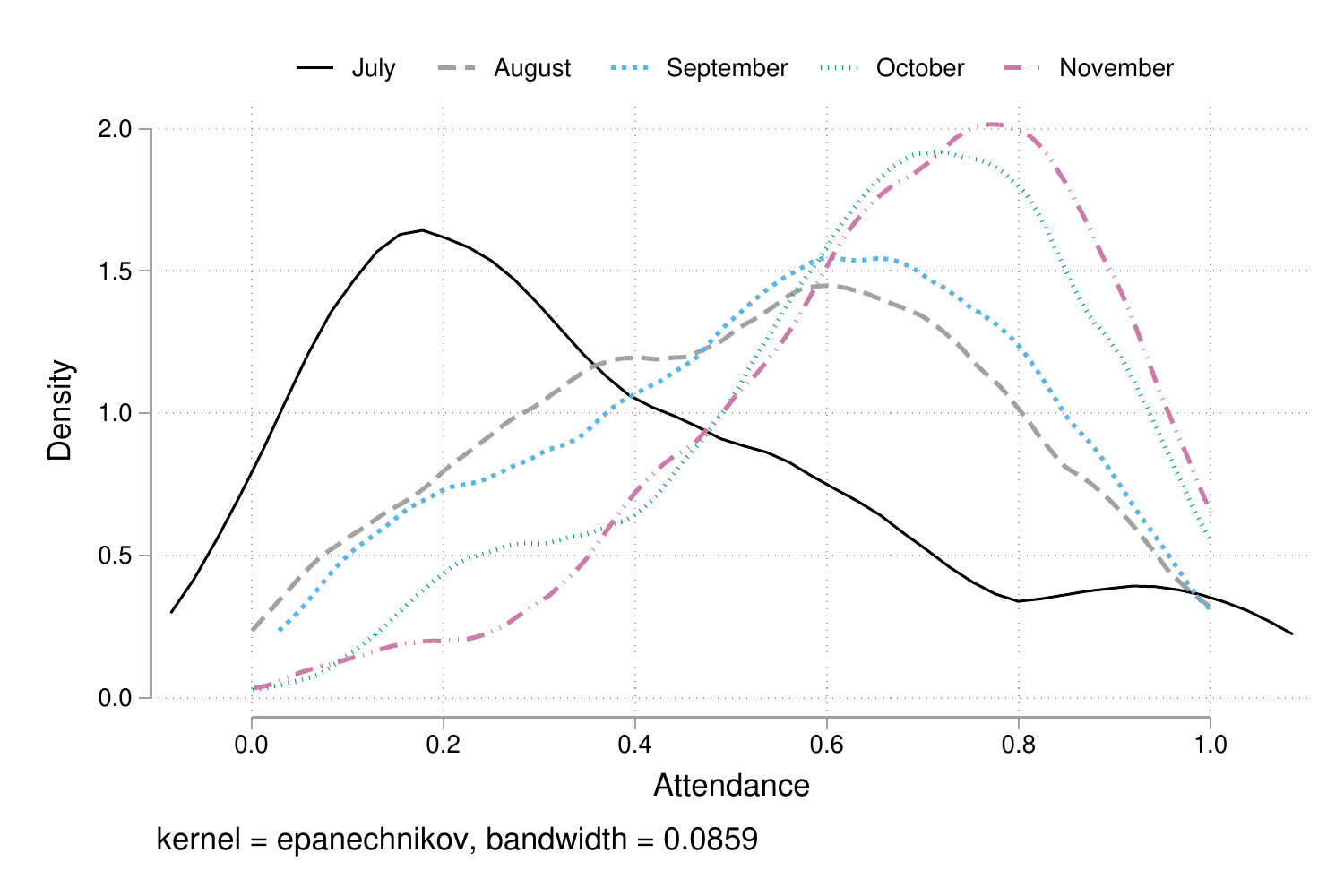}%
}\\
\subfloat[Preschool and primary level]{%
\includegraphics[width=0.5\textwidth]{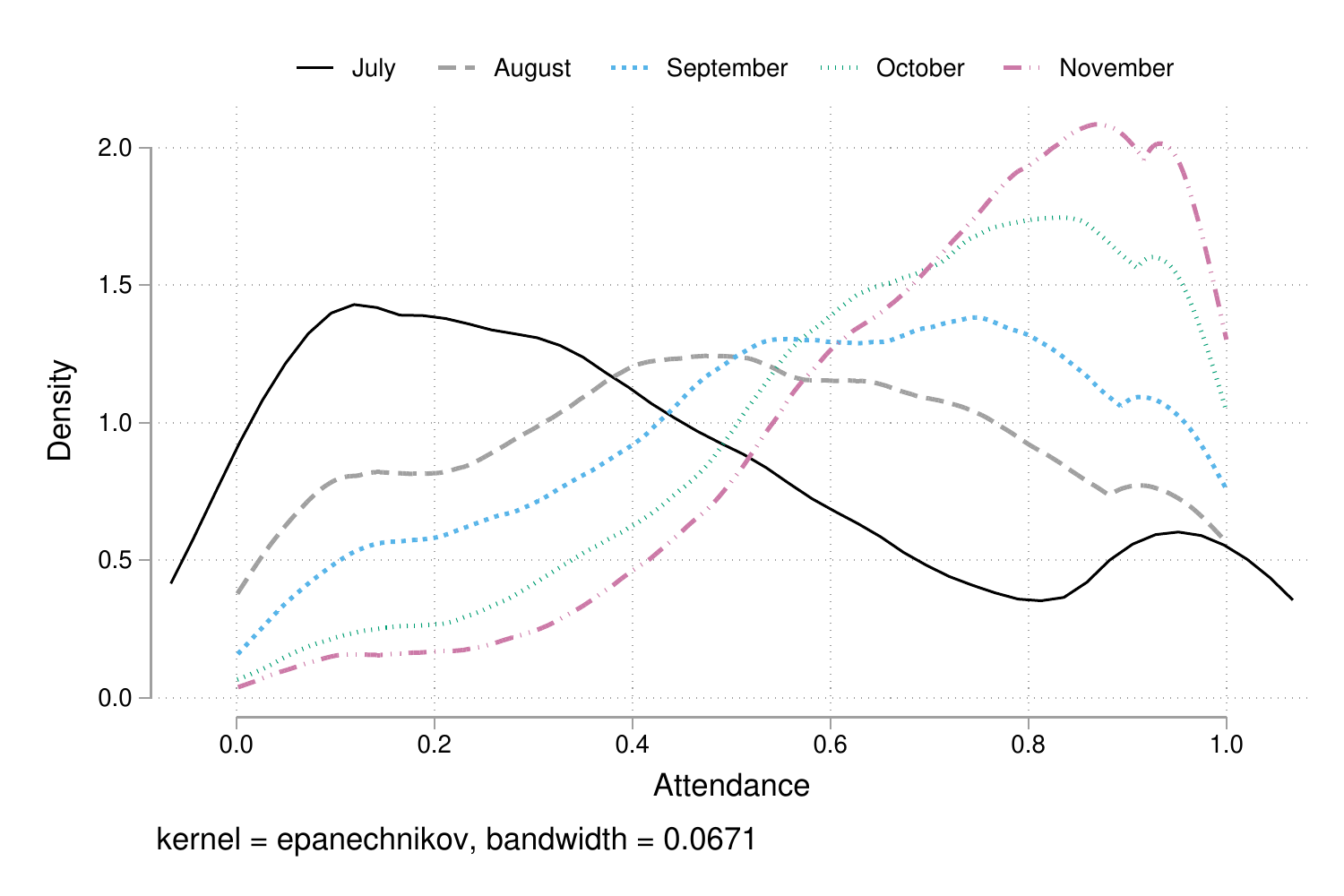}%
}
\subfloat[Less than or equal to secondary level]{%
\includegraphics[width=0.5\textwidth]{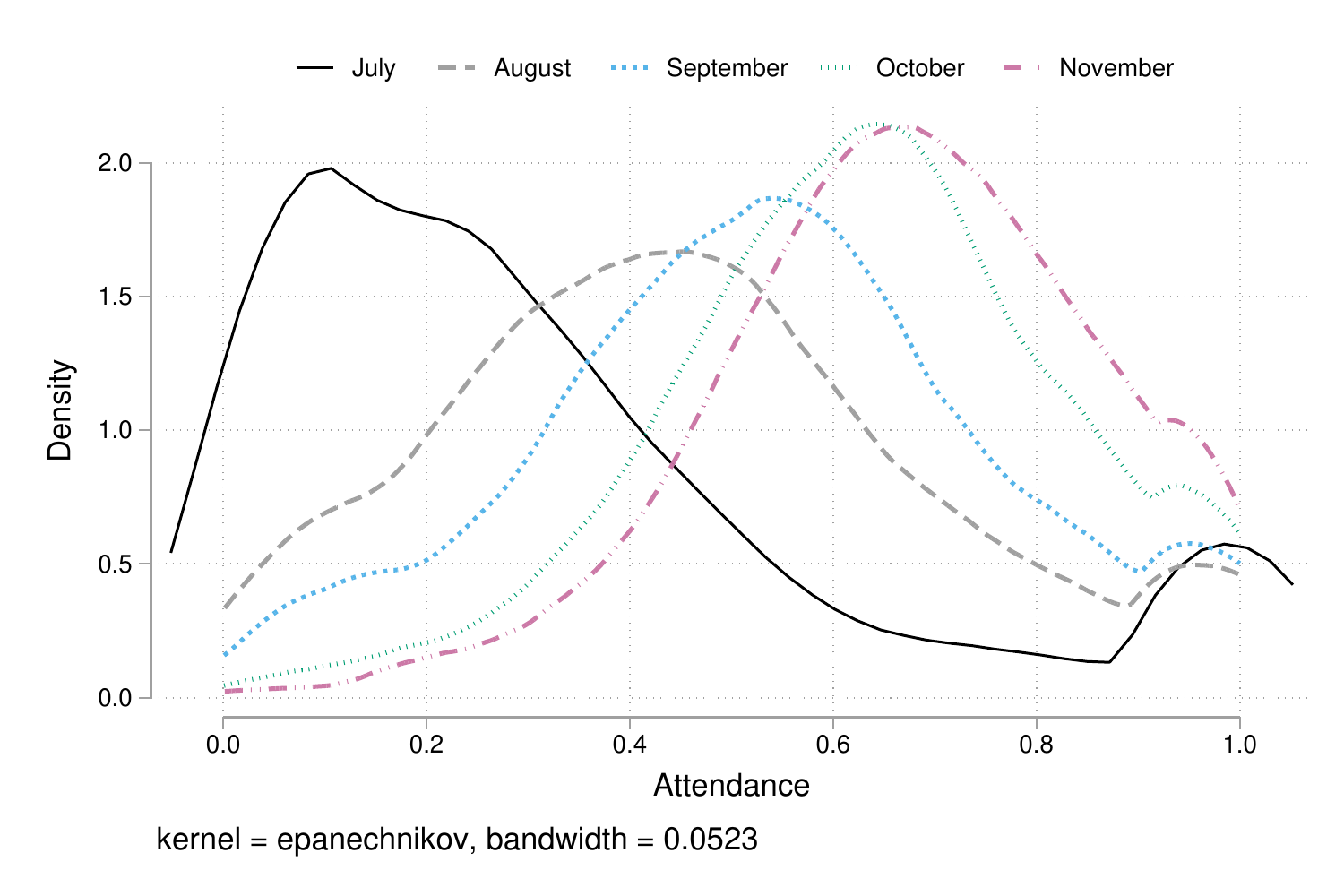}%
}
\end{center}
\floatfoot{\textbf{Notes to Fig.\ \ref{SIfig:AttendanceGroups}}: Refer to notes to Figure \ref{SIfig:AttendanceAll}. Here similar school attendance frequency distributions are documented, however now a single plot is provided for each education level, where densities are observed over all schools of each type in each re-opening month for which attendance data is available (July--December, 2021).}
\end{figure}

\begin{figure}[htpb!]
\begin{center}
\caption{Tests of Differences between Re-opening and Closing in Demographic and Socio-economic Heterogeneity Analyses}
\label{SIfig:heterogeneity3}
\subfloat[School Closure: Intra-family Violence]{%
\includegraphics[width=0.49\textwidth]{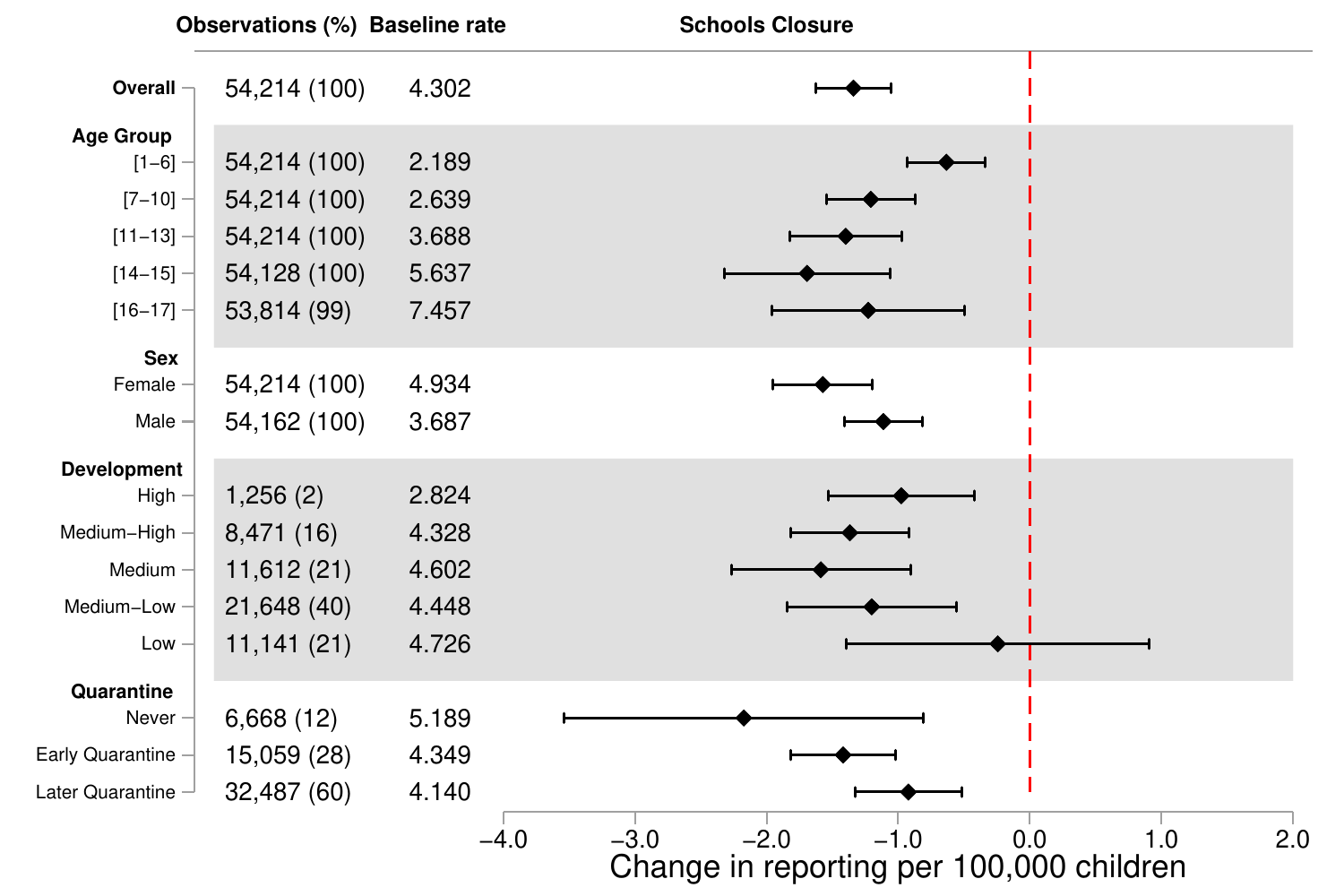}%
}
\subfloat[School Opening: Intra-family Violence]{%
\includegraphics[width=0.49\textwidth]{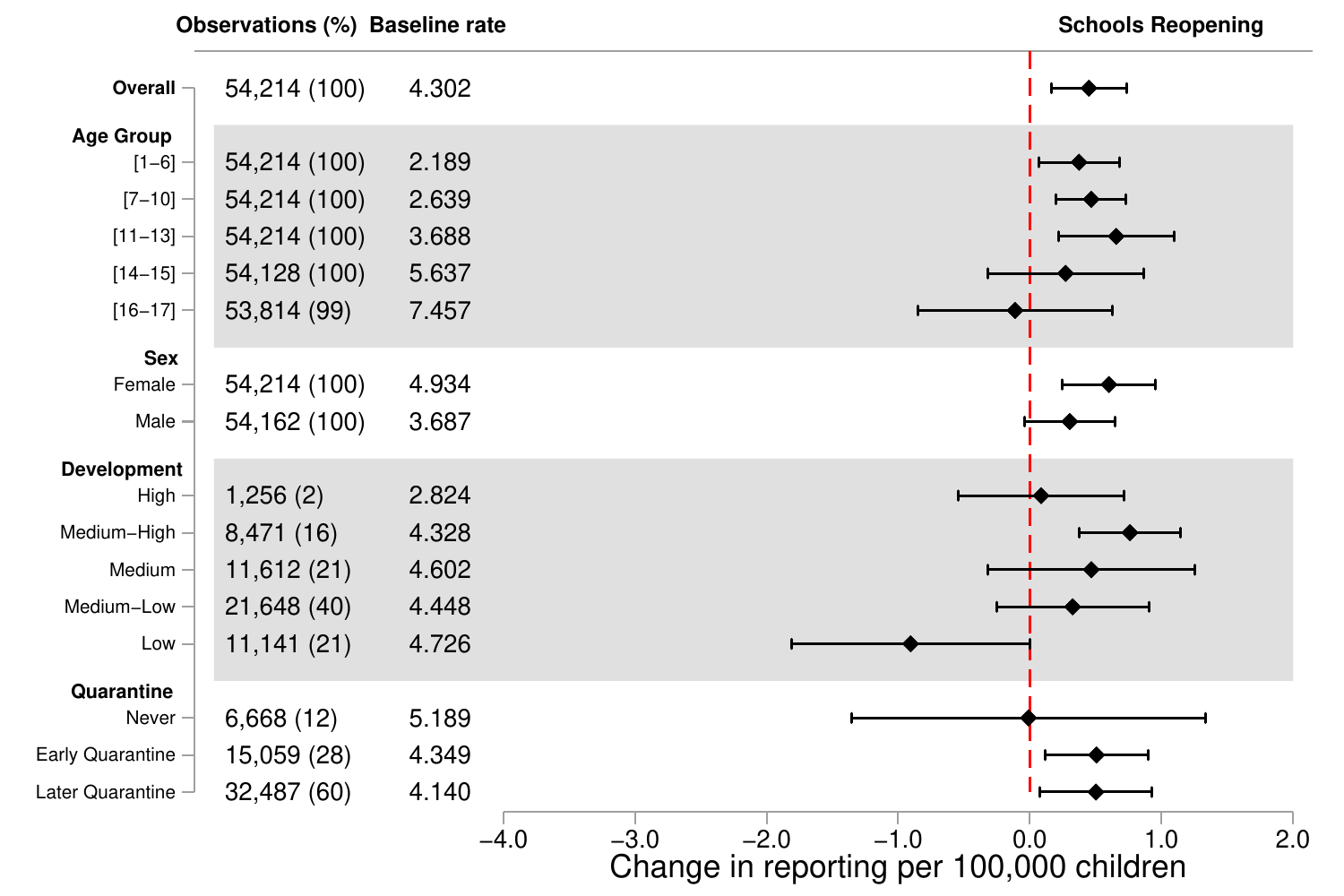}%
}\\
\subfloat[School Closure: Sexual Abuse]{%
\includegraphics[width=0.49\textwidth]{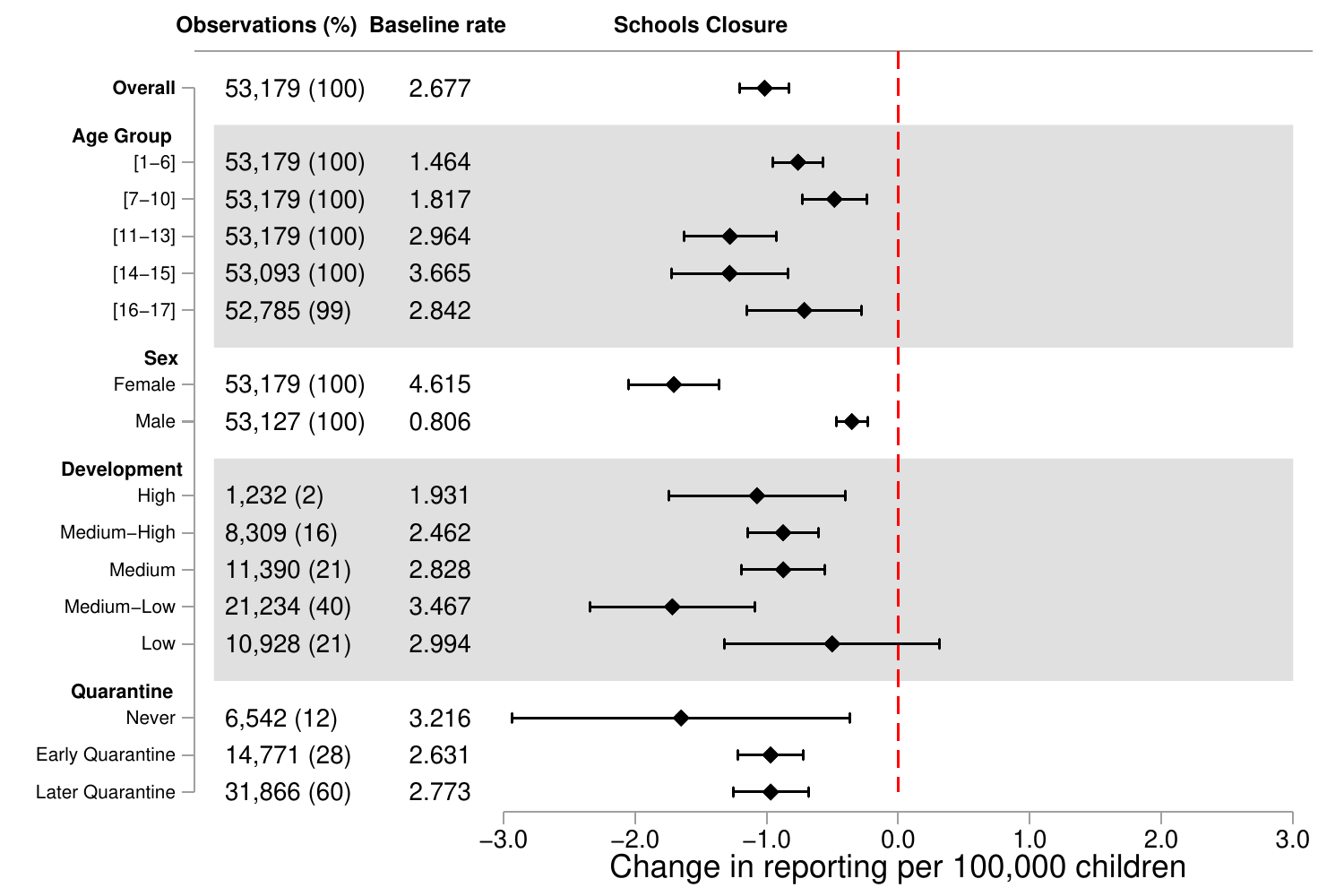}%
}
\subfloat[School Opening: Sexual Abuse]{%
\includegraphics[width=0.49\textwidth]{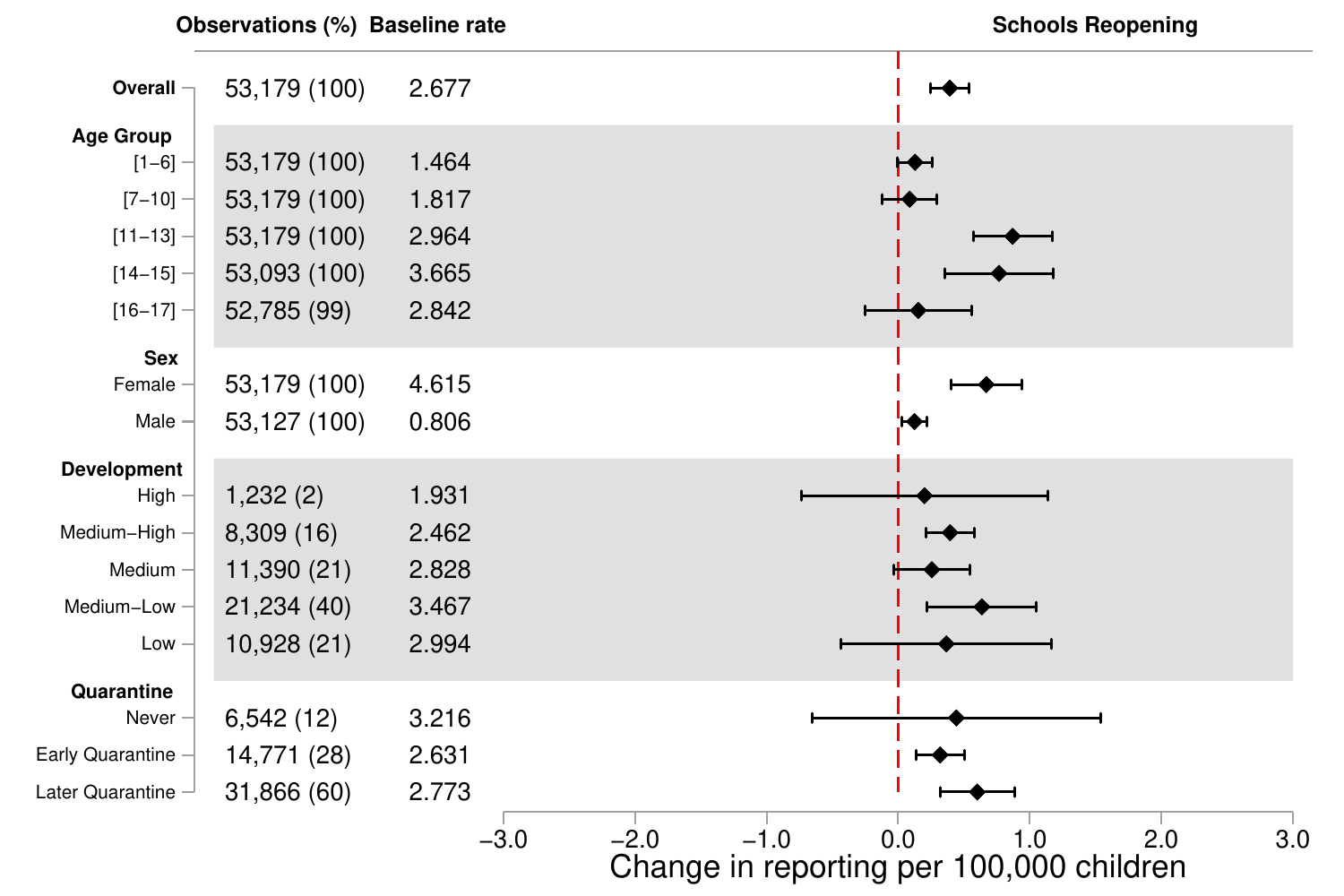}%
}\\
\subfloat[School Closure: Rape]{%
\includegraphics[width=0.49\textwidth]{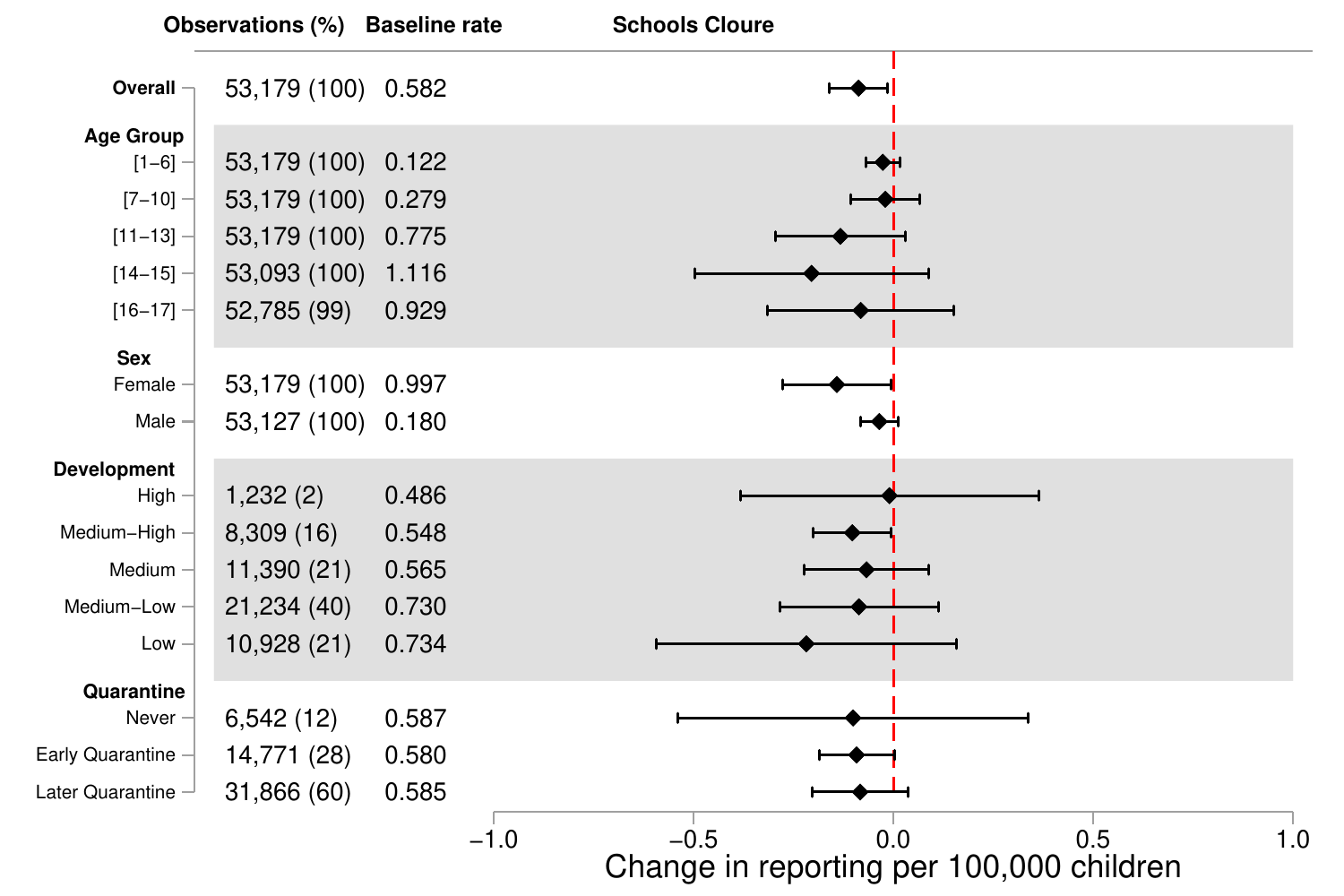}%
}
\subfloat[School Opening: Rape]{%
\includegraphics[width=0.49\textwidth]{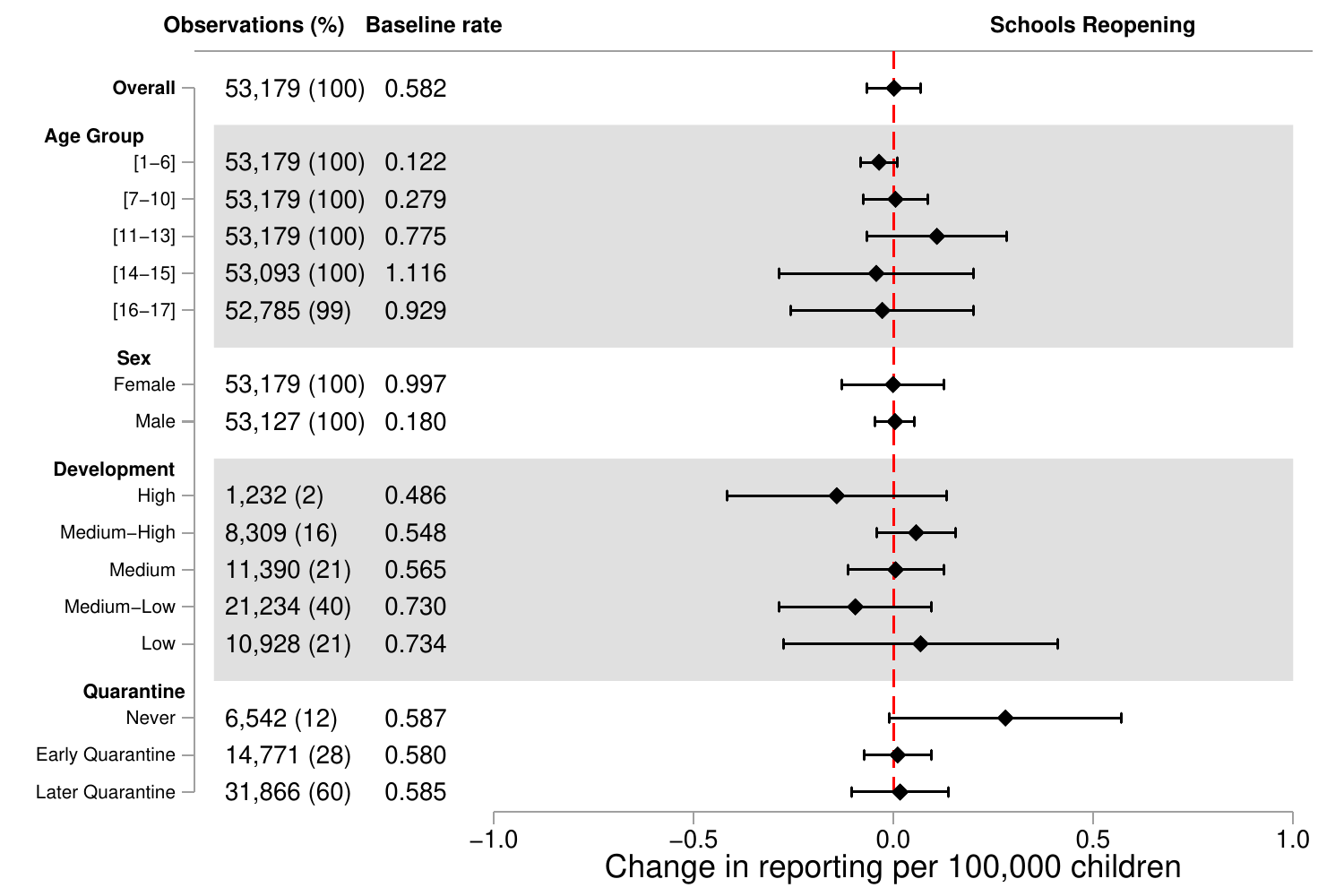}%
}\\
\end{center}
\floatfoot{\textbf{Notes to Fig.\ \ref{SIfig:heterogeneity3}}: Results replicate those of Figure 2 of main analysis, however now rather than comparing each of closure and re-opening to baseline periods, the effect of re-opening is compared to the closure period.  Left-hand panels present the impact of school closure as compared to pre-closure periods (replicating black diamonds and CIs from Figure 2), while right-hand panels present the impact of school re-opening, compared to closure periods.  Thus, if estimates in the right hand panel are above zero, this implies a significant increase in reporting in this group in the post-reopening period, compared with changes observed in the right-hand panel.  All other details follow those in Figure 2.}
\end{figure}

\begin{figure}[t!]
\begin{center}
\caption{Alternative Counterfactual Models -- Including Additional Controls}
\label{SIfig:counterfactualsV4}
\textbf{Panel A: Simple counterfactual (time only)} \\
\subfloat[Intra-family Violence]{%
\includegraphics[width=0.33\textwidth]{./graphs/C1_V_2018_lineal}%
}
\subfloat[Sexual Abuse]{%
\includegraphics[width=0.33\textwidth]{./graphs/C1_SA_2018_cuadratic}%
}
\subfloat[Rape]{%
\includegraphics[width=0.33\textwidth]{./graphs/C1_R_2018_cuadratic}%
}
\\
\textbf{Panel B: Counterfactual (No school channel + Epidemiological controls)} \\
\subfloat[Intra-family Violence]{%
\includegraphics[width=0.33\textwidth]{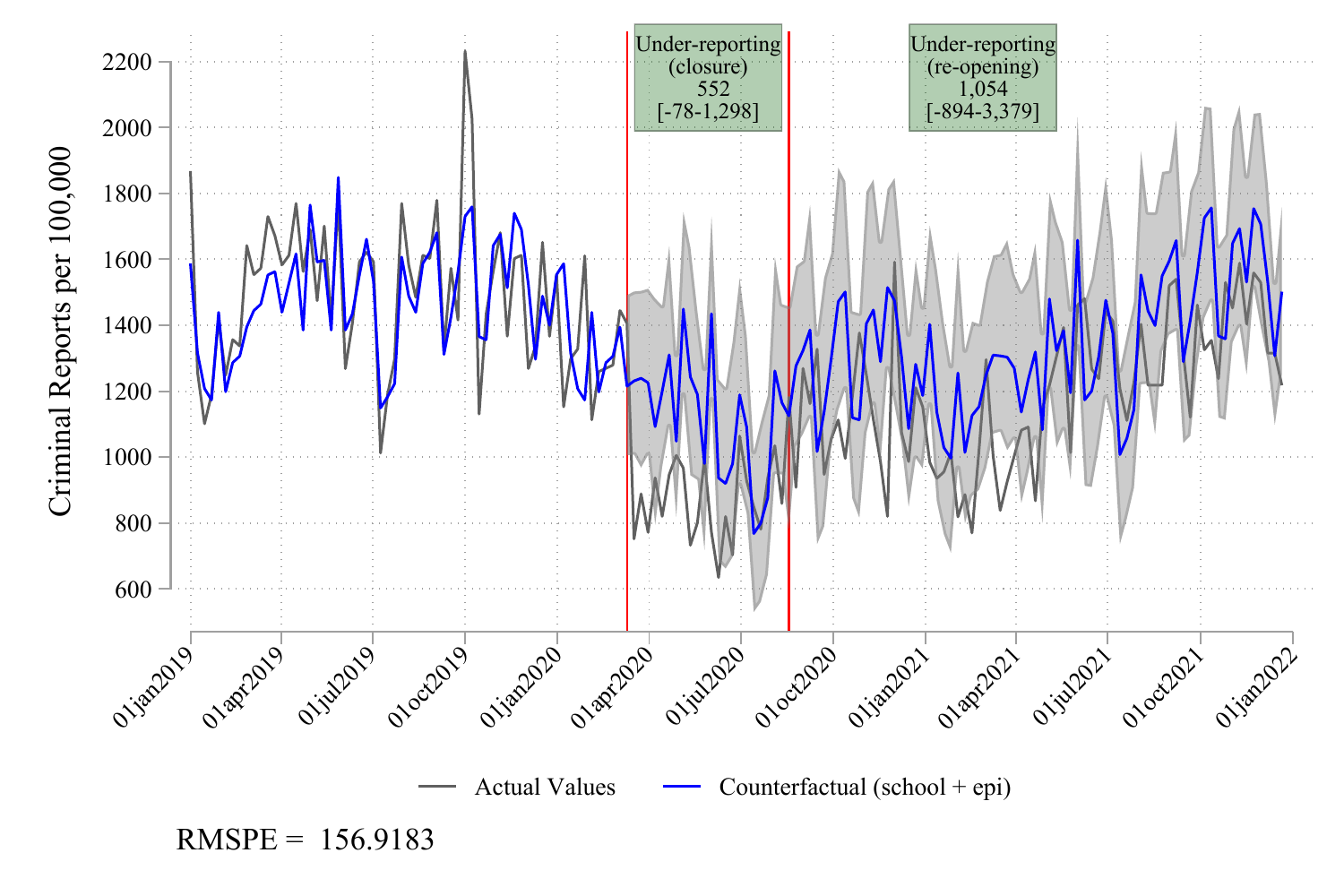}%
}
\subfloat[Sexual Abuse]{%
\includegraphics[width=0.33\textwidth]{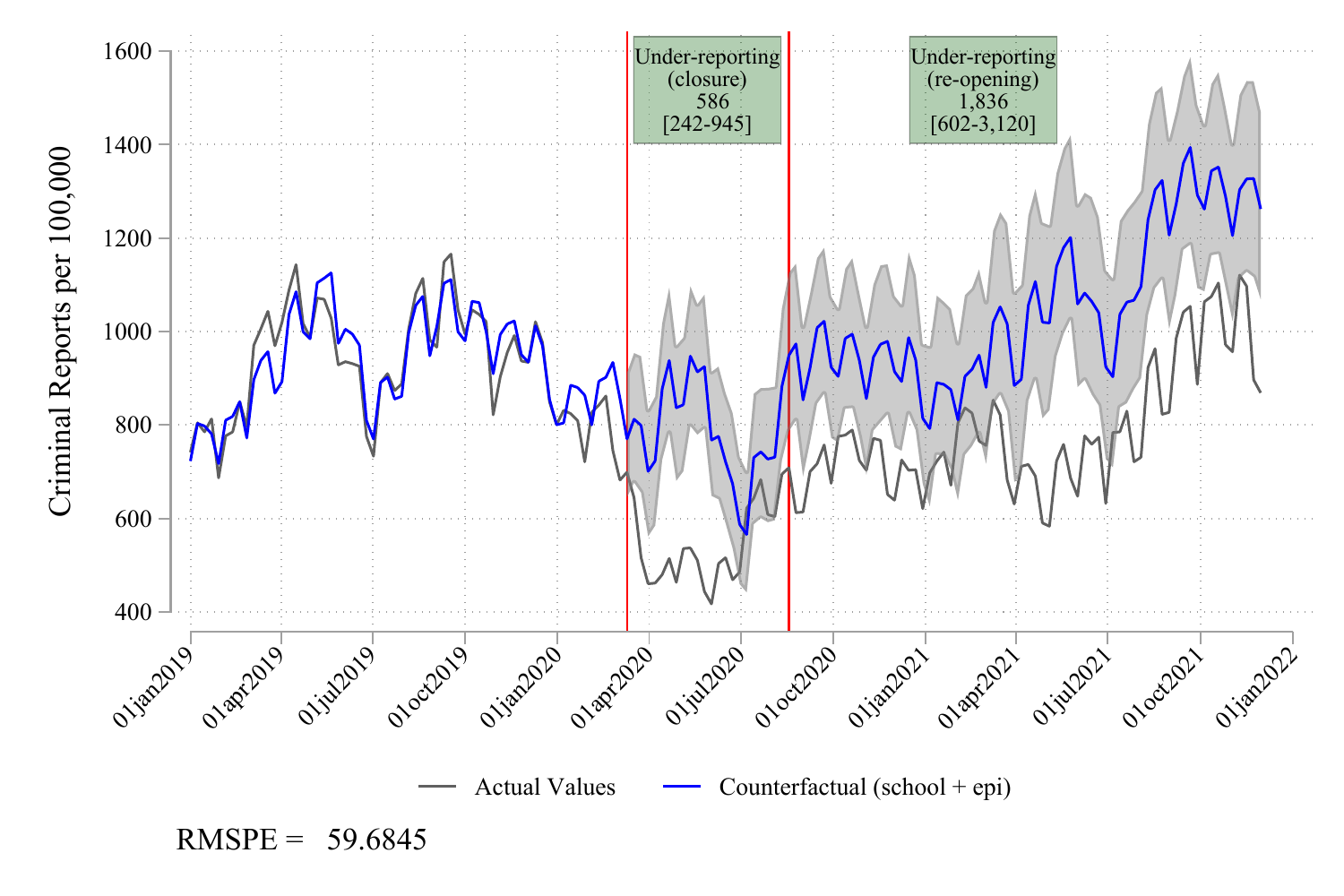}%
}
\subfloat[Rape]{%
\includegraphics[width=0.33\textwidth]{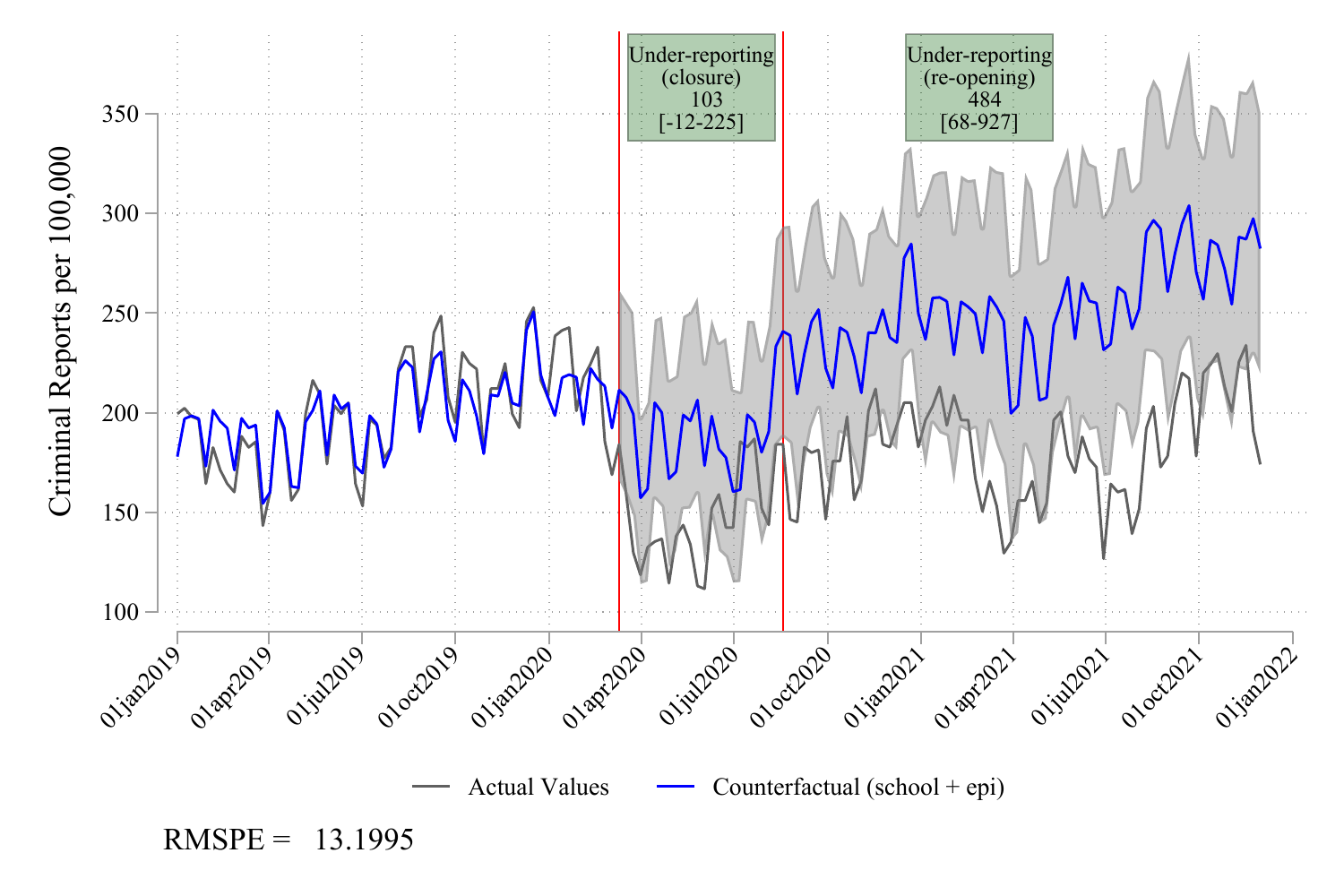}%
}\\
\textbf{Panel C: Projected under-reporting} \\
\subfloat[Intra-family Violence]{%
\includegraphics[width=0.33\textwidth]{./graphs/diff_Count_1_V_2018_lineal}%
}
\subfloat[Sexual Abuse]{%
\includegraphics[width=0.33\textwidth]{./graphs/diff_Count_1_SA_2018_cuadratic}%
}
\subfloat[Rape]{%
\includegraphics[width=0.33\textwidth]{./graphs/diff_Count_1_R_2018_cuadratic}%
}
\end{center}
\floatfoot{\textbf{Notes to Fig.\ \ref{SIfig:counterfactualsV4}}: Results replicate those of Figure 2 of main analysis, however here in Panel B additionally include epidemiological controls used in all main models, as well as controls for the schooling channel documented in Figure 2.  Panels A and C are identical, and are displayed in the interest of comparison.  All other details follow Figure 2.}
\end{figure}

\begin{figure}[t!]
\begin{center}
\caption{Alternative Counterfactual Models -- Optimal Based on 2015 Onwards}
\label{SIfig:counterfactuals2015}
\textbf{Panel A: Simple counterfactual (time only)} \\
\subfloat[Intra-family Violence]{%
\includegraphics[width=0.33\textwidth]{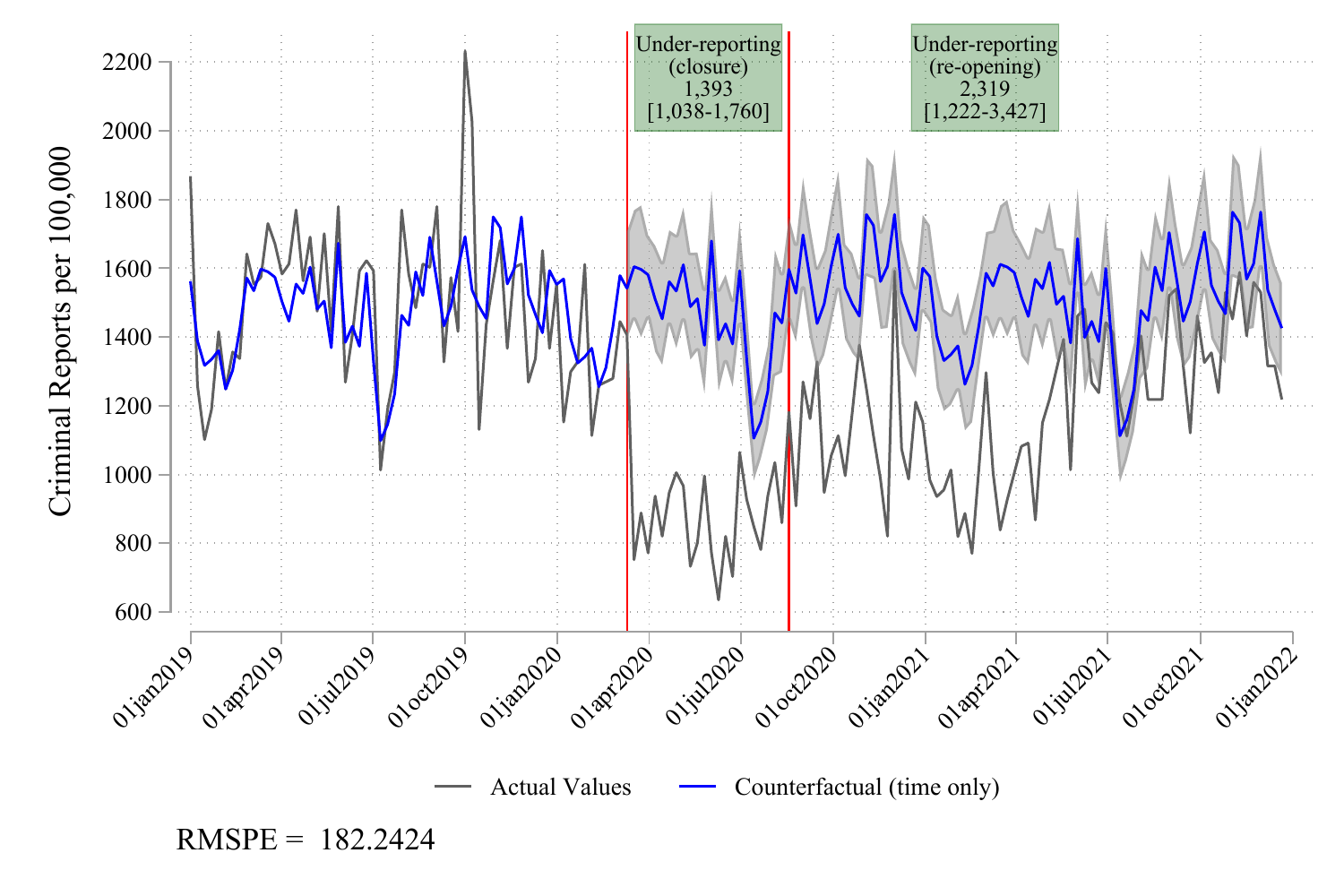}%
}
\subfloat[Sexual Abuse]{%
\includegraphics[width=0.33\textwidth]{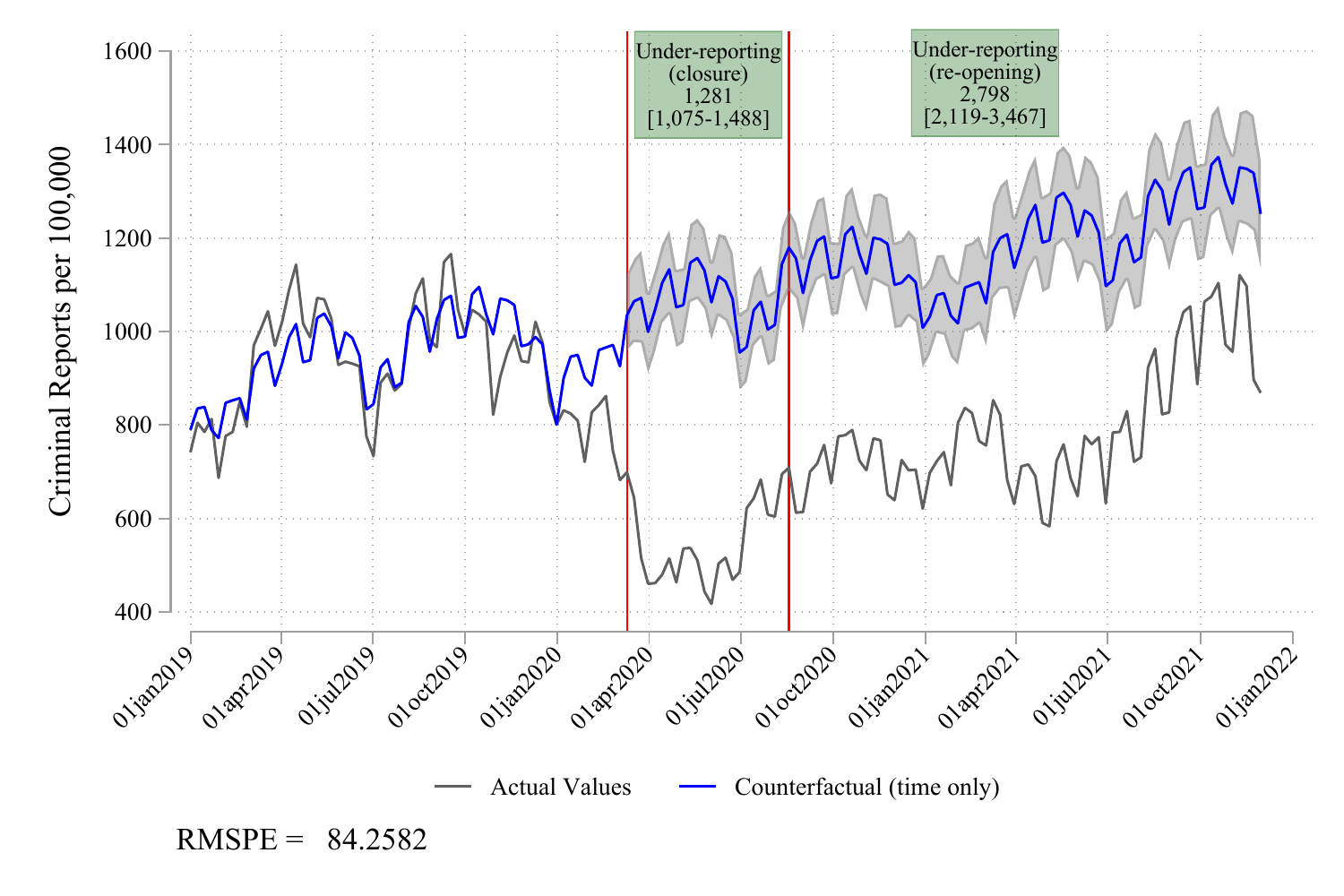}%
}
\subfloat[Rape]{%
\includegraphics[width=0.33\textwidth]{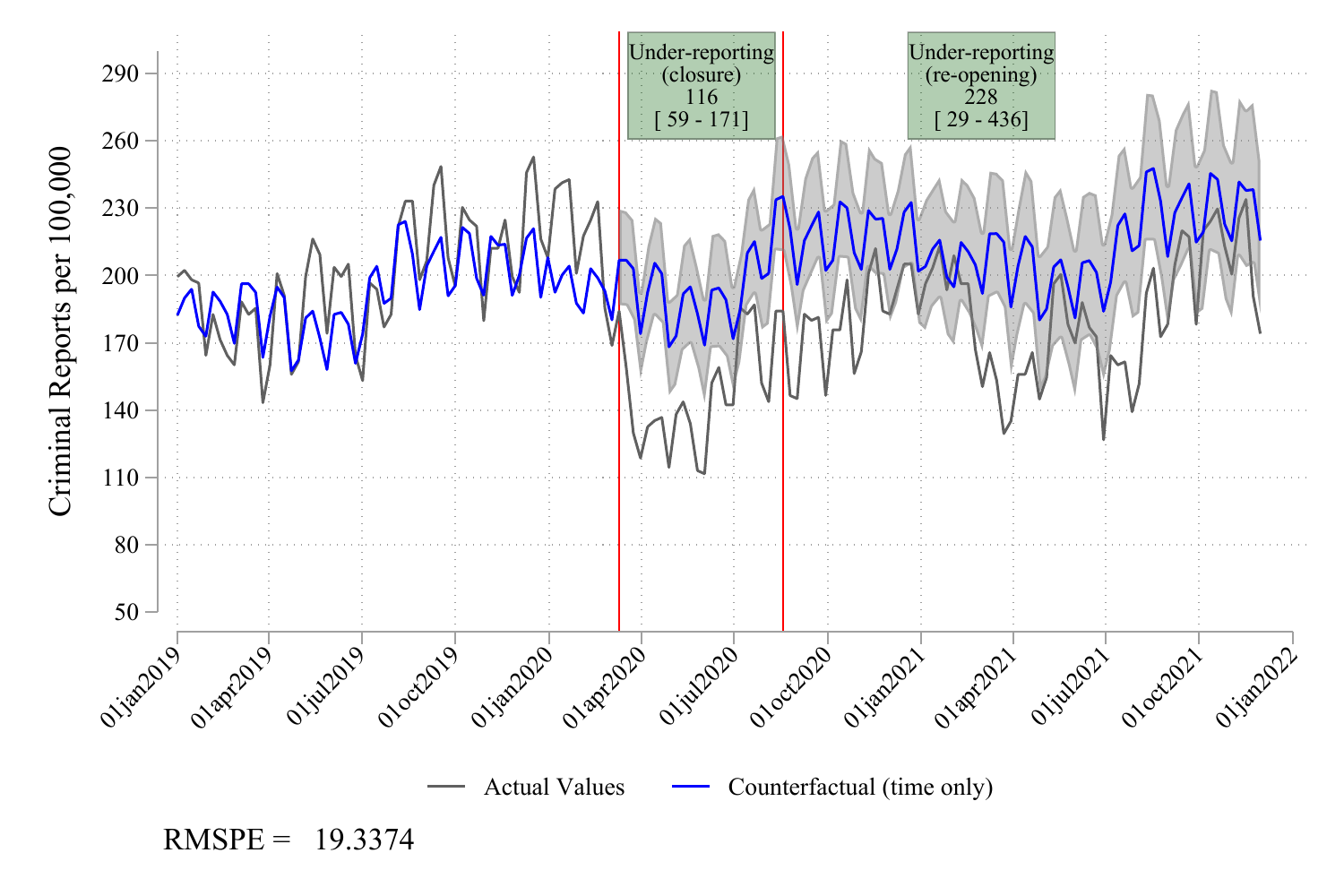}%
}
\\
\textbf{Panel B: Counterfactual (No school channel)} \\
\subfloat[Intra-family Violence]{%
\includegraphics[width=0.33\textwidth]{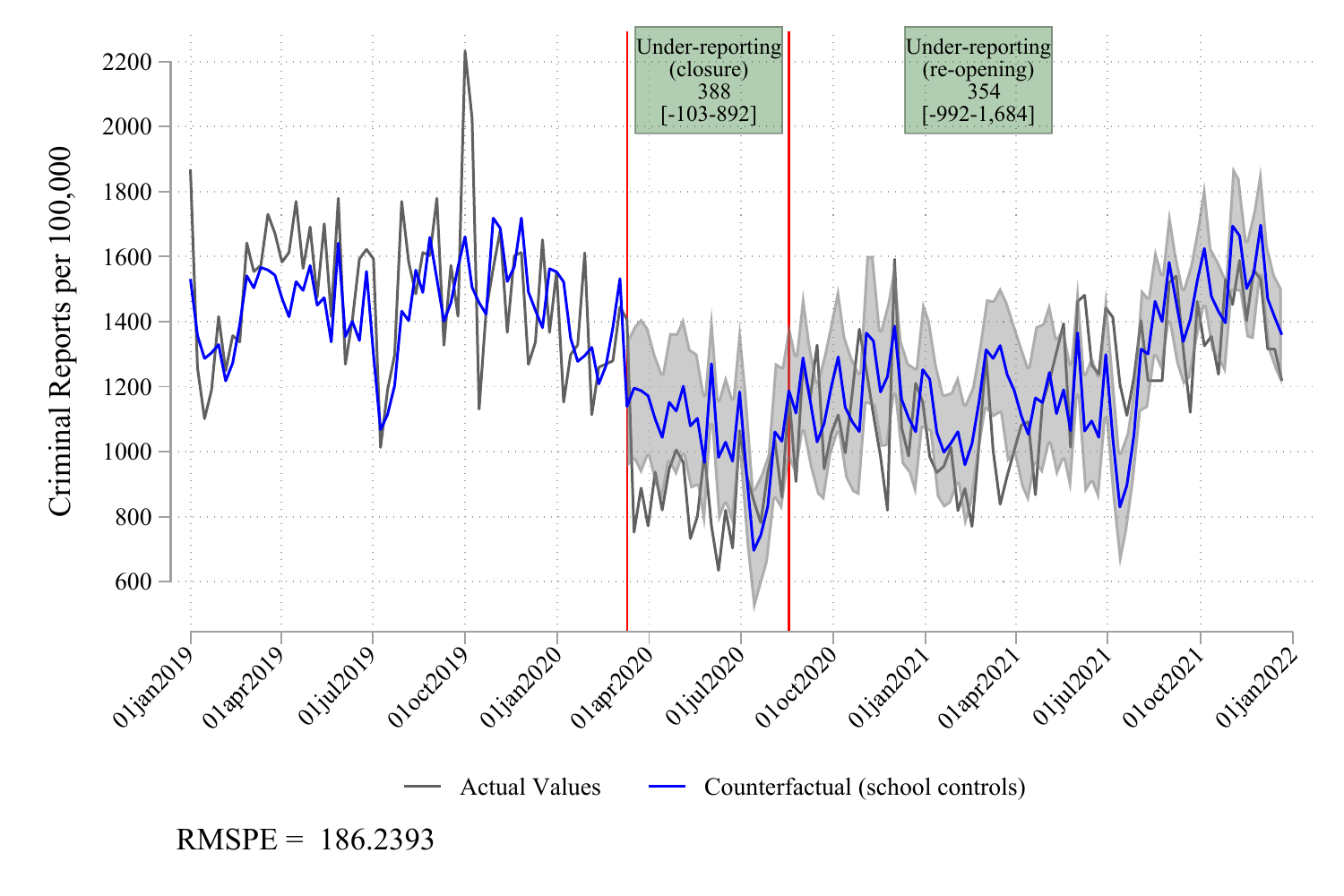}%
}
\subfloat[Sexual Abuse]{%
\includegraphics[width=0.33\textwidth]{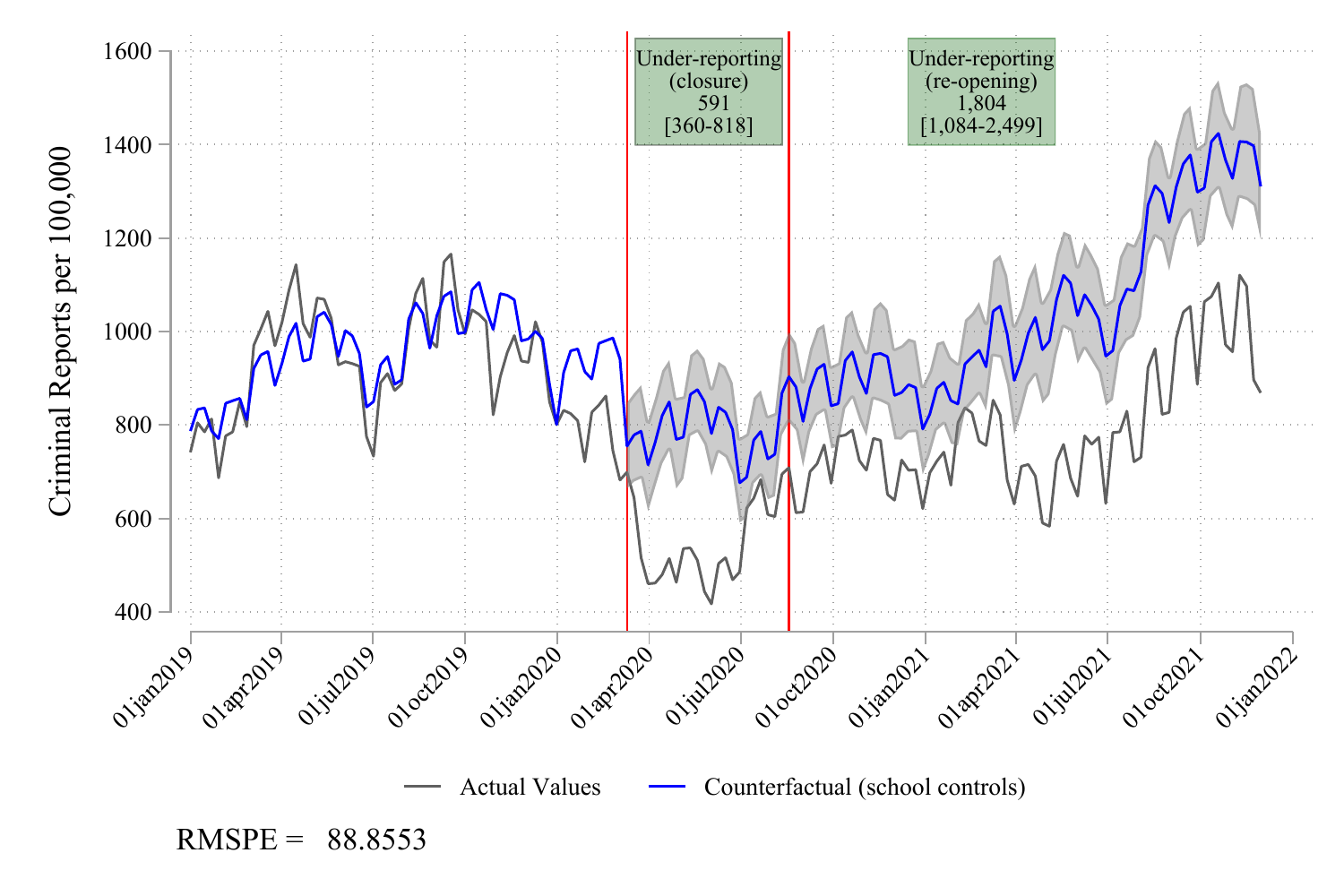}%
}
\subfloat[Rape]{%
\includegraphics[width=0.33\textwidth]{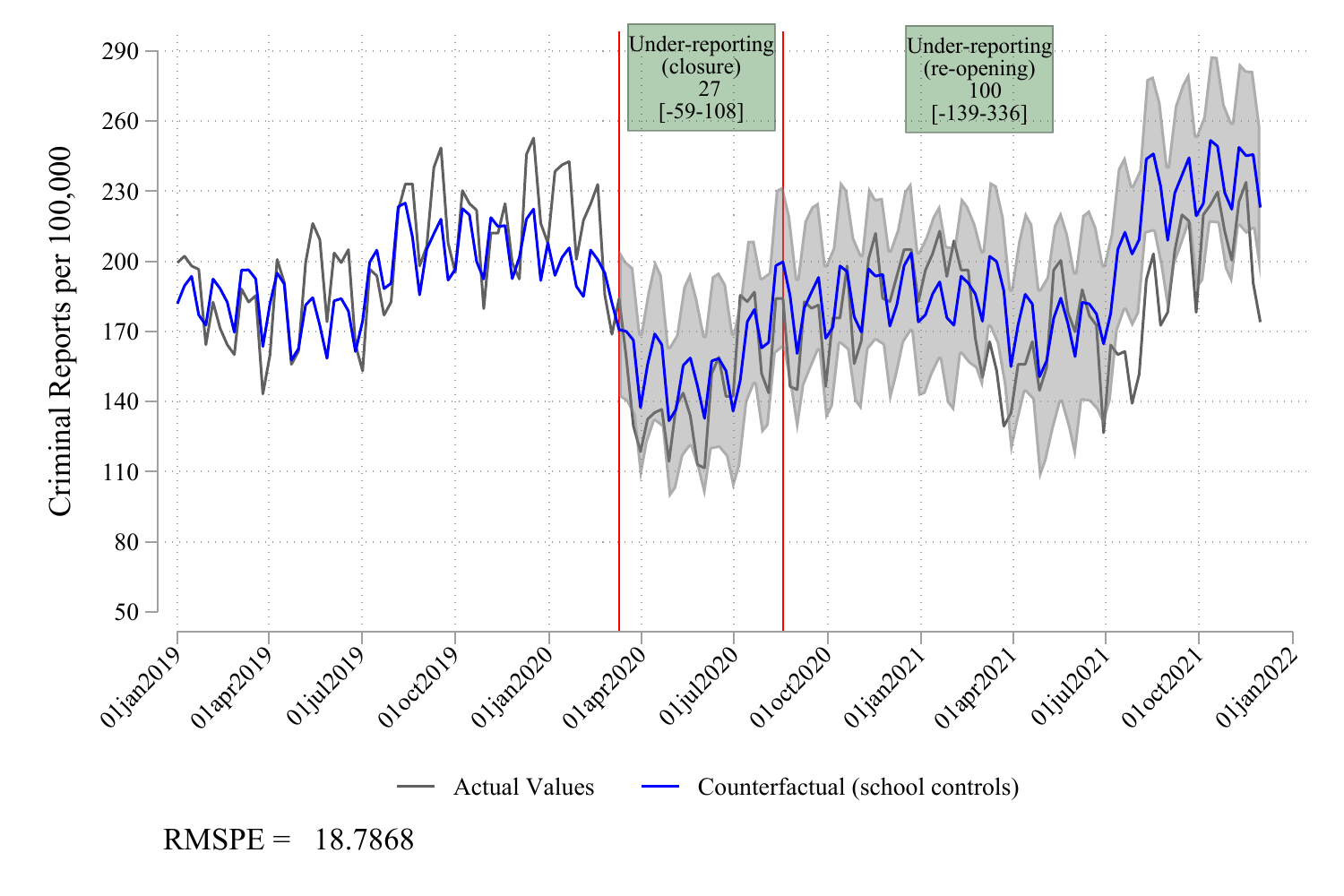}%
}\\
\textbf{Panel C: Projected under-reporting} \\
\subfloat[Intra-family Violence]{%
\includegraphics[width=0.33\textwidth]{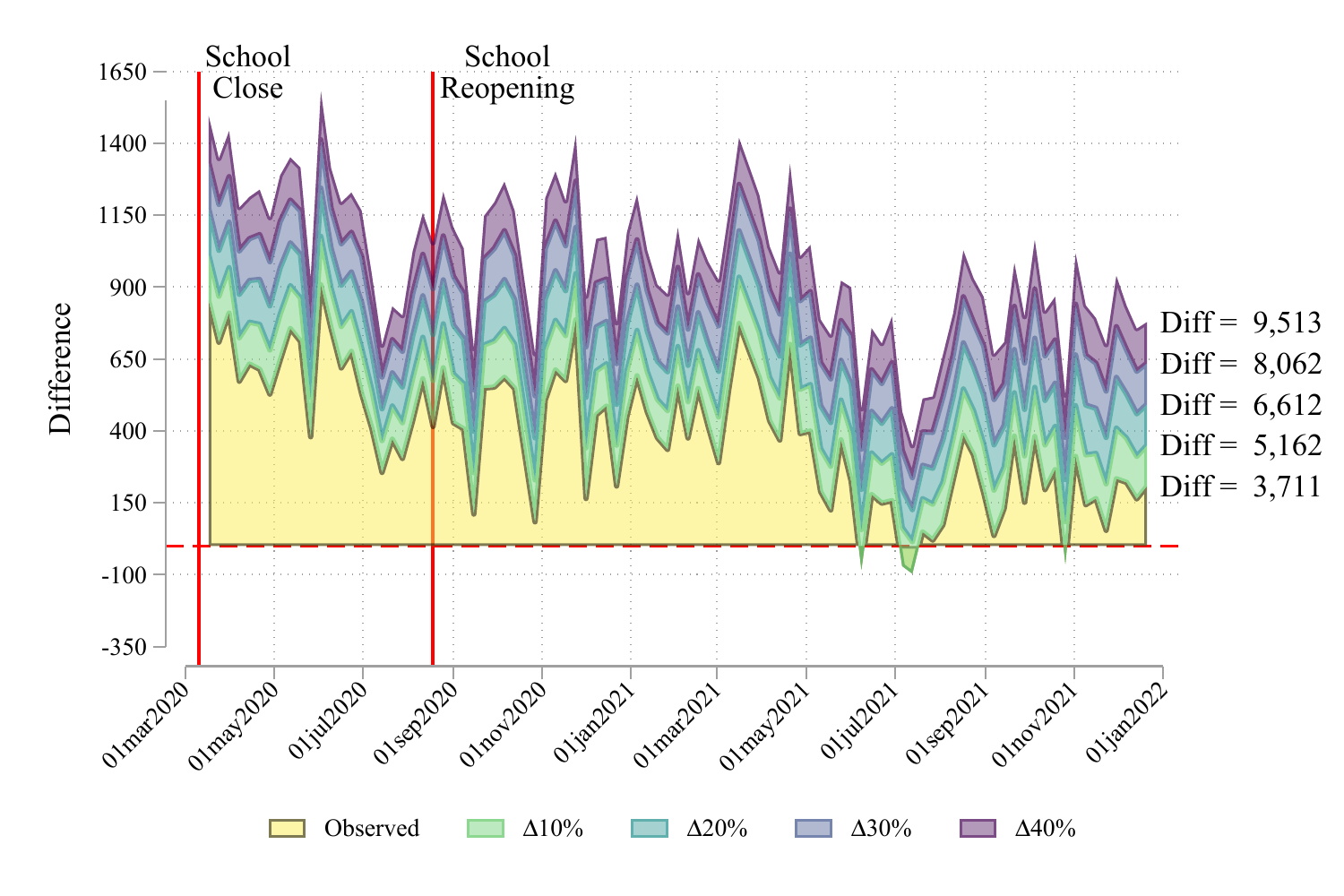}%
}
\subfloat[Sexual Abuse]{%
\includegraphics[width=0.33\textwidth]{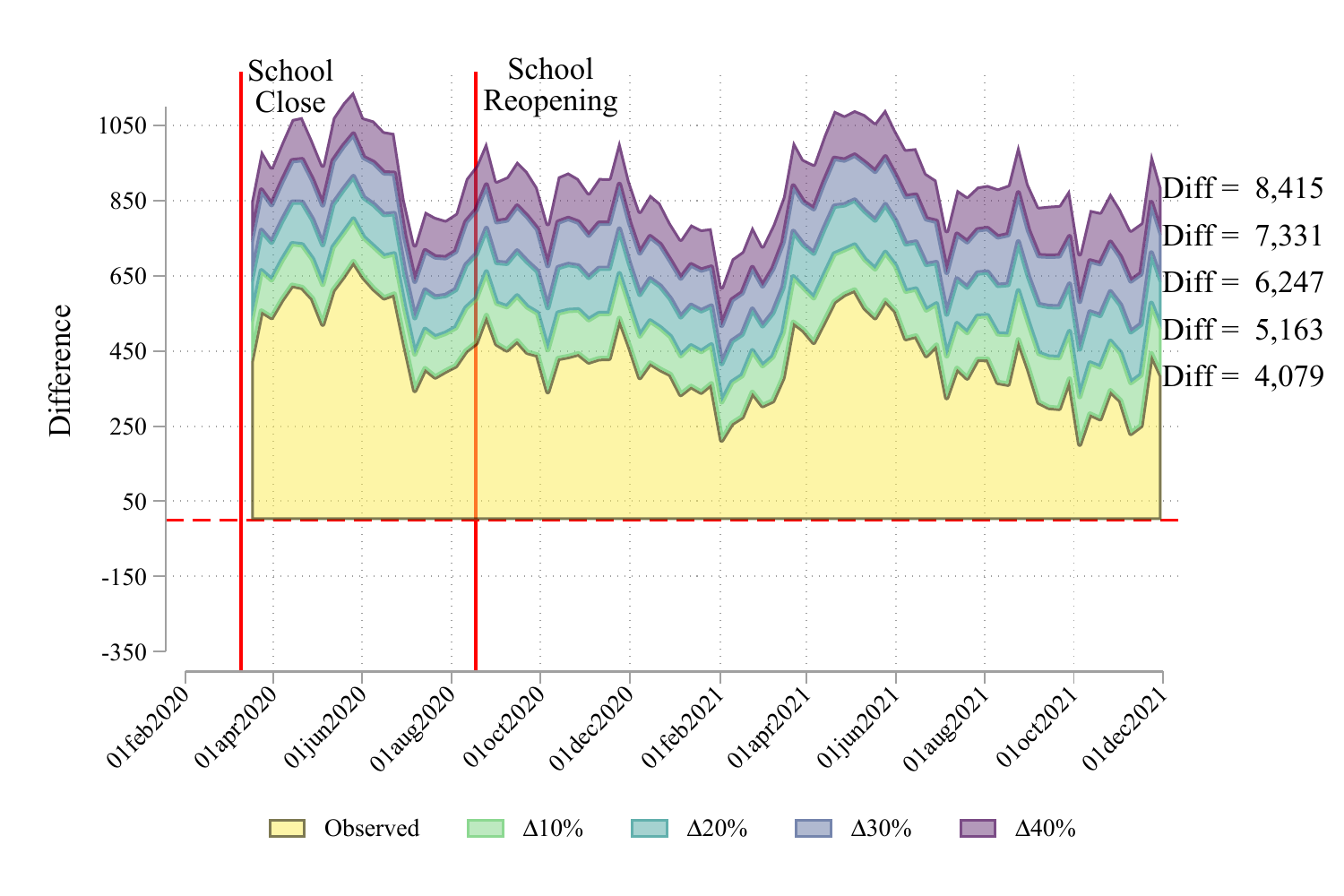}%
}
\subfloat[Rape]{%
\includegraphics[width=0.33\textwidth]{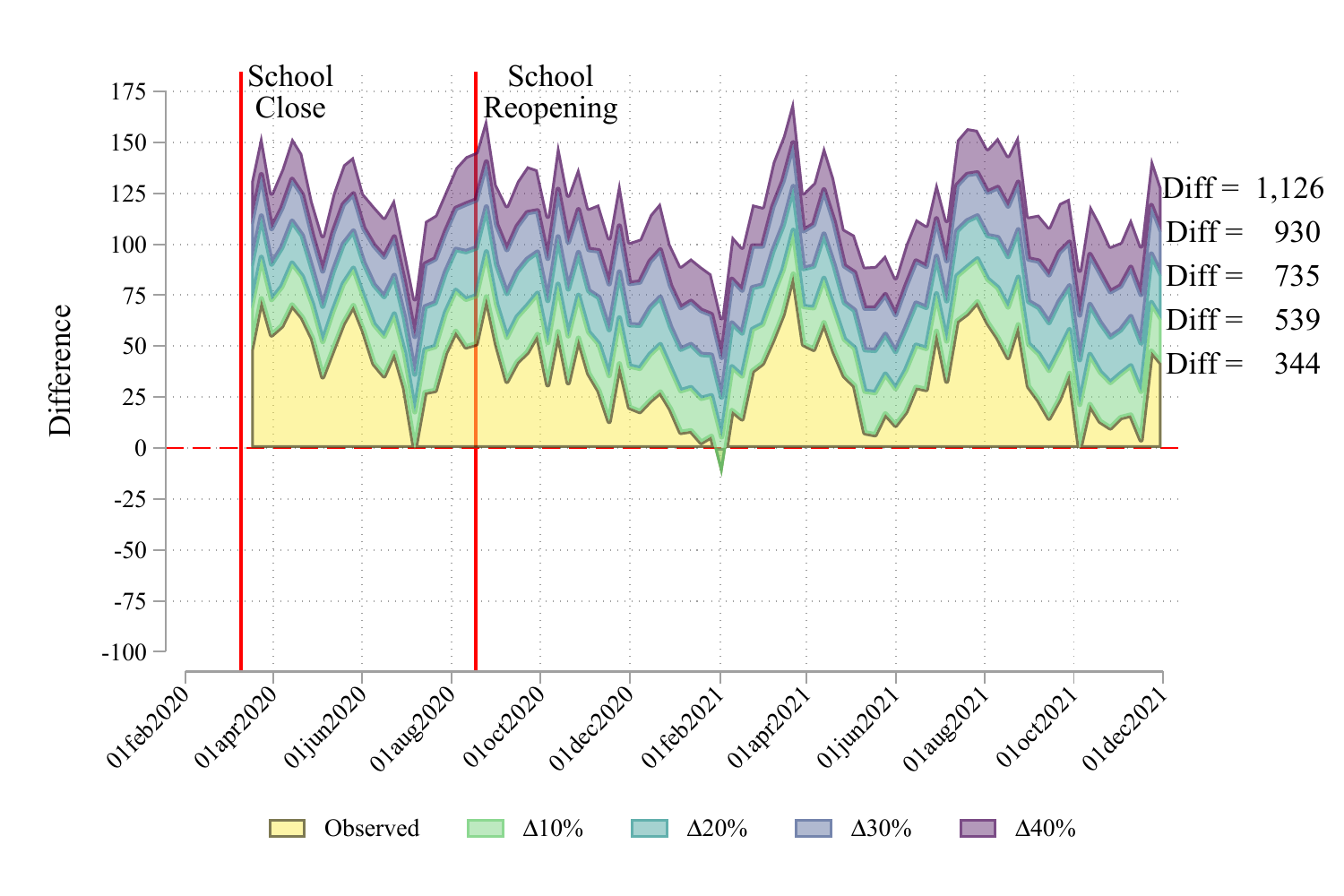}%
}
\end{center}
\floatfoot{\textbf{Notes to Fig.\ \ref{SIfig:counterfactuals2015}}: Alternative projections and reporting differentials are reported, where rather than basing projections off MSE optimal models choosing both the length of pre-COVID prediction years as well as secular trends, optimal secular trends are chosen, in each case using all years from 2015 onwards.  Refer to notes to Figure 3 for further details.}
\end{figure}

\begin{figure}[t!]
\begin{center}
\caption{Alternative Counterfactual Models -- No Trend (2018)}
\label{SIfig:counterfactualsNoTrend}
\textbf{Panel A: Simple counterfactual (time only)} \\
\subfloat[Intra-family Violence]{%
\includegraphics[width=0.33\textwidth]{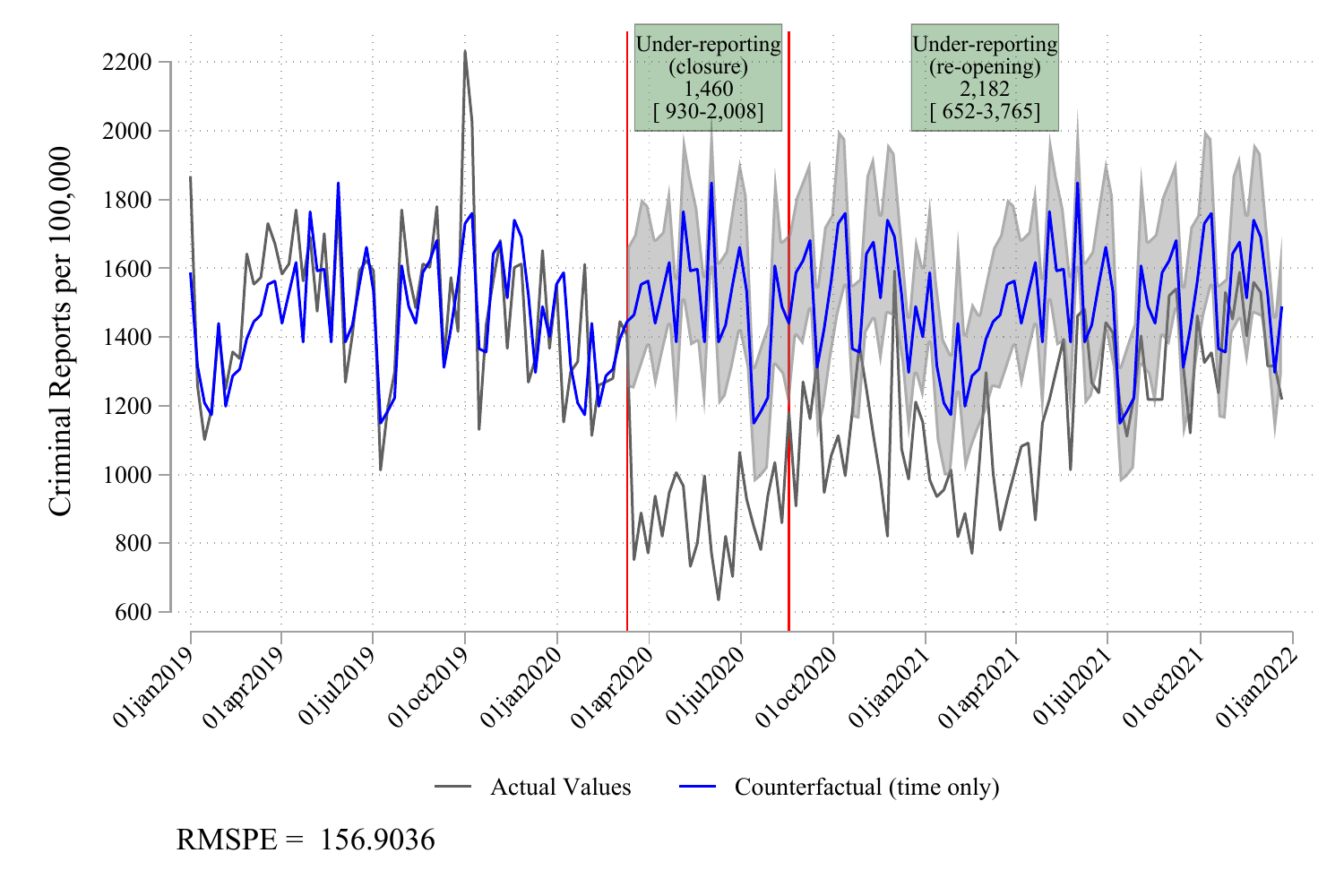}%
}
\subfloat[Sexual Abuse]{%
\includegraphics[width=0.33\textwidth]{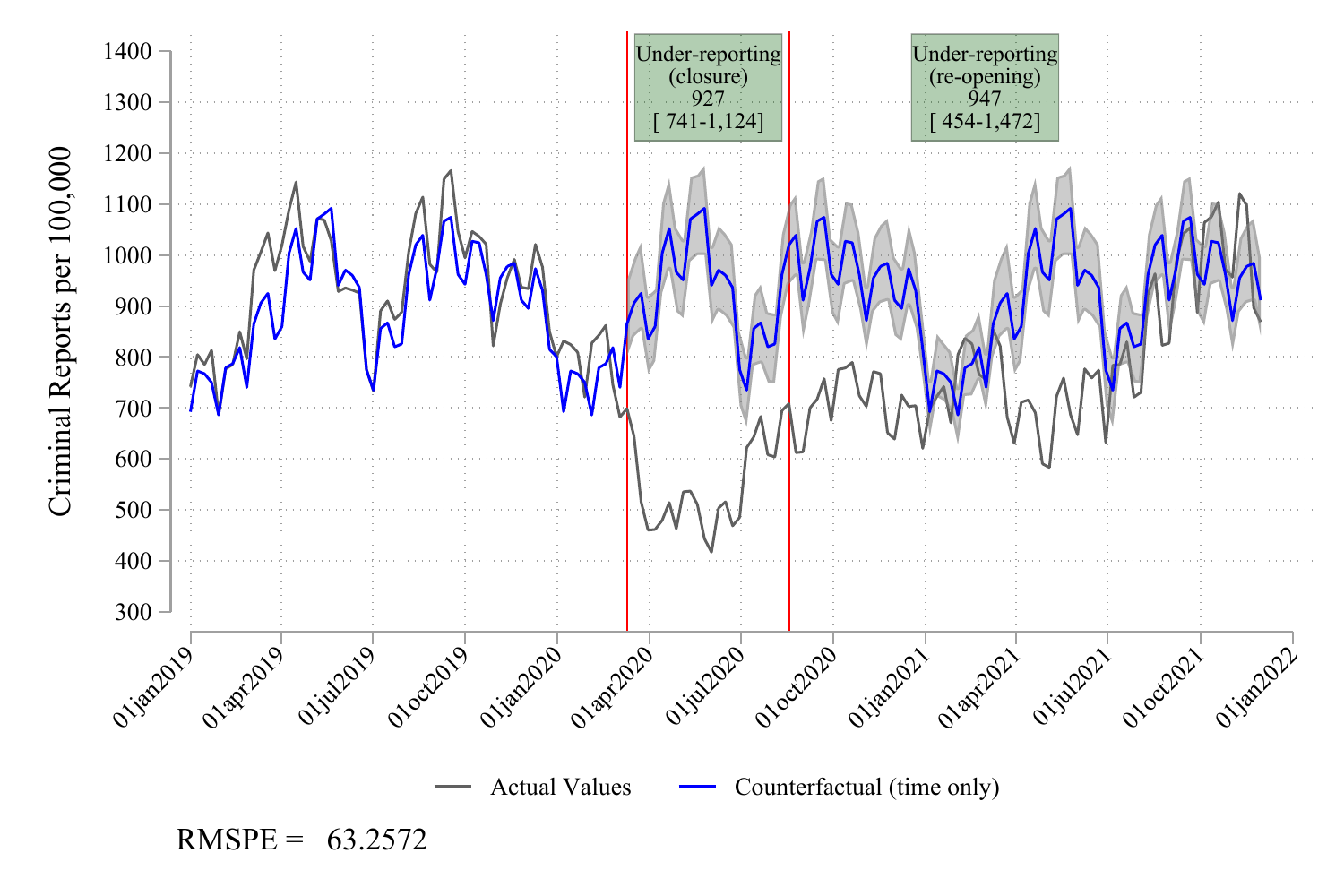}%
}
\subfloat[Rape]{%
\includegraphics[width=0.33\textwidth]{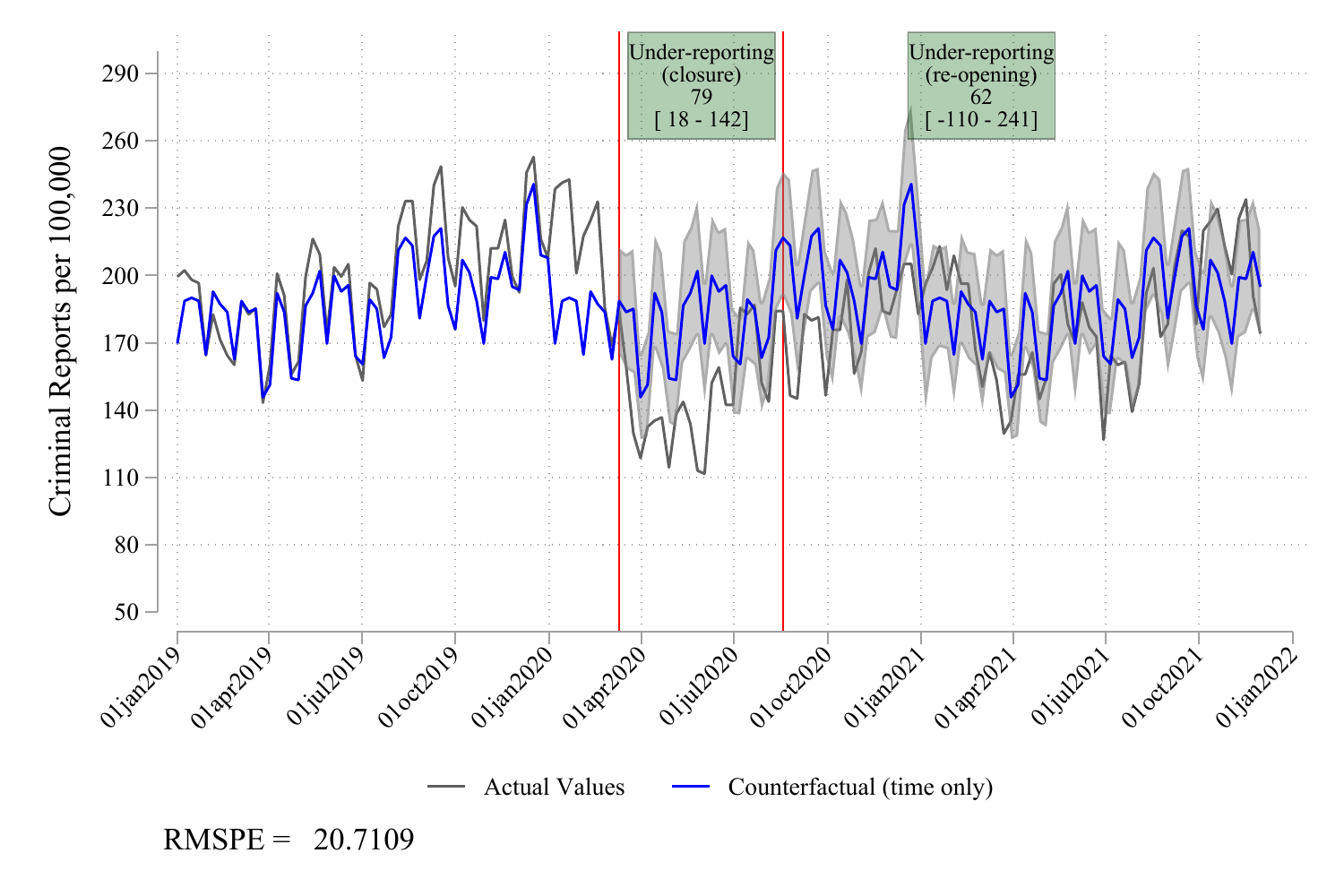}%
}
\\
\textbf{Panel B: Counterfactual (No school channel)} \\
\subfloat[Intra-family Violence]{%
\includegraphics[width=0.33\textwidth]{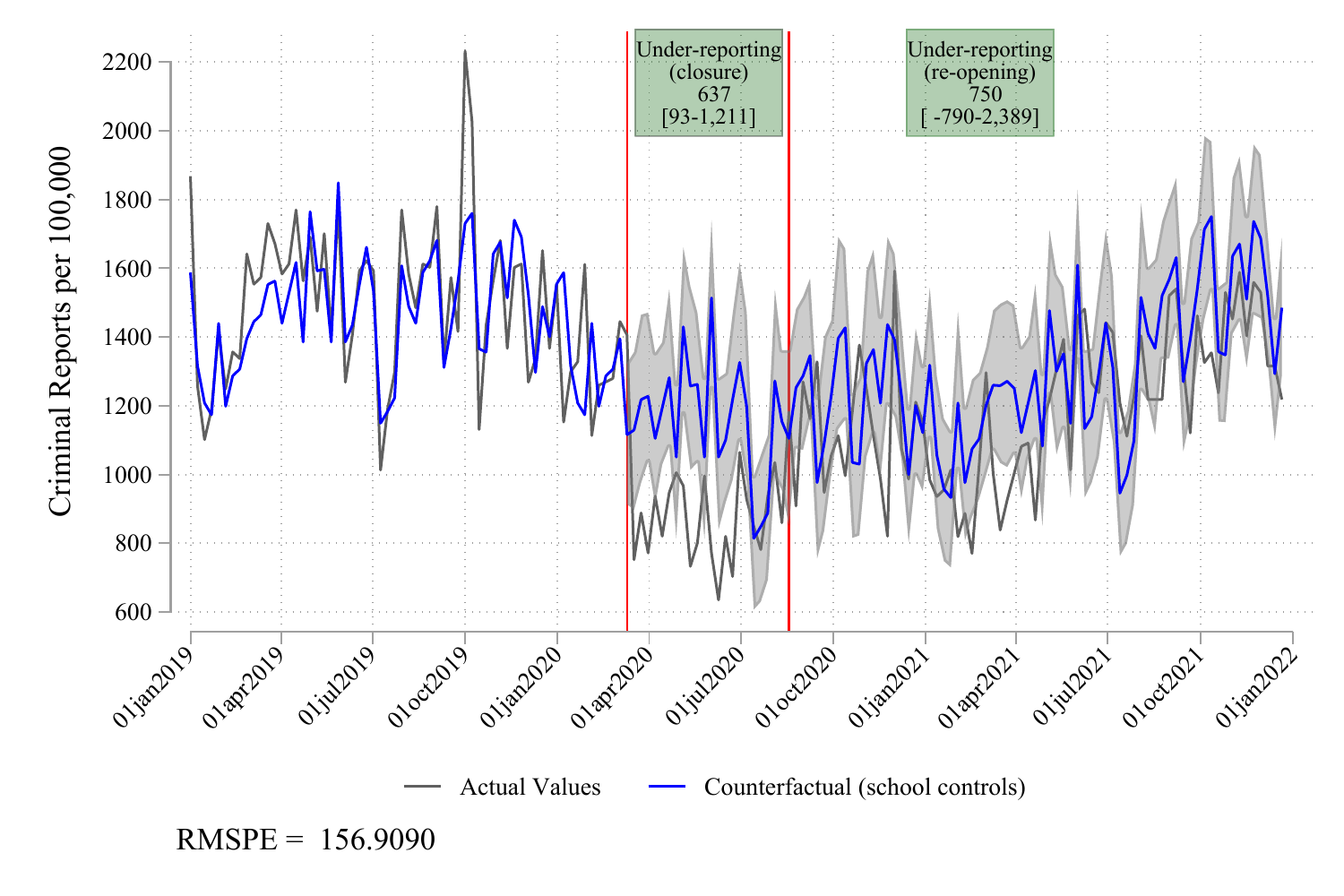}%
}
\subfloat[Sexual Abuse]{%
\includegraphics[width=0.33\textwidth]{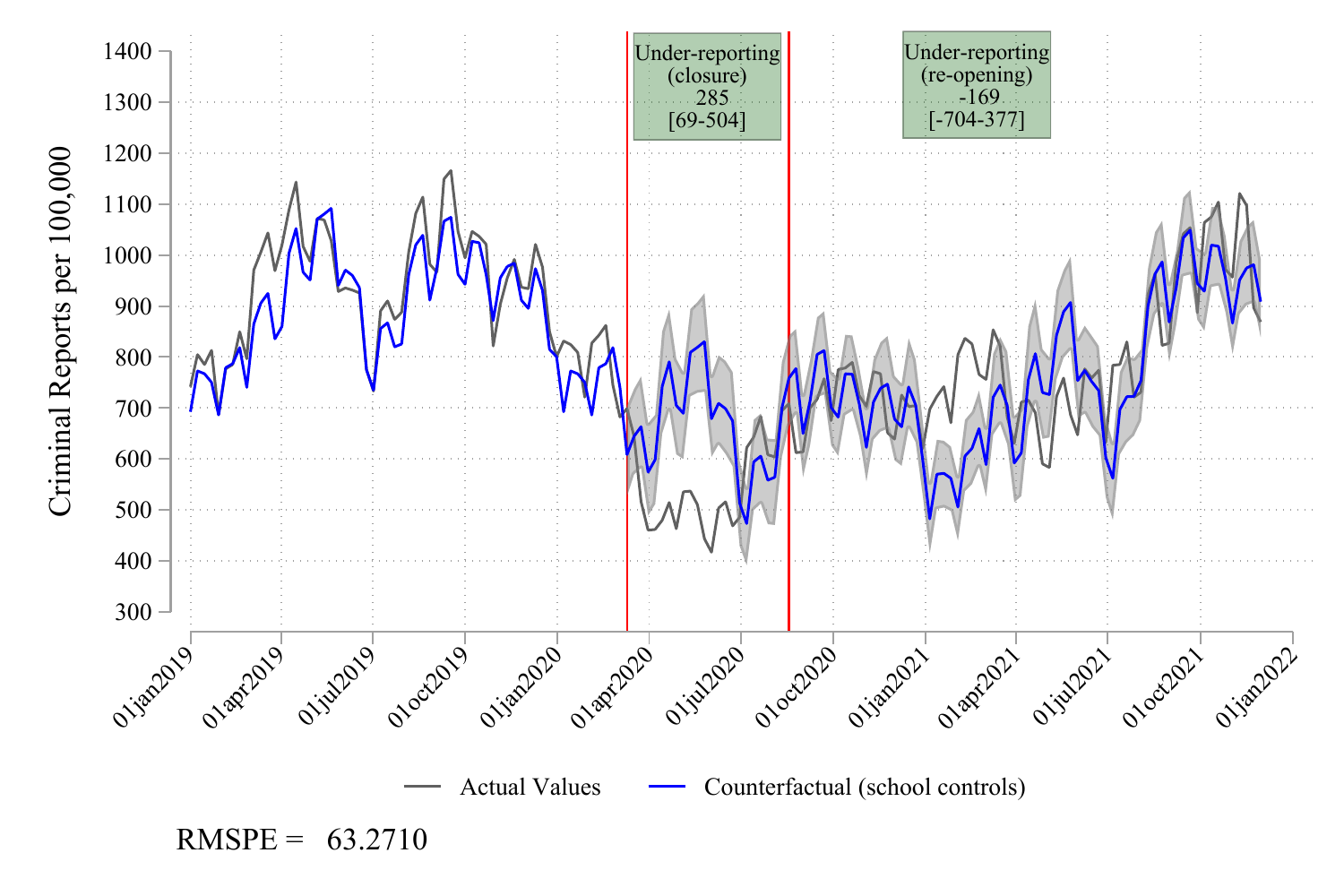}%
}
\subfloat[Rape]{%
\includegraphics[width=0.33\textwidth]{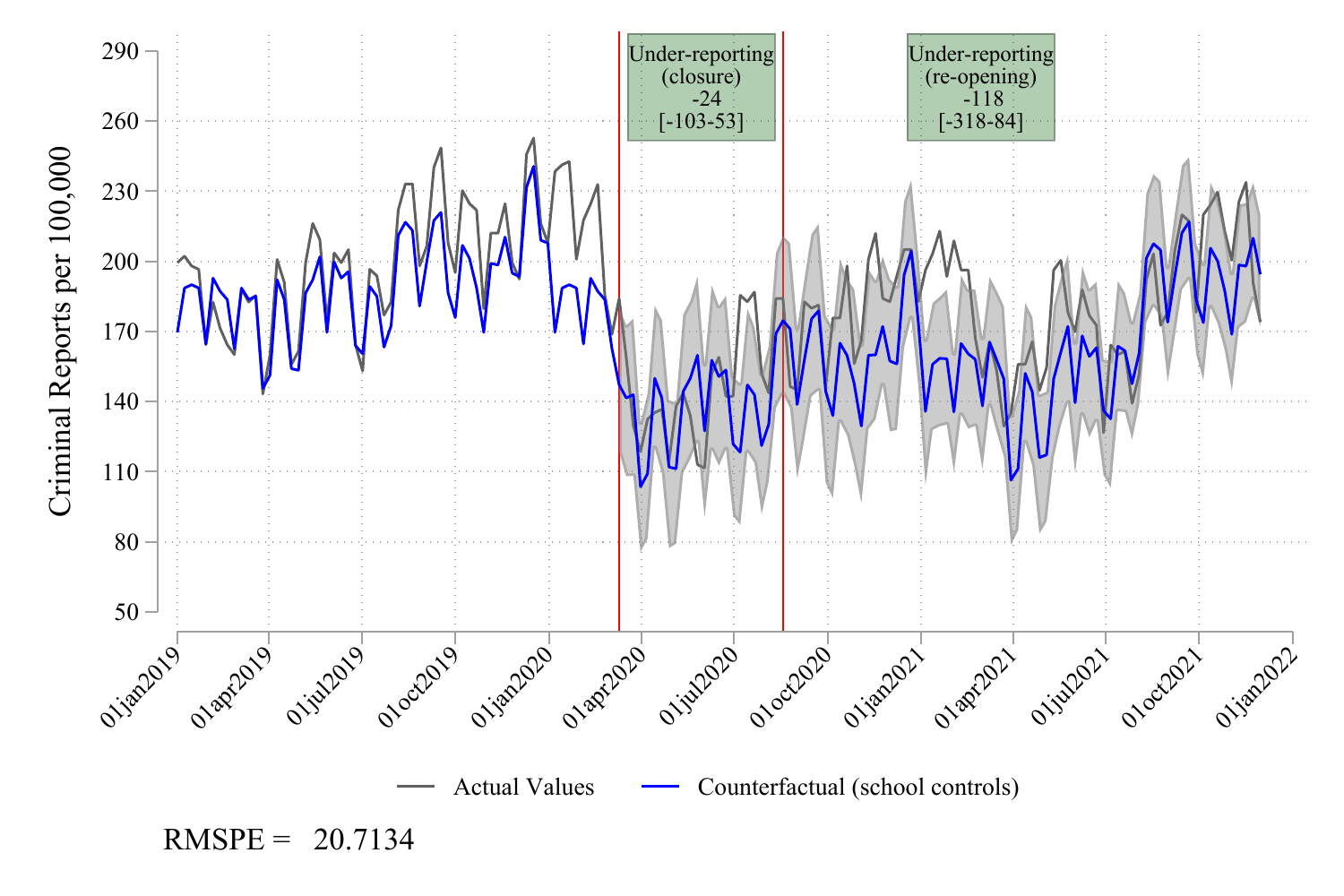}%
}\\
\textbf{Panel C: Projected under-reporting} \\
\subfloat[Intra-family Violence]{%
\includegraphics[width=0.33\textwidth]{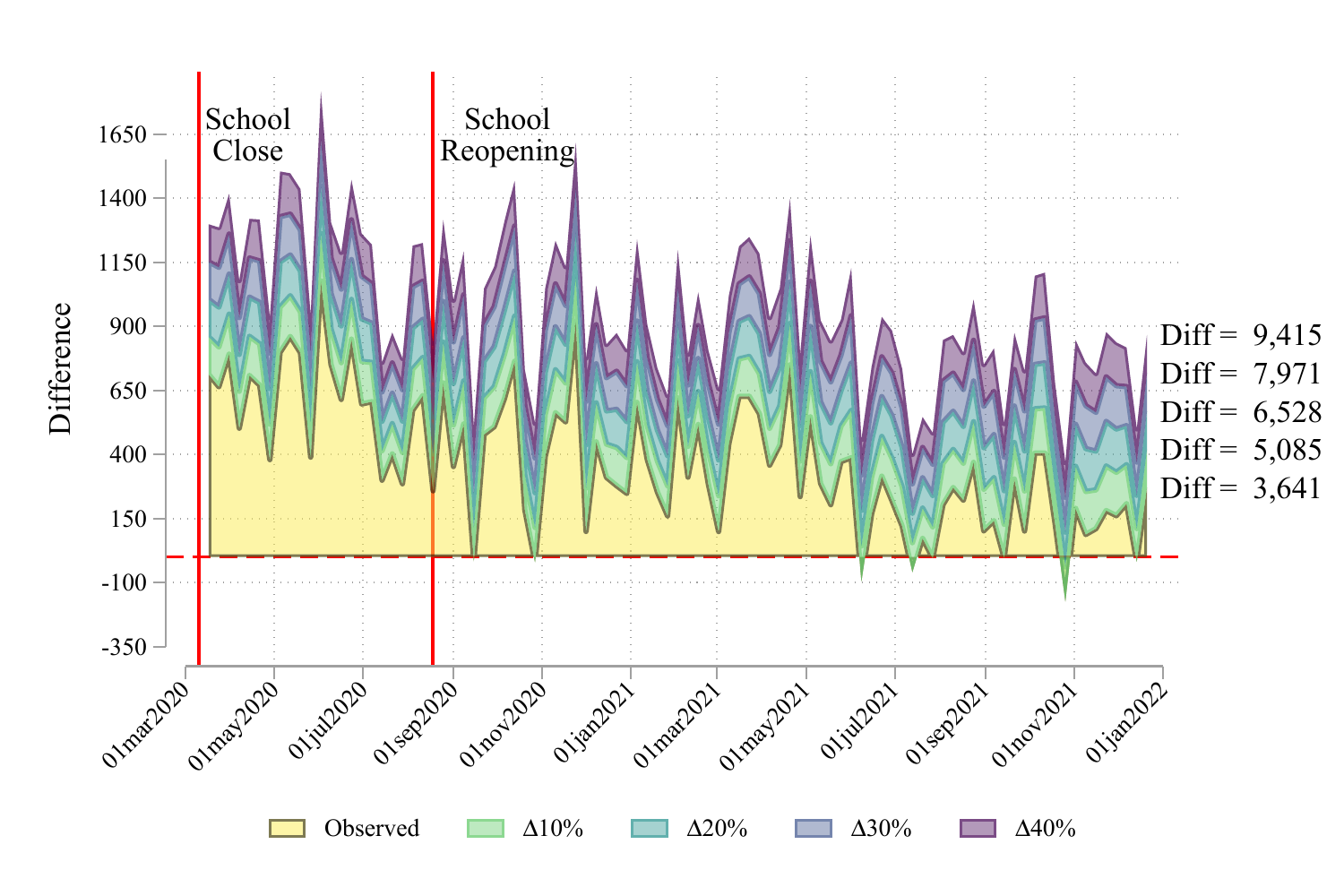}%
}
\subfloat[Sexual Abuse]{%
\includegraphics[width=0.33\textwidth]{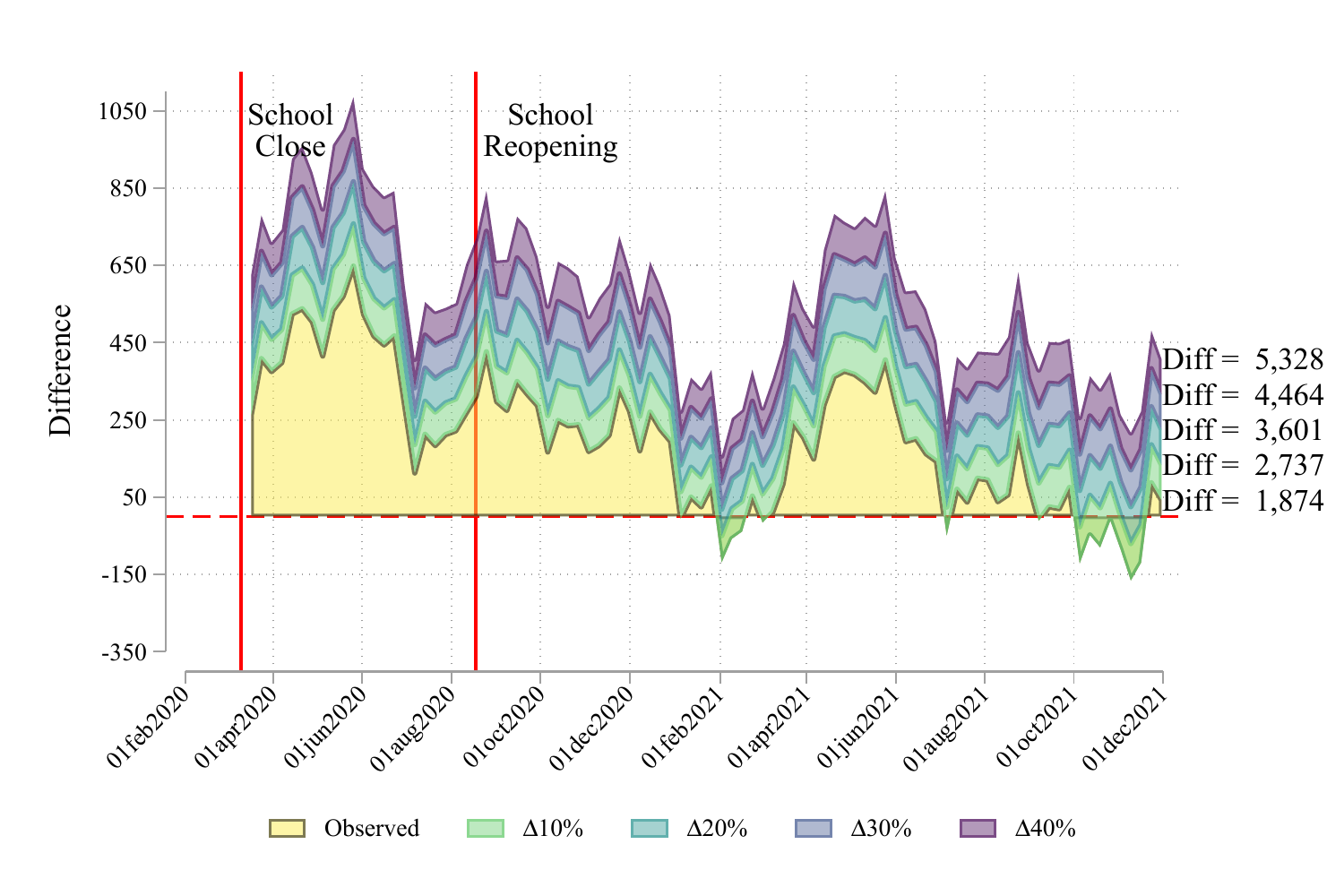}%
}
\subfloat[Rape]{%
\includegraphics[width=0.33\textwidth]{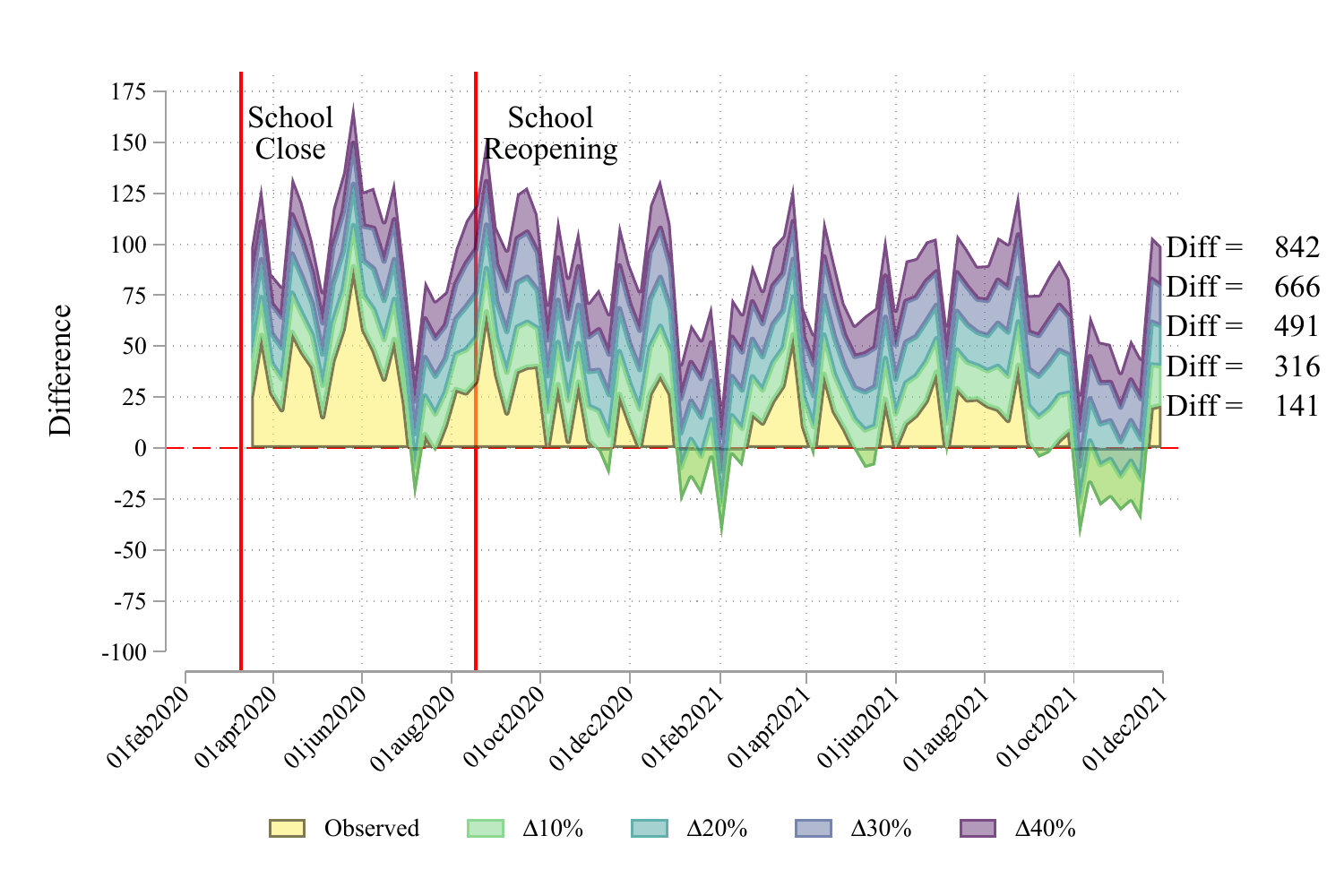}%
}
\end{center}
\floatfoot{\textbf{Notes to Fig.\ \ref{SIfig:counterfactualsNoTrend}}: Alternative projections and reporting differentials are reported, where rather than basing projections off MSE optimal models, projections are based on simple cyclical estimates (week of year fixed effects) based only off year 2018, rather than additionally incorporating a secular trend and choosing pre-periods optimally.  Refer to notes to Figure 3 for further details.}
\end{figure}

\begin{figure}[t!]
\begin{center}
\caption{Alternative Counterfactual Models -- Linear (2018)}
\label{SIfig:counterfactualsLinear}
\textbf{Panel A: Simple counterfactual (time only)} \\
\subfloat[Intra-family Violence]{%
\includegraphics[width=0.33\textwidth]{./graphs/C1_V_2018_lineal}%
}
\subfloat[Sexual Abuse]{%
\includegraphics[width=0.33\textwidth]{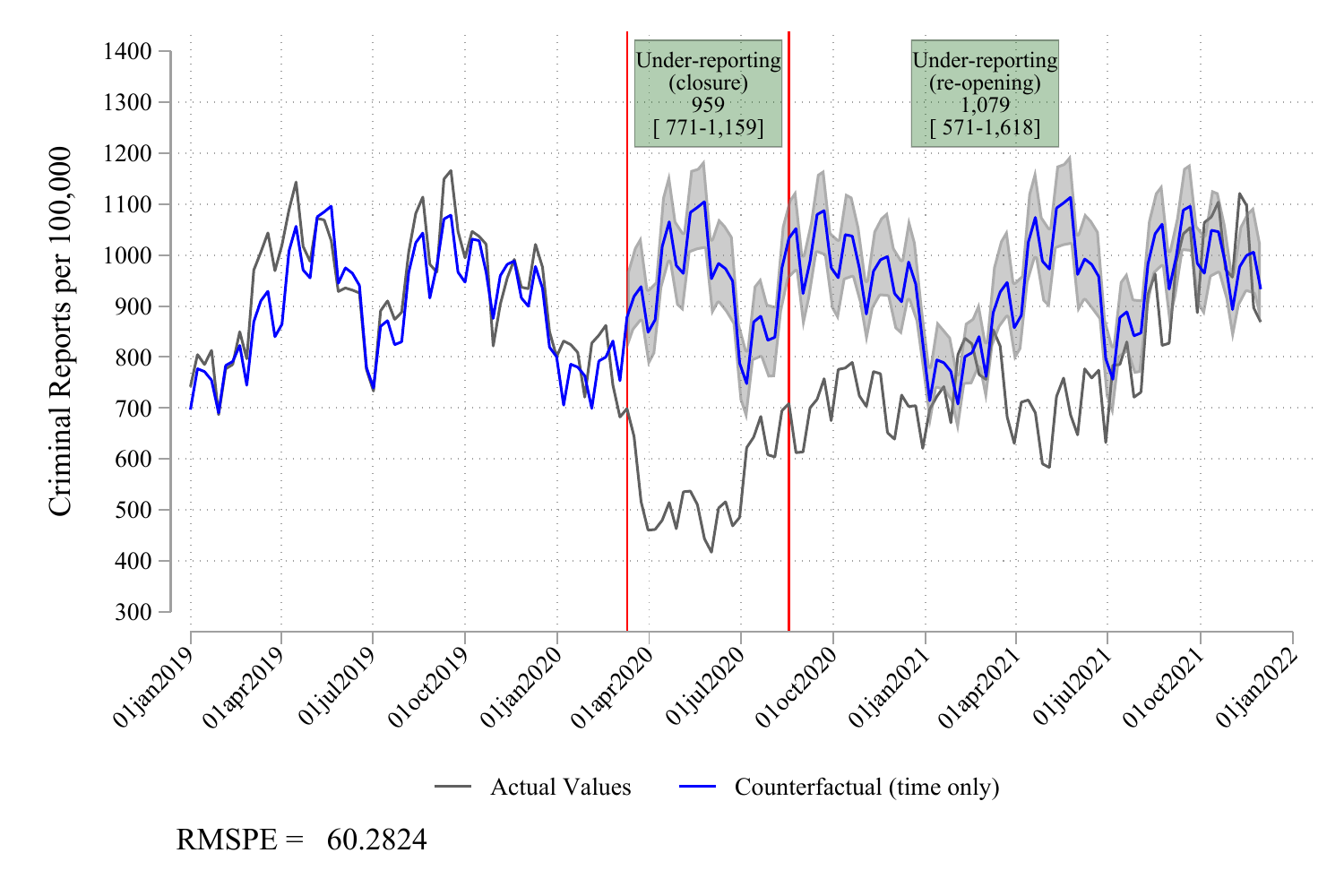}%
}
\subfloat[Rape]{%
\includegraphics[width=0.33\textwidth]{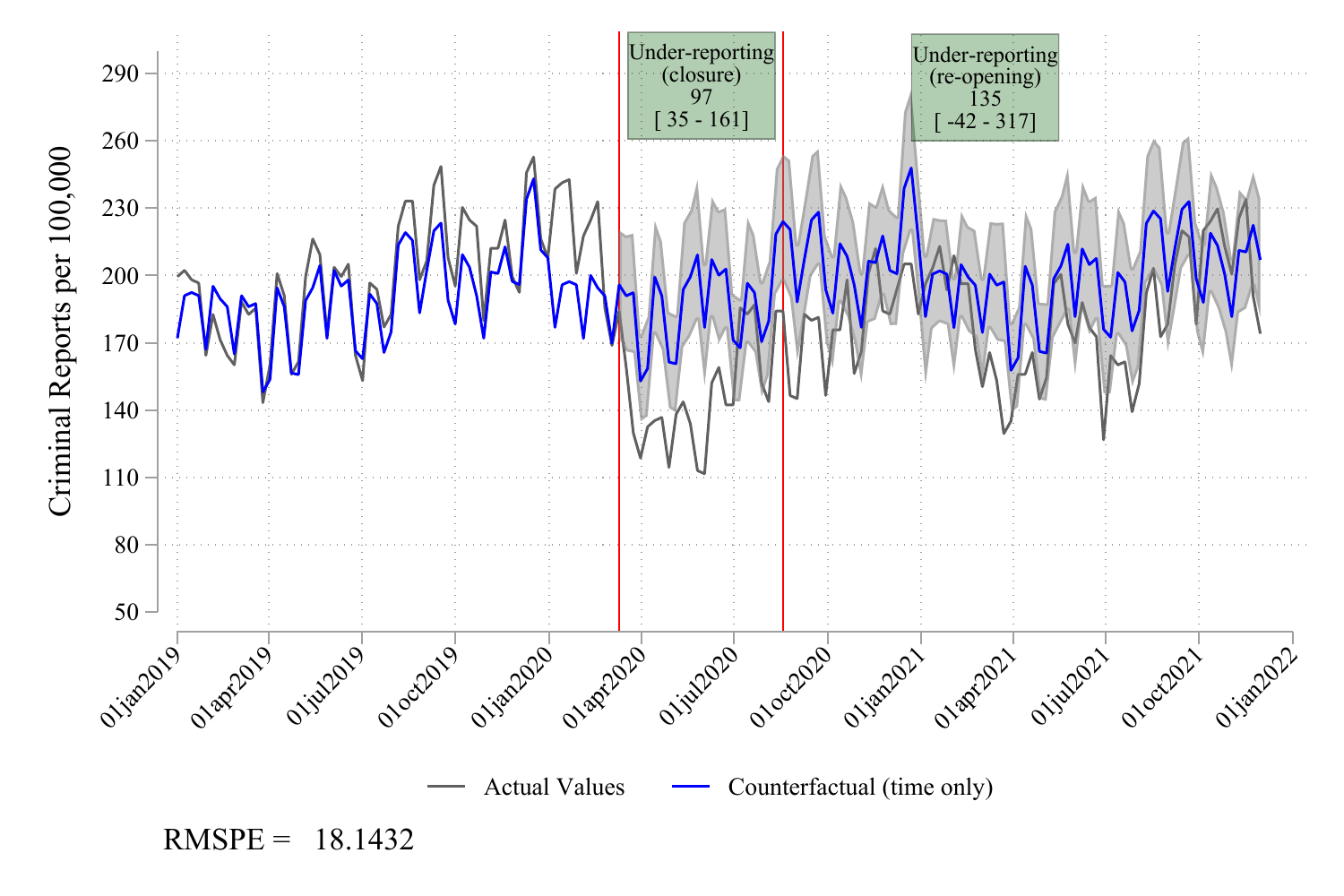}%
}
\\
\textbf{Panel B: Counterfactual (No school channel)} \\
\subfloat[Intra-family Violence]{%
\includegraphics[width=0.33\textwidth]{./graphs/C3_V_2018_lineal}%
}
\subfloat[Sexual Abuse]{%
\includegraphics[width=0.33\textwidth]{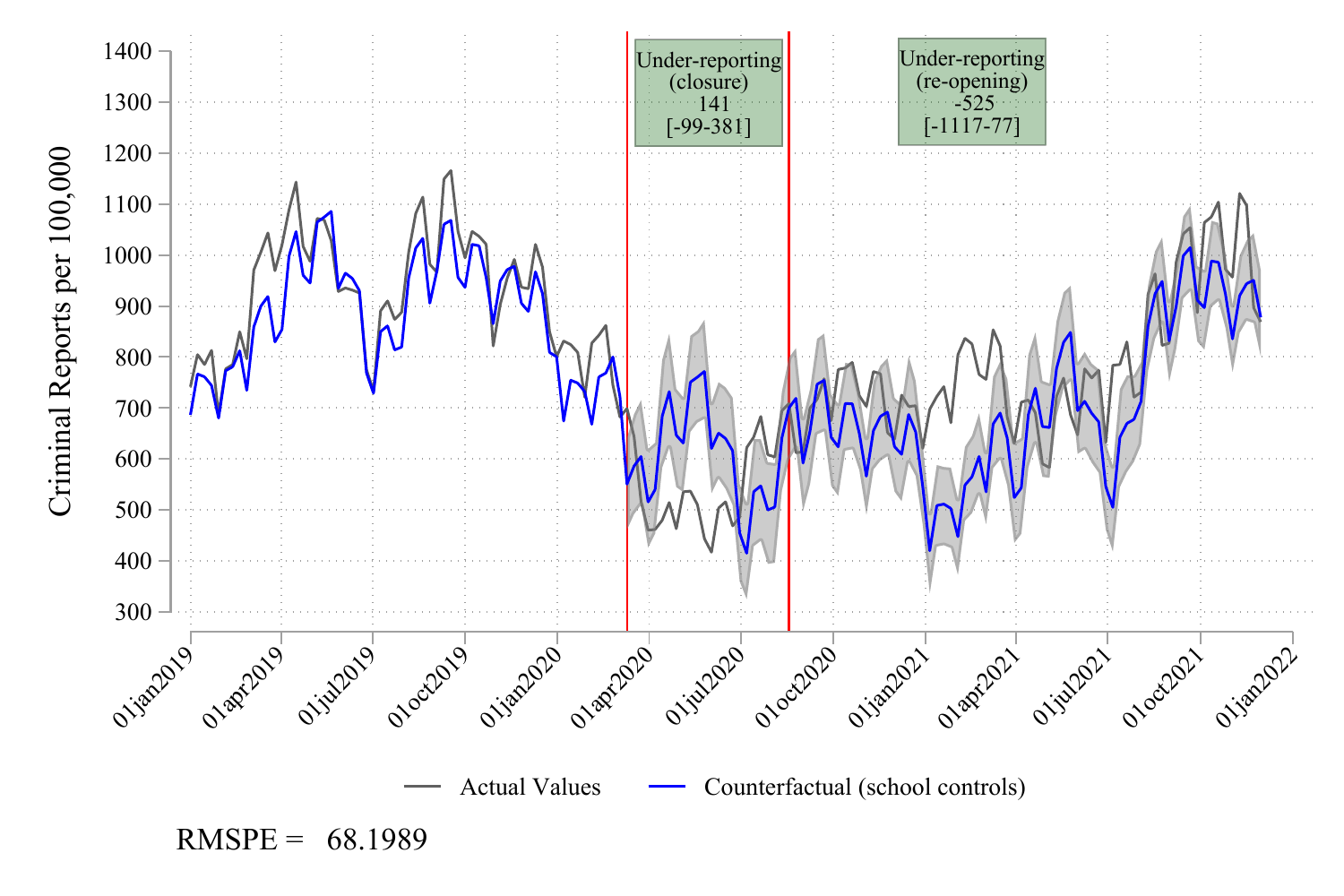}%
}
\subfloat[Rape]{%
\includegraphics[width=0.33\textwidth]{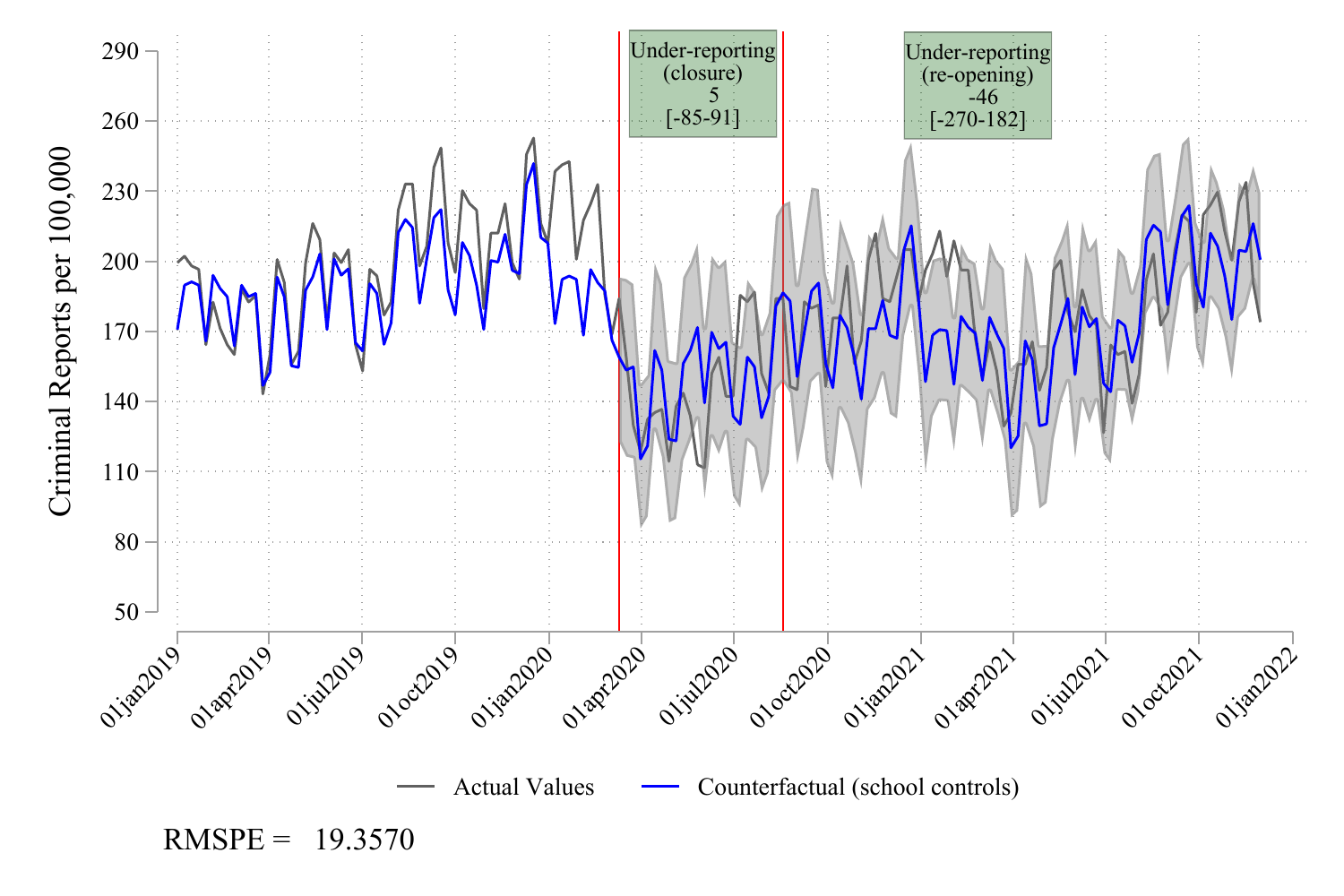}%
}\\
\textbf{Panel C: Projected under-reporting} \\
\subfloat[Intra-family Violence]{%
\includegraphics[width=0.33\textwidth]{./graphs/diff_Count_1_V_2018_lineal}%
}
\subfloat[Sexual Abuse]{%
\includegraphics[width=0.33\textwidth]{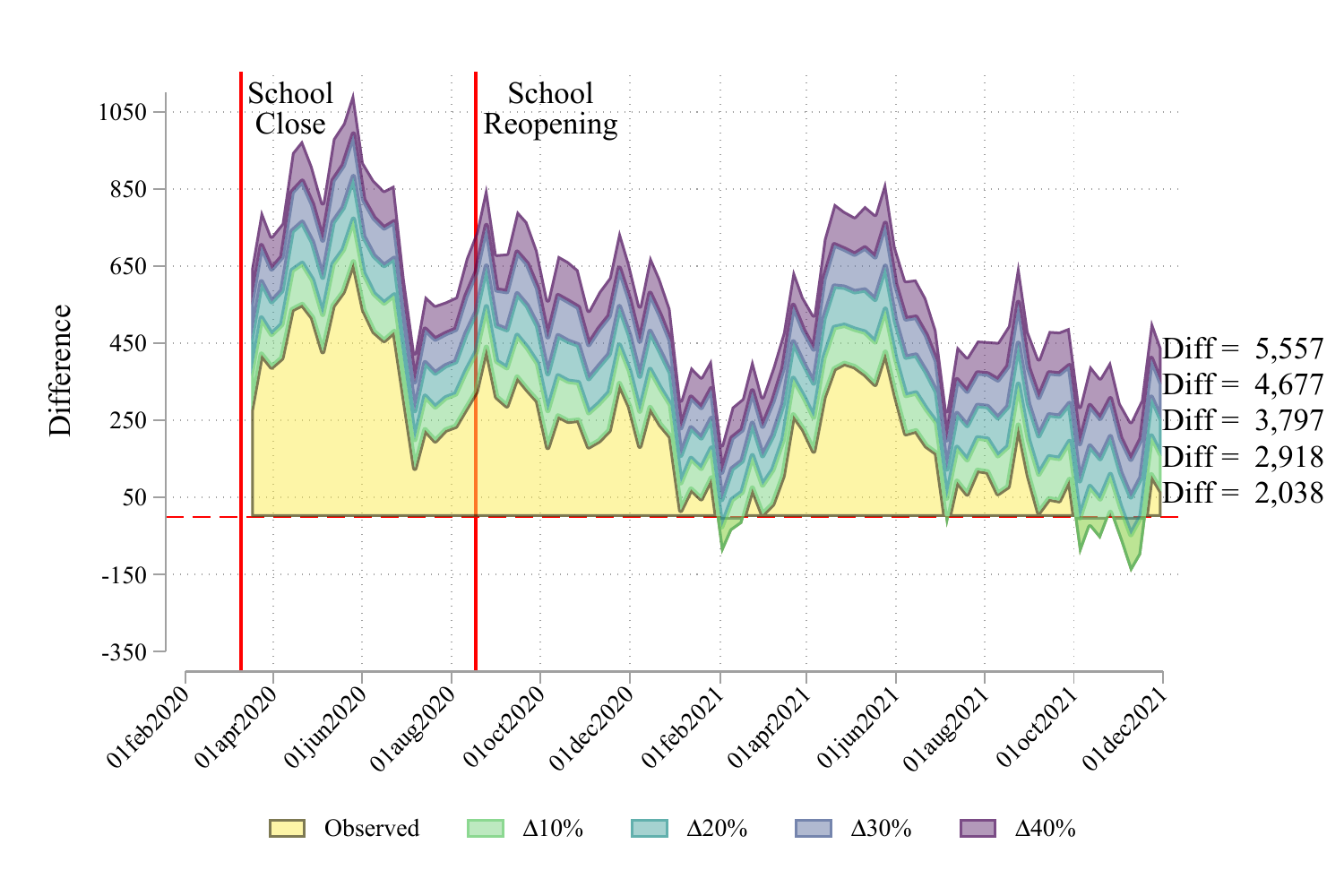}%
}
\subfloat[Rape]{%
\includegraphics[width=0.33\textwidth]{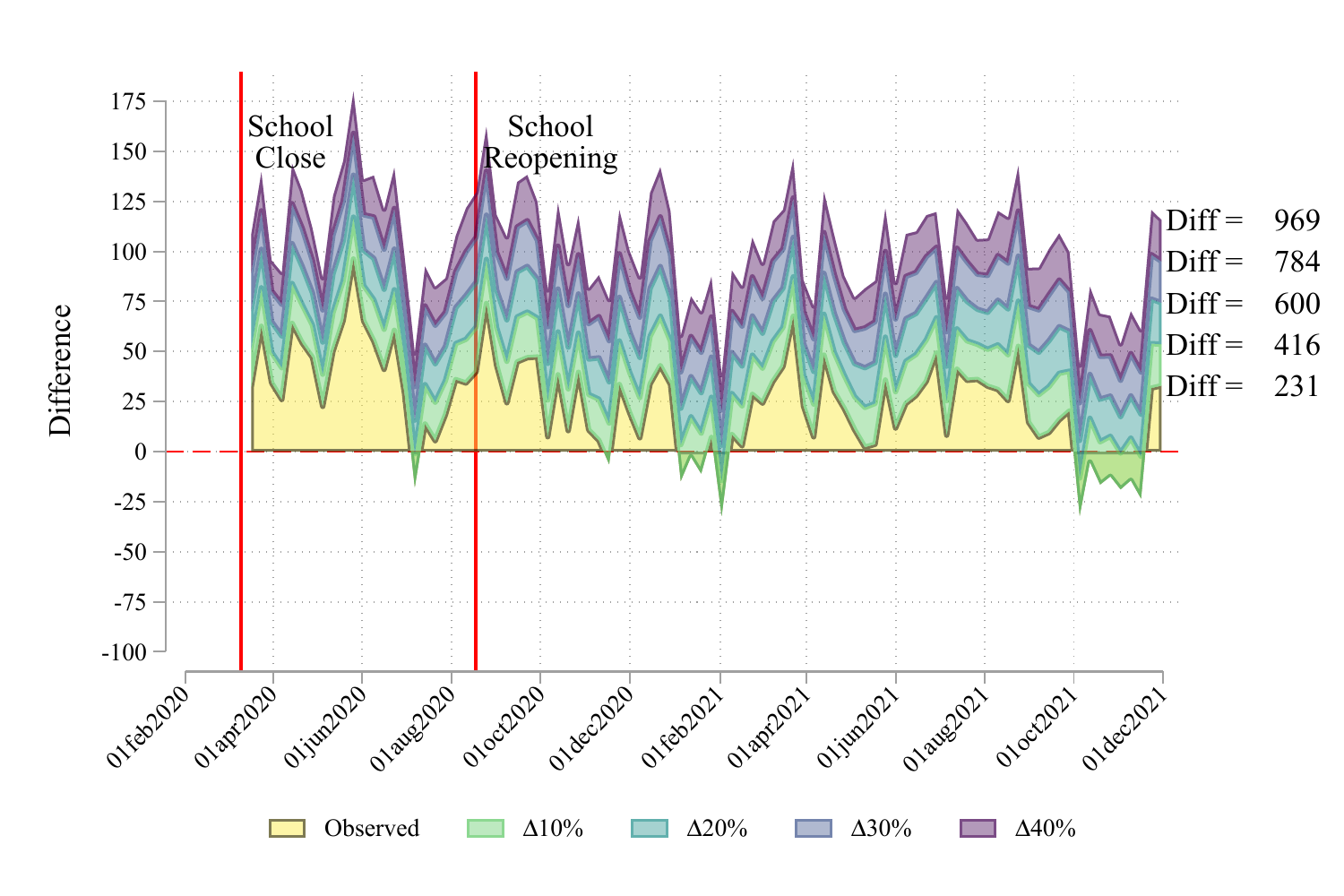}%
}
\end{center}
\floatfoot{\textbf{Notes to Fig.\ \ref{SIfig:counterfactualsLinear}}: Alternative projections and reporting differentials are reported, where rather than basing projections off MSE optimal models, projections are based on simple cyclical estimates (week of year fixed effects) and a linear secular trend, based only off year 2018, rather than optimally choosing parametrization of the secular trend and choosing pre-periods optimally.  Refer to notes to Figure 3 for further details.}
\end{figure}

\begin{figure}[t!]
\begin{center}
\caption{Alternative Counterfactual Models -- Quadratic (2018)}
\label{SIfig:counterfactualsQuadratic}
\textbf{Panel A: Simple counterfactual (time only)} \\
\subfloat[Intra-family Violence]{%
\includegraphics[width=0.33\textwidth]{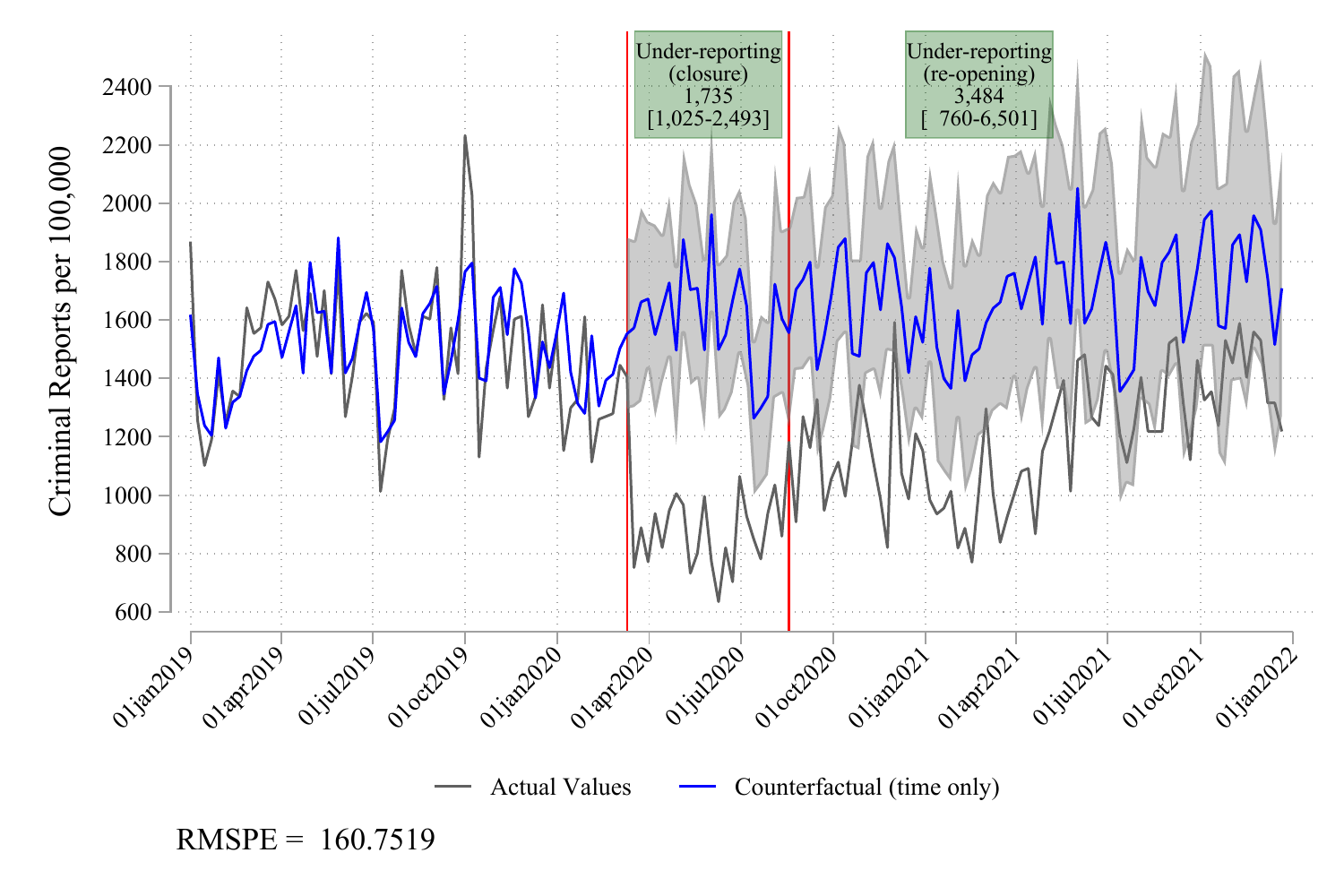}%
}
\subfloat[Sexual Abuse]{%
\includegraphics[width=0.33\textwidth]{./graphs/C1_SA_2018_cuadratic}%
}
\subfloat[Rape]{%
\includegraphics[width=0.33\textwidth]{./graphs/C1_R_2018_cuadratic}%
}
\\
\textbf{Panel B: Counterfactual (No school channel)} \\
\subfloat[Intra-family Violence]{%
\includegraphics[width=0.33\textwidth]{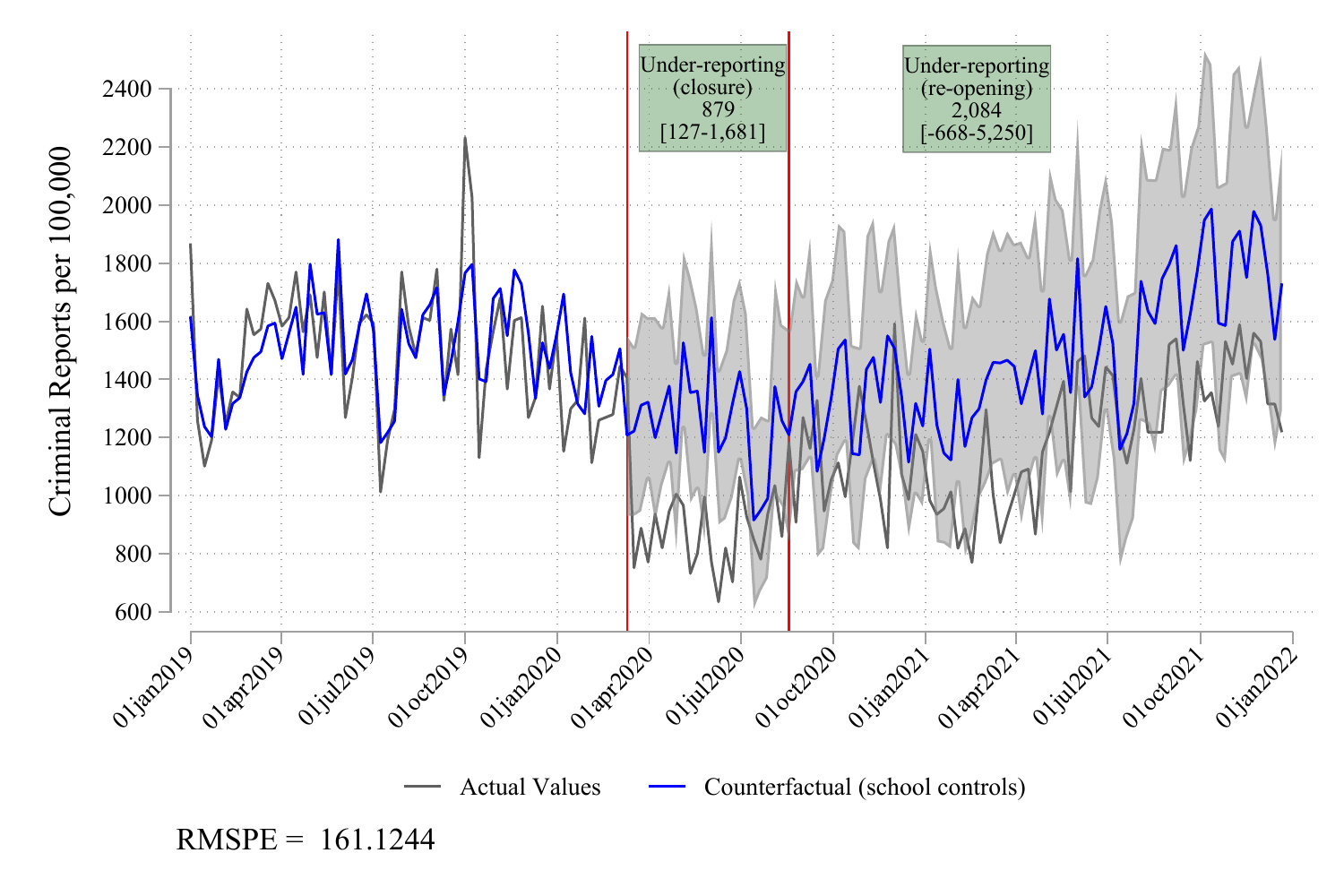}%
}
\subfloat[Sexual Abuse]{%
\includegraphics[width=0.33\textwidth]{./graphs/C3_SA_2018_cuadratic}%
}
\subfloat[Rape]{%
\includegraphics[width=0.33\textwidth]{./graphs/C3_R_2018_cuadratic}%
}\\
\textbf{Panel C: Projected under-reporting} \\
\subfloat[Intra-family Violence]{%
\includegraphics[width=0.33\textwidth]{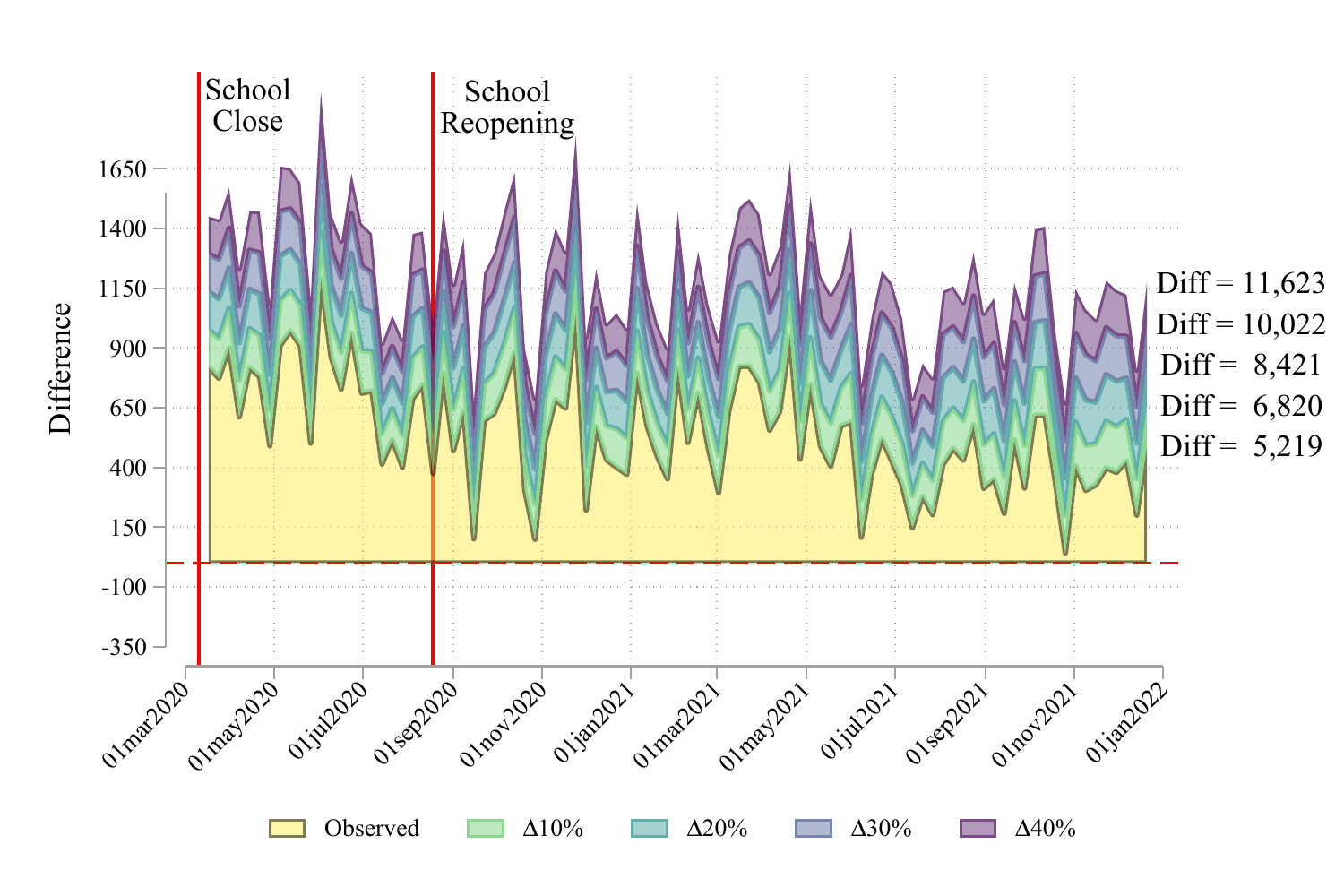}%
}
\subfloat[Sexual Abuse]{%
\includegraphics[width=0.33\textwidth]{./graphs/diff_Count_1_SA_2018_cuadratic}%
}
\subfloat[Rape]{%
\includegraphics[width=0.33\textwidth]{./graphs/diff_Count_1_R_2018_cuadratic}%
}
\end{center}
\floatfoot{\textbf{Notes to Fig.\ \ref{SIfig:counterfactualsQuadratic}}: Alternative projections and reporting differentials are reported, where rather than basing projections off MSE optimal models, projections are based on cyclical estimates (week of year fixed effects) and a quadratic secular trend, based only off of year 2018, rather than optimally choosing parametrization of the secular trend and choosing pre-periods optimally.  Refer to notes to Figure 3 for further details..}
\end{figure}

\begin{figure}[t!]
\begin{center}
\caption{Projected Under-reporting under Various Counterfactual Assumptions (Temporal Projection Only)}
\label{SIfig:Under-reporting}
\textbf{Panel A: Post School Closure and before School Reopening} \\
\subfloat[Intra-family violence (Time only)]{%
\includegraphics[width=0.33\textwidth]{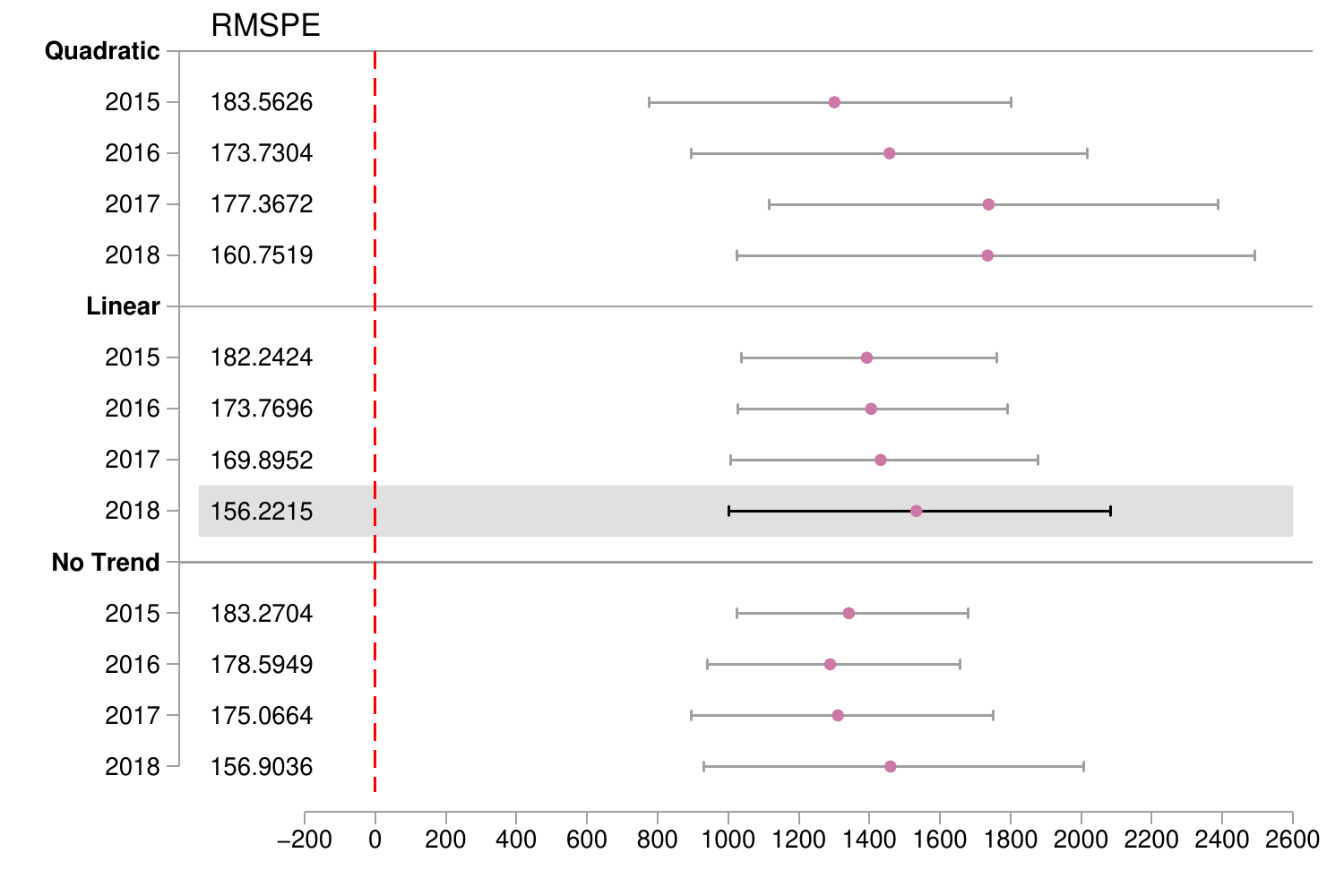}%
}
\subfloat[Sexual Abuse (Time only)]{%
\includegraphics[width=0.33\textwidth]{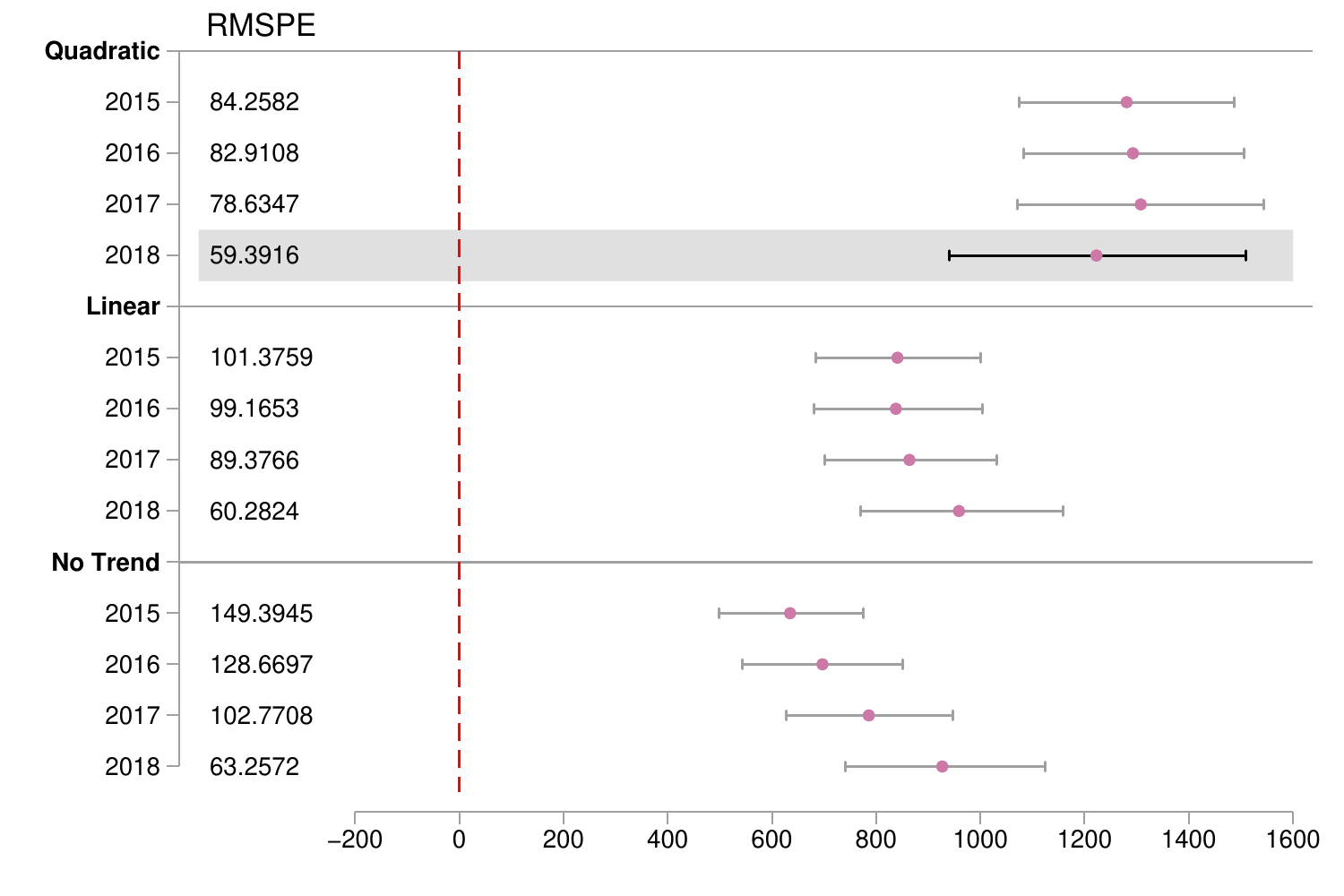}%
}
\subfloat[Rape (Time only)]{%
\includegraphics[width=0.33\textwidth]{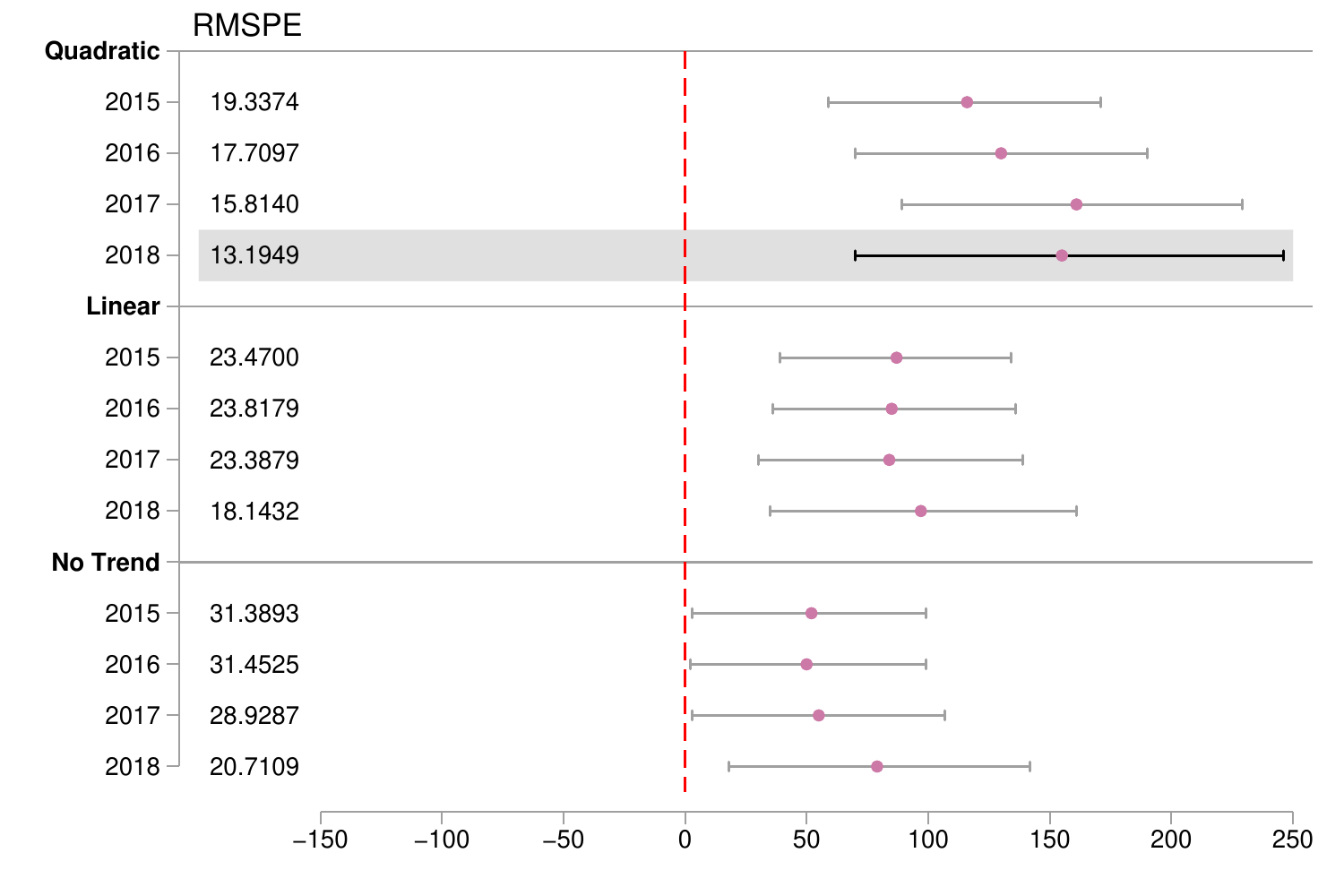}%
}\\
\textbf{Panel B: Post School Reopening} \\
\subfloat[Intra-family violence (Time only)]{%
\includegraphics[width=0.33\textwidth]{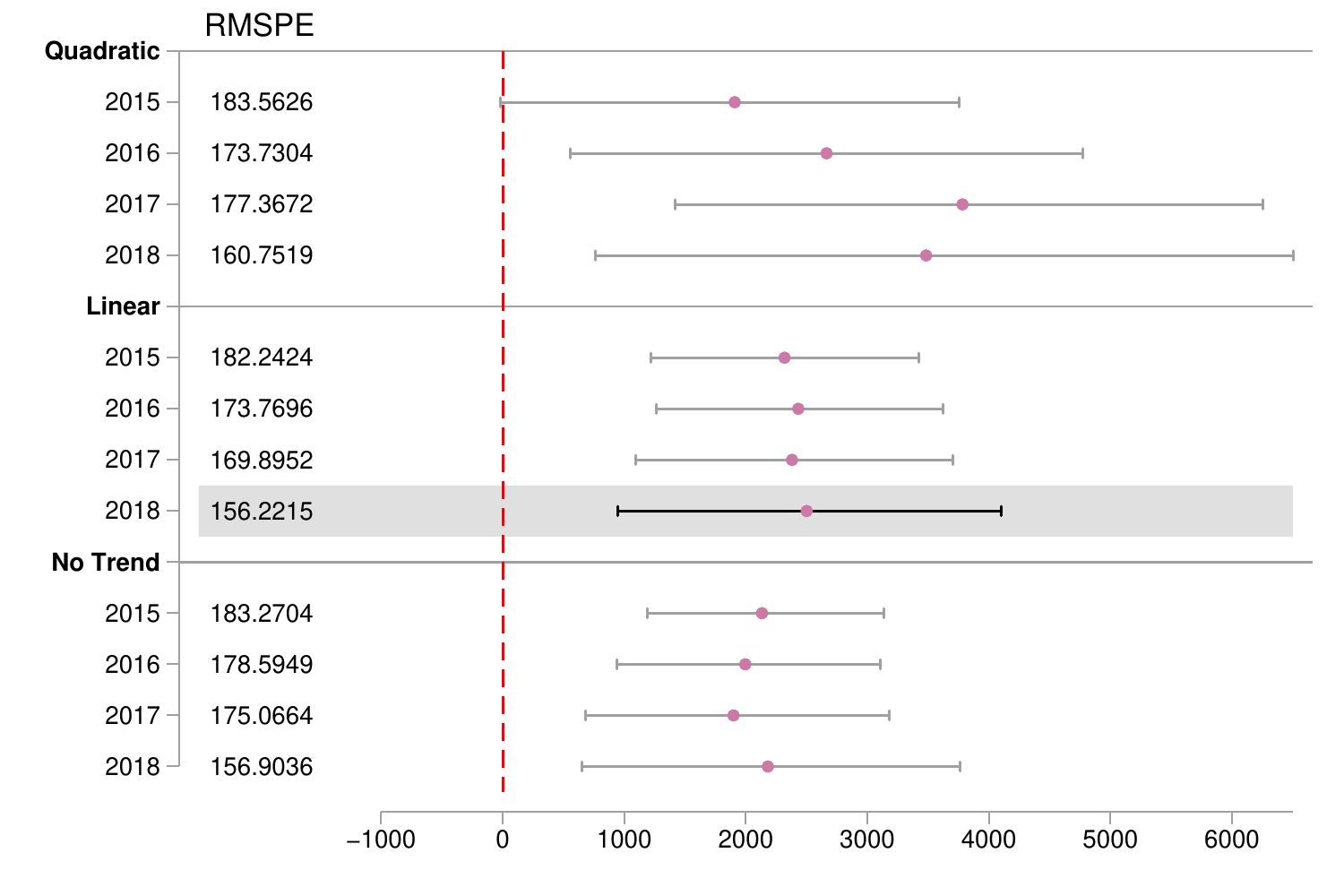}%
}
\subfloat[Sexual Abuse (Time only)]{%
\includegraphics[width=0.33\textwidth]{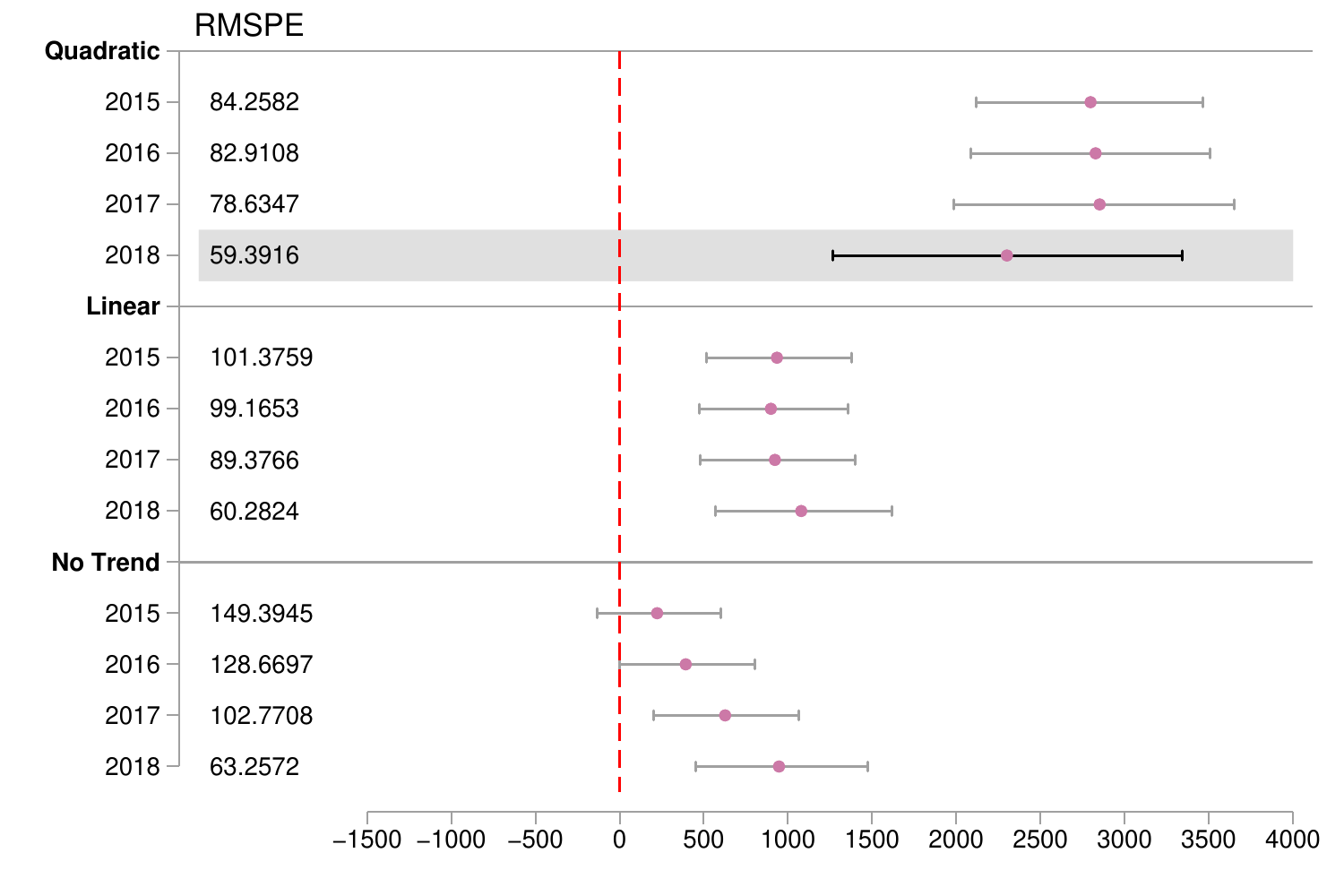}%
}
\subfloat[Rape (Time only)]{%
\includegraphics[width=0.33\textwidth]{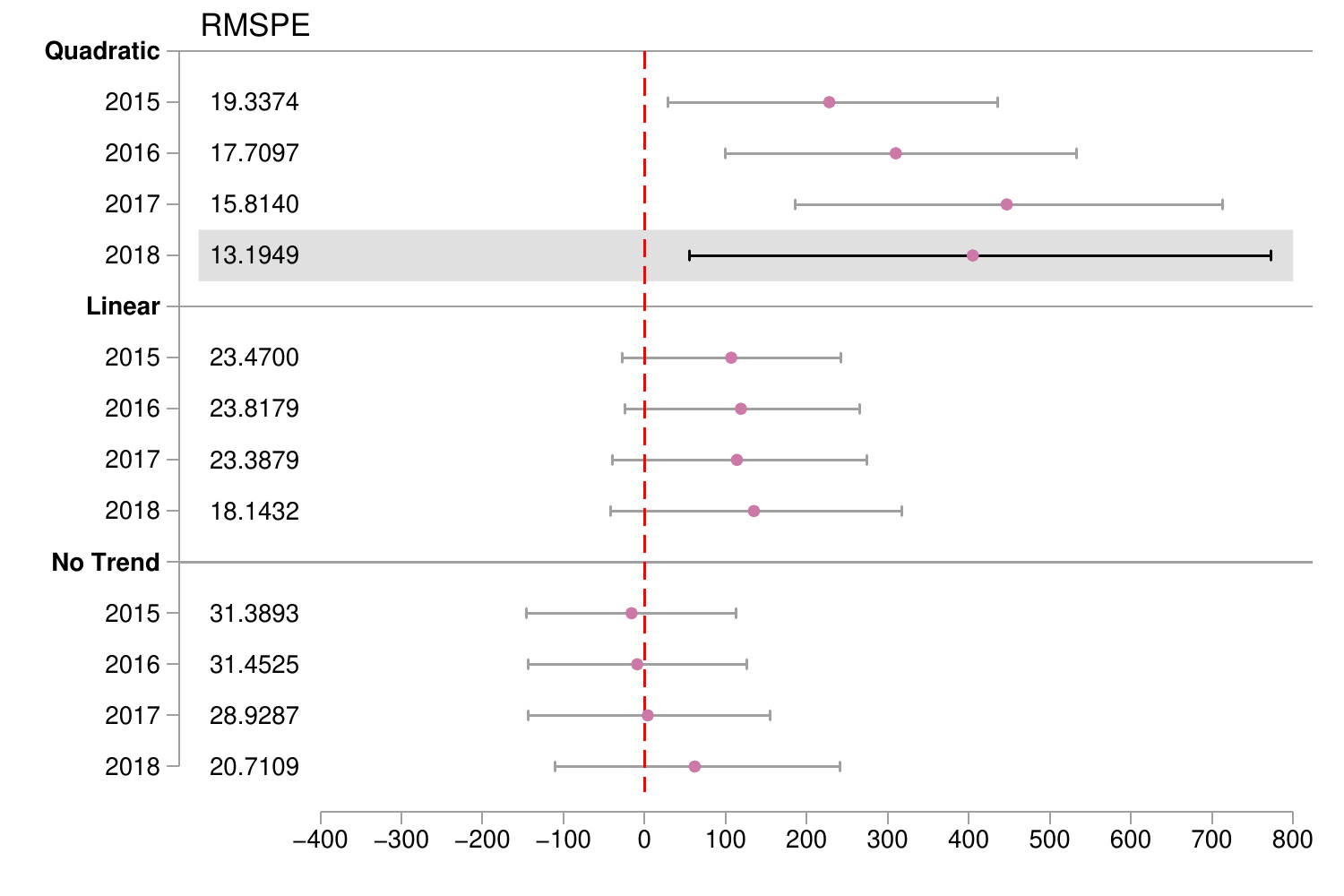}%
}
\end{center}
\floatfoot{\textbf{Notes to Fig.\ \ref{SIfig:Under-reporting}}: Projected values for under-reporting of cases (pink circles), and 95\% CIs (error bars) are displayed for alternative counterfactual models.  In each case, total under-reporting corresponds to the total estimated ``missing cases'' reported in green boxes on Figure 3, however here under alternative modelling assumptions.  MSE optimal models displayed in Figure 3 are shaded and indicated with thicker error bars, and Root-Mean Squared Prediction Errors are displayed in the left hand margin of each figure.  In each case, we consider alternative secular trends (quadratic, linear and no trend), estimated off periods from 2015 onwards, 2016 onwards, 2017 onwards, or 2018 onwards.  Panel A documents under-reporting estimates in periods of full school closure, while Panel B documents under-reporting estimates in periods of school re-opening.}
\end{figure}

\begin{figure}[t!]
\begin{center}
\caption{Projected Under-reporting under Various Counterfactual Assumptions (Projection with no School Channel)}
\label{SIfig:Under-reportingNoSchool}
\textbf{Panel A: Post School Closure and before School Reopening} \\
\subfloat[Intra-family violence (No school channel)]{%
\includegraphics[width=0.33\textwidth]{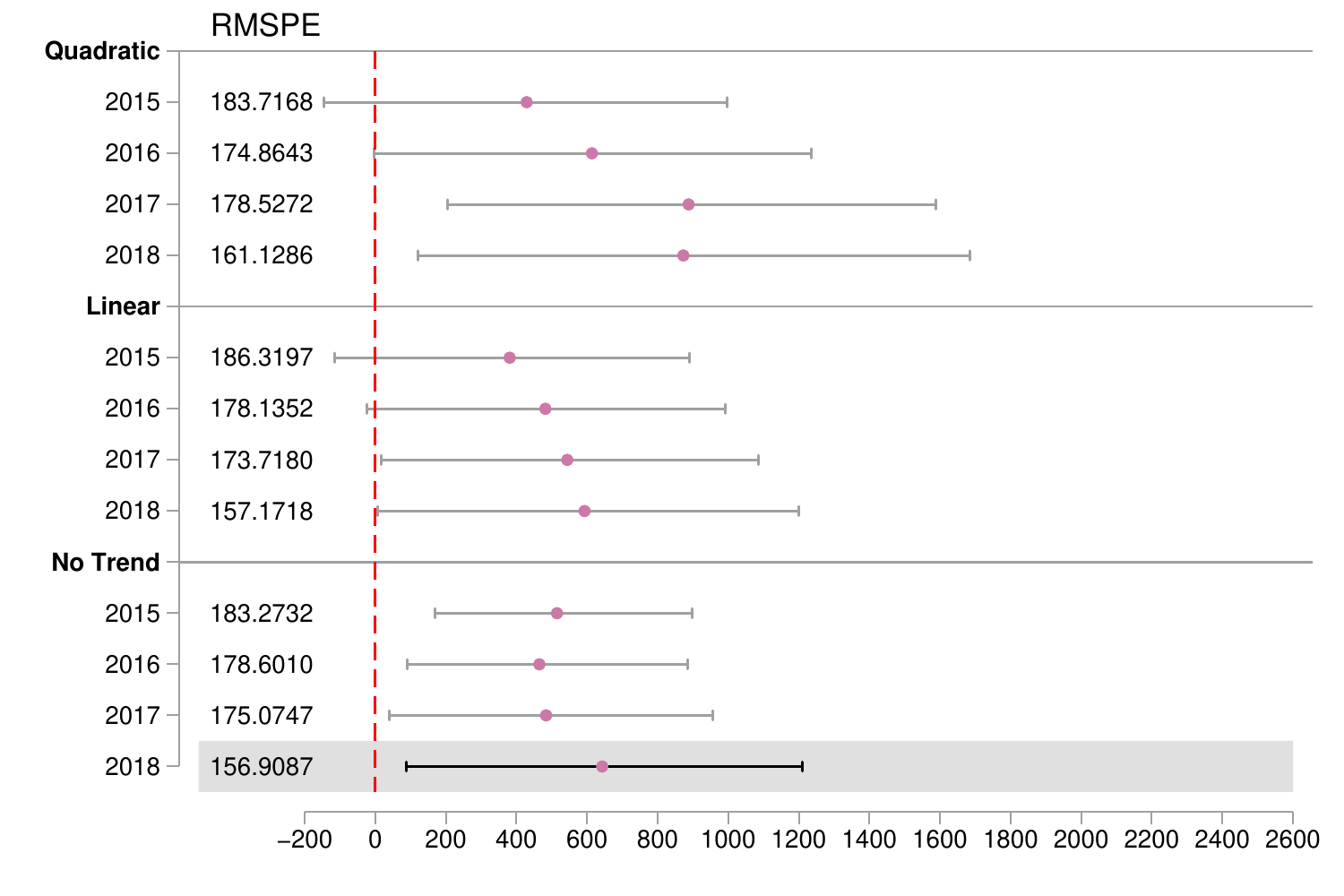}%
}
\subfloat[Sexual Abuse (No school channel)]{%
\includegraphics[width=0.33\textwidth]{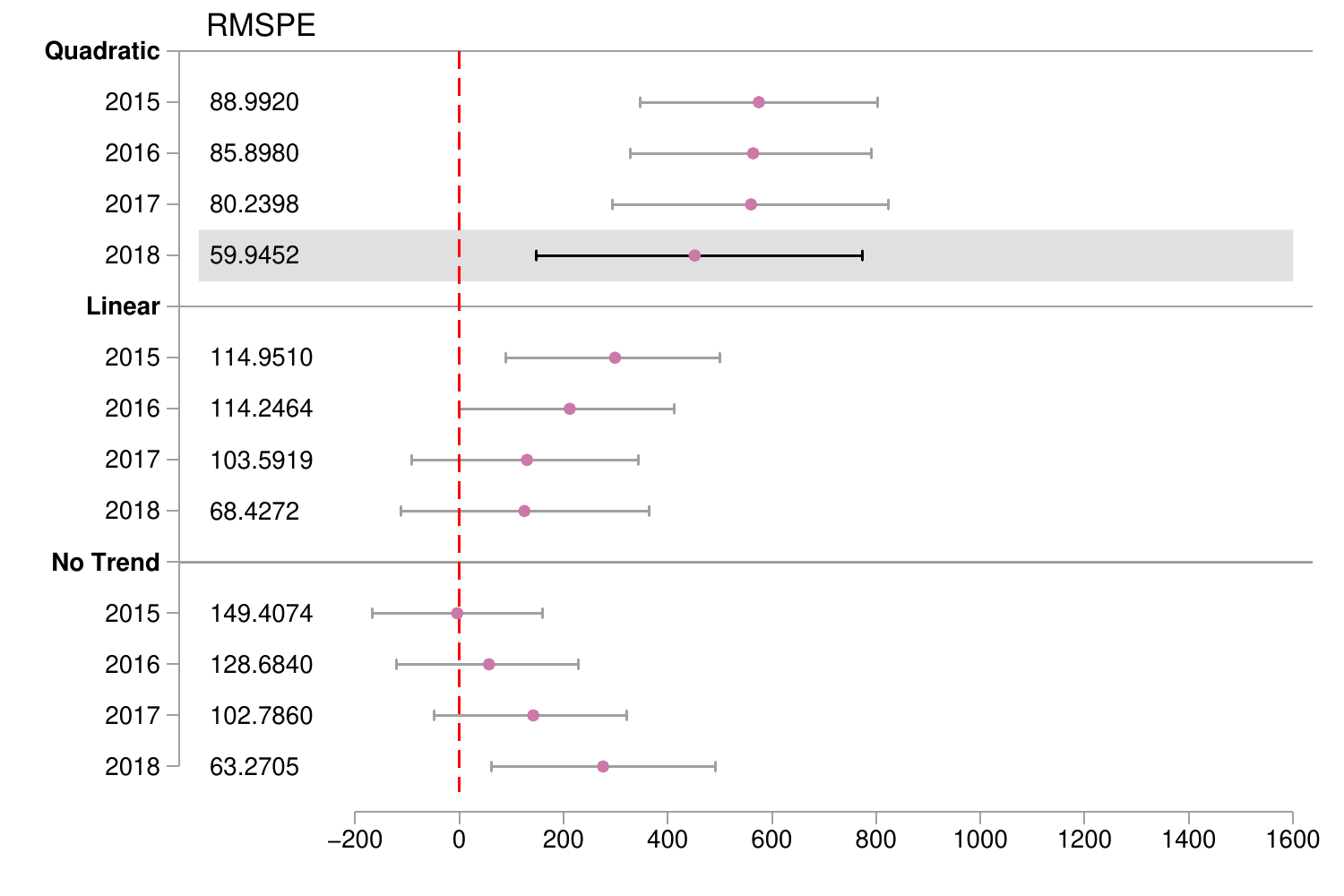}%
}
\subfloat[Rape (No school channel)]{%
\includegraphics[width=0.33\textwidth]{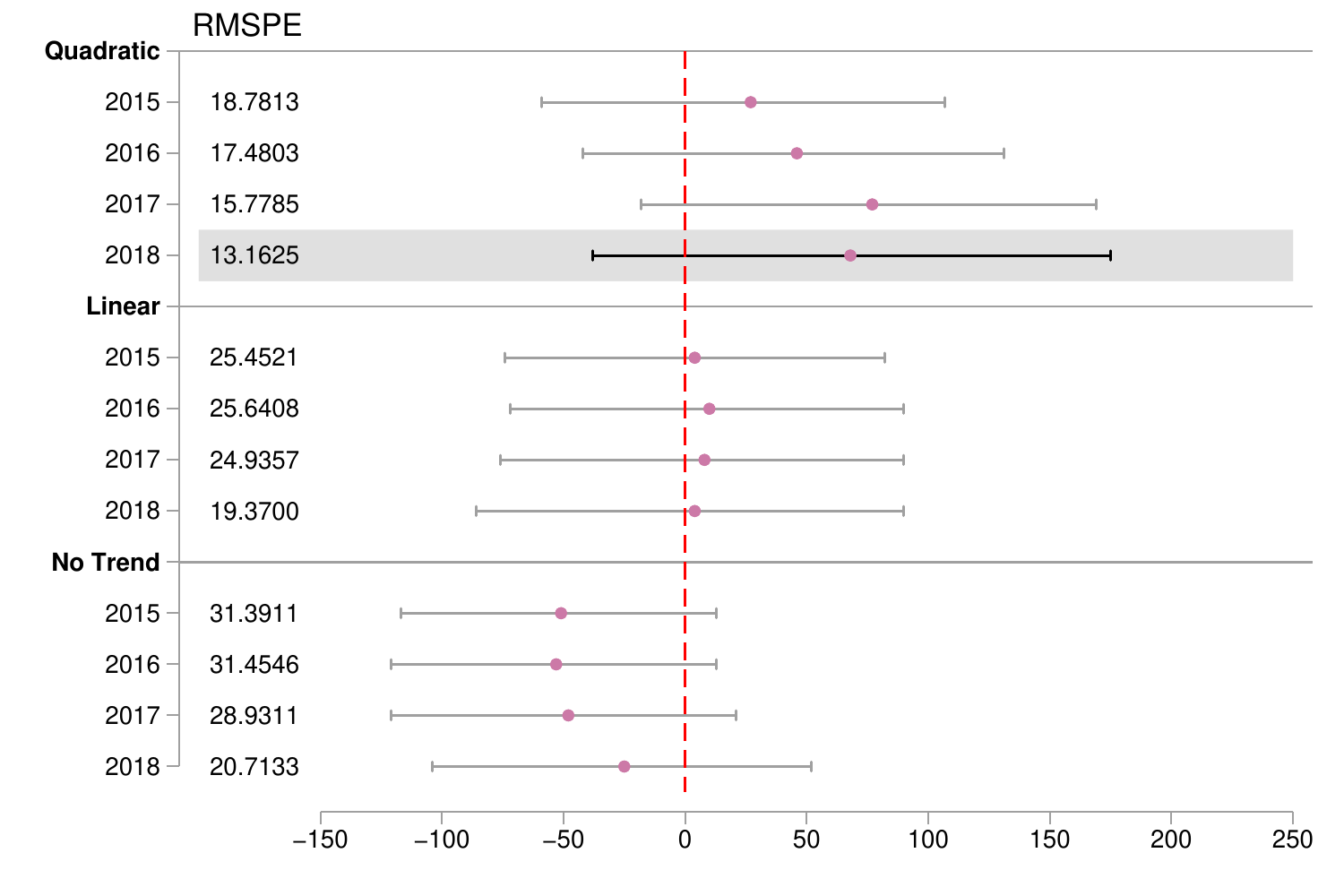}%
}\\
\textbf{Panel B: Post School Reopening} \\
\subfloat[Intra-family violence (No school channel)]{%
\includegraphics[width=0.33\textwidth]{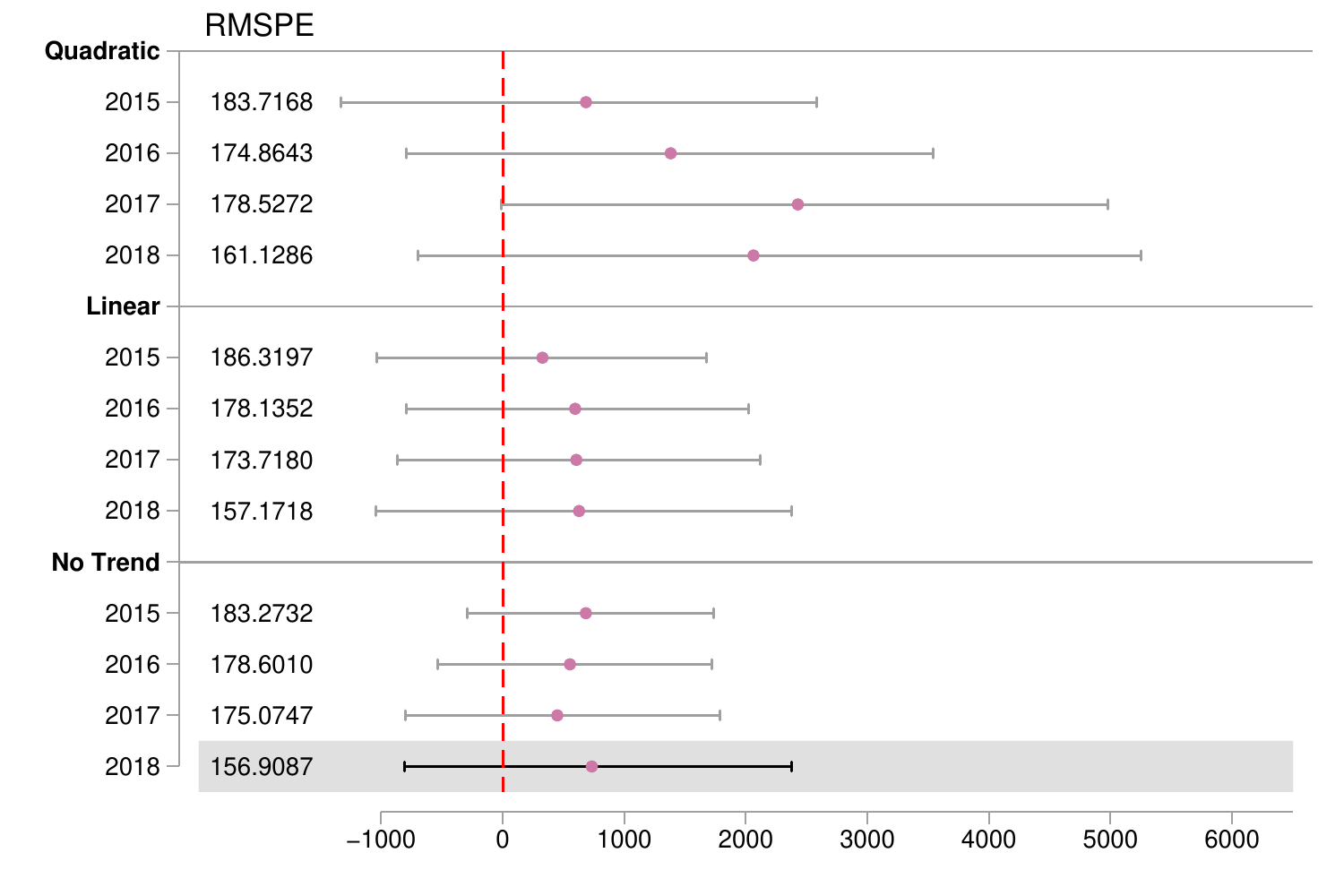}%
}
\subfloat[Sexual Abuse (No school channel)]{%
\includegraphics[width=0.33\textwidth]{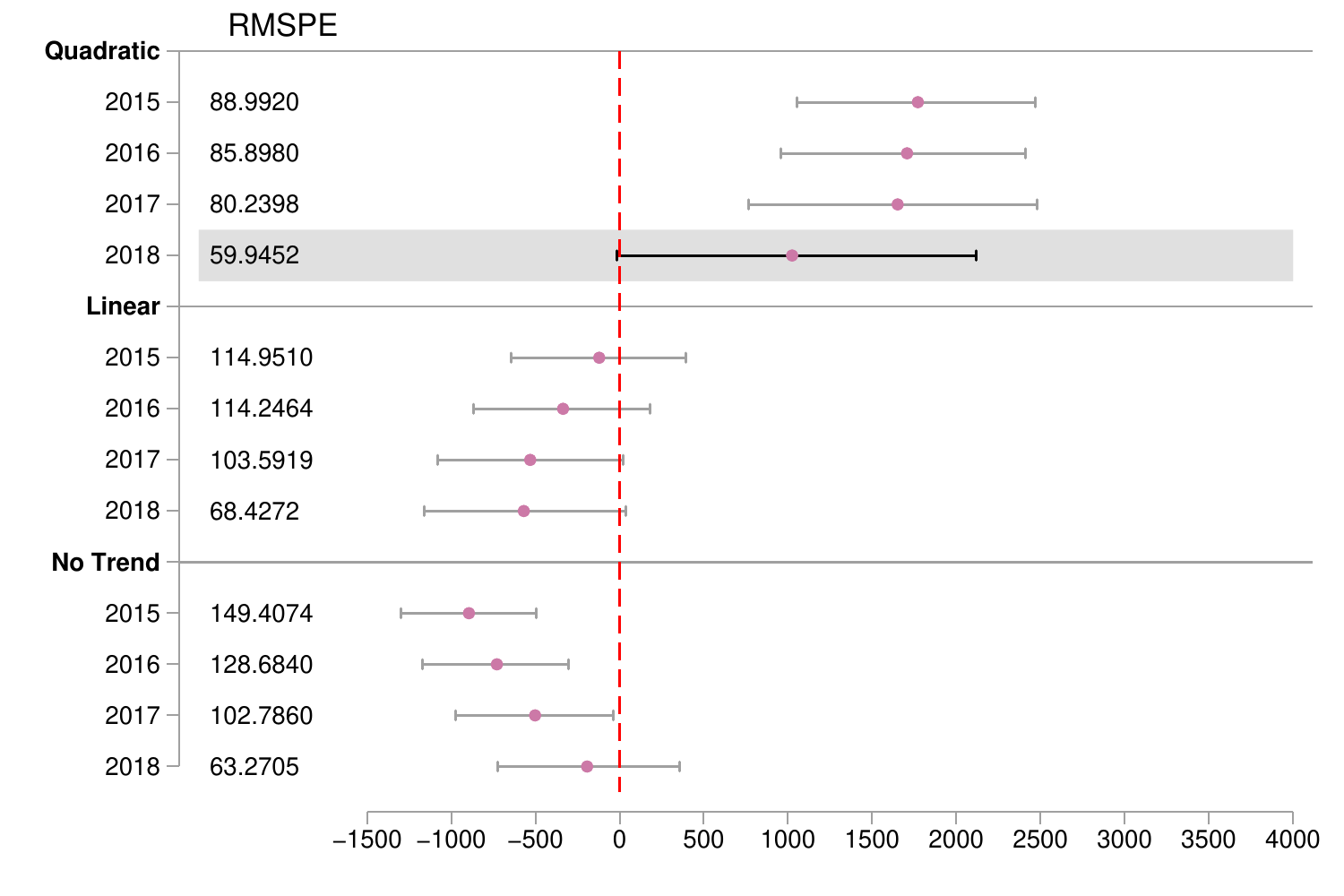}%
}
\subfloat[Rape (No school channel)]{%
\includegraphics[width=0.33\textwidth]{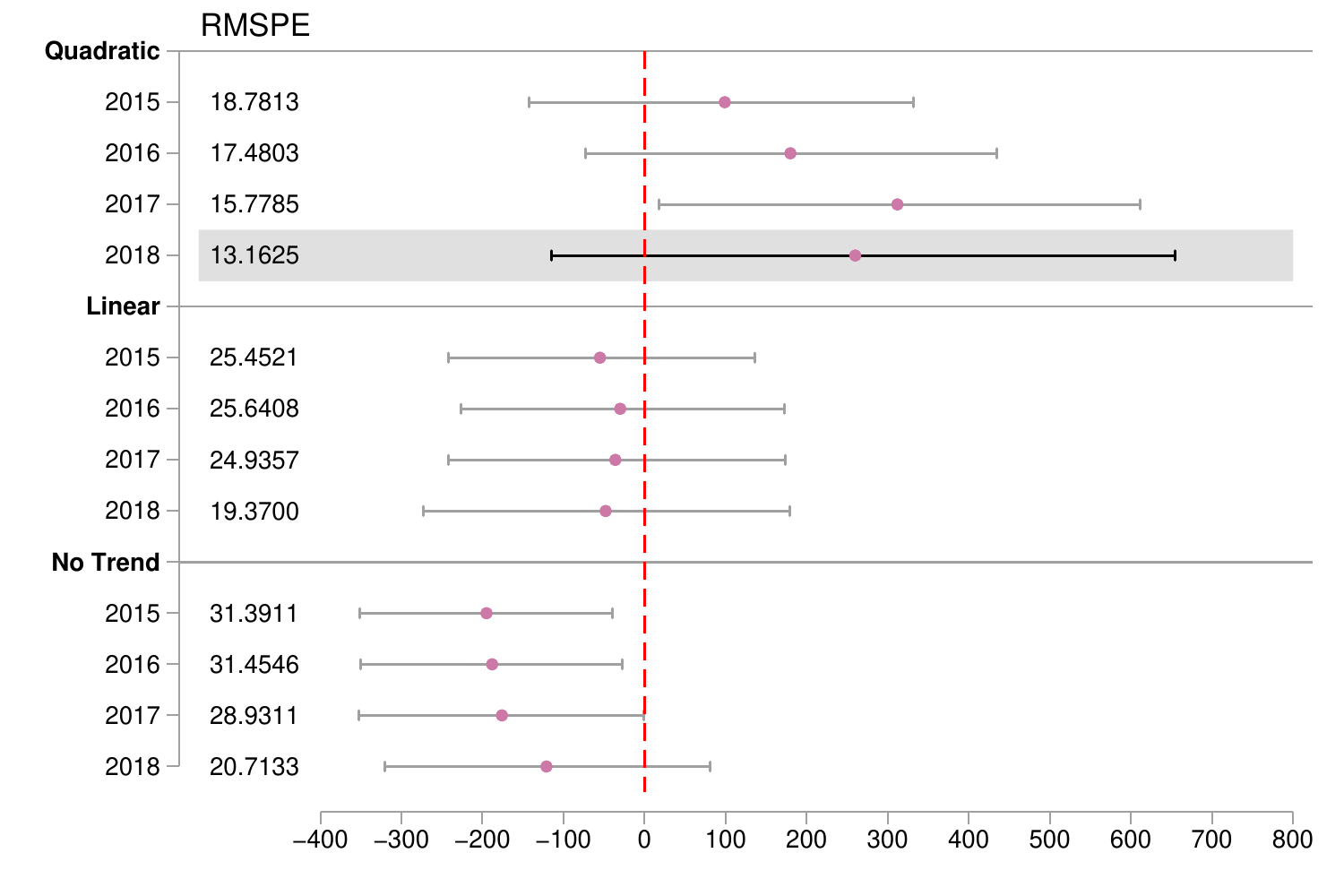}%
}
\end{center}
\floatfoot{\textbf{Notes to Fig.\ \ref{SIfig:Under-reportingNoSchool}}: Refer to Notes to Figure \ref{SIfig:Under-reporting}.  Identical sensitivity analyses are reported, however now for under-reporting displayed in Panel B of Figure 4, where the school closure channel is conditioned out.}
\end{figure}

\begin{figure}[htpb!]
\begin{center}
\caption{Temporal Trends -- Crimes Reported Against Children by Place of Occurrence}
\label{SIfig:partestrends}
\subfloat[Reporting of Intra-family Violence Against Minors]{%
\includegraphics[width=0.33\textwidth]{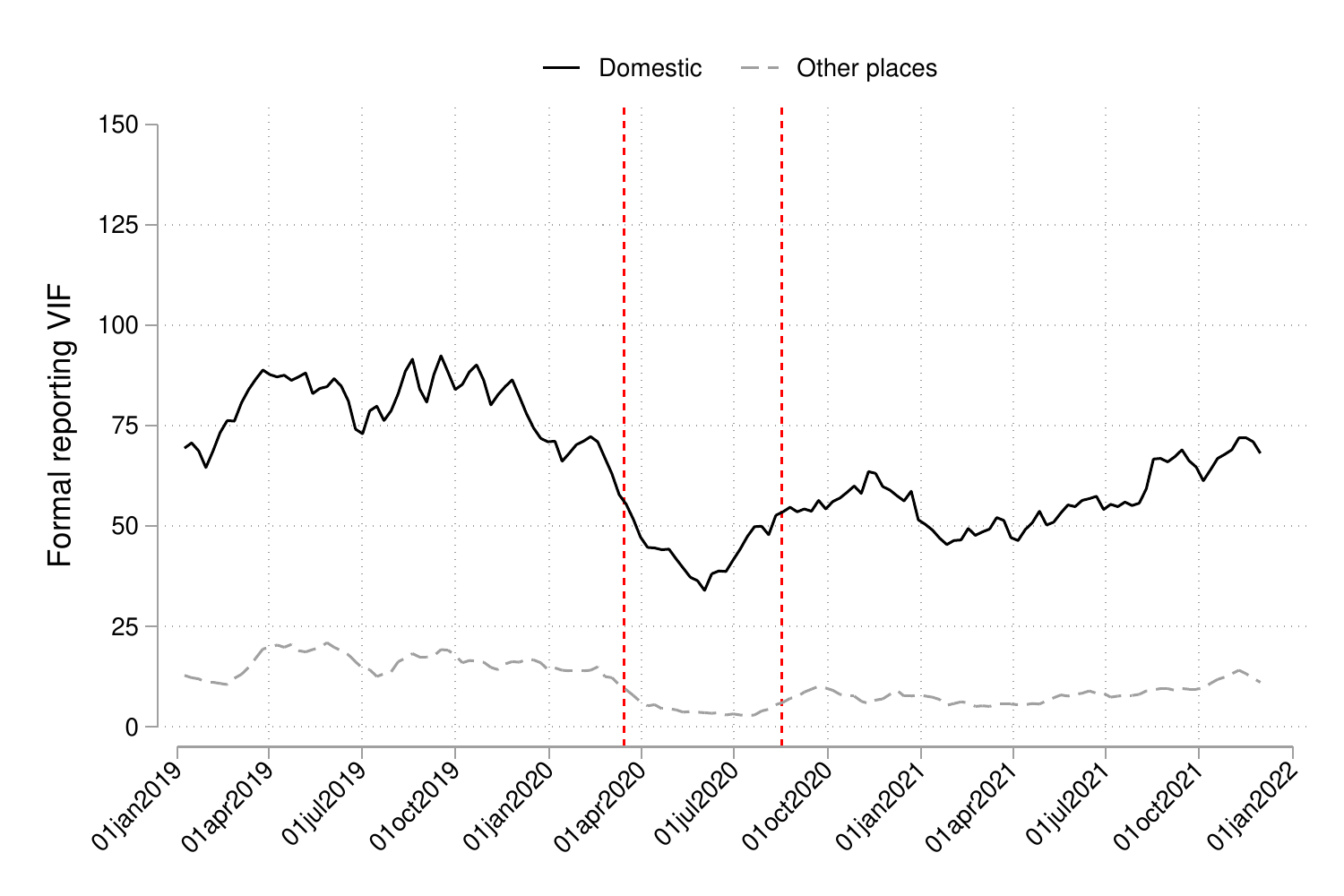}%
}
\subfloat[Reporing of Sexual Assault Against Minors]{%
\includegraphics[width=0.33\textwidth]{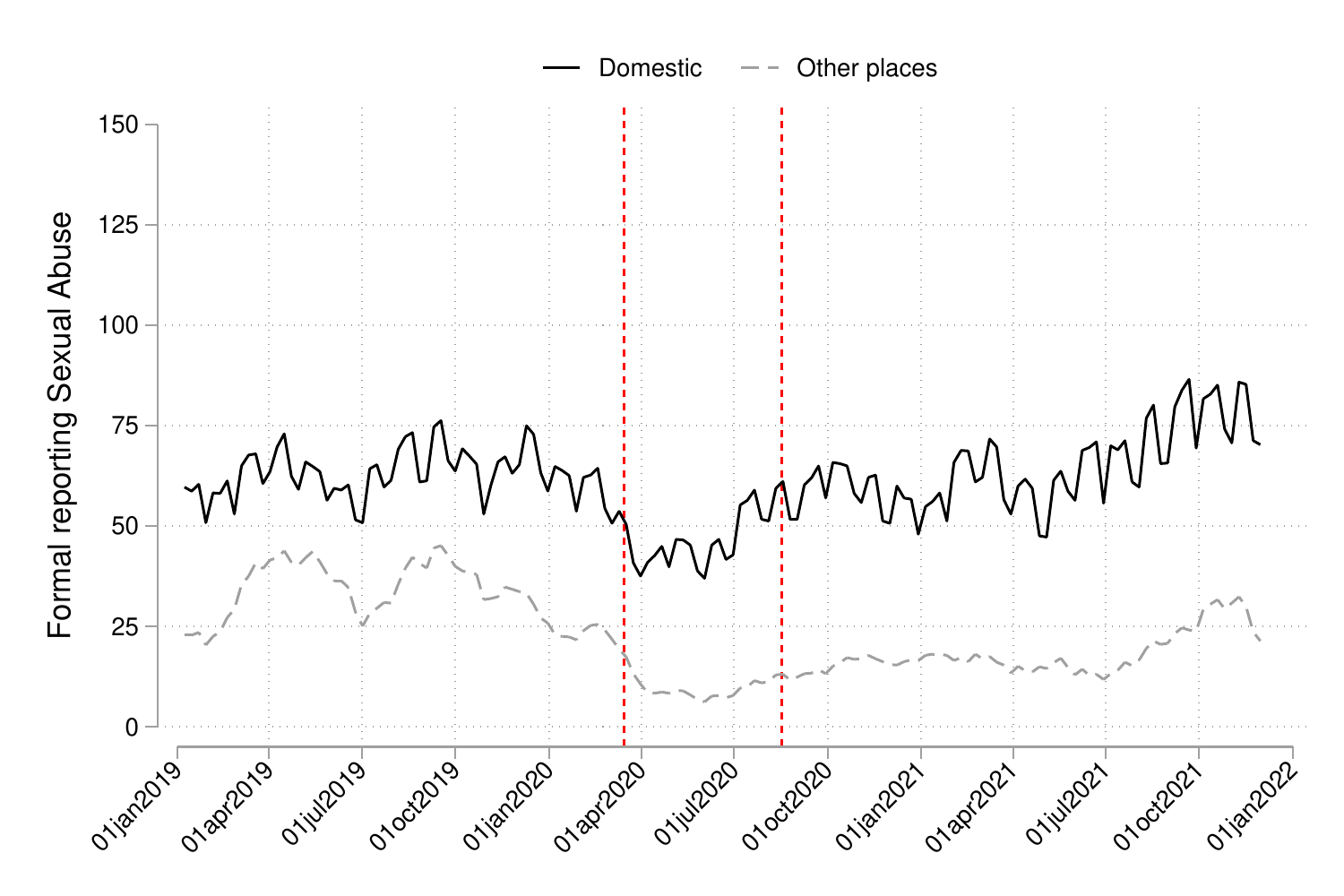}%
}
\subfloat[Reporting of Rape Against Minors]{%
\includegraphics[width=0.33\textwidth]{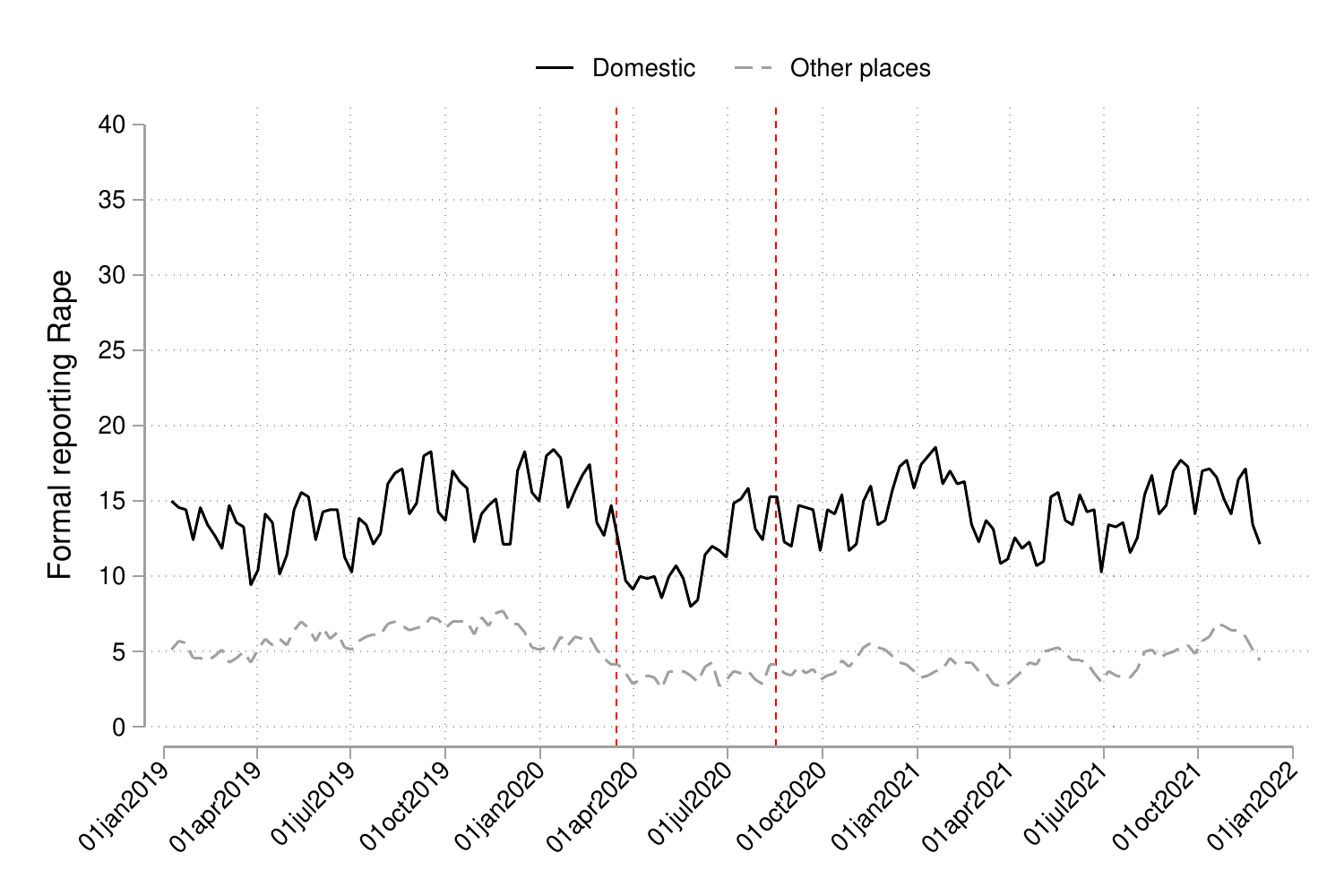}%
}\\
\subfloat[Proportion of Reporting Violence Against Minors]{%
 \includegraphics[width=0.33\textwidth]{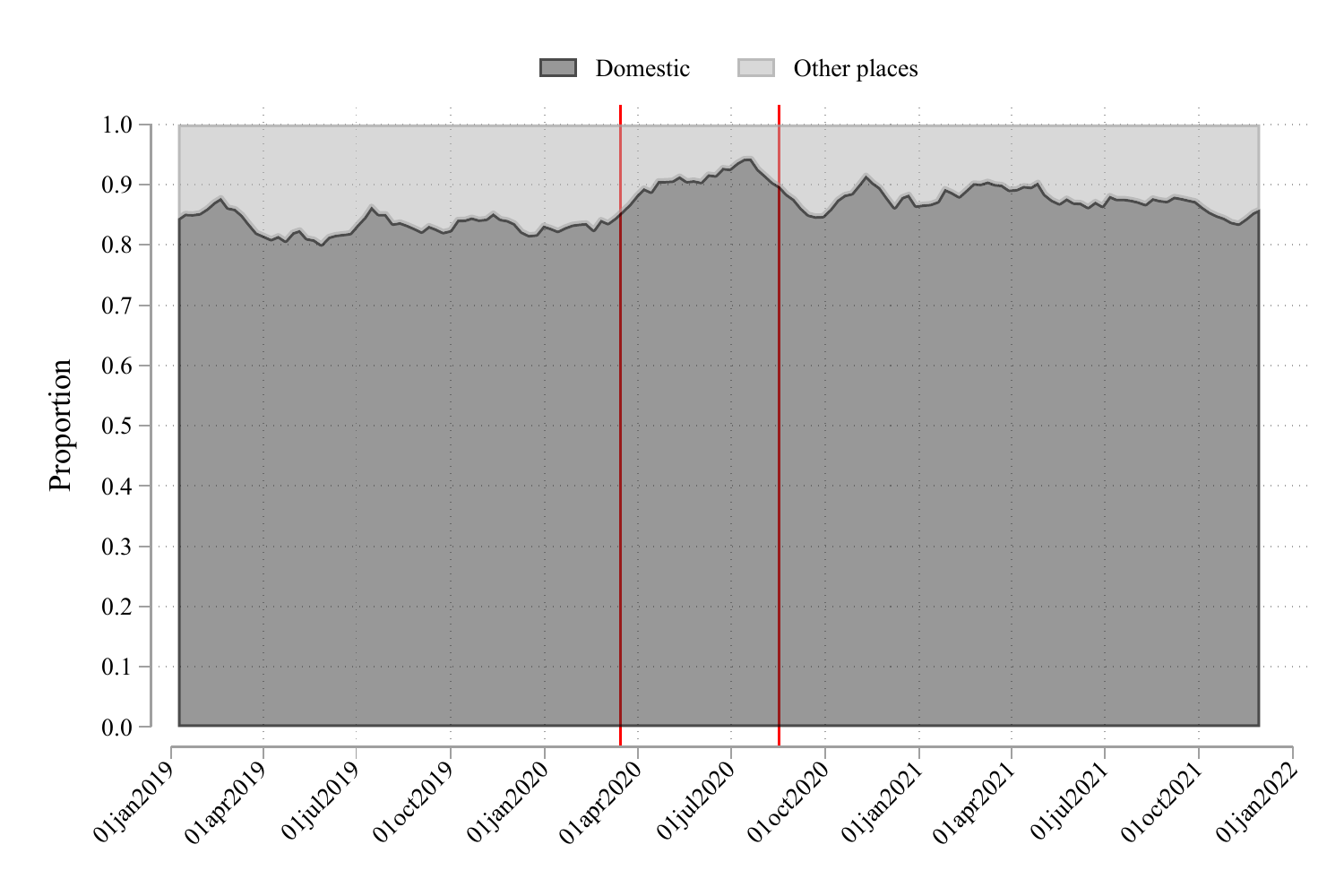}%
}
\subfloat[Proportion of Reporting of Sexual Assault Against Minors]{%
\includegraphics[width=0.33\textwidth]{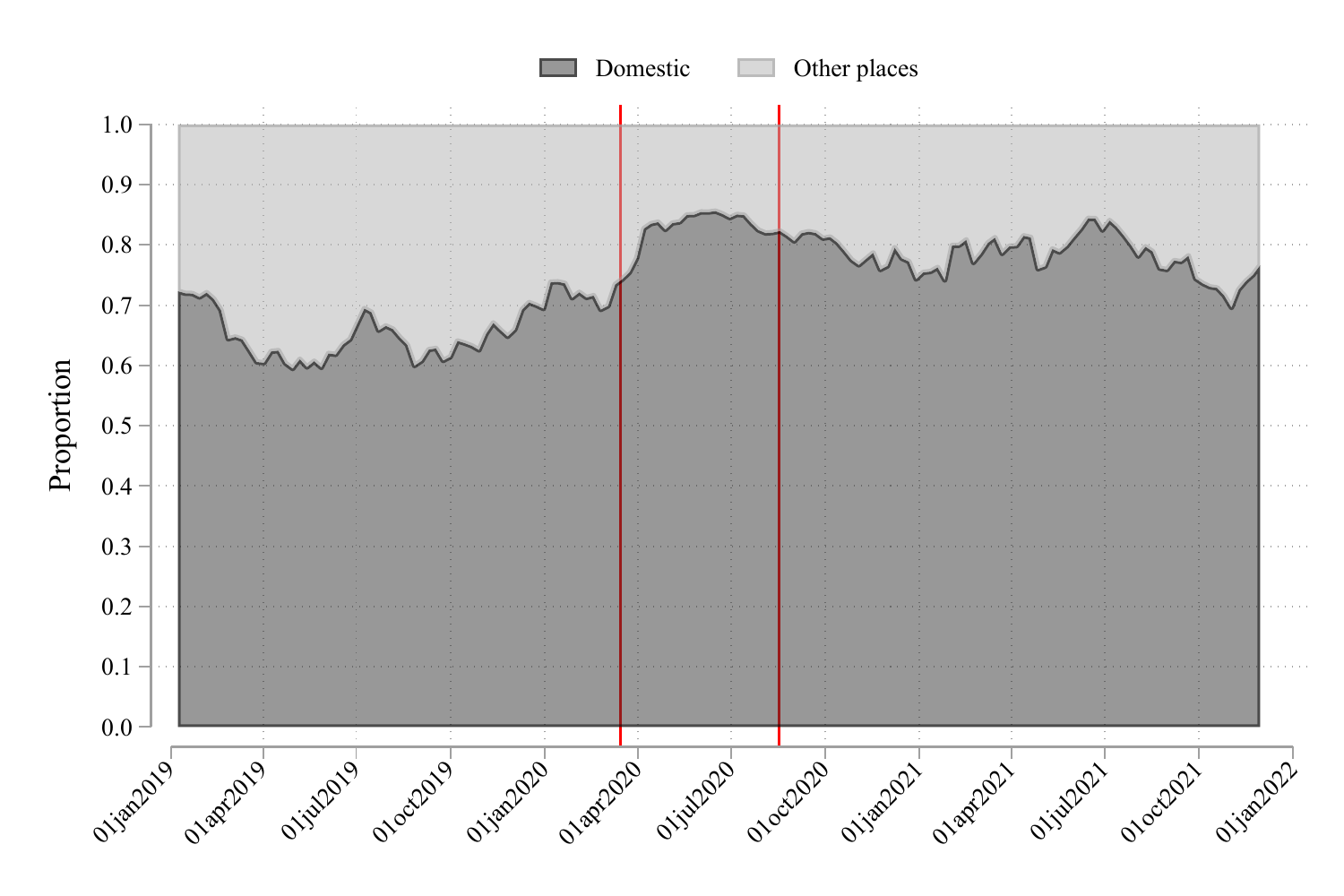}%
}
\subfloat[Proportion of Reporting of Rape Against Minors]{%
 \includegraphics[width=0.33\textwidth]{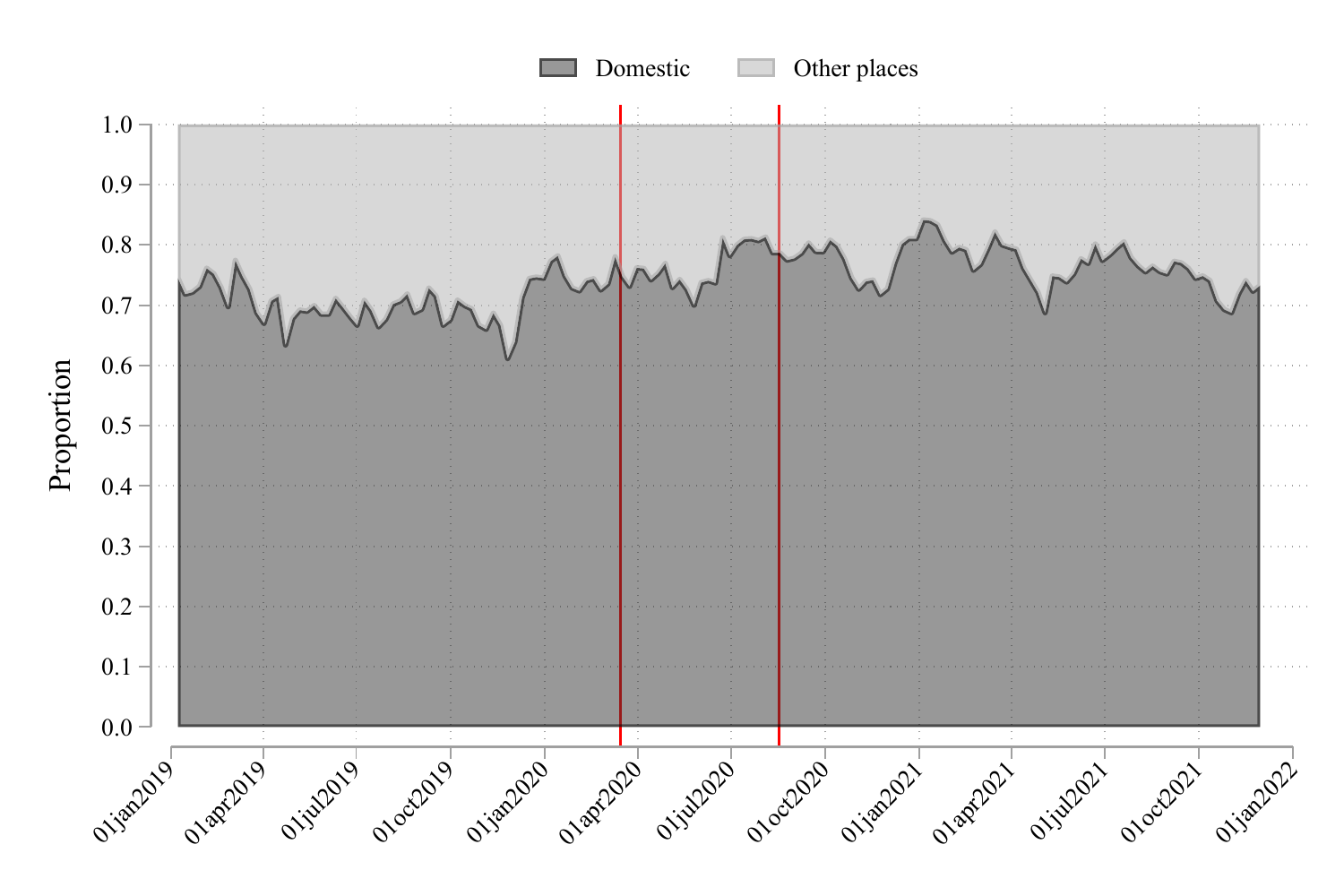}%
}
\end{center}
\floatfoot{\textbf{Notes to Fig.\ \ref{SIfig:partestrends}}: Descriptive trends show the total number of crimes against minors (top panels) and proportion of crimes against minors (bottom panels) classified as occurring in a private home (``Domestic''), or in another place (``Other places'').  The total number and proportion of cases are documented by week for the full period of analysis.  Information on the location of occurrence is not available in data on crime victims, but rather an auxiliary database covering all crimes.  As crimes can have more than 1 victim, these descriptive trends have slightly less crimes than victimization data documented in Figure 1 of the main analysis.  In all cases, crimes are defined as in the main analysis (intra-family violence in panels (a) and (d), sexual assault in panels (b) and (e), and rape in panels (c) and (f)).  Vertical red lines document the data of first school closure, and first re-opening.}
\end{figure}

\begin{figure}[t!]
\begin{center}
\caption{Trends in Other Relevant Factors}
\label{SIfig:othertrends}
\subfloat[Stress and COVID (Web Search Frequency)]{%
\includegraphics[width=0.5\textwidth]{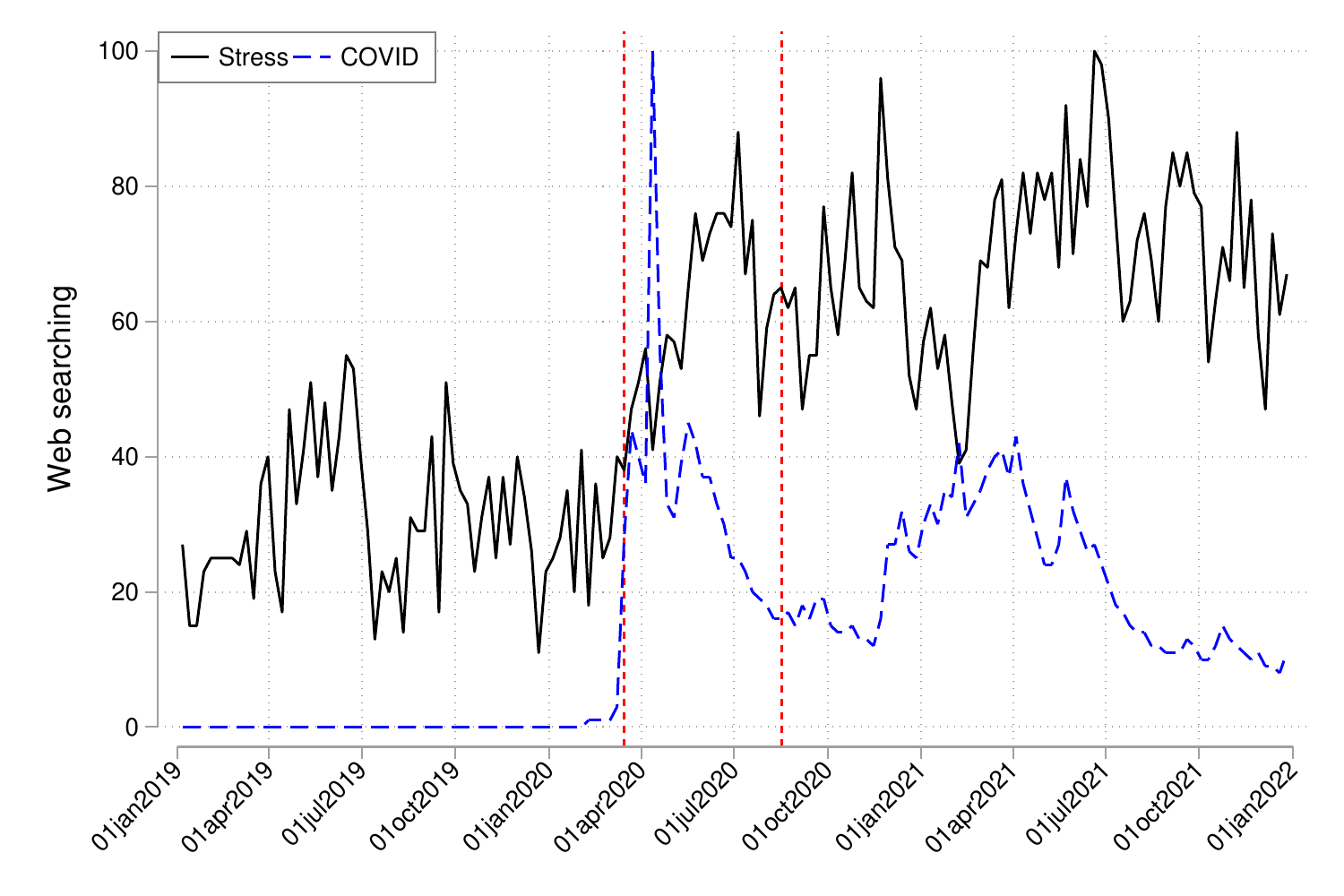}%
}
\subfloat[Calls to \#149]{%
\includegraphics[width=0.5\textwidth]{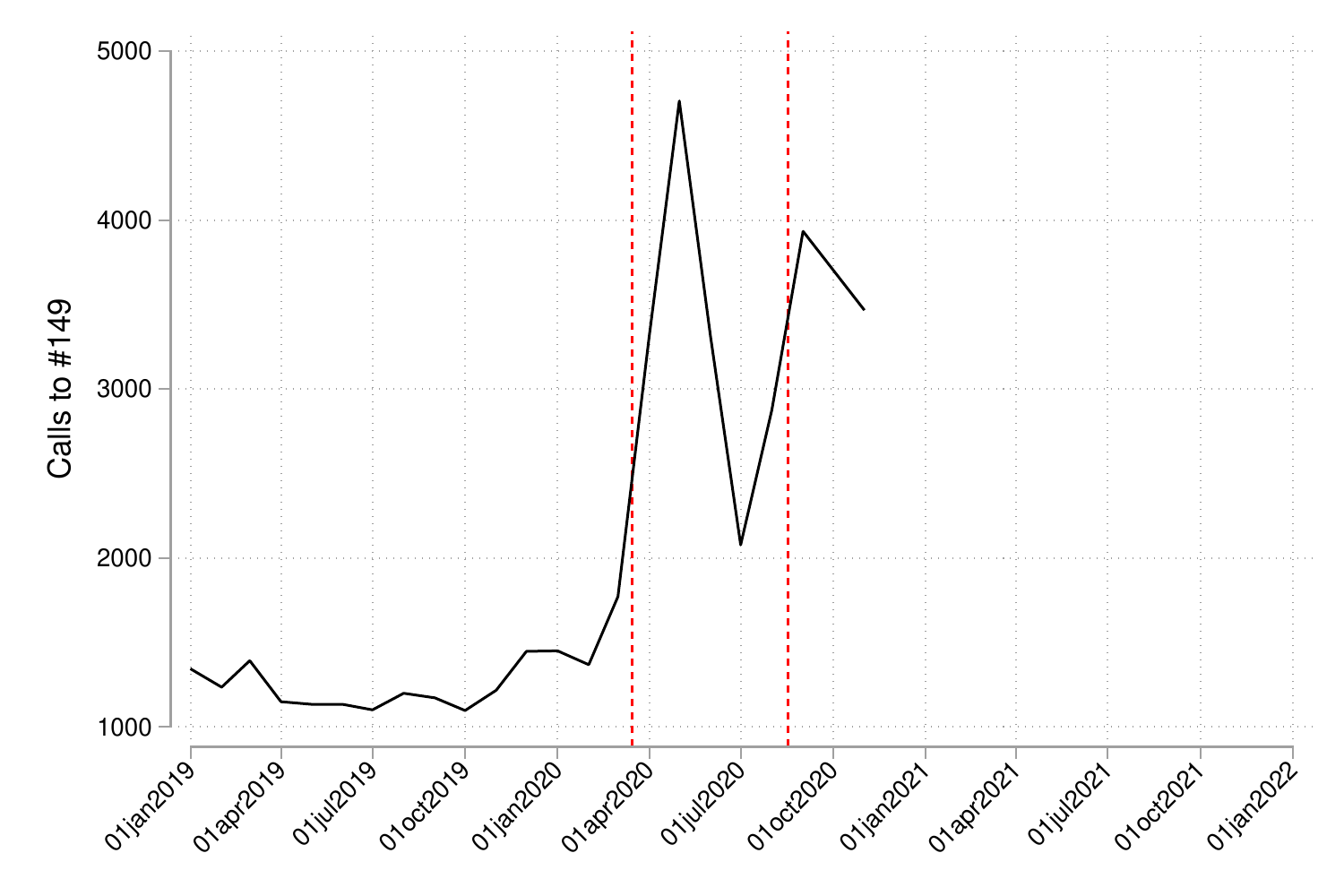}%
}\\
\subfloat[Unemployment]{%
\includegraphics[width=0.5\textwidth]{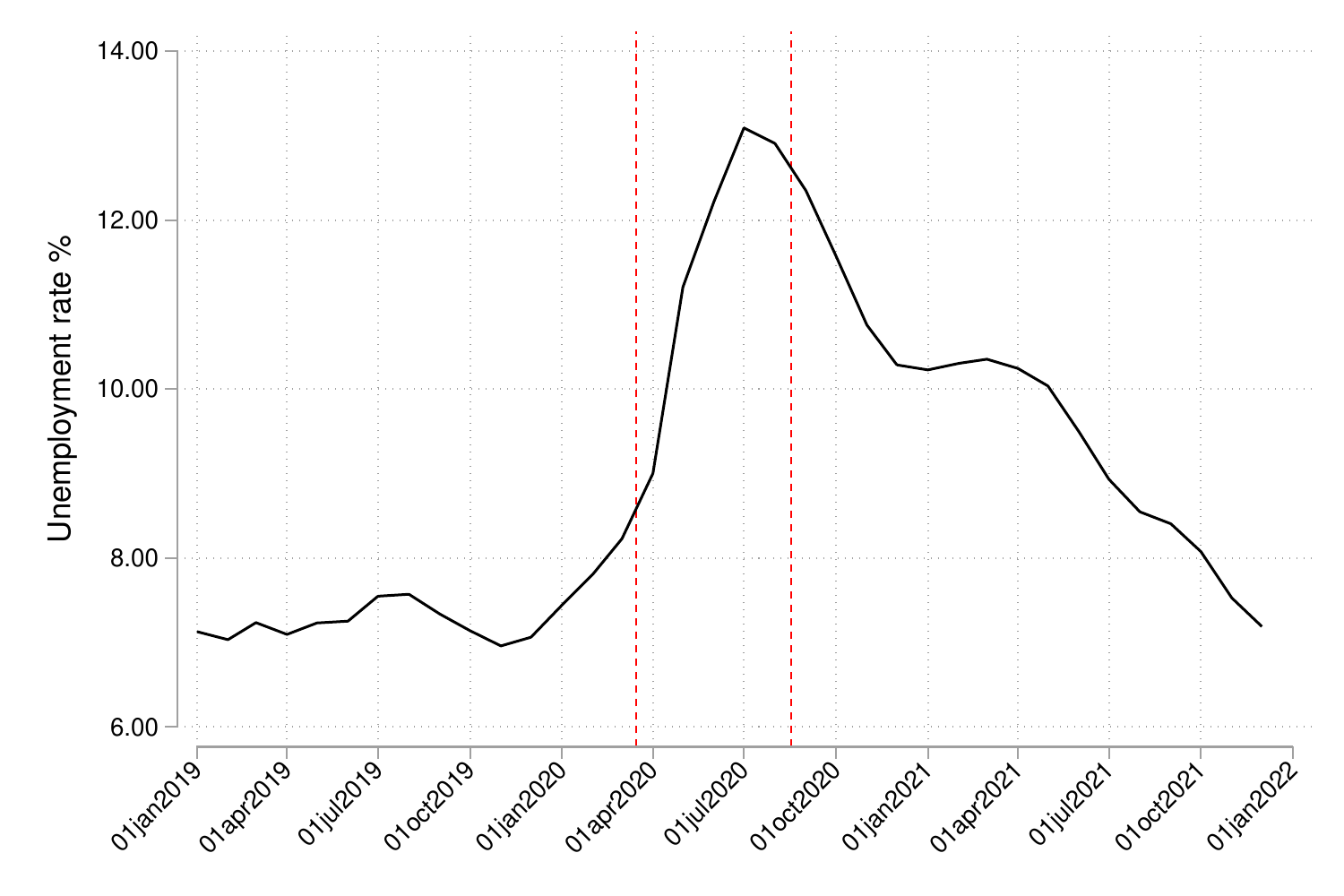}%
}
\subfloat[Home time]{%
\includegraphics[width=0.5\textwidth]{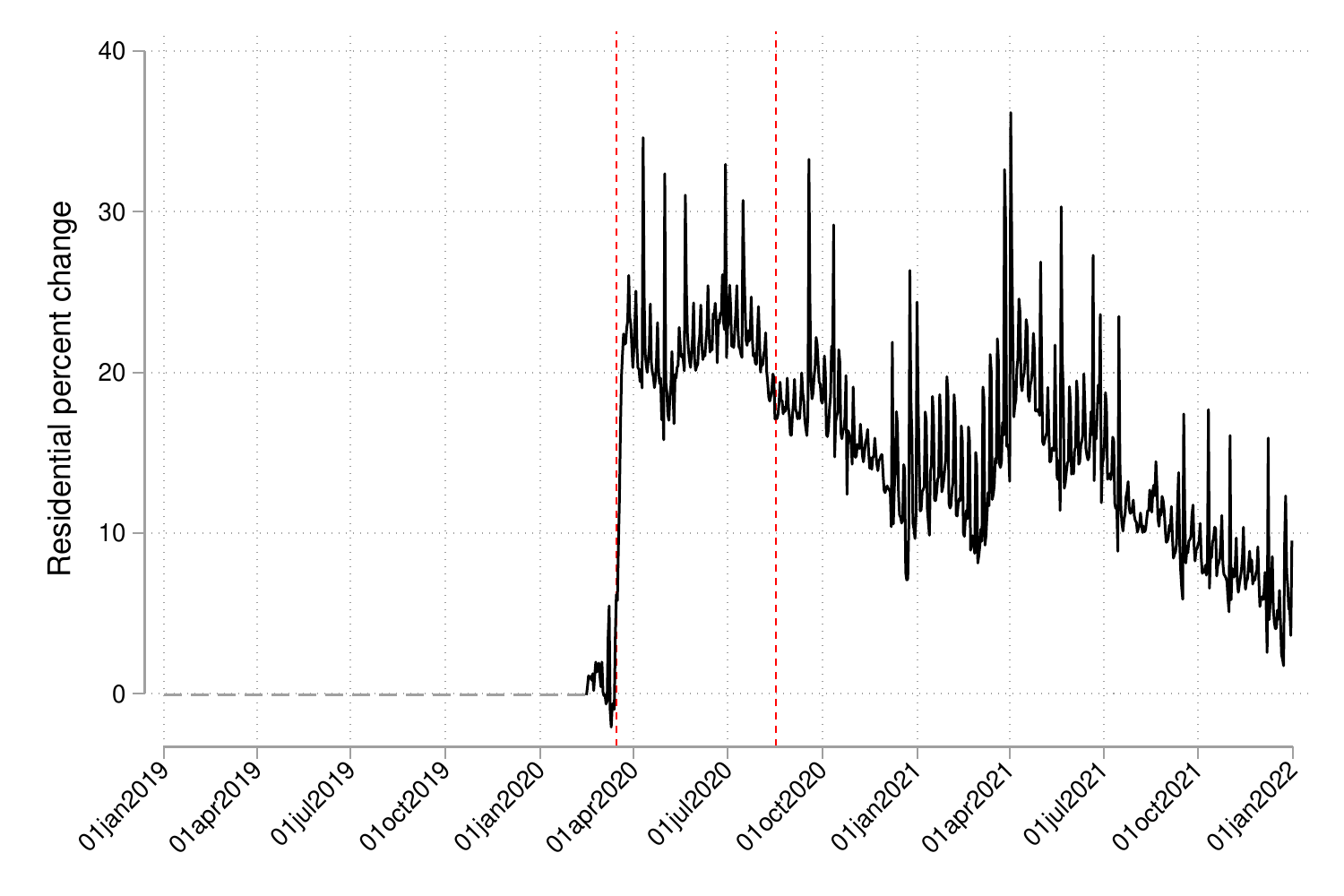}%
}
\end{center}
\floatfoot{\textbf{Notes to Fig.\ \ref{SIfig:othertrends}}: Trends by day (panel (d)), week (panel (a)), or month (panels (b) and (c)) of other relevant factors in Chile are displayed over time.  Panel (a) documents relative frequency of online search based on Google trends data in Chile for the term `stress' (\emph{estr\'es}) and COVID (for comparison), suggesting increases in searches related to stress following the arrival of COVID to the country. Panel (b) documents the monthly quantity of calls to the Police's Family Help Phone line (\emph{fono familia}, \# 149), which are only available until November 2021.  Panel (c) documents monthly unemployment rates as reported by the Central Bank of Chile.  And Panel (d) documents relative changes in the amount of time which individuals are estimated to spend in residential areas based on Google's Community Mobility Reports in the country. Vertical red lines document the data of first school closure, and first re-opening.}
\end{figure}
    
\begin{spacing}{1.3}    
\captionsetup[figure]{list=yes}
\captionsetup[table]{list=yes}
\setcounter{table}{0}
\renewcommand{\thetable}{A\arabic{table}}
\setcounter{figure}{0}
\renewcommand{\thefigure}{A\arabic{figure}}
    
\clearpage

\section*{Appendix Tables}    
\begin{table}[htpb!]
\centering
\caption{Summary Statistics of Principal Variables}
\label{SItab:sumstats}
    \begin{tabular}{lccccc} \\ \toprule
    & Observations & Mean & Std.\ Dev.\ & Min.\ & Max.\ \\ \midrule
    \multicolumn{6}{l}{\textbf{Panel A: Violence Against Children}}\\
    Intra-family Violence&       54321&        3.93&       18.91&        0.00&     2325.58\\
\ \ Physical Violence (serious)&       54321&        0.13&        2.20&        0.00&      116.82\\
\ \ Physical Violence (moderate)&       54321&        2.44&       12.62&        0.00&     1086.96\\
\ \ Psychological Violence&       54321&        1.35&       13.81&        0.00&     2325.58\\
Sexual Abuse        &       53283&        2.94&       14.09&        0.00&     1020.41\\
Rape                &       53283&        0.60&        2.18&        0.00&       80.71\\
\\
    \multicolumn{6}{l}{\textbf{Panel B: Schools}}\\
    School Closure      &       54321&        0.31&        0.46&        0.00&        1.00\\
School Reopening (Binary)&       54214&        0.29&        0.45&        0.00&        1.00\\
School Reopening (Continuous)&       54214&        0.19&        0.35&        0.00&        1.00\\
Attendance (1 day)  &       45802&        0.54&        0.47&        0.00&        1.00\\
Attendance (1-5 days)&       45802&        0.49&        0.48&        0.00&        1.00\\
Attendance (6-10 days)&       45802&        0.49&        0.48&        0.00&        1.00\\
Attendance (10+ days)&       45802&        0.50&        0.48&        0.00&        1.00\\
\\
    \multicolumn{6}{l}{\textbf{Panel C: COVID/Other Measures}}\\
    Quarantine          &       54321&        0.10&        0.30&        0.00&        1.00\\
COVID-19 Cases per 1,000&       54321&        0.68&        1.88&        0.00&      233.58\\
PCR Testing per 1,000&       54321&        8.10&       10.69&        0.00&      179.53\\
PCR Test Positivity &       54321&        5.27&        9.66&        0.00&      100.00\\
Popultaion between 5-18 years&       54321&    10222.90&    15866.86&        9.00&   127390.00\\
\\ \bottomrule
    \multicolumn{6}{p{15cm}}{\footnotesize \textbf{Notes to Table \ref{SItab:sumstats}}: Summary statistics are displayed across all municipal by week cells for the period of January 2019--December 2021. Panel A documents mean outcomes of violence against children measured as weekly reports per 100,000 children in each municipality.  Panel B documents principal measures of school closure and re-opening, as well as attendance figures for periods in which attendance is available.  Panel C documents epidemiological controls.}
    \end{tabular}
\end{table}

\begin{table}[htpb!]
    \caption{Estimated Impacts of School Closure, School Reopening, and Attendance on Intra-family Violence Reports}
    \label{SItab:attendanceDV}
    \centering
    \scalebox{0.9}{
    \begin{tabular}{lcccccc} \toprule
    & \multicolumn{6}{c}{Intra-family Violence}\\ 
    \cmidrule(r){2-7}
    & (1) & (2) & (3) & (4) & (5) & (6) \\ \midrule 
    \multicolumn{7}{l}{\textbf{Panel A:} Binary Re-opening Measure} \\
    School Closure      &      -1.381***&      -1.381***&      -1.533***&      -1.535***&      -1.253***&      -1.258***\\
                    &     (0.105)   &     (0.105)   &     (0.135)   &     (0.136)   &     (0.159)   &     (0.161)   \\
School Reopening    &      -0.356** &      -1.123***&      -0.586***&      -0.702** &      -0.559** &      -0.739** \\
                    &     (0.156)   &     (0.247)   &     (0.182)   &     (0.291)   &     (0.246)   &     (0.362)   \\
School Reopening $\times$ Attendance&               &       1.985***&               &       0.304   &               &       0.451   \\
                    &               &     (0.602)   &               &     (0.656)   &               &     (0.675)   \\
 \\
    Reopening Effect at Percentile 25 of Attendance &&    -0.671         &&    -0.633 &&    -0.637   \\
Reopening Effect at Percentile 50 of Attendance &&    -0.166         &&    -0.555 &&    -0.522   \\
Reopening Effect at Percentile 75 of Attendance &&     0.211         &&    -0.498 &&    -0.436   \\
Reopening Effect at Percentile 90 of Attendance &&     0.540         &&    -0.447 &&    -0.361   \\
Baseline Mean    &     4.302 &     4.302 &     4.302          &     4.302 &     4.302 &     4.302        \\
Observations &    45,802 &    45,802          &    45,802 &    45,802 &    45,802 &    45,802      \\
 \\
    \midrule
    \multicolumn{7}{l}{\textbf{Panel B:} Continuous Re-opening Measure} \\
    School Closure      &      -1.360***&      -1.380***&      -1.514***&      -1.526***&      -1.198***&      -1.229***\\
                    &     (0.102)   &     (0.103)   &     (0.130)   &     (0.132)   &     (0.139)   &     (0.143)   \\
School Reopening    &      -0.331*  &      -1.942***&      -0.655***&      -1.211***&      -0.540** &      -1.209***\\
                    &     (0.181)   &     (0.385)   &     (0.205)   &     (0.397)   &     (0.249)   &     (0.455)   \\
School Reopening $\times$ Attendance&               &       3.504***&               &       1.228   &               &       1.365   \\
                    &               &     (0.828)   &               &     (0.829)   &               &     (0.856)   \\
 \\
    Reopening Effect at Percentile 25 of Attendance &&    -1.144         &&    -0.931 &&    -0.898   \\
Reopening Effect at Percentile 50 of Attendance &&    -0.253         &&    -0.619 &&    -0.551   \\
Reopening Effect at Percentile 75 of Attendance &&     0.412         &&    -0.386 &&    -0.292   \\
Reopening Effect at Percentile 90 of Attendance &&     0.994         &&    -0.182 &&    -0.065   \\
Baseline Mean    &     4.302 &     4.302 &     4.302          &     4.302 &     4.302 &     4.302        \\
Observations &    45,802 &    45,802          &    45,802 &    45,802 &    45,802 &    45,802      \\
 \\
    \midrule
    Municipal \& WoY FEs     &  &  & Y & Y & Y & Y  \\
    Lockdown \& Epidemiological controls &  &  & &  & Y & Y \\
    \bottomrule
    \multicolumn{7}{p{19.9cm}}{{\footnotesize \textbf{Notes to Tab.\ \ref{SItab:attendanceDV}}:  Results replicate those of Table 1 from the main text for the outcome of intra-family violence against children, however here additionally interacting School Reopening measures with the proportion of attendance in each municipality by week cell.  Attendance data is not observed for all periods (refer to SI section \ref{SIscn:data}, and as such, these models are only estimated for periods in which all data are available.  Columns 1, 3 and 5 present baseline models for the period in which attendance data are available, while columns 2, 4 and 6 additionally include the School Reopening by Attendance interaction. In table footers, linear estimated effects on school re-opening are reported at various margins of attendance, namely percentiles 25, 50, 75 and 90 of attendance from observed data.  These percentiles are p25=26.6\% attendance, p50=49.9\% attendance, p75=68.2\% attendance, and p90=84.7\% attendance. All other details follow those laid out in Table 1. $^{***}$ p$<0.01$; $^{**}$ p$<0.05$; $^{*}$ p$<0.10$.}}
    \end{tabular}}
\end{table}

\begin{table*}[htpb!]
    \caption{Estimated Impacts of School Closure, School Reopening, and Attendance on Sexual Assault Reports}
    \label{SItab:attendanceSA}
    \centering
    \scalebox{0.9}{
    \begin{tabular}{lcccccc} \toprule
    & \multicolumn{6}{c}{Sexual Abuse}\\ 
    \cmidrule(r){2-7}
    & (1) & (2) & (3) & (4) & (5) & (6) \\ \midrule 
    \multicolumn{7}{l}{\textbf{Panel A:} Binary Re-opening Measure} \\
    School Closure      &      -0.817***&      -0.817***&      -0.966***&      -0.972***&      -0.995***&      -1.007***\\
                    &     (0.068)   &     (0.068)   &     (0.085)   &     (0.085)   &     (0.098)   &     (0.098)   \\
School Reopening    &       0.025   &      -0.847***&      -0.199** &      -0.536***&      -0.415***&      -0.823***\\
                    &     (0.083)   &     (0.152)   &     (0.095)   &     (0.149)   &     (0.134)   &     (0.180)   \\
School Reopening $\times$ Attendance&               &       2.257***&               &       0.888** &               &       1.022***\\
                    &               &     (0.357)   &               &     (0.365)   &               &     (0.367)   \\
 \\
    Reopening Effect at Percentile 25 of Attendance &&    -0.334         &&    -0.334 &&    -0.590   \\
Reopening Effect at Percentile 50 of Attendance &&     0.240         &&    -0.108 &&    -0.330   \\
Reopening Effect at Percentile 75 of Attendance &&     0.668         &&     0.061 &&    -0.136   \\
Reopening Effect at Percentile 90 of Attendance &&     1.043         &&     0.208 &&     0.033   \\
Baseline Mean    &     2.677 &     2.677 &     2.677          &     2.677 &     2.677 &     2.677        \\
Observations &    45,802 &    45,802          &    45,802 &    45,802 &    45,802 &    45,802      \\
 \\
    \midrule
    \multicolumn{7}{l}{\textbf{Panel B:} Continuous Re-opening Measure} \\
    School Closure      &      -0.803***&      -0.824***&      -0.952***&      -0.968***&      -0.923***&      -0.965***\\
                    &     (0.067)   &     (0.067)   &     (0.083)   &     (0.083)   &     (0.092)   &     (0.092)   \\
School Reopening    &       0.093   &      -1.534***&      -0.189*  &      -0.923***&      -0.314** &      -1.238***\\
                    &     (0.091)   &     (0.226)   &     (0.106)   &     (0.218)   &     (0.139)   &     (0.252)   \\
School Reopening $\times$ Attendance&               &       3.540***&               &       1.622***&               &       1.888***\\
                    &               &     (0.489)   &               &     (0.487)   &               &     (0.494)   \\
 \\
    Reopening Effect at Percentile 25 of Attendance &&    -0.729         &&    -0.554 &&    -0.809   \\
Reopening Effect at Percentile 50 of Attendance &&     0.172         &&    -0.141 &&    -0.329   \\
Reopening Effect at Percentile 75 of Attendance &&     0.843         &&     0.167 &&     0.029   \\
Reopening Effect at Percentile 90 of Attendance &&     1.431         &&     0.436 &&     0.343   \\
Baseline Mean    &     2.677 &     2.677 &     2.677          &     2.677 &     2.677 &     2.677        \\
Observations &    45,802 &    45,802          &    45,802 &    45,802 &    45,802 &    45,802      \\
 \\
    \midrule
    Municipal \& WoY FEs     &  &  & Y & Y & Y & Y  \\
    Lockdown \& Epidemiological controls &  &  & &  & Y & Y \\
    \bottomrule
    \multicolumn{7}{p{19.9cm}}{{\footnotesize \textbf{Notes to Tab.\ \ref{SItab:attendanceSA}}: Refer to Notes to Table \ref{SItab:attendanceDV}.  Identical models are estimated, however here for the outcome of sexual abuse against children. $^{***}$ p$<0.01$; $^{**}$ p$<0.05$; $^{*}$ p$<0.10$.}}
    \end{tabular}}
\end{table*}
\clearpage

\begin{table*}[htpb!]
    \caption{Estimated Impacts of School Closure, School Reopening, and Attendance on Rape Reports}
    \label{SItab:attendanceRape}
    \centering
    \scalebox{0.9}{
    \begin{tabular}{lcccccc} \toprule
    & \multicolumn{6}{c}{Rape}\\ 
    \cmidrule(r){2-7}
    & (1) & (2) & (3) & (4) & (5) & (6) \\ \midrule 
    \multicolumn{7}{l}{\textbf{Panel A:} Binary Re-opening Measure} \\
    School Closure      &      -0.112***&      -0.112***&      -0.097***&      -0.097***&      -0.072*  &      -0.073*  \\
                    &     (0.023)   &     (0.023)   &     (0.026)   &     (0.026)   &     (0.038)   &     (0.038)   \\
School Reopening    &      -0.022   &      -0.160***&      -0.049   &      -0.074   &      -0.088   &      -0.128*  \\
                    &     (0.036)   &     (0.059)   &     (0.039)   &     (0.060)   &     (0.054)   &     (0.074)   \\
School Reopening $\times$ Attendance&               &       0.359***&               &       0.065   &               &       0.100   \\
                    &               &     (0.133)   &               &     (0.140)   &               &     (0.142)   \\
 \\
    Reopening Effect at Percentile 25 of Attendance &&    -0.079         &&    -0.059 &&    -0.105   \\
Reopening Effect at Percentile 50 of Attendance &&     0.013         &&    -0.042 &&    -0.080   \\
Reopening Effect at Percentile 75 of Attendance &&     0.081         &&    -0.030 &&    -0.061   \\
Reopening Effect at Percentile 90 of Attendance &&     0.141         &&    -0.019 &&    -0.044   \\
Baseline Mean    &     0.582 &     0.582 &     0.582          &     0.582 &     0.582 &     0.582        \\
Observations &    45,802 &    45,802          &    45,802 &    45,802 &    45,802 &    45,802      \\
 \\
    \midrule
    \multicolumn{7}{l}{\textbf{Panel B:} Continuous Re-opening Measure} \\
    School Closure      &      -0.109***&      -0.112***&      -0.092***&      -0.093***&      -0.052   &      -0.055   \\
                    &     (0.023)   &     (0.023)   &     (0.026)   &     (0.026)   &     (0.036)   &     (0.037)   \\
School Reopening    &      -0.011   &      -0.265***&      -0.041   &      -0.070   &      -0.053   &      -0.109   \\
                    &     (0.041)   &     (0.091)   &     (0.046)   &     (0.095)   &     (0.059)   &     (0.110)   \\
School Reopening $\times$ Attendance&               &       0.551***&               &       0.065   &               &       0.113   \\
                    &               &     (0.184)   &               &     (0.190)   &               &     (0.195)   \\
 \\
    Reopening Effect at Percentile 25 of Attendance &&    -0.139         &&    -0.056 &&    -0.083   \\
Reopening Effect at Percentile 50 of Attendance &&     0.001         &&    -0.039 &&    -0.054   \\
Reopening Effect at Percentile 75 of Attendance &&     0.105         &&    -0.027 &&    -0.033   \\
Reopening Effect at Percentile 90 of Attendance &&     0.197         &&    -0.016 &&    -0.014   \\
Baseline Mean    &     0.582 &     0.582 &     0.582          &     0.582 &     0.582 &     0.582        \\
Observations &    45,802 &    45,802          &    45,802 &    45,802 &    45,802 &    45,802      \\
 \\
    \midrule
    Municipal \& WoY FEs     &  &  & Y & Y & Y & Y  \\
    Lockdown \& Epidemiological controls &  &  & &  & Y & Y \\
    \bottomrule
    \multicolumn{7}{p{19.9cm}}{{\footnotesize \textbf{Notes to Tab.\ \ref{SItab:attendanceRape}}: Refer to Notes to Table \ref{SItab:attendanceDV}.  Identical models are estimated, however here for the outcome of rape against children. $^{***}$ p$<0.01$; $^{**}$ p$<0.05$; $^{*}$ p$<0.10$.}}
    \end{tabular}}
\end{table*}
\clearpage

\begin{landscape}
\begin{table}[htpb!]
    \caption{Modelled Impacts of School Closure and Re-opening on DV Against Children (no January and February)} 
    \label{SItab:novacations}
    \centering
    \begin{tabular}{lccccccccc} \toprule
    & \multicolumn{3}{c}{Intra-family Violence} &  \multicolumn{3}{c}{Sexual Abuse} &  \multicolumn{3}{c}{Rape} \\ \cmidrule(r){2-4}\cmidrule(r){5-7}\cmidrule(r){8-10}
    & (1) & (2) & (3) & (4) & (5) & (6) & (7) & (8) & (9)  \\ \midrule 
    \multicolumn{10}{l}{\textbf{Panel A:} Binary Re-opening Measure} \\
    School Closure      &      -1.561***&      -1.582***&      -1.311***&      -0.974***&      -1.010***&      -1.069***&      -0.115***&      -0.104***&      -0.083** \\
                    &     (0.120)   &     (0.131)   &     (0.149)   &     (0.082)   &     (0.085)   &     (0.099)   &     (0.025)   &     (0.026)   &     (0.036)   \\
School Reopening    &      -0.782***&      -0.767***&      -0.818***&      -0.407***&      -0.413***&      -0.694***&      -0.052*  &      -0.059** &      -0.083*  \\
                    &     (0.144)   &     (0.146)   &     (0.214)   &     (0.082)   &     (0.082)   &     (0.119)   &     (0.029)   &     (0.030)   &     (0.045)   \\
 \\
    Test of $\beta=\gamma$ (p-value) &     0.000 &     0.000 &     0.002  &     0.000 &     0.000 &     0.000 &     0.014  &     0.117  &     0.995   \\
Observations   &    45,572 &    45,572 &    45,572  &    44,537 &    44,537 &    44,537 &    44,537  &    44,537  &    44,537 \\
Baseline Mean   &     4.472 &     4.472 &     4.472  &     2.825 &     2.825 &     2.825  &     0.581  &     0.581 &     0.581  \\
\\
    \midrule
    \multicolumn{10}{l}{\textbf{Panel B:} Continuous Re-opening Measure} \\
    School Closure      &      -1.383***&      -1.435***&      -0.983***&      -0.811***&      -0.892***&      -0.721***&      -0.100***&      -0.089***&      -0.042   \\
                    &     (0.108)   &     (0.122)   &     (0.119)   &     (0.072)   &     (0.077)   &     (0.077)   &     (0.023)   &     (0.024)   &     (0.030)   \\
School Reopening    &      -0.630***&      -0.649***&      -0.306   &      -0.085   &      -0.197** &      -0.041   &      -0.029   &      -0.035   &      -0.008   \\
                    &     (0.186)   &     (0.175)   &     (0.208)   &     (0.099)   &     (0.096)   &     (0.117)   &     (0.039)   &     (0.040)   &     (0.048)   \\
 \\
    Test of $\beta=\gamma$ (p-value) &     0.000 &     0.000 &     0.001  &     0.000 &     0.000 &     0.000 &     0.051  &     0.171  &     0.448   \\
Observations   &    45,572 &    45,572 &    45,572  &    44,537 &    44,537 &    44,537 &    44,537  &    44,537  &    44,537 \\
Baseline Mean   &     4.472 &     4.472 &     4.472  &     2.825 &     2.825 &     2.825  &     0.581  &     0.581 &     0.581  \\
\\
    \midrule
    Municipal \& WoY FEs     &  & Y & Y &  & Y & Y &  & Y & Y \\
    Lockdown \& Epidemiological controls    &  &   & Y &  &   & Y &  &   & Y \\
    \bottomrule
    \multicolumn{10}{p{23.6cm}}{{\footnotesize \textbf{Notes to Tab.\ \ref{SItab:novacations}}: Refer to notes to Tab.\ 1 of the main text.  Identical specifications are estimated, however here consistently removing all observations in months January-February of all years, when schools are closed for summary vacations.   $^{***}$ p$<0.01$; $^{**}$ p$<0.05$; $^{*}$ p$<0.10$.}}
    \end{tabular}
\end{table}
\end{landscape}
\clearpage

\begin{table*}[htpb!]
    \caption{Modelled Impacts of School Closure and Re-opening on Sexual Violence (Unadjusted Variables)}
    \label{SItab:unadjusted}
    \centering
    \begin{tabular}{lcccccc} \toprule
     &  \multicolumn{3}{c}{Sexual Abuse} &  \multicolumn{3}{c}{Rape} \\ \cmidrule(r){2-4}\cmidrule(r){5-7}
    & (4) & (5) & (6) & (7) & (8) & (9)  \\ \midrule 
    \multicolumn{7}{l}{\textbf{Panel A:} Binary Re-opening Measure} \\
    School Closure      &      -0.878***&      -0.966***&      -1.031***&      -0.130***&      -0.098***&      -0.092** \\
                    &     (0.069)   &     (0.082)   &     (0.099)   &     (0.025)   &     (0.028)   &     (0.040)   \\
School Reopening    &      -0.381***&      -0.377***&      -0.655***&      -0.065** &      -0.047*  &      -0.078*  \\
                    &     (0.070)   &     (0.073)   &     (0.116)   &     (0.026)   &     (0.028)   &     (0.046)   \\
 \\
    Test of $\beta=\gamma$ (p-value) &     0.000 &     0.000 &     0.000 &     0.012  &     0.096  &     0.706   \\
Observations   &    54,214 &    54,214 &    54,214 &    54,214  &    54,214  &    54,214 \\
Baseline Mean   &     2.705 &     2.705 &     2.705  &     0.588  &     0.588 &     0.588  \\
\\
    \midrule
    \multicolumn{7}{l}{\textbf{Panel B:} Continuous Re-opening Measure} \\
    School Closure      &      -0.752***&      -0.886***&      -0.743***&      -0.121***&      -0.095***&      -0.071** \\
                    &     (0.064)   &     (0.079)   &     (0.086)   &     (0.023)   &     (0.026)   &     (0.034)   \\
School Reopening    &      -0.156*  &      -0.288***&      -0.213*  &      -0.072*  &      -0.062   &      -0.068   \\
                    &     (0.090)   &     (0.093)   &     (0.122)   &     (0.037)   &     (0.039)   &     (0.048)   \\
 \\
    Test of $\beta=\gamma$ (p-value) &     0.000 &     0.000 &     0.000 &     0.173  &     0.412  &     0.945   \\
Observations   &    54,214 &    54,214 &    54,214 &    54,214  &    54,214  &    54,214 \\
Baseline Mean   &     2.705 &     2.705 &     2.705  &     0.588  &     0.588 &     0.588  \\
\\
    \midrule
    Municipal \& WoY FEs      &  & Y & Y &  & Y & Y \\
    Lockdown \& Epidemiological controls &  &   & Y &  &   & Y \\
    \bottomrule
    \multicolumn{7}{p{18.6cm}}{{\footnotesize \textbf{Notes to Tab.\ \ref{SItab:unadjusted}}: Refer to notes to Tab.\ 1 of the main text.  Identical specifications are estimated for models where the outcome is sexual abuse or rape, however using original un-smoothed data, where over-reporting occurs on the first day of each month, rather than smoothed data re-assigning excess reporting uniformly across the month.  All other details follow those in Tab.\ 1.  Column numbers (4)-(9) are used here for sake of comparison with column numbers in Tab.\ 1.   $^{***}$ p$<0.01$; $^{**}$ p$<0.05$; $^{*}$ p$<0.10$.}}
    \end{tabular}
\end{table*}

\begin{landscape}
\begin{table}[htpb!]
    \caption{Modelled Impacts of School Closure and Re-opening on Reporting of Violence Against Children by Type of Intra-family Violence}
    \label{SItab:subDV}
    \centering
    \begin{tabular}{lccccccccc} \toprule
    & \multicolumn{3}{c}{Physical Violence (Serious)} &  \multicolumn{3}{c}{Physical Violence (Moderate)} &  \multicolumn{3}{c}{Psychological Violence} \\ \cmidrule(r){2-4}\cmidrule(r){5-7}\cmidrule(r){8-10}
    & (1) & (2) & (3) & (4) & (5) & (6) & (7) & (8) & (9)  \\ \midrule 
    \multicolumn{10}{l}{\textbf{Panel A:} Binary Re-opening Measure} \\
    School Closure      &      -0.031** &      -0.029** &      -0.005   &      -1.016***&      -1.149***&      -0.947***&      -0.345***&      -0.410***&      -0.390***\\
                    &     (0.013)   &     (0.013)   &     (0.017)   &     (0.073)   &     (0.086)   &     (0.105)   &     (0.057)   &     (0.067)   &     (0.078)   \\
School Reopening    &      -0.018   &      -0.024   &       0.008   &      -0.745***&      -0.798***&      -0.760***&      -0.007   &      -0.022   &      -0.139   \\
                    &     (0.014)   &     (0.015)   &     (0.019)   &     (0.073)   &     (0.077)   &     (0.119)   &     (0.083)   &     (0.083)   &     (0.109)   \\
 \\
    Test of $\beta=\gamma$ (p-value) &     0.423 &     0.754 &     0.514 &     0.000  &     0.000  &     0.038  &     0.000  &     0.000  &     0.003  \\
Observations   &    54,214 &    54,214 &    54,214 &    54,214  &    54,214  &    54,214 &    54,214  &    54,214  &    54,214 \\
Baseline Mean   &     0.136 &     0.136 &     0.136  &     2.783  &     2.783 &     2.783 &     1.383  &     1.383 &     1.383  \\
\\
    \midrule
    \multicolumn{10}{l}{\textbf{Panel B:} Continuous Re-opening Measure} \\
    School Closure      &      -0.024** &      -0.024*  &      -0.002   &      -0.868***&      -1.026***&      -0.652***&      -0.310***&      -0.382***&      -0.286***\\
                    &     (0.012)   &     (0.013)   &     (0.016)   &     (0.066)   &     (0.081)   &     (0.085)   &     (0.049)   &     (0.060)   &     (0.061)   \\
School Reopening    &      -0.005   &      -0.016   &       0.023   &      -0.673***&      -0.787***&      -0.375***&       0.122   &       0.077   &       0.091   \\
                    &     (0.020)   &     (0.021)   &     (0.023)   &     (0.101)   &     (0.104)   &     (0.126)   &     (0.104)   &     (0.099)   &     (0.110)   \\
 \\
    Test of $\beta=\gamma$ (p-value) &     0.371 &     0.722 &     0.328 &     0.049  &     0.019  &     0.019  &     0.000  &     0.000  &     0.000  \\
Observations   &    54,214 &    54,214 &    54,214 &    54,214  &    54,214  &    54,214 &    54,214  &    54,214  &    54,214 \\
Baseline Mean   &     0.136 &     0.136 &     0.136  &     2.783  &     2.783 &     2.783 &     1.383  &     1.383 &     1.383  \\
\\
    \midrule
    Municipal \& WoY FEs     &  & Y & Y &  & Y & Y &  & Y & Y \\
    Lockdown \& Epidemiological controls    &  &   & Y &  &   & Y &  &   & Y \\
    \bottomrule
    \multicolumn{10}{p{23.8cm}}{{\footnotesize \textbf{Notes to Tab.\ \ref{SItab:subDV}}: Refer to notes to Tab.\ 1 of the main text.  Identical specifications are estimated, however here rather than considering total reports of intra-family violence against children (as reported in columns (1)-(3) of Tab.\ 1), reports of intra-family violence against children for each sub-class of intra-family violence are reported.  These classifications are generated from police reports, and each case of intra-family violence from Tab.\ 1 is classified as one (and only one) of the three classes displayed here.  All other details follow those in Tab.\ 1. $^{***}$ p$<0.01$; $^{**}$ p$<0.05$; $^{*}$ p$<0.10$.}}
    \end{tabular}
\end{table}
\end{landscape}
\clearpage

\begin{table*}[htpb!]
\begin{center}
\caption{Weights in Time-Varying Adoption Two-Way Fixed Effect Estimate on School Reopening}
\label{SItab:2WayWeights}
\begin{tabular}{lcc} \toprule
Comparison Group & Weights & Average DD Estimate \\ \midrule
\multicolumn{3}{l}{\textbf{Panel A:} Intra-family violence} \\
Earlier Treatment vs.\ Later Control &0.788&-0.637 \\
Later Treatment vs.\ Earlier Control &0.212&0.127 \\ \\
\multicolumn{3}{l}{\textbf{Panel B:} Sexual abuse} \\
Earlier Treatment vs.\ Later Control &0.788&-0.544 \\
Later Treatment vs.\ Earlier Control &0.212&-1.368 \\ \\
\multicolumn{3}{l}{\textbf{Panel C:} Rape} \\
Earlier Treatment vs.\ Later Control &0.788&-0.074 \\
Later Treatment vs.\ Earlier Control &0.212&-0.001 \\ \bottomrule
\multicolumn{3}{p{12.4cm}}{{\footnotesize \textbf{Notes to Tab \ref{SItab:2WayWeights}}: Aggregate values for two-way FE weights are reported (following \cite{GB2021,GBetal2019}) for estimates of impacts of time-varying school closure on outcomes indicated in each panel.  These are aggregate values across all points documented in Figure \ref{SIfig:GB}.  Weights refer to the total proportion of estimates based on each comparison type (earlier treatment vs.\ later adopter, or later adopter vs.\ earlier treated), and ``Average DD Estimate'' refers to the average difference between these groups in a 2$\times$2 DD setting. For all models, the proportion of negative weights following \cite{dCDH2020} is 0.}}
\end{tabular}
\end{center}
\end{table*}
\clearpage

\begin{table}[htpb!]
    \caption{Actual Violence Reporting Channels Reported by a Single Child Protection Office -- Temporal Differences}
    \label{SItab:OPDdesc}
    \centering
    \begin{tabular}{lcccccccc} \toprule
   & \multicolumn{2}{c}{Jan-Feb}  &\multicolumn{2}{c}{Mar-Sep}  &\multicolumn{2}{c}{Oct-Dec}  & \multicolumn{2}{c}{Aug-Dec} \\
   & 2019&2020&2019&2020&2019&2020&2019&2021\\ \midrule 
Total Reporting & 57 & 46 & 227 & 163 & 83 & 43 & 138 & 120  \\ 
    Total Reporting by Schools & 4 & 4 & 57 & 8 & 32 & 4 & 48 & 25  \\
    Percentage of Reporting by Schools & 0.07 & 0.09 & 0.25 & 0.05 & 0.39 & 0.09 & 0.35 & 0.21 \\  
    Percentage of Schools Open & --& -- & 1.00 & 0.00 & 1.00&0.08&1.00&0.99\\    
  \bottomrule
    \multicolumn{9}{p{15.4cm}}{{\footnotesize \textbf{Notes to Tab. \ref{SItab:OPDdesc}}: Descriptive values document official recorded channels of violence reporting received by a single child protection office in a large municipality in Santiago.  Here year by year comparisons are documented of reports received via schools and total reports received in various periods.  Jan-Feb is vacations in both years.  Mar-Sep covers periods where schools returned in 2019, but were closed in 2020.  Oct-Dec covers periods where schools were completely open in 2019, and only very partially open in 2020.  Aug-Dec covers periods in which schools sere completely open in 2019, and nearly entirely open in 2021.}}
    \end{tabular}
\end{table}
\clearpage

\begin{table}[htpb!]
    \caption{Percentage of Violence Reporting by Age Group and Channel in a Single Child Protection Office}
    \label{SItab:channelsOPD}
    \centering
    \begin{tabular}{rcccc} \toprule
           &\multicolumn{4}{c}{Reporting Channel} \\ \cmidrule(r){2-5}
Age group  & Schools&Courts&Health Centers&Others\\ \midrule 
\multicolumn{5}{l}{\textbf{Panel A: 2019-2021}} \\
$[1-6]$ & 0.12 & 0.41 & 0.13 & 0.34\\ 
$[7-10]$ & 0.29 & 0.42 & 0.08 &0.21\\
$[11-13]$&  0.21 &  0.49 & 0.08 &0.22 \\  
$[14-15]$&  0.11  &0.57 & 0.08 & 0.24\\  
$[16-17]$ &  0.14 &0.58 & 0.08 &0.20\\ \midrule
\multicolumn{5}{l}{\textbf{Panel B: 2019 Only}} \\
$[1-6]$   & 0.24 & 0.35 & 0.15 & 0.26 \\ 
$[7-10]$  & 0.38 & 0.35 & 0.09 & 0.18 \\
$[11-13]$ & 0.33 & 0.44 & 0.10 & 0.13 \\  
$[14-15]$ & 0.12 & 0.61 & 0.04 & 0.23 \\  
$[16-17]$ & 0.17 & 0.54 & 0.08 & 0.21 \\ 
\bottomrule
\multicolumn{5}{p{9.6cm}}{{\footnotesize \textbf{Notes to Tab. \ref{SItab:channelsOPD}}: Descriptive values are reported documenting official recorded channels of violence reporting received by a single child protection office in a large municipality in Santiago.  Here channels are separated as entering via schools, courts, health centres, or other sources, and relative proportions of each type of reporting channel by age of the victim is documented. Panel A reports values over all years in which data was provided (2019--2021), while panel B reports values only in the period entirely preceding COVID (year 2019).}}
    \end{tabular}
\end{table}

\end{spacing}
\end{document}